\documentclass{report}
\usepackage{suthesis-2e}
\usepackage{amsfonts}
\usepackage{amsmath}
\usepackage{amssymb}
\usepackage{graphicx}
\usepackage{kbordermatrix}
\usepackage{ccaption}
\usepackage[hyphens]{url}
\usepackage{hyperref}
\usepackage[numbers]{natbib}
\usepackage{subcaption, multirow}
\usepackage{booktabs} 
\usepackage{diagbox}
\usepackage{array}
\usepackage{mathtools}
\usepackage{bbm}
\newcolumntype{L}[1]{>{\raggedright\let\newline\\\arraybackslash\hspace{0pt}}m{#1}}
\newcolumntype{C}[1]{>{\centering\let\newline\\\arraybackslash\hspace{0pt}}m{#1}}
\newcolumntype{R}[1]{>{\raggedleft\let\newline\\\arraybackslash\hspace{0pt}}m{#1}}
\newcommand{\STAB}[1]{\begin{tabular}{@{}c@{}}#1\end{tabular}}

\newcommand{\bc}{\boldsymbol{c}}
\newcommand{\bk}{\boldsymbol{k}}
\newcommand{\bp}{\boldsymbol{p}}
\newcommand{\bu}{\boldsymbol{u}}
\newcommand{\bv}{\boldsymbol{v}}
\newcommand{\bx}{\boldsymbol{x}}
\newcommand{\bz}{\boldsymbol{z}}
\newcommand{\bX}{\boldsymbol{X}}
\newcommand{\btheta}{\boldsymbol{\theta}}
\newcommand{\R}{\mathbb{R}}
\renewcommand{\L}{\mathcal{L}}

\newcommand{\indicator}{\mathbbm{1}}

\newcommand{\animalA}{\mathsf{A}}
\newcommand{\animalB}{\mathsf{B}}
\DeclareMathOperator*{\median}{median}
\DeclareMathOperator{\corr}{\mathsf{Corr}}
\DeclareMathOperator{\rsa}{\mathsf{RSA}}
\DeclareMathOperator{\spbrsa}{\widetilde{\mathsf{RSA}}}
\DeclareMathOperator{\rdm}{\mathsf{RDM}}
\DeclareMathOperator{\train}{\mathsf{train}}
\DeclareMathOperator{\test}{\mathsf{test}}
\DeclareMathOperator{\trueA}{t^{\animalA}_{\train}}
\DeclareMathOperator{\trueB}{t^{\animalB}_{\train}}
\DeclareMathOperator{\trueBtest}{t^{\animalB}_{\test}}

\DeclareMathOperator{\trueAid}{t^{\animalA}}
\DeclareMathOperator{\trueBid}{t^{\animalB}}
\DeclareMathOperator{\sfAtrain}{s^{\animalA}_{1,\train}}
\DeclareMathOperator{\ssAtrain}{s^{\animalA}_{2,\train}}
\DeclareMathOperator{\sfBtrain}{s^{\animalB}_{1,\train}}
\DeclareMathOperator{\ssBtrain}{s^{\animalB}_{2,\train}}

\DeclareMathOperator{\sfBtest}{s^{\animalB}_{1,\test}}
\DeclareMathOperator{\ssBtest}{s^{\animalB}_{2,\test}}
\DeclareMathOperator{\sfAid}{s^{\animalA}_{1}}
\DeclareMathOperator{\ssAid}{s^{\animalA}_{2}}

\DeclareMathOperator{\sfBid}{s^{\animalB}_{1}}
\DeclareMathOperator{\ssBid}{s^{\animalB}_{2}}

\programthesis{\dept{Neuroscience}}

\setcitestyle{square}
\setcitestyle{citesep={,}}

\begin{document}
\title{A Goal-Driven Approach to Systems Neuroscience}
\author{Aran Nayebi}
\submitdate{March 2022}
\copyrightyear{2022}
\principaladviser{Daniel L.K. Yamins}
\coprincipaladviser{Surya Ganguli}
\firstreader{Stephen A. Baccus}
\secondreader{Shaul Druckmann}
\thirdreader{David Sussillo} 
 
\beforepreface
\prefacesection{Abstract}
Humans and animals exhibit a range of interesting behaviors in dynamic environments, and it is unclear how our brains actively reformat this dense sensory information to enable these behaviors.
Experimental neuroscience is undergoing a revolution in its ability to record and manipulate hundreds to thousands of neurons while an animal is performing a complex behavior.
As these paradigms enable unprecedented access to the brain, a natural question that arises is how to distill these data into interpretable insights about how neural circuits give rise to intelligent behaviors.
The classical approach in systems neuroscience has been to ascribe well-defined operations to individual neurons and provide a description of how these operations combine to produce a circuit-level theory of neural computations.
While this approach has had some success for small-scale recordings with simple stimuli, designed to probe a particular circuit computation, often times these ultimately lead to disparate descriptions of the same system across stimuli.
Perhaps more strikingly, many response profiles of neurons are difficult to succinctly describe in words, suggesting that new approaches are needed in light of these experimental observations.
In this thesis, we offer a different definition of interpretability that we show has promise in yielding unified structural and functional models of neural circuits, and describes the evolutionary constraints that give rise to the response properties of the neural population, including those that have previously been difficult to describe individually.
Specifically, our approach is to ``reverse engineer'' neural circuits by simulating the evolutionary process to build \emph{in silico} neural networks that are subject to the combined interaction of three types of constraints: the \emph{task}, expressed as an \emph{objective function} to be maximized or minimized given a \emph{data stream}; the network \emph{architecture}, expressed as the connections through which inputs flow; and the \emph{learning rule}, expressed as synaptic weight updates.
This joint set of constraints is an interpretable hypothesis for the evolutionary design principles that enable a biological circuit to perform its computations, and crucially, the set of combinations of these constraints gives rise to multiple hypotheses that will be quantitatively evaluated against high-throughput neural and behavioral recordings.
We demonstrate the utility of this framework across multiple brain areas and species to study the roles of recurrent processing in the primate ventral visual pathway; mouse visual processing; heterogeneity in rodent medial entorhinal cortex; and facilitating biological learning.

\prefacesection{Acknowledgments}
First and foremost, I would like to thank my advisors, Daniel Yamins and Surya Ganguli.
Dan, I have learned so much from you -- from showing me firsthand how to approach scientific problems with clarity, to aligning figures pristinely in Illustrator, you taught me to always stay focused on the question, keeping the big picture in clear view but never letting the details slide.
Thank you for always challenging me and pushing me; any success I am fortunate to experience in my career will be in large part because I learned to do science from you.
Surya, since the day I took your theoretical neuroscience course in college, I have been inspired by your breadth of knowledge, curiosity, and drive to find the gems in the scientific haystack.
Thank you for being a supportive co-mentor and role model all these years.

I would also like to thank my thesis committee members, Steve Baccus, David Sussillo, and Shaul Druckmann, for all of your input and guidance over the years.
I am grateful I got to share my science with you, for asking insightful questions during my committee meetings and one-on-one, and for offering unwavering support and career guidance along the way.
Thank you to Lisa Giocomo for chairing my defense and for being an incredibly supportive collaborator!
In addition, I am grateful to Kalai Diamond, Tony Ricci, Marrium Fatima, Elise Kleeman, Jay McClelland, and Nisa Cao, for providing ancillary support that enabled me to focus on my research.
I am especially grateful to Jay McClelland for spearheading the Mind, Brain, Computation, and Technology (MBCT) program and for always making time to meet with me to discuss my vaguely posed scientific questions.
Finally, I am grateful to my fellow cohort of ``Blebs'' and the wider Stanford Neurosciences community for being such a supportive environment.

During my time in graduate school, the Yamins and Ganguli labs were a constant source of intellectual stimulation and camaraderie.
Thank you to my Ganguli lab mates, Lane McIntosh, Niru Maheswaranathan, Ben Poole, Sarah Harvey, Kiah Hardcastle, Alex Williams, Sam Ocko, St\'{e}phane Deny, Subhy Lahiri, Jonathan Kadmon, Hidenori Tanaka, Dan Kunin, Gabriel Mel, Ben Sorscher, Chris Stock, Linnie Jiang, Mansheej Paul, Brett Larsen, and Brandon Benson for all of the deep scientific discussions and heart-to-hearts over the years.
Thank you to my official (and honorary!) Yamins lab mates, Kevin Feigelis, Eshed Margalit, Nathan Kong, Josh Melander, Chengxu Zhuang, Dan \& Mona Bear (plus Zane), Dawn Finzi, Damian Mrowca, Tyler Bonnen, Violet Xiang, Javier Sagastuy-Brena, Elias Wang, Nick Haber, and Judy Fan, for making each day bright and fulfilling -- it wouldn't be possible with you.

Stanford has been my intellectual home for 11 years, and I am grateful to those who nurtured my scientific interests early on. 
Steve Marsheck, my high school math teacher, made the subject light-hearted and exciting at the same time. 
Sol Feferman, Grisha Mints, and Dana Scott provided the encouragement for a starry-eyed college freshman to work independently on research questions.
Bill Newsome first introduced me to Dayan \& Abbott's textbook my senior year, and opened my eyes to the relevance of a quantitative background to problems in systems neuroscience.
Steve Baccus was the first neuroscientist who took a chance on me.
I learned so much from his clarity of thought and use of illustrative, yet deceptively simple examples by which to approach difficult problems.
I am also fortunate to have known and worked with his students, Lane McIntosh and Niru Maheswaranathan, who inspired me with their creativity, rigor, and generosity.
I am grateful to David Kastner and Pablo Jadzinsky for their mentorship during this time as well.

I am forever indebted to the friends and relationships I have leaned on for support during my time at Stanford.
Dominic Becker, Alison Nguyen, and Matt Vitelli were the first friendships I made in high school and college, and they have been great friends all these years since.
I am grateful to have fortuitously befriended Swetava Ganguli in CS 228 during my master's -- your generosity, intellectual, and emotional depth never cease to amaze me.
Katherine Hermann, thank you for all the laughs and great conversations over the years -- you're someone I know I can always count on.
Eshed Margalit, your work ethic and rigor will always be an inspiration to me.
Thank you for the creative levity you add to every situation, and for always helping me when I needed it.
Nathan Kong, thank you for being an amazing collaborator and one of the kindest people I've ever met. 
I'm blessed to have you in my life.
Kevin Feigelis, thank you for always being there for me through thick and thin.
Talking to you feels like I am conversing with my innermost voice -- I have and will always feel that we were cut from the same cloth.

None of this would be possible without the love and support of my family.
I would like to thank my brilliant, dedicated wife, Heather, for being my partner through it all.
As I transitioned from theory to applied science, you gave me the encouragement to continue learning programming, and you were always there for debugging and hugging me when I needed it.
Thank you for introducing me to the world of cats -- without you, we wouldn't have Shira and Zoe, who make the simple moments so precious.
Your love, support, and encouragement has sustained me, and I eternally cherish the opportunity to have grown up and to grow old with you.
I am also grateful to my brothers, Ryan and Sean, for their love and support.
Finally, I would like to thank my parents, Mehrdad and Floura, who nurtured my love of science from an early age.
I am deeply grateful for their undying love, support, and guidance.
My first exposure to the wonders of science was through my father, and his passion for it has been fueling my curiosity ever since.
This thesis simply would not exist without them, and I dedicate this thesis to my parents.

\afterpreface

\chapter{Introduction}
\label{intro}
Humans and animals exhibit a range of interesting behaviors in dynamic environments, and it is unclear how our brains actively reformat this dense sensory information to enable these behaviors.
Technical advances in experimental neuroscience are enabling us to manipulate neural circuits at unprecedented scale, where recording hundreds to thousands of neurons is becoming more standard, even while the animal is performing increasingly complex behaviors~\citep{rajalingham2018large, steinmetz2018challenges,stringer2019spontaneous,steinmetz2021neuropixels}.
If our goal is to ultimately understand how the brain gives rise to these behaviors, how might we effectively leverage this data in order to do so?

As a starting example, we consider the visual system, which must discover meaningful patterns in a complex physical world~\citep{james1890principles}.
Within 200\emph{ms}, primates can quickly identify objects despite changes in position, pose, contrast, background, foreground, and many other factors from one occasion to the next: a behavior known as ``core object recognition''~\citep{Pinto:2008ua, DiCarlo_2012}.
It is known that the ventral visual stream (VVS) underlies this ability by transforming the retinal image of an object into a new internal representation, in which high-level properties, such as object identity and category, are more explicit~\citep{DiCarlo_2012}.
The seminal work of Hubel and Weisel~\citep{hubel1959receptive} provided evidence that object recognition behavior is generated by a series of hierarchically connected cortical areas.
For early visual cortex (V1), these experiments more specifically suggested an interpretable mathematical structure resembling Gabor wavelet filters, of different frequencies and orientations.
In fact, models with hand-tuned Gabor filters lead to some initial success towards explaining V1 response patterns~\citep{jones1987evaluation}.
The inclusion of thresholding nonlinearities and normalizations further improved these models~\citep{movshon1978spatial,carandini2005we}.

\begin{figure}
  \centering
  \includegraphics[width=1.0\columnwidth]{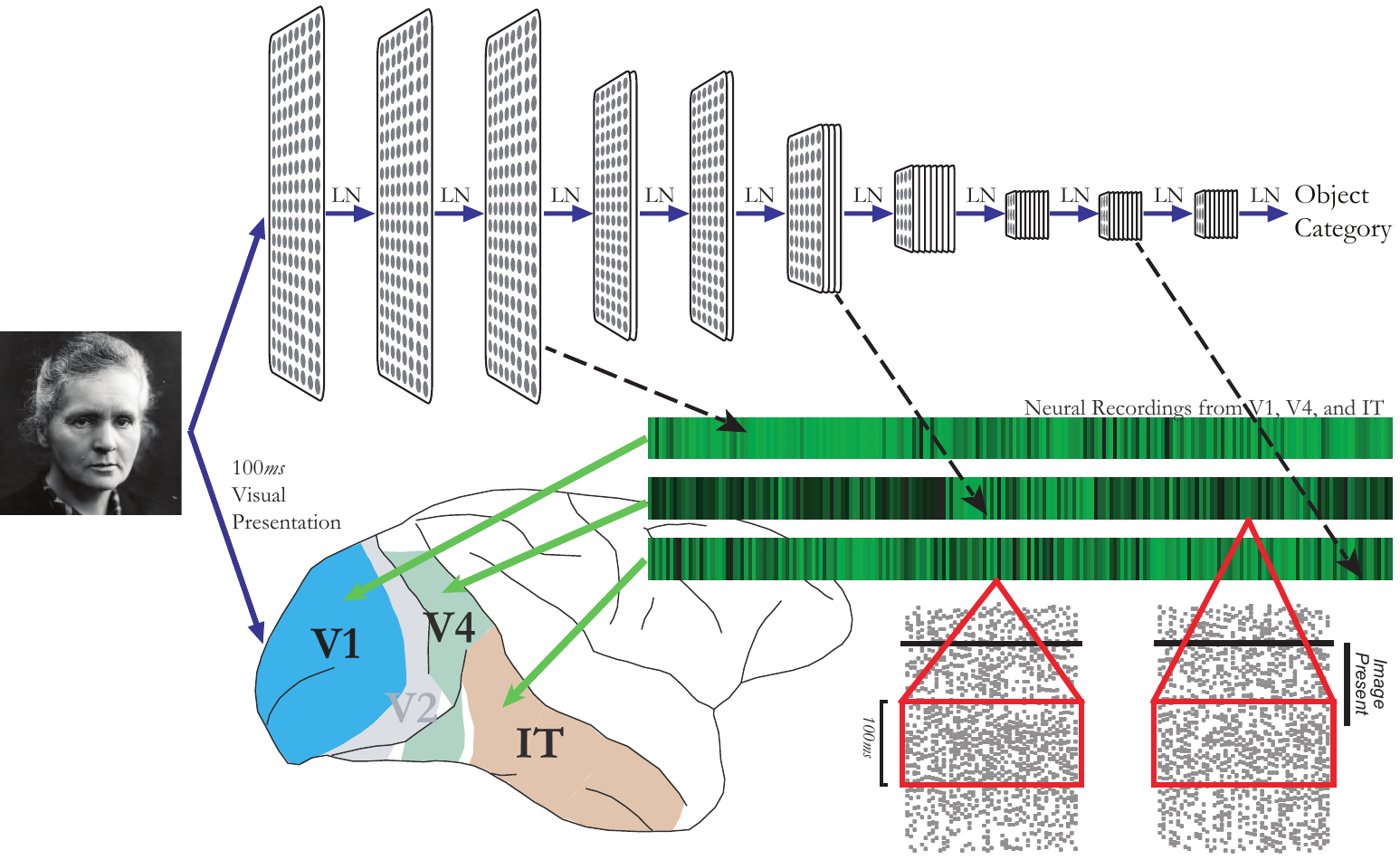}
  \caption[Goal-driven convolutional neural networks (CNNs) as models of sensory cortex]{\textbf{Goal-driven convolutional neural networks (CNNs) as models of sensory cortex.}
  The ventral visual stream is a well-studied sensory cascade, consisting of a series of hierarchically connected cortical areas (macaque brain pictured), encoding the initial stimulus into patterns of neural activity and enabling core object recognition behaviors.
  Convolutional neural networks (CNNs) are a model of this encoding step, composed of multiple linear-nonlinear (LN) layers, each consisting of simple operations such as filtering, thresholding, pooling, and normalization.
  These networks are trained on a challenging, ethologically-relevant object categorization task, before being compared to neural and behavioral data.
  \emph{Comparison to neural response patterns.}
  Each stimulus is presented for 100\emph{ms}.
  We stipulate that units from each multi-unit array must be fit by features from a single model layer.
  Model features produce a stimulus-computable output that can be compared to neural responses across visual areas (in this example, V1, V4, and IT), as denoted by the dotted arrows.
  } 
\label{intro:fig0}
\end{figure}

\section{Background}
How might we extend these observations to downstream visual areas, such as V4 or IT?
A starting point would be to compose these nonlinearities, producing deep ``artificial'' neural network architectures~\citep{fukushima1980self,riesenhuber1999hierarchical}, known as convolutional neural networks (CNNs)~\citep{lecun1995convolutional}.
CNNs (Figure~\ref{intro:fig0}) are comprised of repeated linear-nonlinear motifs, which take dot products of local patches of the input with a given filter template (analogous to Gabor wavelets of differing frequencies), followed by a pointwise nonlinearity to perform thresholding.
Additionally, we can include an aggregation of these local values via (mean or max) pooling, along with normalization operations to put the output within a given range.

A natural approach then is to directly apply this hand-tuned approach hierarchically.
However, the inputs to these higher cortical areas are more challenging to study analytically, making the application of the hand-tuned approach limited in this multi-layer setting.
Of course, one should attempt an analytical solution when possible, but ultimately we want solutions that are applicable across the range of naturalistic stimuli the visual system operates over.
Thus, one option is to try to acquire suitable representations through optimization, most directly on the neural response data that has been collected from these areas.
This approach can work well for shallow networks mapped to V1 and A1~\citep{decharms2000neural}.
It can also work for subcortical areas where we can collect a lot of data from, such as the retina.
In this case, somewhat deeper networks (of depth comparable to the retina itself) can be fit on the responses of a small fraction of the outermost (ganglion) cells of this circuit~\citep{mcintosh2016deep}, and end up producing responses well-matched to intermediate cell types~\citep{maheswaranathan2018deep} and circuit mechanisms on unseen stimuli~\citep{tanaka2019deep}, without ever being directly optimized for either of them.

It turns out that directly fitting to the observed stimulus-response relationship is less successful for higher visual areas such as V4 and IT, as multi-layered networks tend to have many more parameters, resulting in overfitting of the data, failing to generalize on novel test stimuli~\citep{gallant1996neural}.
Besides the technical issue of overfitting, there is perhaps a more fundamental reason why this approach is not generally desirable.
Namely, we have effectively replaced one neural circuit with another; albeit, the latter one is more amenable to analysis and can generate predictions for new circuit mechanisms~\citep{tanaka2019deep}.
However, we do not possess any normative insight into \emph{what} gave rise to the response variability observed in the circuit in the first place.
Instead, feedforward CNNs trained on high-variation tasks (``goals'') are recently the most accurate models of V1, V4, and IT responses~\citep{yamins2014performance,khaligh2014deep,cadena2019deep}, along with the auditory pathway~\citep{kell2018task}, as well as recurrent networks for the motor system~\citep{sussillo2015neural, michaels2020goal}.
This accomplishment is quite remarkable, since for the first time we have models that can start to perform the behaviors under consideration, offering a new way forward.

\section{Goal-driven modeling}
These results are examples of a more general ``goal-driven approach'' (surveyed in e.g.~\citep{Yamins2016,richards2019deep}), motivated by the perspective that neural circuits were evolved to enable the organism to support a range of behaviors in order to survive.
Specifically, rather than optimize a neural network directly to the neural data, we instead optimize these network parameters for behavioral task(s) that the circuit we have data for supports.
Therefore, we aim to ``reverse engineer'' neural circuits towards this end goal in a hypothesis-driven manner, by simulating the evolutionary process via \emph{in silico} neural network models that toggle four main components:
\begin{itemize}
\item \textbf{Data Stream:} This is perhaps the most basic component, but nonetheless crucial.
Prior descriptions of the goal-driven approach tend to combine this with ``objective function'' (see next item below), into a single component called the ``task''.
However, it is important to individuate this aspect of the task into its own component, since having a stream of inputs that does not match those of the neural circuit in question, despite having correctly identified the remaining components, can limit the explanatory power of the resultant model.
For example, suppose one trains a CNN to categorize only apples -- it would be not very surprising if it failed to recognize faces!
The fact the inputs are a consideration implies that our models will be ``input computable'', and should accept arbitrary inputs within the domain of interest.
For example, for visual models, pixels are the most common input format, but for higher-order areas, the putative outputs of other areas can considered potential inputs as well.
The type of input format is itself an important hypothesis about the functional role of the circuit.
\item \textbf{Objective Function:} These formalize the behavioral goals of the system, and are functions of the network and task.
The idea that neural circuits change their properties in order to improve objective function(s) that define their role is motivated by observations that human behavior can approach optimality in domains such as movement and energy consumption~\citep{todorov2002optimal,kording2007decision,taylor2011does}.
In the context of sensory processing, the most well-known recent example is the cross entropy loss, which assesses categorization performance:
\begin{equation}\label{celoss}
L(y,\hat{y}) := -\sum_{i=1}^C y_i\log(\hat{y}_i),
\end{equation}
where $C$ is the total number of categories, $\hat{y} \in \mathbb{R}^C$ are the predicted category probabilities for the given stimuli, and $y \in \mathbb{R}^C$ are the ground-truth category probabilities (typically a one-hot vector with 1 in the correct category and 0 everywhere else).
Of course, this objective function may seem unlikely to be instantiated in a neural circuit, given the explicit dependence on category labels.
However, we can view this objective function as a ``proxy'' for more ethologically relevant unsupervised functions, such as contrastive representation learning~\citep{Wu2018,Zhuang2021}, which we will explore further in this thesis (Chapter~\ref{ch:mouse}).
Furthermore, we can imagine that there are a multitude of objective functions distributed across brain areas, each of which can be considered a specialized subsystem~\citep{minksy1977plain,marblestone2016toward}.
This idea will become important in the later parts of this thesis as we apply goal-driven modeling to multiple organisms and non-sensory domains.
\item \textbf{Architecture:} These describe the ways in which units are arranged in an artifical network and their corresponding operations.
The McCulloh-Pitt neuron~\citep{mcculloh1943logical} is one such possible design choice, whereby each artificial neuron is a linear thresholding function of its inputs.
We are free to additionally consider spiking units and more biophysically realistic neurons, but the architectures we consider here are all comprised of these simple units, and can fall into largely two types: \emph{feedforward}~\citep{haykin1994neural,lecun1995convolutional} and \emph{recurrent}~\citep{schmidhuber2015deep}.
Both consist mainly of iterated linear-nonlinear transformations\footnote{The simplest instantiation being a single layer perceptron~\citep{rosenblatt1958perceptron,minsky1969introduction}.}, the former being physically through depth, and the latter being temporally through time (with the crucial difference being that the synaptic weights are reused).
These classes of networks need not be considered separately, and in fact can be combined -- an application of which we will describe in Chapter~\ref{ch:convrnn} of this thesis.
One can consider additional operations (e.g. ``convolution'' and ``gating''), and these inductive biases play an important role in enabling effective learning of representations, which we turn to next.
\item \textbf{Learning Rule:} So far we have provided an objective function to minimize through an architecture which is a function of its inputs.
How can we go about effectively minimizing this objective function?
One approach is to learn one iteration at a time, and update the network parameters in the opposite (negative) direction of the error gradient (defined via the chain rule from differential calculus), until training converges.
This is the basic notion behind the backpropagation of errors algorithm~\citep{linnainmaa1970representation,rumelhart1986learning}, the most effective learning algorithm to date.
The hyperparameters of learning rate, batch size, and initialization scheme are all salient in this component.
We are of course free to use any learning algorithm, even those directly inspired from biological observations (e.g. Hebbian learning~\citep{hebb1949organization}), but these have been difficult to achieve performance (comparable at least to humans and animals) on tasks with.
Moreover, while we may view backpropagation as a ``proxy'' by which to efficiently learn representations that are consistent with neural and behavioral data, a literal interpretation of backpropagation as a learning rule instantiated in neural circuits has biological plausibility issues.
However, these issues do not rule out the possibility that various approximations to backpropagation may still be implemented in the brain; although identifying the specific learning rule in any given system is by far more experimentally difficult than identifying the objective function (from witnessing behavior) or architecture (from observing anatomy).
In this thesis, we will demonstrate performant approximations to backpropagation (Chapter~\ref{ch:lrperf}), as well as identify experimental observables that we can use to help separate various hypotheses about what learning rules might be operative in a given neural circuit (Chapter~\ref{ch:lrobs}).
\end{itemize}

\section{What constitutes understanding?}
At the end of this procedure, we obtain \emph{functional} and \emph{structural} models that satisfy three criteria~\citep{Yamins2016}: \emph{input computable} from an arbitrary stimulus (not necessarily the one it was trained with), \emph{structural/mappable} in that the components of the model correspond to anatomically well-defined brain areas, and \emph{functional/predictive} in that the model provides predictions of the mapped components on a per input basis.
These criteria enable the model to be compared directly to neural and behavioral recordings.

A natural question then is -- suppose we have a model that can quantitatively explain the data collected from a given brain area better than alternatives, what have we gained?
The difference is that we have designed the model according to the four design principles above.
Each of those design principles has a concise, word-level summary that jointly describes a quantitatively accurate representation of the system in question.
It directly impinges on Marr's levels, namely as to how a system's computational goals give rise to algorithmic and implementation level mechanisms~\citep{marr2010vision}.
Furthermore, the access to every unit in the model enables ``virtual electrophysiology'', which can be used identify optimal stimuli for actual recorded neurons, enabling neural population control beyond that which was possible with more hand-designed, classical stimuli~\citep{bashivan2019neural,ponce2019evolving}.

How exactly then do we go about making these quantitative comparisons between models and data?
We will largely deal with two types of metrics in this thesis: those derived from neural recordings, and those derived from behavioral measurements.
Part of the quantitative assessment of models to data in this thesis will be the development of new neural and behavioral metrics, with the primary consideration being how to map either model units or behavior in a comparable manner to those from the animal.
For example, \emph{behavioral metrics} examine the consistency of patterns of explicitly decodable information available to support potential behavioral tasks. 
We then analyze both the model neural population and the data with identical decoding procedures (typically linear classifiers for object recognition, as these may represent potential downstream decoding circuits~\citep{rajalingham2018large,kar2019evidence}).
Crucially, we have a pattern of response choices for the model and the neural population, which can then be compared to one another on a stimulus-by-stimulus response level, resulting in the consistency measure.
For \emph{neural metrics}, the mapping between model units to recorded neurons is the same as the mapping we use between animals for the same brain area.
The neural predictivity (explained variance) across animals under this mapping defines the \emph{inter-animal consistency}, which forms a minimal upper bound when comparing models to neural responses.
For most brain areas, it is reasonable to suppose that they are relatable up to linear transform between animals, as this would suggest they are equivalent ``bases''\footnote{If the mapping were to be nonlinear, then we may be skeptical that the areas really are equivocal between animals.}.
Therefore, the problem of mapping models to animals reduces to the problem of finding the appropriate linear transform between animals, which will be addressed in Chapters~\ref{ch:convrnn},~\ref{ch:mouse}, and~\ref{ch:mec} of this thesis.

\section{Overview}
This thesis is organized so that each chapter corresponds to papers for which I am a first author\footnote{Other papers during the time of my PhD, for which I am an author, include the first multi-layer network models (feedforward and recurrent) of the retinal response to natural scenes~\citep{mcintosh2016deep,maheswaranathan2018deep}, distilling circuit mechanisms from these models~\citep{tanaka2019deep}, recurrent models of object recognition~\citep{cornets2019} and their application to 3D visual scenes~\citep{bear2020learning}, as well as unsupervised models of the ventral visual pathway~\citep{zhuang2019self, Zhuang2021}. They will be excluded from the discussion here, though are all related to aspects of goal-driven modeling.}.
I have grouped some of these papers thematically into a single chapter, and list below a high-level summary of each.

\textbf{Chapter~\ref{ch:convrnn}} contains work from two publications~\citep{nayebi2018task, nayebi2022} studying the role of recurrent connections in the primate ventral visual pathway during core object recognition.
The main finding is that through the design of new convolutional recurrent networks that better explain neural response dynamics and high-throughput temporal behaviors over traditional feedforward networks, we find evidence for the role of recurrent connections to mediate a tradeoff between task performance and network size, enabling a deeper network to be placed in the ventral pathway by extending a shallower network through time.
In the larger context of goal-driven modeling, one can view this chapter as primarily operating within the ``architecture'' component.

\textbf{Chapter~\ref{ch:mouse}} contains work from one paper~\citep{nayebi2021unsupervised} which shows that the most quantitatively accurate description of mouse visual cortex is a low-resolution, shallow network that makes best use of the mouse’s limited resources to create a light-weight, general-purpose visual system – in contrast to the deep, high-resolution, and more object-recognition-specific visual system of primates.
One can view this chapter as related to showing the impact of each component of goal-driven modeling, with an emphasis in particular on ``architecure'', ``objective function'', and ``data stream''. 

\textbf{Chapter~\ref{ch:mec}} contains work from one publication~\citep{nayebi2021explaining} applying goal-driven modeling to the non-sensory area of medial entorhinal cortex (MEC), a brain area that plays a key role in navigation and memory.
I build neural network models that explain the full diversity of neural responses in MEC, explaining practically all the response variability in a wide spectrum of experimental data.
The main implication is that specific processes of biological performance optimization may have directly shaped the neural mechanisms in MEC as a whole, and provides a path for enlarging the study of MEC beyond overly-restrictive response stereotypes, which the field has traditionally focused on.
Specifically, this work suggests the existence not of a specialized class of heterogeneous cells that is functionally segregated from classic cell types, but rather a continuum of cells within a single unified network that naturally encompasses grid, border, and heterogeneous cells.
This chapter is therefore primarily related to the ``architecture'', ``objective function'', and ``data stream'' aspects of goal-driven modeling.

\textbf{Chapter~\ref{ch:lrperf}} contains work from a single publication~\citep{tworoutes2020} demonstrating relaxations of backpropagation that could feasibly be implemented in a biological circuit and maintains competitive performance on ImageNet with deep CNNs, along with a ``language'' for parametrizing the larger space of learning circuits.
Effectively searching in this space yields one of the first biologically plausible versions of backpropagation that does not have performance deficits on large-scale tasks with CNNs of depth comparable to what we expect in the primate ventral visual pathway.
One can view this chapter as mainly related to the ``learning rule'' component of goal-driven modeling.

\textbf{Chapter~\ref{ch:lrobs}} contains work from two publications~\citep{melander2021distinct,nayebi2020identifying} related to the problem of extracting dynamical rules from observations related to synaptic plasticity.
The first publication~\citep{melander2021distinct} analyzes the dynamics of hundreds of synaptic weights \emph{in-vivo} over the course of approximately a month.
The main finding is that the changes are multiplicative in nature, but that dynamics of excitatory synapses onto inhibitory interneurons exhibit a strong additive component, providing the first description of shaft excitatory synapses in inhibitory interneurons.
The second publication~\citep{nayebi2020identifying} examines the necessity of measuring synaptic strengths (which is generally experimentally difficult to do), and demonstrates that across neural network architectures and tasks, one can reliably identify the operative learning rule from statistics derived from the network's activations alone, without needing to resort to perfect (and difficult to obtain) measurements of the synaptic weights over time.
One can view this chapter as mainly related to the ``learning rule'' component of goal-driven modeling.

\chapter{Recurrent Connections in the Primate Ventral Stream}
\label{ch:convrnn}
\section{Chapter Abstract}
The computational role of the abundant feedback connections in the ventral visual stream (VVS) is unclear, enabling humans and non-human primates to effortlessly recognize objects across a multitude of viewing conditions.
Prior studies have augmented feedforward convolutional neural networks (CNNs) with recurrent connections to study their role in visual processing; however, often these recurrent networks are optimized directly on neural data or the comparative metrics used are undefined for standard feedforward networks that lack these connections.
In this work, we develop \emph{task-optimized} convolutional recurrent (ConvRNN) network models that more correctly mimic the timing and gross neuroanatomy of the ventral pathway.
Properly chosen intermediate-depth ConvRNN circuit architectures, which incorporate mechanisms of feedforward bypassing and recurrent gating, can achieve high performance on a core recognition task, comparable to that of much deeper feedforward networks.
We then develop methods that allow us to compare both CNNs and ConvRNNs to fine-grained measurements of primate categorization behavior and neural response trajectories across thousands of stimuli.
We find that high performing ConvRNNs provide a better match to this data than feedforward networks of any depth, predicting the precise timings at which each stimulus is behaviorally decoded from neural activation patterns.
Moreover, these ConvRNN circuits consistently produce quantitatively accurate predictions of neural dynamics from V4 and IT across the entire stimulus presentation.
In fact, we find that the highest performing ConvRNNs, which best match neural and behavioral data, also achieve a strong Pareto-tradeoff between task performance and overall network size.
Taken together, our results suggest the functional purpose of recurrence in the ventral pathway is to fit a high performing network in cortex, attaining computational power through temporal rather than spatial complexity.

\section{Introduction}
\label{convrnn:section:intro}
The visual system of the brain must discover meaningful patterns in a complex physical world~\citep{james1890principles}.
Within 200\emph{ms}, primates can quickly identify objects despite changes in position, pose, contrast, background, foreground, and many other factors from one occasion to the next: a behavior known as ``core object recognition''~\citep{Pinto:2008ua, DiCarlo_2012}.
It is known that the ventral visual stream (VVS) underlies this ability by transforming the retinal image of an object into a new internal representation, in which high-level properties, such as object identity and category, are more explicit~\citep{DiCarlo_2012}.

Non-trivial dynamics result from a ubiquity of recurrent connections in the VVS, including synapses that facilitate or depress, dense local recurrent connections within each cortical region, and long-range connections between different regions, such as feedback from higher to lower visual cortex~\citep{Gilbert2013}.
Furthermore, the behavioral roles of recurrence and dynamics in the visual system are not well understood.
Several conjectures are that recurrence ``fills in'' missing data, \citep{Spoerer2017, Michaelis2018, Rajaei2019, Linsley2018} such as object parts occluded by other objects; that it ``sharpens'' representations by top-down attentional feature refinement, allowing for easier decoding of certain stimulus properties or performance of certain tasks \citep{Gilbert2013, Lindsay2015, Mcintosh2018, Li2018, kar2019evidence}; that it allows the brain to ``predict'' future stimuli (such as the frames of a movie) \citep{Rao1999, Lotter2017, Issa2018}; or that recurrence ``extends'' a feedforward computation, reflecting the fact that an unrolled recurrent network is equivalent to a deeper feedforward network that conserves on neurons (and learnable parameters) by repeating transformations several times \citep{Liao2016, Zamir2017, Li2018, Rajaei2019, cornets2019, Spoerer2020}.
Formal computational models are needed to test these hypotheses: if optimizing a model for a certain task leads to accurate predictions of neural dynamics, then that task may be a primary reason those dynamics occur in the brain.

\begin{figure}
  \centering
  \includegraphics[width=1.0\columnwidth]{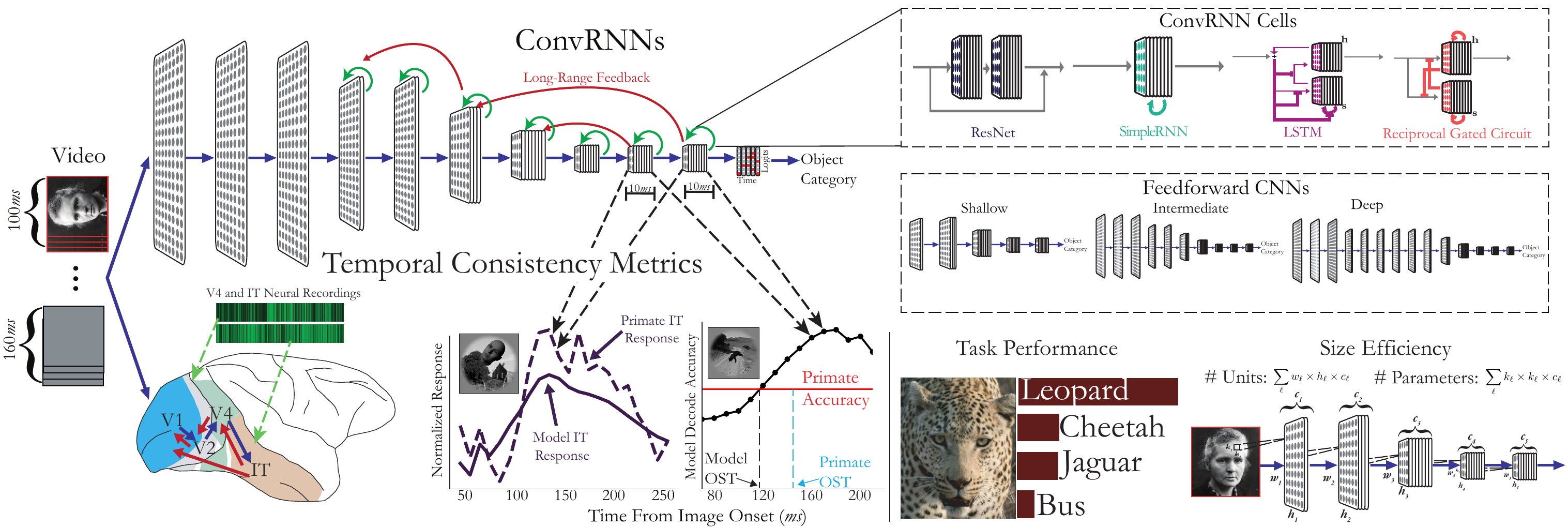}
  \caption[ConvRNNs as models of the primate ventral visual stream] {\textbf{ConvRNNs as models of the primate ventral visual stream.} \emph{Performance-optimized recurrence.}
  Convolutional recurrent networks (ConvRNNs) have a combination of local recurrent circuits (green) and long-range feedback connections (red) added on top of a feedforward CNN ``BaseNet'' backbone (blue).
  Feedforward CNNs are therefore a special case of ConvRNNs, and we consider a variety of CNNs of varying depths, trained on the ImageNet categorization task.
  We also perform large-scale evolutionary searches over the local and long-range feedback connections.
  In addition, we consider particular choices of ``light-weight'' (in terms of parameter count) decoding strategy that determines the final object category of that image.
  In our implementation displayed on the top, propagation along each arrow takes one time step (10\emph{ms}) to mimic conduction delays between cortical layers.
  \emph{Measurements.}
  From each network class, we measure categorization performance and its size in terms of its parameter and neuron count.
  \emph{Comparison to neural and behavioral data.}
  Each stimulus was presented for 100\emph{ms}, followed by a mean gray stimulus interleaved between images, lasting a total of 260\emph{ms}.
  All images were presented to the models for 10 time steps (corresponding to 100\emph{ms}), followed by a mean gray stimulus for the remaining 15 time steps, to match the image presentation to the primates.
  We stipulated that units from each multi-unit array must be fit by features from a single model layer, detailed in Section~\ref{convrnn:sss:methods-neural-fitting}.
  Model features produce a temporally-varying output that can be compared to primate neural dynamics in V4 and IT, as well as temporally-varying behaviors in the form of object solution times (OST).
  } 
\label{convrnn:fig0}
\end{figure}

We therefore broaden the method of goal-driven modeling from solving tasks with feedforward CNNs~\citep{Yamins2016} or RNNs~\citep{Mante2013} to explain dynamics in the primate visual system, building convolutional recurrent neural networks (ConvRNNs), depicted in Figure~\ref{convrnn:fig0}.
There has been substantial prior work in this domain~\citep{Liao2016,Mcintosh2018,Zamir2017,cornets2019,Kietzmann2019,Spoerer2020}, which we go beyond in several important ways.  

We show that with a novel choice of layer-local recurrent circuit and long-range feedback connectivity pattern, ConvRNNs can match the performance of much deeper feedforward CNNs on ImageNet but with far fewer units and parameters, as well as a more anatomically consistent number of layers, by extending these computations through time.
In fact, such ConvRNNs most accurately explain neural dynamics from V4 and IT across the entirety of stimulus presentation with a temporally-fixed linear mapping, compared to alternative recurrent circuits. 
Furthermore, we find these suitably-chosen ConvRNN circuit architectures provide a better match to primate behavior in the form of object solution times, compared to feedforward CNNs. 
We observe that ConvRNNs that attain high task performance but have small overall network size, as measured by number of units, are most consistent with this data -- while even the highest-performing but biologically-implausible deep feedforward models are overall a \emph{less consistent} match.
In fact, we find a strong Pareto-tradeoff between network size and performance, with ConvRNNs of biologically-plausible intermediate-depth occupying the sweet spot with high performance and a (comparatively) small overall network size.
Because we do not fit neural networks end-to-end to neural data (c.f. \cite{Kietzmann2019}), but instead show that these outcomes emerge naturally from task performance, our approach enables a normative interpretation of the structural and functional design principles of the model.

Our work is also the first to develop large-scale task-optimized ConvRNNs with biologically-plausible temporal unrolling. 
Unlike most studies of combinations of convolutional and recurrent networks, which posit a recurrent subnetwork concatenated onto the end of a convolutional backbone~\citep{Mcintosh2018}, we model local recurrence implanted within ConvRNN layers, and allow long-range feedback between layers.  
Moreover, we treat each connection in the network -- whether feedforward or feedback -- as a real temporal object with a biophysical conduction delay (set at $\sim$10\emph{ms}), rather than the typical procedure (e.g. as in~\cite{Mcintosh2018,Zamir2017,cornets2019}) in which the feedforward component of the network (no matter now deep) operates in one timestep.
As a result, our networks can be directly compared with neural and behavioral trajectories at a fine-grained scale limited only by the conduction delay itself.

This level of realism is especially important for establishing what we have found appears to be the main real quantitative advantage of ConvRNNs as biological models, as compared to very deep feedforward networks.  
In particular, we can define an improved metric for assessing the correctness of the match between a ConvRNN network -- thought of as a dynamical system -- and the actual trajectories of real neurons. 
By treating such feedforward networks as ConvRNNs with recurrent connections set to 0, we can map these networks to temporal trajectories as well.
As a result, we can directly ask, how much of the neural-behavioral trajectory of real brain data is explicable by very deep feedforward networks? 
This is an important question because implausibly deep networks have been shown in the literature not only to achieve the highest categorization performance~\citep{He2016} but also achieve competitive results on matching static (temporally-averaged) neural responses~\citep{Schrimpf2018}.
Due to non-biological temporal unrolling, previous work with comparing such networks to temporal trajectories in neural data~\citep{cornets2019} has been forced to unfairly score feedforward networks as total failures, with temporal match score artificially set at 0.
With our improved realism, we find (see results section below) that deep feedforward networks actually make quite nontrivial temporal predictions that do explain \emph{some} of the reliable temporal variability of real neurons.
In this context, our finding that plausibly-deep ConvRNNs in turn meaningfully outperform these deep feedforward networks on this more fair metric is a strong and nontrivial signal of the actually-better biological match of ConvRNNs as compared to deep feedforward networks.

\section{Results}
\label{convrnn:section:results}
\subsection{An evolutionary architecture search yields specific layer-local recurrent circuits and long-range feedback connectivity patterns that improve task performance and maintain small network size.}
\label{convrnn:ss:results-performance}
\begin{figure}
  \centering
  \includegraphics[width=1.0\columnwidth]{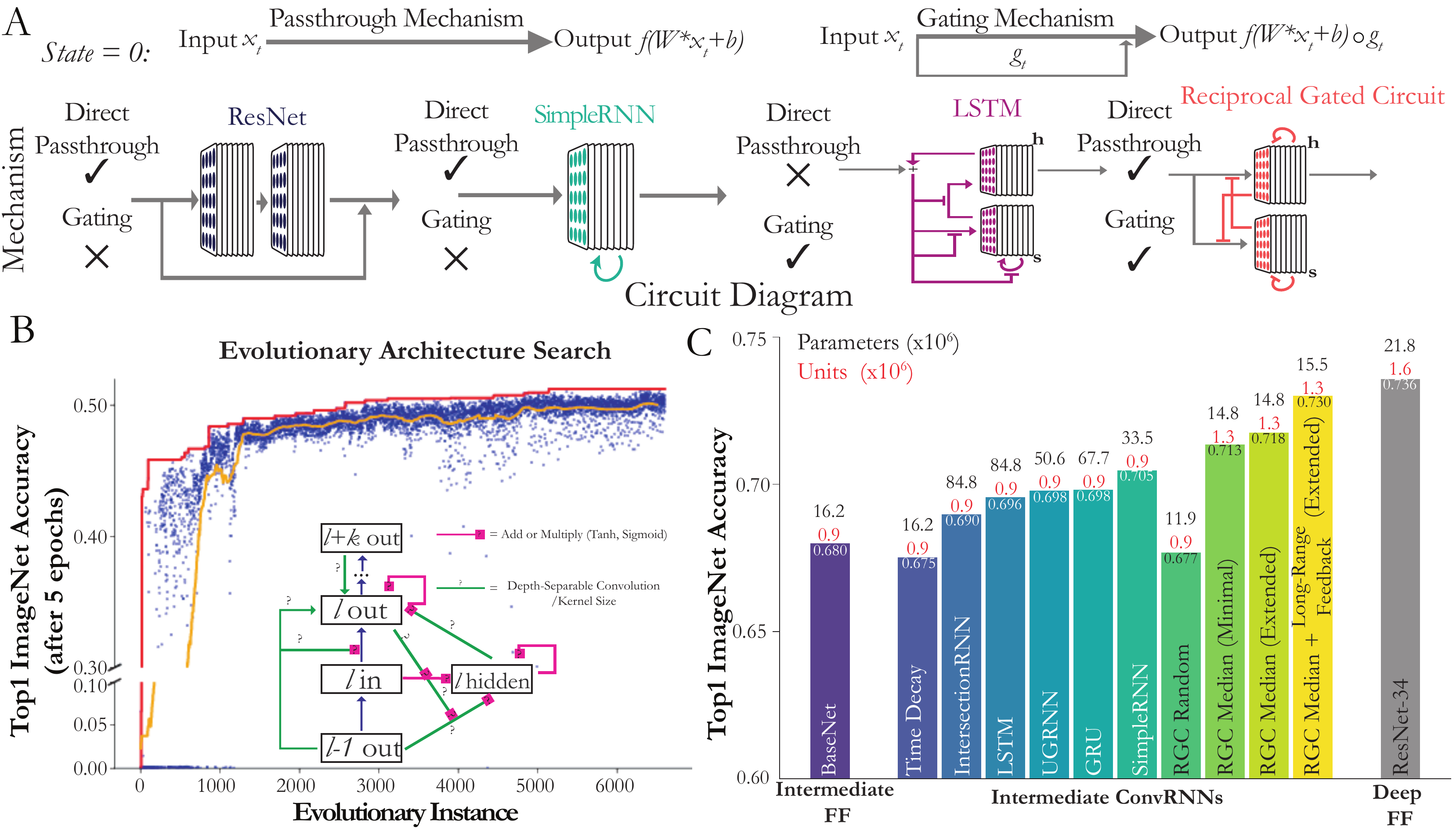}
  \caption[Suitably-chosen intermediate ConvRNN circuits can match the object recognition performance of much deeper feedforward models]{\textbf{Suitably-chosen intermediate ConvRNN circuits can match the object recognition performance of much deeper feedforward models.} \textbf{(a) Architectural differences between ConvRNN circuits.} Standard ResNet blocks and SimpleRNN circuits have direct passthrough but not gating.
  Namely, on the first timestep, the output of a given ConvRNN layer is directly a single linear-nonlinear function of its input, equivalent to that of a feedforward CNN layer (namely, $f(W*x_t + b)$, where $f$ is a nonlinear function such as ELU/ReLU and $x_t$ is the input).
  The LSTM circuit has gating, denoted by T-junctions, but not direct passthrough. 
  The Reciprocal Gated Circuit has both.
  \textbf{(b) ConvRNN circuit search.} Each blue dot represents a model, sampled from hyperparameter space, trained for five epochs.
  The orange line is the average performance of the last 50 models up to that time.
  The red line denotes the top performing model at that point in the search.
  \emph{Search space schematic:} 
  Question marks denote optional connections, which may be conventional or depth-separable convolutions with a choice of kernel size.
  \textbf{(c) Performance of models fully trained on ImageNet.} We compared the performance of an 11-layer feedforward base model (``BaseNet'') modeled after ResNet-18, a control ConvRNN model with trainable time constants (``Time Decay''), along with various other common RNN architectures implanted into this BaseNet, as well as the median Reciprocal Gated Circuit (RGC) model from the search (``RGC Median'') with or without global feedback connectivity, and its minimally-unrolled control (see the first table in Section~\ref{convrnn:ss:methods-rnn} for the exact timestep values). 
  The ``RGC Random'' model was selected randomly from the initial, random phase of the model search. 
  Parameter and unit counts (total number of neurons in the output of each layer) in millions are shown on top of each bar.}
\label{convrnn:fig2}
\end{figure}

We first tested whether augmenting CNNs with standard RNN circuits from the machine learning community, SimpleRNNs and LSTMs, could improve performance on ImageNet object recognition (Figure~\ref{convrnn:fig2}a).
We found that these recurrent circuits added a small amount of accuracy when introduced into the convolutional layers of a shallow, 6-layer feedforward backbone (``FF'' in Figure~\ref{convrnn:supp:minextend}) based off of the AlexNet~\citep{Krizhevsky2012} architecture, which we will refer to as a ``BaseNet'' (see Section~\ref{convrnn:ss:methods-rnn} for architecture details).
However, there were two problems with these resultant recurrent architectures: first, these ConvRNNs did not perform substantially better than parameter-matched, minimally unrolled controls -- defined as the minimum number of timesteps after the initial feedforward pass whereby all recurrence connections were engaged at least once.
This control comparison suggested that the observed performance gain was due to an increase in the number of unique parameters added by the implanted ConvRNN circuits rather than temporally-extended recurrent computation.
Second, making the feedforward model wider or deeper yielded an even larger performance gain than adding these standard RNN circuits, but with fewer parameters.
This suggested that standard RNN circuits, although well-suited for a range of temporal tasks, are less well-suited for inclusion within deep CNNs to solve challenging object recognition tasks.

We speculated that this was because standard circuits lack a combination of two key properties, each of which on their own have been successful either purely for RNNs or for feedforward CNNs: (1) \textbf{Direct passthrough}, where at the first timestep, a zero-initialized hidden state allows feedforward input to pass on to the next layer as a single linear-nonlinear layer just as in a standard feedforward CNN layer (Figure~\ref{convrnn:fig2}a; top left); and (2) \textbf{Gating}, in which the value of a hidden state determines how much of the bottom-up input is passed through, retained, or discarded at the next time step (Figure~\ref{convrnn:fig2}a; top right).
For example, LSTMs employ gating, but not direct passthrough, as their inputs must pass through several nonlinearities to reach their output; whereas SimpleRNNs do passthrough a zero-initialized hidden state, but do not gate their input (Figure~\ref{convrnn:fig2}a; see Section~\ref{convrnn:ss:methods-rnn} for cell equations).
Additionally, each of these computations have direct analogies to biological mechanisms -- ``direct passthrough'' would correspond to feedforward processing in time, and ``gating'' would correspond to adaptation to stimulus statistics across time~\citep{Hosoya2005,mcintosh2016deep}.

We thus implemented recurrent circuits with both features to determine whether they function better than standard circuits within CNNs. 
One example of such a structure is the ``Reciprocal Gated Circuit'' (RGC)~\citep{nayebi2018task}, which passes through its zero-initialized hidden state and incorporates gating (Figure~\ref{convrnn:fig2}a, bottom right; see Section~\ref{convrnn:sss:methods-rgc} for the circuit equations).
Adding this circuit to the 6-layer BaseNet (``FF'') increased accuracy from 0.51 and 0.53 (``RGC Minimal'', the minimally unrolled, parameter-matched control version) to 0.6 (``RGC Extended'').
Moreover, the RGC used substantially fewer parameters than the standard circuits to achieve greater accuracy (Figure~\ref{convrnn:supp:minextend}).

However, it has been shown that different RNN structures can succeed or fail to perform a given task because of differences in trainability rather than differences in capacity~\citep{collins2017}.
Therefore, we designed an evolutionary search to jointly optimize over both discrete choices of recurrent connectivity patterns as well as continuous choices of learning hyperparameters and weight initializations (search details in Section~\ref{convrnn:ss:methods-hyperopt}).
While a large-scale search over thousands of convolutional LSTM architectures did yield a better purely gated LSTM-based ConvRNN (``LSTM Opt''), it did not eclipse the performance of the smaller RGC ConvRNN.
In fact, applying the same hyperparameter optimization procedure to the RGC ConvRNNs equally increased that architecture class's performance and further reduced its parameter count (Figure~\ref{convrnn:supp:minextend}, ``RGC Opt'').

Therefore, given the promising results with shallower networks, we turned to embedding recurrent circuit motifs into intermediate-depth feedforward networks at scale, whose number of feedforward layers corresponds to the timing of the ventral stream~\citep{DiCarlo_2012}.
We then performed an evolutionary search over these resultant intermediate-depth recurrent architectures (Figure~\ref{convrnn:fig2}b).
If the primate visual system uses recurrence in lieu of greater network depth to perform object recognition, then a shallower recurrent model with a suitable form of recurrence should achieve recognition accuracy equal to a deeper feedforward model, albeit with temporally-fixed parameters~\citep{Liao2016}.
We therefore tested whether our search had identified such well-adapted recurrent architectures by fully training a representative ConvRNN, the model with the median (across 7000 sampled models) validation accuracy after five epochs of ImageNet training.
This median model (``RGC Median'') reached a final ImageNet top-1 validation accuracy nearly equal to a ResNet-34 model with nearly twice as many layers, even though the ConvRNN used only $\sim 75\%$ as many parameters.
The fully unrolled model from the random phase of the search (``RGC Random'') did not perform substantially better than the BaseNet, though the minimally unrolled control did (Figure~\ref{convrnn:fig2}c).
We suspect the improvement of the base intermediate feedforward model over using shallow networks (as in Figure~\ref{convrnn:supp:minextend}) diminishes the difference between the minimal and extended versions of the RGC compared to suitably chosen long-range feedback connections. 
However, compared to alternative choices of ConvRNN circuits, even the minimally extended RGC significantly outperforms them with fewer parameters and units, indicating the importance of this circuit motif for task performance.
This observation suggests that our evolutionary search strategy yielded effective recurrent architectures beyond the initial random phase of the search.

We also considered a control model (``Time Decay'') that produces temporal dynamics by learning time constants on the activations independently at each layer, rather than by learning connectivity between units.
In this ConvRNN, unit activations have exponential rather than immediate falloff once feedforward drive ceases.
These dynamics could arise, for instance, from single-neuron biophysics (e.g. synaptic depression) rather than interneuronal connections.
However, this model did not perform any better than the feedforward BaseNet, implying that ConvRNN performance is not a trivial result of outputting a dynamic time course of responses.
We further implanted other more sophisticated forms of ConvRNN circuits into the BaseNet, and while this improved performance over the Time Decay model, it did not outperform the RGC Median ConvRNN despite having many more parameters (Figure~\ref{convrnn:fig2}c).
Together, these results demonstrate that the RGC Median ConvRNN uses recurrent computations to subserve object recognition behavior and that particular motifs in its recurrent architecture (Figure~\ref{convrnn:supp:fig1}), found through an evolutionary search, are required for its improved accuracy.
Thus, given suitable local recurrent circuits and patterns of long-range feedback connectivity, a physically more compact, temporally-extended ConvRNN can do the same challenging object recognition task as a deeper feedforward CNN.

\subsection{ConvRNNs better match temporal dynamics of primate behavior than feedforward models.}
\label{convrnn:ss:results-dynamics-ost}
To address whether recurrent processing is engaged during core object recognition behavior, we turn to behavioral data collected from behaving primates.
There is a growing body of evidence that current feedforward models fail to accurately capture primate behavior~\citep{rajalingham2018large, kar2019evidence}.
We therefore reasoned that if recurrence is critical to core object recognition behavior, then recurrent networks should be more consistent with suitable measures of primate behavior compared to the feedforward model family.
Since the identity of different objects is decoded from the IT population at different times, we considered the first time at which the IT neural decoding accuracy reaches the (pooled) primate behavioral accuracy of a given image, known as the ``object solution time (OST)''~\citep{kar2019evidence}.
Given that our ConvRNNs also have an output at each 10\emph{ms} timebin, the procedure for computing the OST for these models is computed from its ``IT-preferred'' layers, and we report the ``OST consistency'', which we define as the Spearman correlation between the model OSTs and the IT population's OSTs on the common set of images solved by the given model and IT under the \emph{same} stimulus presentation (see Sections~\ref{convrnn:sss:methods-neural-data} and ~\ref{convrnn:ss:methods-ost} for more details).

\begin{figure}
  \centering
  \includegraphics[width=1.0\columnwidth]{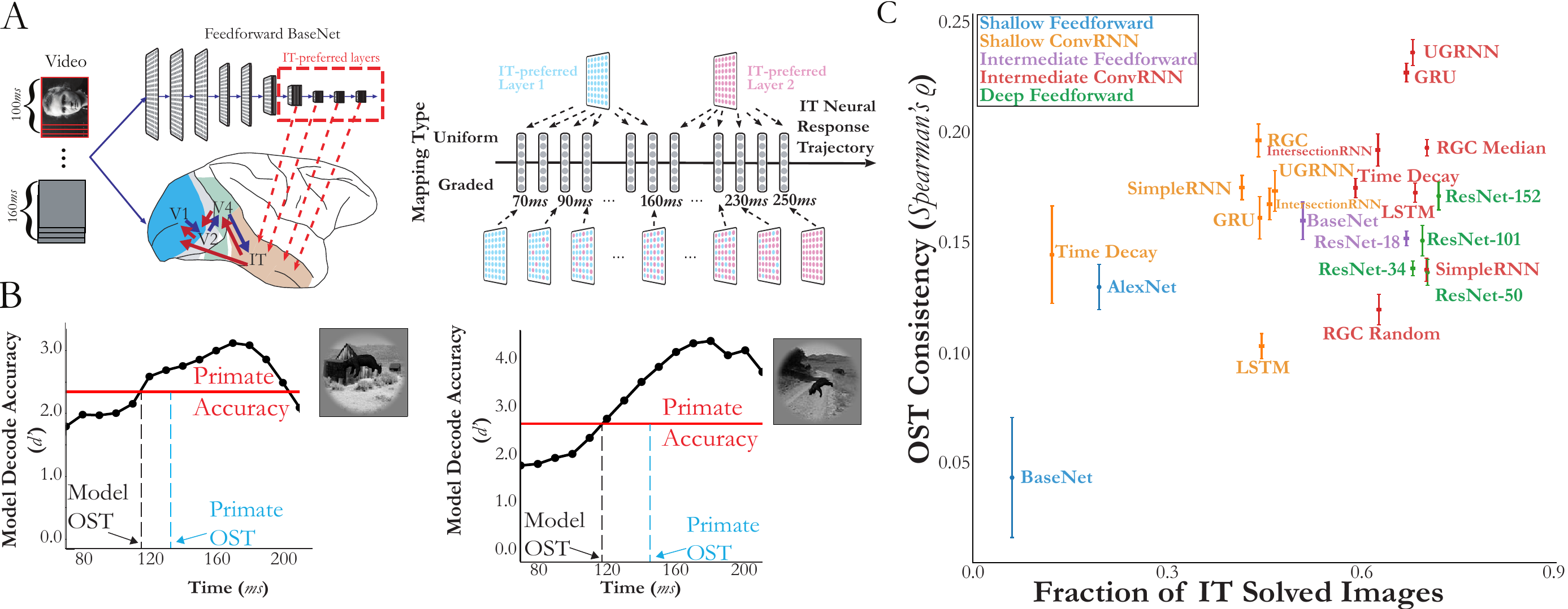}
  \caption[Intermediate ConvRNNs best explain the object solution times (OST) of IT across images]{\textbf{Intermediate ConvRNNs best explain the object solution times (OST) of IT across images.}
  \textbf{(a) Comparing to primate OSTs.}
  \emph{Mapping model layers to timepoints.}
  In order to compare to primate IT object solution times, namely the first time at which the neural decode accuracy for each image reached the level of the (pooled) primate behavioral accuracy, we first need to define object solution times for models.
  This procedure involves identification of the ``IT-preferred'' layer(s) via a standard linear mapping to temporally averaged IT responses.
  \emph{Choosing a temporal mapping gradation.}
  These ``IT-preferred'' model layer(s) are then mapped to 10\emph{ms} timebins from 70-260\emph{ms} in either a uniform or graded fashion, if the model is feedforward.
  For ConvRNNs, this temporal mapping is always one-to-one with these 10\emph{ms} timebins.
  \textbf{(b) Defining model OSTs.}
  Once the temporal mapping has been defined, we train a linear SVM at each 10\emph{ms} model timebin and compute the classifier's $d^{'}$ (displayed in each of the black dots for a given example image).
  The first timebin at which the model $d^{'}$ matches the primate's accuracy is defined as the ``Model OST'' for that image (obtained via linear interpolation), which is the same procedure previously used~\citep{kar2019evidence} to determine the ground truth IT OST (``Primate OST'' vertical dotted line).
  \textbf{(c) Proper choices of recurrence best match IT OSTs.}
  Mean and s.e.m. are computed across train/test splits ($N=10$) when that image (of 1320 images) was a test-set image, with the Spearman correlation computed with the IT object solution times (analogously computed from the IT population responses) across the imageset solved by both the given model and IT, constituting the ``Fraction of IT Solved Images'' on the $x$-axis.
  We start with either a shallow base feedforward model consisting of 5 convolutional layers and 1 layer of readout (``BaseNet'' in blue) as well as an intermediate-depth variant with 10 feedforward layers and 1 layer of readout (``BaseNet'' in purple), detailed in Section~\ref{convrnn:sss:methods-basenet}.
  From these base feedforward models, we embed recurrent circuits, resulting in either ``Shallow ConvRNNs'' or ``Intermediate ConvRNNs'', respectively.
}
 \label{convrnn:ostfig}
\end{figure}

Unlike our ConvRNNs, which exhibit more biologically plausible temporal dynamics, evaluating the temporal dynamics in feedforward models poses an immediate problem.
Given that recurrent networks repeatedly apply nonlinear transformations across time, we can analogously map the layers of a feedforward network to timepoints, observing that a network with $k$ distinct layers can produce $k$ distinct OSTs in this manner.
Thus, the most direct definition of a feedforward model's OST is to uniformly distribute the timebins between 70-260\emph{ms} across its $k$ layers.
For very deep feedforward networks such as ResNet-101 and ResNet-152, this number of distinct layers will be as fine-grained as the 10\emph{ms} timebins of the IT responses; however, for most other shallower feedforward networks this will be much coarser.
Therefore to enable these feedforward models to be maximally temporally expressive, we \emph{additionally} randomly sample units from consecutive feedforward layers to produce a more graded temporal mapping, depicted in Figure~\ref{convrnn:ostfig}a.
This graded mapping is ultimately what we use for the feedforward models in Figure~\ref{convrnn:ostfig}c, providing the highest OST consistency for that model class\footnote{Mean OST difference $0.0120$ and s.e.m. $0.0045$, Wilcoxon test on uniform vs. graded mapping OST consistencies across feedforward models, $p < 0.001$; see also Figure~\ref{convrnn:supp:fig4}.}.
Note that for ConvRNNs and very deep feedforward models (ResNet-101 and ResNet-152) whose number of ``IT-preferred'' layers matches the number of timebins, then the uniform and graded mappings are equivalent, whereby the earliest (in the feedforward hierarchy) layer is matched to the earliest 10\emph{ms} timebin of 70\emph{ms}, and so forth.

With model OST defined across both model families, we compared various ConvRNNs and feedforward models to the IT population's OST in Figure~\ref{convrnn:ostfig}c.
Among shallower and deeper models, we found that ConvRNNs were generally able to better explain IT's OST than their feedforward counterparts.
Specifically, we found that ConvRNN circuits without \emph{any} multi-unit interaction such as the Time Decay ConvRNN only marginally, and not always significantly, improved the OST consistency over its respective BaseNet model\footnote{Paired $t$-test with Bonferroni correction: shallow Time Decay vs. ``BaseNet'' in blue, mean OST difference $0.101$ and s.e.m. $0.0313$, $t(9) \approx 3.23, p < 0.025$; intermediate Time Decay vs. ``BaseNet'' in purple, mean OST difference $0.0148$ and s.e.m. $0.00857$, $t(9) \approx 1.73, p\approx 0.11$.}.
On the other hand, ConvRNNs with multi-unit interactions generally provided the greatest match to IT OSTs than even the deepest feedforward models\footnote{Paired $t$-test with Bonferroni correction: shallow RGC vs. ``BaseNet'' in blue, mean OST difference $0.153$ and s.e.m. $0.0252$, $t(9) \approx 6.08, p < 0.001$; intermediate UGRNN vs. ResNet-152, mean OST difference $0.0652$ and s.e.m. $0.00863$, $t(9) \approx 7.55, p < 0.001$; intermediate GRU vs. ResNet-152, mean OST difference $0.0559$ and s.e.m. $0.00725$, $t(9) \approx 7.71, p < 0.001$; RGC Median vs. ResNet-152, mean OST difference $0.0218$ and s.e.m. $0.00637$, $t(9) \approx 3.44, p < 0.01$.}, where the best feedforward model (ResNet-152) attains a mean OST consistency of 0.173 and the best ConvRNN (UGRNN) attains an OST consistency of 0.237.

Consistent with our observations in Figure~\ref{convrnn:fig2} that different recurrent circuits with multi-unit interactions were not all equally effective when embedded in CNNs (despite outperforming the simple Time Decay model), we similarly found that this observation held for the case of matching IT's OST.
Given recent observations~\citep{Kar2021} that inactivating parts of macaque ventrolateral PFC (vlPFC) results in behavioral deficits in IT for late-solved images, we reasoned that additional decoding procedures employed at the categorization layer during task optimization might meaningfully impact the model's OST consistency, in addition to the choice of recurrent circuit used.
We designed several decoding procedures (defined in Section~\ref{convrnn:ss:methods-framework-decoders}), motivated by prior observations of accumulation of relevant sensory signals during decision making in primates~\citep{Shadlen2001}.
Overall, we found that ConvRNNs with different decoding procedures, but with the \emph{same} layer-local recurrent circuit (RGC Median) and long-range feedback connectivity patterns, yielded significant differences in final consistency with the IT population OST (Figure~\ref{convrnn:supp:decoder-ostfig}; Friedman test, $p < 0.05$).
Moreover, the simplest decoding procedure of outputting a prediction at the last timepoint, a strategy commonly employed by the computer vision community, had a lower OST consistency than each of the more nuanced Max Confidence\footnote{Paired $t$-test with Bonferroni correction, mean OST difference $0.0195$ and s.e.m. $0.00432$, $t(9) \approx -4.52, p < 0.01$.} and Threshold decoding procedures\footnote{Paired $t$-test with Bonferroni correction, mean OST difference $0.0279$ and s.e.m. $0.00634$, $t(9) \approx -4.41, p < 0.01$.} that we considered.
Taken together, our results suggest that the type of multi-unit layer-wise recurrence \emph{and} downstream decoding strategy are important features for OST consistency with IT, suggesting that specific, non-trivial connectivity patterns further downstream of the ventral visual pathway may be important to core object recognition behavior over timescales of a couple hundred milliseconds.

\begin{figure}
  \centering
  \includegraphics[width=1.0\columnwidth]{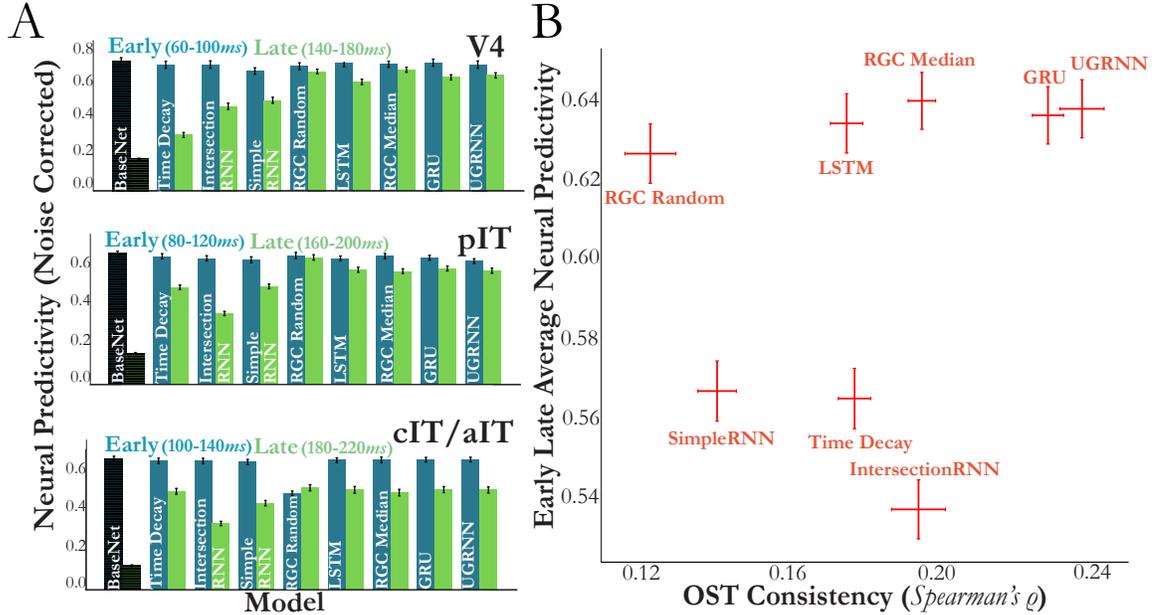}
  \caption[Suitably-chosen intermediate ConvRNN circuits provide consistent predictions of primate ventral stream neural dynamics]{\textbf{Suitably-chosen intermediate ConvRNN circuits provide consistent predictions of primate ventral stream neural dynamics.}
  \textbf{(a)} The $y$-axis indicates the median across neurons of the explained variance between predictions and ground-truth responses on held-out images divided by the square root of the internal consistencies of the neurons, defined in Section~\ref{convrnn:sss:methods-neural-metrics}. 
  Error bars indicates the s.e.m across neurons ($N=88$ for V4, $N=88$ for pIT, $N=80$ for cIT/aIT) averaged across 10\emph{ms} timebins ($N=4$ each for the ``Early'' and ``Late'' designations).
  As can be seen, the intermediate-depth feedforward BaseNet model (first bars) is a poor predictor of the subset of late responses that are beyond the feedforward pass, but certain types of ConvRNN circuits (such as ``RGC Median'', ``UGRNN'', and ``GRU'') added to the BaseNet are overall best predictive across visual areas at late timepoints (Wilcoxon test (with Bonferroni correction) with feedforward BaseNet, $p < 0.001$ for each visual area).
  See Figure~\ref{convrnn:supp:neuralfig} for the full timecourses at the resolution of 10\emph{ms} bins.
  \textbf{(b)} For each ConvRNN circuit, we compare the average neural predictivity (averaged per neuron across early and late timepoints) averaged across areas, to the OST consistency.
  The ConvRNNs that have the best average neural predictivity also best match the OST consistency (``RGC Median'', ``UGRNN'', and ``GRU'').}
 \label{convrnn:neuralfig}
\end{figure}

\subsection{Neural dynamics differentiate ConvRNN circuits.}
\label{convrnn:ss:results-dynamics-neural}
ConvRNNs naturally produce a dynamic time series of outputs given an unchanging input stream, unlike feedforward networks.
While these recurrent dynamics could be used for tasks involving time, here we optimized the ConvRNNs to perform the ``static'' task of object classification on ImageNet.
It is possible that the primate visual system is optimized for such a task, because even static images produce reliably dynamic neural response trajectories at temporal resolutions of tens of milliseconds~\citep{Issa2018}.
The object content of some images becomes decodable from the neural population significantly later than the content of other images, even though animals recognize both object sets equally well.
Interestingly, late-decoding images are not well characterized by feedforward CNNs, raising the possibility that they are encoded in animals through recurrent computations~\citep{kar2019evidence}.
If this were the case, we reason then that recurrent networks trained to perform a difficult, but static object recognition task might explain neural \emph{dynamics} in the primate visual system, just as feedforward models explain time-averaged responses~\citep{yamins_ventralneural, khaligh2014deep}.

Prior studies~\citep{Kietzmann2019} have \emph{directly} fit recurrent parameters to neural data, as opposed to optimizing them on a task.
While it is natural to try to fit recurrent parameters to the temporally-varying neural responses directly, this approach naturally has a loss of normative explanatory power.
In fact, we found that this approach suffers from a fundamental overfitting issue to the particular image statistics of the neural data collected.
Specifically, we directly fit these recurrent parameters (implanted into the task-optimized feedforward BaseNet) to the dynamic firing rates of primate neurons recorded during encoding of visual stimuli.
However, while these non-task optimized dynamics generalized to held-out images and neurons (Figure~\ref{convrnn:supp:fig2}a,b), they had no longer retained performance to the original object recognition task that the primate itself is able to perform (Figure~\ref{convrnn:supp:fig2}c).
Therefore, to avoid this issue, we instead asked whether \emph{fully} task-optimized ConvRNN models (including the ones introduced in Section~\ref{convrnn:ss:results-performance}) could predict these dynamic firing rates from multi-electrode array recordings from the ventral visual pathway of rhesus macaques~\citep{Majaj2015}.

We began with the feedforward BaseNet and added a variety of ConvRNN circuits, including the RGC Median ConvRNN and its counterpart generated at the random phase of the evolutionary search (``RGC Random'').
All of the ConvRNNs were presented with the same images shown to the primates, and we collected the time series of features from each model layer.
To decide which layer should be used to predict which neural responses, we fit linear models from each feedforward layer's features to the neural population and measured where explained variance on held-out images peaked (see Section~\ref{convrnn:ss:methods-neural} for more details).
Units recorded from distinct arrays -- placed in the successive V4, posterior IT (pIT), and central/anterior IT (cIT/aIT) cortical areas of the macaque -- were fit best by the successive layers of the feedforward model, respectively.
Finally, we measured how well ConvRNN features from these layers predicted the dynamics of each unit.
In contrast with feedforward models fit to temporally-averaged neural responses, the linear mapping in the temporal setting must be \emph{fixed} at all timepoints.
The reason for this choice is that the linear mapping yields ``artificial units'' whose activity can change over time (just like the real target neurons), but the identity of these units should not change over the course of 260\emph{ms}, which would be the case if instead a separate linear mapping was fit at each 10\emph{ms} timebin.
This choice of a temporally-fixed linear mapping therefore maintains the physical relationship between real neurons and model neurons.

As can be seen from Figure~\ref{convrnn:neuralfig}a, with the exception of the RGC Random ConvRNN, the ConvRNN feature dynamics fit the neural response trajectories as well as the feedforward baseline features on early phase responses (Wilcoxon test $p$-values in Table~\ref{convrnn:tab:pval-basenet-early}) and better than the feedforward baseline features for late phase responses (Wilcoxon test with Bonferroni correction $p < 0.001$), across V4, pIT, and cIT/aIT on held-out images.
For the early phase responses, the ConvRNNs that employ direct passthrough are elaborations of the baseline feedforward network, although the ConvRNNs which only employ gating are still a nonlinear function of their input, similar to a feedforward network.
For the late phase responses, any feedforward model exhibits similar ``square wave'' dynamics as its 100\emph{ms} visual input, making it a poor predictor of the subset of late responses that are beyond the initial feedforward pass (Figure~\ref{convrnn:supp:neuralfig}, purple lines).
In contrast, the activations of ConvRNN units have persistent dynamics, yielding predictions of the \emph{entire} neural response trajectories.

Crucially, these predictions result from the task-optimized nonlinear dynamics from ImageNet, as both models are fit to neural data with the same form of temporally-fixed linear model with the \emph{same} number of parameters.
Since the initial phase of neural dynamics was well-fit by feedforward models, we asked whether the later phase could be fit by a much simpler model than any of the ConvRNNs we considered, namely the Time Decay ConvRNN with ImageNet-trained time constants at convolutional layers.
If the Time Decay ConvRNN were to explain neural data as well as the other ConvRNNs, it would imply that interneuronal recurrent connections are not needed to account for the observed dynamics; however, this model did not fit the late phase dynamics of intermediate areas (V4 and pIT) as well as the other ConvRNNs\footnote{Wilcoxon test with Bonferroni correction $p < 0.001$ for each ConvRNN vs. Time Decay, except for the SimpleRNN $p\approx 0.46$ for pIT.}.
The Time Decay model did match the other ConvRNNs for cIT/aIT, which may indicate some functional differences in the temporal processing of this area versus V4 and pIT.
Thus, the more complex recurrence found in ConvRNNs is generally needed both to improve object recognition performance over feedforward models \emph{and} to account for neural dynamics in the ventral stream, even when animals are only required to fixate on visual stimuli.
However, not all forms of complex recurrence are equally predictive of temporal dynamics.
As depicted in Figure~\ref{convrnn:neuralfig}b, we found among these that the RGC Median, UGRNN, and GRU ConvRNNs attained the highest median neural predictivity for each visual area in both early and late phases, but in particular significantly outperformed the SimpleRNN ConvRNN at the late phase dynamics of these areas\footnote{Wilcoxon test with Bonferroni correction between each of these ConvRNNs vs. the SimpleRNN on late phase dynamics, $p < 0.001$ per visual area.}, and these models in turn were among the best matches to IT object solution times (OST) from Section~\ref{convrnn:ss:results-dynamics-ost}.

A natural follow-up question to ask is whether a \emph{lack} of recurrent processing is the reason for the prior observation that there is a drop in explained variance for feedforward models from early to late timebins~\citep{kar2019evidence}.
In short, we find that this is not the case, and that this drop likely has to do with task-orthogonal dynamics specific to individual primates, which we examine below.

It is well-known that recurrent neural networks can be viewed as very deep feedforward networks with weight sharing across layers that would otherwise be recurrently connected~\citep{Liao2016}.
Thus, to address this question, we compare feedforward models of varying depths to ConvRNNs across the entire temporal trajectory under a \emph{varying} linear mapping at each timebin, in contrast to the above.
This choice of linear mapping allows us to evaluate how well the model features are at explaining early compared to late time dynamics without information from the early dynamics influencing the later dynamics, and also more crucially, to allow the feedforward model features to be independently compared to the late dynamics.
Specifically, as can be seen in Figure~\ref{convrnn:supp:fig3}a, we observe a drop in explained variance from early (130-140\emph{ms}) to late (200-210\emph{ms}) timebins for the BaseNet and ResNet-18 models, across multiple neural datasets.
Models with increased feedforward depth (such as ResNet-101 or ResNet-152), along with our performance-optimized RGC Median ConvRNN, exhibit a similar drop in median population explained variance as the intermediate feedforward models.
The benefit of model depth with respect to increased explained variance of late IT responses might be only noticeable while comparing shallow models ($< 7$ nonlinear transforms) to much deeper ($> 15$ nonlinear transforms) models~\citep{kar2019evidence}.
Our results suggest that the amount of variance explained in the late IT responses is not a monotonically increasing function of model depth.

As a result, an alternative hypothesis is that the drop in explained variance from early to late timebins could instead be attributed to task-orthogonal dynamics specific to an individual primate as opposed to iterated nonlinear transforms, resulting in variability unable to be captured by any task-optimized model (feedforward or recurrent).
To explore this possibility, we examined whether the model's neural predictivity at these early and late timebins was relatively similar in ratio to that of one primate's IT neurons mapped to that of another primate (see Section~\ref{convrnn:ss:methods-interanimal} for more details, where we derive a novel measure of the the neural predictivity between animals, known as the ``inter-animal consistency'').

As shown in Figure~\ref{convrnn:supp:fig3}b, across various hyperparameters of the linear mapping, we observe a ratio close to one between the neural predictivity (of the target primate neurons) of the feedforward BaseNet to that of the source primate mapped to the same target primate.
Therefore, as it stands, \emph{temporally-varying} linear mappings to neural responses collected from an animal during rapid visual stimulus presentation (RSVP) may not sufficiently separate feedforward models from recurrent models any better than one animal to another -- though more investigation is needed to ensure tight estimates of the inter-animal consistency measure we have introduced here with neural data recorded from more primates.
Nonetheless, this observation further motivates our earlier result of additionally turning to temporally-varying \emph{behavioral} metrics (such as the OST consistency), in order to be able to separate these model classes beyond what is currently achievable by neural response predictions.

\subsection{ConvRNNs mediate a tradeoff between task performance and network size.}
\label{convrnn:ss:ost-statistics}
\begin{figure}
  \centering
  \includegraphics[width=1.0\columnwidth]{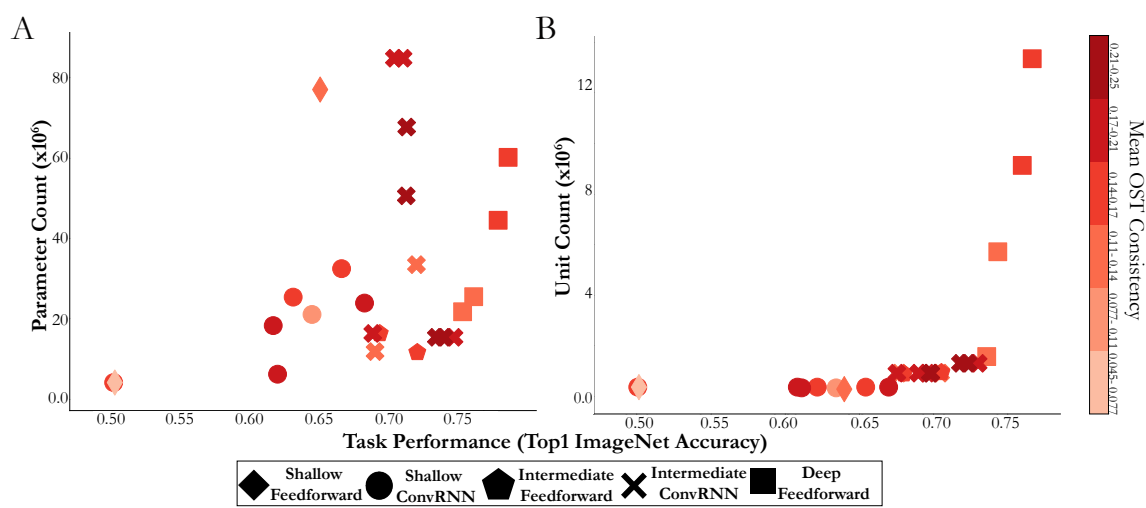}
  \caption[Intermediate ConvRNN circuits with highest OST consistency conserve on network size while maintaining task performance]{\textbf{Intermediate ConvRNN circuits with highest OST consistency conserve on network size while maintaining task performance.}
  Across all models considered, the intermediate ConvRNNs (denoted by ``x'') that attain high categorization performance ($x$-axis) while maintaining a low unit count (panel B) rather than parameter count (panel A) for their given performance level, achieve the highest mean OST consistency (Spearman correlation with IT population OST, averaged across $N=10$ train/test splits).
  The colorbar indicates this mean OST consistency (monotonically increasing from purple to red), binned into 6 equal ranges.
  Models with a larger network size at a fixed performance level are \emph{less consistent} with primate object recognition behavior (e.g. deep feedforward models, denoted by boxes), with recurrence maintaining a fundamental tradeoff between network size and task performance.
}
 \label{convrnn:oststats}
\end{figure}

\begin{figure}[tb]
  \centering
  \includegraphics[width=\columnwidth]{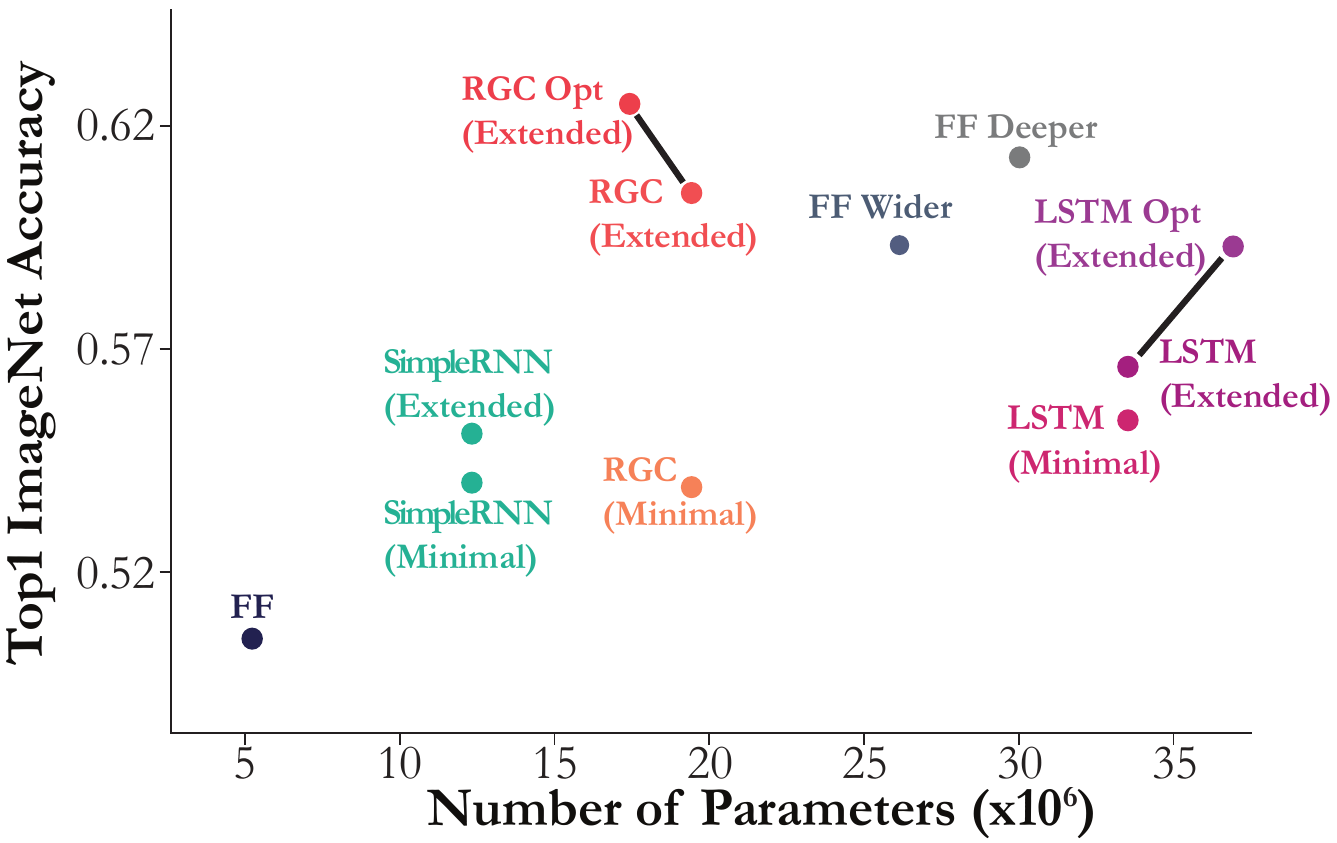}
  \caption[Performance of shallow ConvRNN and feedforward models as a function of number of parameters]{\textbf{Performance of shallow ConvRNN and feedforward models as a function of number of parameters.} 
  Colored points incorporate the respective ConvRNN circuit into the shallow, 6-layer feedforward BaseNet architecture (``FF'').
  ``Minimal'' is defined as the minimum number of timesteps (7) after the initial feedforward pass whereby all recurrence connections were engaged at least once, which the model was trained with.
  ``Extended'' is a greater number of timesteps (16) that the model was trained for given optimization and memory constraints.
  Hyperparameter-optimized versions of the LSTM (``LSTM Opt'') and Reciprocal Gated Circuit ConvRNNs (``RGC Opt'') are connected to their non-optimized versions by black lines.
  Note that the feedforward (FF) models are already optimized for the relevant hyperparameters of batch size, learning rate, and L2 regularization.
  The SimpleRNN is also hyperparameter optimized since unlike the more sophisticated ConvRNN circuit architectures of the LSTM and RGC, it is unable to train otherwise -- with layer normalization being an important factor (see Section~\ref{convrnn:sss:methods-simplernn} for more details).
}
 \label{convrnn:supp:minextend}
\end{figure}

\begin{figure}[tb]
  \centering
  \includegraphics[width=1.0\columnwidth]{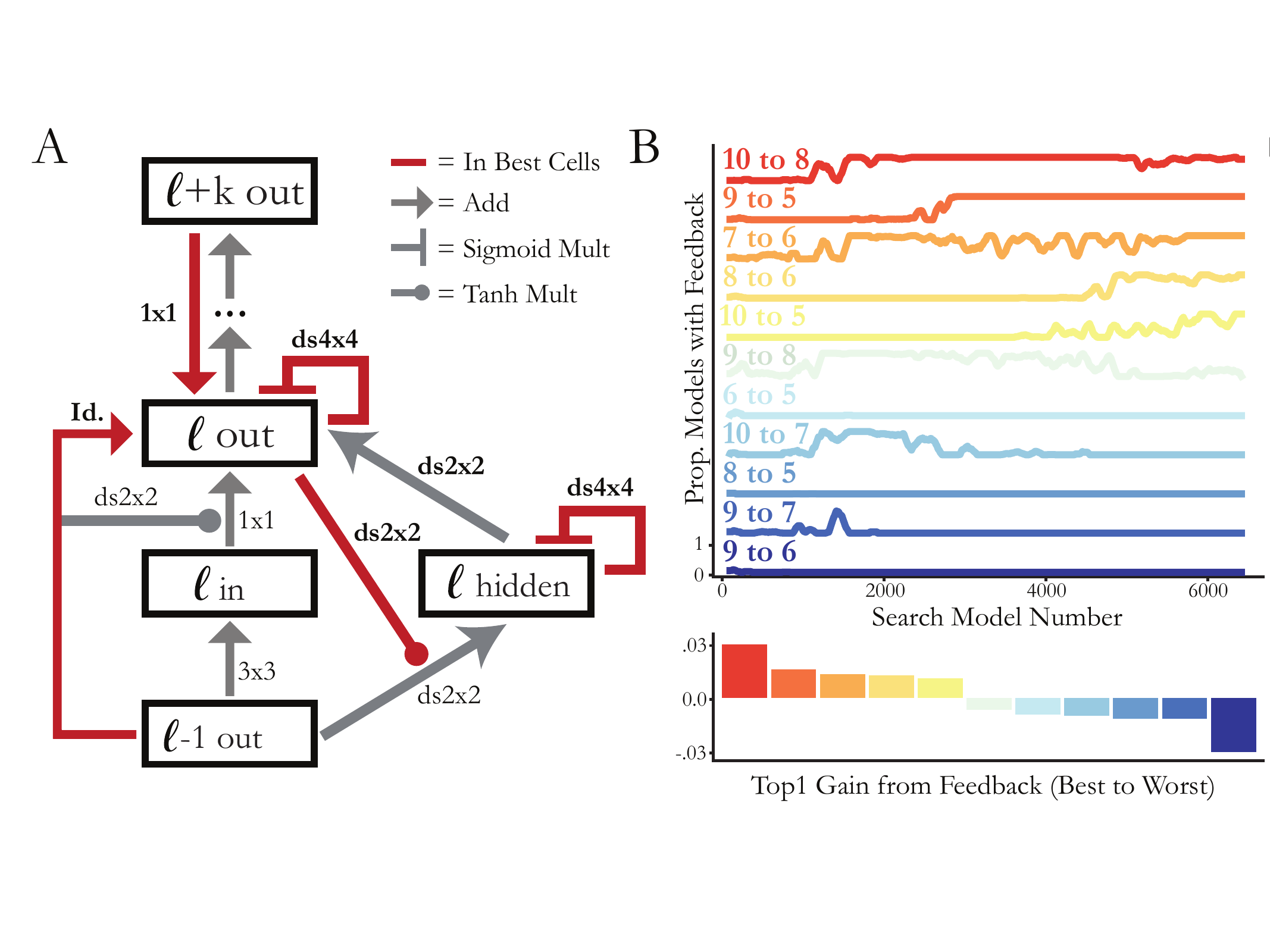}
 \caption[Optimal local recurrent circuit motif and global feedback connectivity]{\textbf{Optimal local recurrent circuit motif and global feedback connectivity.} \textbf{(a) RNN circuit structure from the top-performing search model.} Red lines indicate that this hyperparameter choice (connection and filter size) was chosen in each of the top unique models from the search. $K\times K$ denotes a convolution and ds$K\times K$ denotes a depth-separable convolution with filter size $K\times K$.
 \textbf{(b) Long-range feedback connections from the search.} (Top) Each trace shows the proportion of models in a 100-sample window that have a particular feedback connection. (Bottom) Each bar indicates the difference between the median performance of models with a given feedback and the median performance of models without that feedback. Colors correspond to the same feedback connectivity as above.}
 \label{convrnn:supp:fig1}
\end{figure}

Why might a suitably shallower feedforward network with temporal dynamics be desirable for the ventral visual stream?
We reasoned that recurrence mediates a tradeoff between network size and task performance; a tradeoff that the ventral stream also maintains.
To examine this possibility, in Figure~\ref{convrnn:oststats}, we compared each network's task performance versus its size, measured either by parameter count or unit count.
Across models, we found unit count (related to the number of neurons) to be more consistent with task performance than parameter count (related to the number of synapses).
In fact, there are many models with a large parameter count but not very good task performance, indicating that adding synapses is not necessarily as useful for performance as adding neurons.
For shallow recurrent networks, task performance seemed to be more strongly associated with OST consistency than network size.
This tradeoff became more salient for deeper feedforward models and the intermediate ConvRNNs, as the very deep ResNets (ResNet-34 and deeper) attained an overall \emph{lower} OST consistency compared to the intermediate ConvRNNs, using both much more units and parameters compared to small relative gains in task performance.
Similarly, intermediate ConvRNNs with high task performance and minimal \emph{unit} count, such as the UGRNN, GRU, and RGCs attained both the highest OST consistency overall (Figures~\ref{convrnn:ostfig} and~\ref{convrnn:oststats}) along with providing the best match to neural dynamics among ConvRNN circuits across visual areas (Figure~\ref{convrnn:neuralfig}b).
This observation indicates that suitably-chosen recurrence can provide a means for maintaining this fundamental tradeoff.

Given our finding that specific forms of task-optimized recurrence are more consistent with IT's OST than iterated feedforward transformations (with unshared weights), we asked whether it was possible to approximate the effect of recurrence with a feedforward model.
This approximation would allow us to better describe the additional ``action'' that recurrence is providing in its improved OST consistency.
In fact, one difference between this metric and the explained variance metric evaluated on neural responses in the prior section is that the latter uses a linear transform from model features to neural responses, whereas the former operates directly on the original model features.
Therefore, a related question is whether the (now standard) use of a linear transform for mapping from model units to neural responses can potentially \emph{mask} the behavioral improvement that suitable recurrent processing has over deep feedforward models in their original feature space.

To address these questions, we trained a separate linear mapping (PLS regression) from each model layer to the corresponding IT response at the given timepoint, on a set of images distinct from those on which the OST consistency metric is evaluated on (see Section~\ref{convrnn:sss:methods-ost-neural} for more details).
The outputs of this linear mapping were then used in place of the original model features for both the uniform and graded mappings, constituting ``PLS Uniform'' and ``PLS Graded'', respectively.
Overall, as depicted in Figure~\ref{convrnn:supp:fig4}, we found that models with \emph{less} temporal variation in their source features (namely, those under a uniform mapping with less ``IT-preferred'' layers than the total number of timebins) had significantly \emph{improved} OST consistency with their linearly transformed features under PLS regression (Wilcoxon test, $p < 0.001$; mean OST difference $0.0458$ and s.e.m. $0.00399$).
On the other hand, the linearly transformed intermediate feedforward models were \emph{not} significantly different from task-optimized ConvRNNs that achieved high OST consistency\footnote{Paired $t$-test with Bonferroni correction: RGC Median vs. PLS Uniform BaseNet, mean OST difference $-0.0052$ and s.e.m. $0.0061$, $t(9)\approx -0.86,p\approx 0.41$; RGC Median with Threshold Decoder vs. PLS Uniform ResNet-18, mean OST difference $0.00697$ and s.e.m. $0.0085$, $t(9)\approx 0.82,p\approx 0.43$; RGC Median with Max Confidence Decoder vs. PLS Uniform ResNet-34, mean OST difference $0.0001$ and s.e.m. $0.0079$, $t(9)\approx 0.02,p\approx 0.99$.}, suggesting that the action of suitable task-optimized recurrence approximates that of a shallower feedforward model with linearly induced ground-truth neural dynamics.

\section{Discussion}
The overall goal of this study is to determine what role recurrent circuits may have in the execution of core object recognition behavior in the ventral stream.
By broadening the method of goal-driven modeling from solving tasks with feedforward CNNs to ConvRNNs that include layer-local recurrence and feedback connections, we first demonstrate that appropriate choices of these recurrent circuits which incorporate specific mechanisms of ``direct passthrough'' and ``gating'' lead to matching the task performance of much deeper feedforward CNNs with fewer units and parameters, even when minimally unrolled.
This observation suggests that the recurrent circuit motif plays an important role even during the initial timepoints of visual processing.
Moreover, unlike very deep feedforward CNNs, the mapping from the early, intermediate, and higher layers of these ConvRNNs to corresponding cortical areas is neuroanatomically consistent and reproduces prior quantitative properties of the ventral stream.
In fact, ConvRNNs with high task performance but small network size (as measured by number of neurons rather than synapses) are most consistent with the temporal evolution of primate IT object identity solutions.
We further find that these task-optimized ConvRNNs can reliably produce quantitatively accurate dynamic neural response trajectories at temporal resolutions of tens of milliseconds throughout the ventral visual hierarchy.

Taken together, our results suggest that recurrence in the ventral stream extends feedforward computations by mediating a tradeoff between task performance and neuron count during core object recognition, suggesting that the computer vision community's solution of stacking more feedforward layers to solve challenging visual recognition problems approximates what is compactly implemented in the primate visual system by leveraging additional nonlinear temporal transformations to the initial feedforward IT response.
This work therefore provides a quantitative prescription for the next generation of dynamic ventral stream models, addressing the call to action in a recent previous study~\citep{kar2019evidence} for a change in architecture from feedforward models.

Many hypotheses about the role of recurrence in vision have been put forward, particularly in regards to overcoming certain challenging image properties~\citep{Spoerer2017, Michaelis2018, Rajaei2019, Linsley2018, Gilbert2013, Lindsay2015, Mcintosh2018, Li2018, kar2019evidence, Rao1999, Lotter2017, Issa2018}. 
We believe this is the first work to address the role of recurrence at scale by connecting novel \emph{task-optimized} recurrent models to temporal metrics defined on high-throughput neural and behavioral data, to provide evidence for recurrent connections extending feedforward computations.
Moreover, these metrics are well-defined for feedforward models (unlike prior work~\citep{cornets2019}) and therefore meaningfully demonstrate a separation between the two model classes.

Though our results help to clarify the role of recurrence during core object recognition behavior, many major questions remain.
Our work addresses why the visual system may leverage recurrence to subserve visually challenging behaviors, replacing a physically implausible architecture (deep feedforward CNNs) with one that is ubiquitously consistent with anatomical observations (ConvRNNs).
However, our work does not address gaps in understanding either the loss function or the learning rule of the neural network.
Specifically, we intentionally implant layer-local recurrence and long-range feedback connections into feedforward networks that have been useful for supervised learning via backpropagation on ImageNet.
A natural next step would be to connect these ConvRNNs with unsupervised objectives, as has been done for feedforward models of the ventral stream in concurrent work~\citep{Zhuang2021}.
The question of biologically plausible learning targets is similarly linked to biologically plausible mechanisms for learning such objective functions.
Recurrence could play a separate role in implementing the propagation of error-driven learning, obviating the need for some of the issues with backpropagation (such as weight transport), as has been recently demonstrated at scale~\citep{Akrout2019, tworoutes2020}.
Therefore, building ConvRNNs with unsupervised objective functions optimized with biologically-plausible learning rules would be essential towards a more complete goal-driven theory of visual cortex.

Additionally, high-throughput experimental data will also be critical to further separate hypotheses about recurrence.
While we see evidence of recurrence as mediating a tradeoff between network size and task performance for core object recognition, it could be that recurrence plays a more task-specific role under more temporally dynamic behaviors.
Not only would it be an interesting direction to optimize ConvRNNs on more temporally dynamic visual tasks than ImageNet, but to compare to neural and behavioral data collected from such stimuli, potentially over longer timescales than 260\emph{ms}.
While the RGC motif of gating and direct passthrough gave the highest task performance among ConvRNN circuits, the circuits that maintain a tradeoff between number of units and task performance (RGC Median, GRU, and UGRNN) had the best match to the current set of neural and behavioral metrics, even if some of them do not employ passthroughs.
However, it could be the case that with the same metrics we develop here but used in concert with such stimuli over potentially longer timescales, that we can better differentiate these three ConvRNN circuits.
Therefore, such models and experimental data would synergistically provide great insight into how rich visual behaviors proceed, while also inspiring better computer vision algorithms.

\section{Methods}
\label{convrnn:section:methods}
\subsection{Model framework}
\label{convrnn:ss:methods-framework}

\subsubsection{Software package}
\label{convrnn:sss:methods-framework-package}
To explore the architectural space of ConvRNNs and compare these models with the primate visual system, we used the Tensorflow library~\citep{Abadi2016} to augment standard CNNs with both local and long-range recurrence (Figure~\ref{convrnn:fig0}). Conduction from one area to another in visual cortex takes approximately 10\emph{ms}~\citep{mizuseki2009}, with signal from photoreceptors reaching IT cortex at the top of the ventral stream by 70-100\emph{ms}.
Neural dynamics indicating potential recurrent connections take place over the course of 100-260\emph{ms}~\citep{Issa2018}.
A single feedforward volley of responses thus cannot be treated as if it were instantaneous relative to the timescale of recurrence and feedback.
Hence, rather than treating each entire feedforward pass from input to output as one integral time step, as is normally done with RNNs~\citep{Spoerer2017}, each time step in our models corresponds to a single feedforward layer processing its input and passing it to the next layer.
This choice required an unrolling scheme different from that used in the standard Tensorflow RNN library, the code for which, including trained model weights, can be found on our Github repository: \url{https://github.com/neuroailab/convrnns}.

\subsubsection{Defining ConvRNNs}
\label{convrnn:sss:methods-framework-convrnn}
Within each ConvRNN layer, feedback inputs from higher layers are resized to match the spatial dimensions of the feedforward input to that layer. Both types of input are processed by standard 2D convolutions.
If there is any local recurrence at that layer, the output is next passed to the recurrent circuit as input.
Feedforward and feedback inputs are combined within the recurrent circuit by spatially resizing the feedback inputs (via bilinear interpolation) and concatenating these with the feedforward input across the channel dimension.
We let $\oplus$ denote this concatenation along the channel dimension with appropriate resizing to align spatial dimensions.
Finally, the output of the circuit is passed through any additional nonlinearities, such as max-pooling.
The generic update rule for the discrete-time trajectory of such a network is thus:
\begin{equation}\label{convrnn:eq:generic-update}
\begin{split}
& h^{\ell}_{t} = C_{\ell}\left(F_{\ell}\left(\bigoplus_{j \ne \ell}r_t^{j}\right), h_{t-1}^{\ell}\right)\\
& r^{\ell}_t = A_{\ell}(h_t^{\ell}), 
\end{split}
\end{equation}
where $r^{\ell}_t$ is the output of layer $\ell$ at time $t$, $h_{t-1}^{\ell}$ is the hidden state of the locally recurrent circuit $C_{\ell}$ at time $t-1$, and
$A_{\ell}$ is the activation function and any pooling post-memory operations.
The learned parameters of such a network consist of $F_{\ell}$, comprising any feedforward and feedback connections coming into layer $\ell=1,\ldots, L$, and any of the learned parameters associated with the local recurrent circuit $C_{\ell}$.

In this work, all forms of recurrence add parameters to the feedforward base model.
Because this could improve task performance for reasons unrelated to recurrent computation, we trained two types of control model to compare to ConvRNNs: 
\begin{enumerate}
\item Feedforward models with more convolution filters (``wider'') or more layers (``deeper'') to approximately match the number of parameters in a recurrent model.

\item Replicas of each ConvRNN model unrolled for a \emph{minimal} number of time steps, defined as the number that allows all model parameters to be used at least once.
A minimally unrolled model has exactly the same number of parameters as its fully unrolled counterpart, so any increase in performance from unrolling longer can be attributed to recurrent computation.
Fully and minimally unrolled ConvRNNs were trained with identical learning hyperparameters.
\end{enumerate}

\subsubsection{Training Procedure}
\label{convrnn:sss:methods-training}
All models (both feedforward and ConvRNN) used the standard ResNet preprocessing provided by TensorFlow here: \url{https://github.com/tensorflow/tpu/blob/master/models/official/resnet/resnet_preprocessing.py}.
Furthermore, they were trained on 224 pixel ImageNet with stochastic gradient descent with momentum (SGDM)~\citep{Sutskever2013}, using a momentum value of 0.9.

We allowed the base learning rate, batch size, and L2 regularization strength to vary for each model, depending on what was optimal in terms of top-1 validation accuracy for that model.
All models (except for AlexNet) used the ResNet training schedule~\cite{He2016}, whereby the base learning rate is decayed by $90\%$ at 30, 60, and 80 epochs, training for 90 epochs total.
The AlexNet had its base learning rate of 0.01 subsequently decayed to 0.005, 0.001, and 0.0005, at 30, 60, and 80 epochs, respectively.
We list these values for each model in the table below:\newline

\begin{tabular}{ | l | c | c | r | }
    \hline
 \textbf{Model Class} & \textbf{Base Learning Rate} & \textbf{Batch Size} & \textbf{L2 Regularization} \\ \hline
 AlexNet & 0.01 & 1024 & $5\times 10^{-4}$ \\ \hline
 6-layer BaseNet & 0.01 & 256 & $1\times 10^{-4}$ \\ \hline
 Shallow ConvRNNs & 0.01 & 256 & $1\times 10^{-4}$ \\ \hline
 11-layer BaseNet & 0.0025 & 64 & $1\times 10^{-4}$ \\ \hline
 ResNets & 0.025 & 64 & $1\times 10^{-4}$ \\ \hline
 Intermediate ConvRNNs & 0.0025 & 64 & $1\times 10^{-4}$ \\ \hline
\end{tabular}\newline

The only exceptions to the above are the models that are the result of the large-scale hyperparameter searches, detailed in Section~\ref{convrnn:ss:methods-hyperopt}.
Here the learning rate and batch size are allowed to vary, and the L2 regularization is not uniform across the model, but is also allowed to vary for both the feedforward backbone and each layer's ConvRNN circuit.
We list the learning rates and batch sizes for these models below:\newline

\begin{tabular}{ | l | c | r | }
    \hline
 \textbf{Model} & \textbf{Base Learning Rate} & \textbf{Batch Size} \\ \hline
 Shallow LSTM (``LSTM Opt'' in Figure~\ref{convrnn:supp:minextend}) & $7.587\times 10^{-3}$ & 64 \\ \hline
 RGC Random & $5.184\times 10^{-3}$ & 64 \\ \hline
 RGC Median & $6.736\times 10^{-3}$ & 64 \\ \hline
\end{tabular}\newline

Since these model hyperparameters are non-standard, we manually drop the learning rate (using the same decay factor of $90\%$) once the top-1 validation accuracy saturates at that given learning rate.

\subsection{Feedforward model architectures}
\label{convrnn:ss:methods-cnn}
\subsubsection{BaseNet architectures}
\label{convrnn:sss:methods-basenet}
Here we provide the architectures of the feedforward CNNs we developed in this paper, referred to as ``BaseNet'' when they are later implanted with ConvRNN circuits.
For all of these architectures, we use ELU nonlinearities~\citep{Clevert2016}. 

The 6-layer BaseNet (into which we implanted ConvRNN circuits to form the orange ``Shallow ConvRNN'' model class in Figure~\ref{convrnn:ostfig}c), referenced as ``FF'' in Figure~\ref{convrnn:supp:minextend}, referred to as ``BaseNet'' among the blue ``Shallow Feedforward'' models in Figure~\ref{convrnn:ostfig}c, and ``Feedforward'' in Figure~\ref{convrnn:supp:fig2}c, had the following architecture:\newline

\begin{tabular}{ | l | c | c | c | r | }
    \hline
 \textbf{Layer} & \textbf{Kernel Size} & \textbf{Channels} & \textbf{Stride} & \textbf{Max Pooling} \\ \hline
 1 & $7\times 7$ & 64 & 2 & $2\times 2$ \\ \hline
 2 & $3 \times 3$ & 128 & 1 & $2\times 2$ \\ \hline
 3 & $3 \times 3$ & 256 & 1 & $2\times 2$ \\ \hline
 4 & $3 \times 3$ & 256 & 1 & $2\times 2$ \\ \hline
 5 & $3 \times 3$ & 512 & 1 & $2\times 2$ \\ \hline
 6 & $2 \times 2$ & 1000 & 1 & No \\ \hline
\end{tabular}\newline

The 11-layer BaseNet used for the ``Intermediate ConvRNNs'' (red models in Figure~\ref{convrnn:ostfig}c) and modeled after ResNet-18~\citep{He2016} (but using MaxPooling rather than stride-2 convolutions to perform downsampling) is given below: \newline

\begin{tabular}{ | l | c | c | c | c | r | }
    \hline
 \textbf{Block} & \textbf{Kernel Size} & \textbf{Depth} & \textbf{Stride} & \textbf{Max Pooling} & \textbf{Repeat} \\ \hline
 1 & $7\times 7$ & 64 & 2 & $2\times 2$ & $\times 1$\\ \hline
 2 & $3 \times 3$ & 64 & 1 & None & $\times 2$\\ \hline
 3 & $3 \times 3$ & 64 & 1 & None & $\times 2$\\ \hline
 4 & $3 \times 3$ & 128 & 1 & $2\times 2$ & $\times 2$\\ \hline
 5 & $3 \times 3$ & 128 & 1 & None & $\times 2$ \\ \hline
 6 & $3 \times 3$ & 256 & 1 & $2\times 2$ & $\times 2$\\ \hline
 7 & $3 \times 3$ & 256 & 1 & $2\times 2$ & $\times 2$\\ \hline
 8 & $3 \times 3$ & 512 & 1 & None & $\times 2$\\ \hline
 9 & $3 \times 3$ & 512 & 1 & None & $\times 2$\\ \hline
 10 & $3 \times 3$ & 512 & 1 & $2\times 2$ & $\times 2$ \\ \hline
 11 & None (Avg. Pool FC) & 1000 & None & None & $\times 1$\\ \hline
\end{tabular}\newline

This is the BaseNet among the purple ``Intermediate Feedforward'' models in Figure~\ref{convrnn:ostfig}c, and used in Figures~\ref{convrnn:neuralfig},~\ref{convrnn:supp:fig4},~\ref{convrnn:supp:neuralfig}, and~\ref{convrnn:supp:fig3}.

The variant of the above 6-layer feedforward CNN, referenced in Figure~\ref{convrnn:supp:minextend} as ``FF Wider'' is given below:\newline

\begin{tabular}{ | l | c | c | c | r | }
    \hline
 \textbf{Layer} & \textbf{Kernel Size} & \textbf{Channels} & \textbf{Stride} & \textbf{Max Pooling} \\ \hline
 1 & $7\times 7$ & 128 & 2 & $2\times 2$ \\ \hline
 2 & $3 \times 3$ & 512 & 1 & $2\times 2$ \\ \hline
 3 & $3 \times 3$ & 512 & 1 & $2\times 2$ \\ \hline
 4 & $3 \times 3$ & 512 & 1 & $2\times 2$ \\ \hline
 5 & $3 \times 3$ & 1024 & 1 & $2\times 2$ \\ \hline
 6 & $2 \times 2$ & 1000 & 1 & None \\ \hline
\end{tabular}\newline

The ``FF Deeper'' model referenced in Figure~\ref{convrnn:supp:minextend} is given below:\newline

\begin{tabular}{ | l | c | c | c | r | }
    \hline
 \textbf{Layer} & \textbf{Kernel Size} & \textbf{Depth} & \textbf{Stride} & \textbf{Max Pooling} \\ \hline
 1 & $7\times 7$ & 64 & 2 & $2\times 2$ \\ \hline
 2 & $3 \times 3$ & 64 & 1 & None \\ \hline
 3 & $3 \times 3$ & 64 & 1 & None\\ \hline
 4 & $3 \times 3$ & 128 & 1 & $2\times 2$\\ \hline
 5 & $3 \times 3$ & 128 & 1 & None\\ \hline
 6 & $3 \times 3$ & 256 & 1 & $2\times 2$\\ \hline
 7 & $3 \times 3$ & 256 & 1 & $2\times 2$\\ \hline
 8 & $3 \times 3$ & 512 & 1 & None\\ \hline
 9 & $3 \times 3$ & 512 & 1 & None\\ \hline
 10 & $3 \times 3$ & 512 & 1 & $2\times 2$\\ \hline
 11 & None (Avg. Pool FC) & 1000 & None & None\\ \hline
\end{tabular}\newline

\subsubsection{AlexNet}
\label{convrnn:sss:methods-alexnet}
We use the standard AlexNet architecture, which uses local response normalization~\citep{Krizhevsky2012}.
We note that we are able to attain a higher than reported top-1 validation accuracy of 63.9\% (compared to 57\% accuracy) by using the ResNet preprocessing mentioned in Section~\ref{convrnn:sss:methods-training}.

\subsubsection{ResNet Architectures}
\label{convrnn:sss:methods-resnet}
For the ResNet architectures, we used the original v1 versions~\citep{He2016} for ResNet-18 and ResNet-34.
For deeper ResNets (ResNet-50, ResNet-101, and ResNet-152), we used the v2 variants of ResNets, as this gave them a slightly higher increase in top-1 ImageNet validation accuracy.
Specifically, the v2 variants of ResNets use the pre-activation of the weight layers rather than the post-activation used in the original versions.
Furthermore, the v2 variants of ResNets apply batch normalization~\citep{Ioffe2015} and ReLU to the input \emph{prior} to the convolution, whereas the original variants apply these operations after the convolution.
We use the TensorFlow Slim implementations for these two variants provided here: \url{https://github.com/tensorflow/models/tree/master/research/slim}.

\subsection{ConvRNN Circuit Equations}
\label{convrnn:ss:methods-rnn}
Here we provide the explicit update equations for each of the ConvRNN circuits referenced in the barplot in Figure~\ref{convrnn:fig2}c ($C_{\ell}$ in \eqref{convrnn:eq:generic-update}), where $\sigma$ denotes the sigmoid function.

Throughout these sections, we let $\circ$ denote Hadamard (elementwise) product, let $*$ denote convolution, let $h^{\ell}_t$ denote the output of the circuit, let $s^{\ell}_t$ denote the propagated memory of the circuit (also known as the hidden state), and let $x^{\ell}_t = \bigoplus_{j \ne \ell}r_t^{j}$ denote the input to the circuit at layer $\ell$ (this is the concatenation of feedforward and feedback inputs to layer $\ell$, defined in Section~\ref{convrnn:sss:methods-framework-convrnn}).

In the following table, we provide the number of timesteps the ConvRNNs were unrolled for during training (``Fully Unrolled''), what the corresponding minimally unrolled timesteps would be to engage recurrent connections once for each model class, and the number of timesteps for evaluation when comparing to neural and behavioral data:\newline

\begin{tabular}{ | l | c | c | c | r | }
    \hline
 \textbf{Model Class} & \textbf{Minimally Unrolled} & \textbf{Fully Unrolled} & \textbf{Evaluation} \\ \hline
 Shallow ConvRNNs & 7 & 16 & 26 \\ \hline
 Intermediate ConvRNNs & 12 & 17 & 26 \\ \hline
 RGC Random & 12 & 26 & 26 \\ \hline
\end{tabular}\newline

We also list the timestep at which the image presentation was replaced by a mean gray stimulus during model training and model evaluation:\newline

\begin{tabular}{ | l | c | r | }
    \hline
 \textbf{Model Class} & \textbf{Training} & \textbf{Evaluation} \\ \hline
 Shallow ConvRNNs & 12 & 10 \\ \hline
 Intermediate ConvRNNs & 12 & 10 \\ \hline
 RGC Random & 10 & 10 \\ \hline
\end{tabular}\newline

The above training parameters were chosen based on what yielded high performance for that model class and also what was able to feasibly fit into TPU memory for training (more unroll timesteps requires more memory, but can also lead to instability during training, as is common with training RNNs~\citep{Bengio1994}).

For the ``Shallow ConvRNNs'', ConvRNN circuits were implanted into convolutional layers 3, 4, and 5 of the 6-layer BaseNet.
For the ``Intermediate ConvRNNs'', ConvRNN circuits were implanted into convolutional layers 4, 5, 6, 7, 8, 9, and 10 of the 11-layer BaseNet.

\subsubsection{Time Decay}
\label{convrnn:sss:methods-timedecay}
This is the simplest form of recurrence that we consider and has a discrete-time trajectory given by
\begin{equation}\label{convrnn:eq:time-decay}
\begin{split}
& s^{\ell}_{t} = F_{\ell}\left(x^{\ell}_t\right) + \tau_{\ell} s_{t-1}^{\ell}\\
& h^{\ell}_t = s^{\ell}_t,
\end{split}
\end{equation}
 where $\tau_{\ell}$ is the learned time constant at a given layer $\ell$.
This model is intended to be a control for simplicity, where the time constants could model synaptic facilitation and depression in a cortical layer.

For the TensorFlow implementation of this circuit, see the \texttt{GenFuncCell()} class in the \texttt{utils.cells.py} file on our Github repository.

\subsubsection{SimpleRNN}
\label{convrnn:sss:methods-simplernn}
The update equations in this case are given by:
\begin{equation}\label{convrnn:eq:srnn}
\begin{split}
& a_{t}^{\ell} = W_{s}^{\ell}*s_{t-1}^{\ell} + b_{s}^{\ell} \\
& i_{t}^{\ell} = W_{i}^{\ell}*x_t^{\ell} + b_{i}^{\ell} \\
& s_{t}^{\ell} = \text{elu}(\text{LN}(i_{t}^{\ell}+a_{t}^{\ell})) \\
& h_{t}^{\ell} = s_{t}^{\ell},
\end{split}
\end{equation}
where $\text{LN}$ denotes the layer normalization operation~\citep{Ba2016} with offset parameter $\beta$ initialized to 0 and scale parameter $\gamma$ initialized to 1.
For the shallow SimpleRNN (among the orange ``Shallow ConvRNN'' models in Figure~\ref{convrnn:ostfig}c), we use layer normalization but omit its usage in the intermediate ConvRNN as it was not able to train with that operation.

For the TensorFlow implementation of this circuit, see the \texttt{ConvNormBasicCell()} class in the \texttt{utils.cells.py} file on our Github repository.

\subsubsection{GRU}
\label{convrnn:sss:methods-gru}
We adapt the standard GRU circuit~\citep{Cho2014} to the convolutional setting:
\begin{equation}\label{convrnn:eq:gru}
\begin{split}
& {r}_{t}^{\ell} = \sigma(W_{r}^{\ell}*x_t^{\ell} + U_{r}^{\ell}*s_{t-1}^{\ell} + b_{r}^{\ell} + 1) \\
& {u}_{t}^{\ell} = \sigma(W_{u}^{\ell}*x_t^{\ell} + U_{u}^{\ell}*s_{t-1}^{\ell} + b_{u}^{\ell}) \\
& c_t^{\ell} = \tanh(W_{c}^{\ell}*x_t^{\ell} + U_{c}^{\ell}*(r_{t}^{\ell}\circ s_{t-1}^{\ell}) + b_{c}^{\ell}) \\
& s_t^{\ell} = u_{t}^{\ell}\circ s_{t-1}^{\ell} + (1-u_{t}^{\ell})\circ c_t^{\ell} \\
& h_t^{\ell} = s_{t}^{\ell}.
\end{split}
\end{equation}

For the TensorFlow implementation of this circuit, see the \texttt{ConvGRUCell()} class in the \texttt{utils.cells.py} file on our Github repository.

\subsubsection{LSTM}
\label{convrnn:sss:methods-lstm}
We adapt the standard LSTM circuit~\citep{Hochreiter1997} to the convolutional setting, with some slight modifications such as added layer normalization for stability in training.

We first make the gates convolutional as follows:
\begin{equation}\label{convrnn:eq:lstm-gates}
\begin{split}
& i_t^{\ell} = LN(W_{i}^{\ell}*x_t^{\ell} + U_{i}^{\ell}*h_{t-1}^{\ell} + b_{i}^{\ell}) \\
& j_t^{\ell} = LN(W_{j}^{\ell}*x_t^{\ell} + U_{j}^{\ell}*h_{t-1}^{\ell} + b_{j}^{\ell}) \\
& f_t^{\ell} = LN(W_{f}^{\ell}*x_t^{\ell} + U_{f}^{\ell}*h_{t-1}^{\ell} + b_{f}^{\ell}) \\
& o_t^{\ell} = LN(W_{o}^{\ell}*x_t^{\ell} + U_{o}^{\ell}*h_{t-1}^{\ell} + b_{o}^{\ell}), \\
\end{split}
\end{equation}
where $\text{LN}$ denotes the layer normalization operation~\citep{Ba2016} with offset parameter $\beta$ initialized to 0 and scale parameter $\gamma$ initialized to 1.

Next, the LSTM update equations are as follows:
\begin{equation}\label{convrnn:eq:lstm-update}
\begin{split}
& s^{\ell}_t = LN(s^{\ell}_{t-1} \circ \sigma(f^{\ell}_t + f^{\ell}_b) + \sigma(i^{\ell}_t) \circ \tanh(j^{\ell}_t)) \\
& h^{\ell}_t = \tanh(s^{\ell}_t) \circ \sigma(o^{\ell}_t),
\end{split}
\end{equation}
where $f^{\ell}_b$ is the forget gate bias, typically set to 1, as recommended by others~\citep{Jozefowicz2015}.
When peephole connections~\citep{Gers2002} are allowed, these update equations are augmented to become:
\begin{equation}\label{convrnn:eq:lstm-peepholes}
\begin{split}
& s^{\ell}_t = LN(s^{\ell}_{t-1} \circ \sigma(f^{\ell}_t + f^{\ell}_b + V_f^{\ell} \circ s^{\ell}_{t-1}) + \sigma(i^{\ell}_t + V_i^{\ell} \circ s^{\ell}_{t-1}) \circ \tanh(j^{\ell}_t)) \\
& h^{\ell}_t = \tanh(s^{\ell}_t) \circ \sigma(o^{\ell}_t + V_o^{\ell} \circ s^{\ell}_{t-1}).
\end{split}
\end{equation}

In the shallow LSTM (among the orange ``Shallow ConvRNN'' models in Figure~\ref{convrnn:ostfig}c), we use peepholes and layer normalization, as that was found in the LSTM search for shallow models (described in Section~\ref{convrnn:sss:lstm-search}) to be useful for performance.
We found, however, that neither of these augmentations are needed in the deeper variant (among the red ``Intermediate ConvRNNs'' in Figure~\ref{convrnn:ostfig}c) in order to achieve high top-1 validation accuracy on ImageNet.

For the TensorFlow implementation of this circuit, see the \texttt{ConvLSTMCell()} class in the \texttt{utils.cells.py} file on our Github repository.

\subsubsection{UGRNN}
\label{convrnn:sss:methods-ugrnn}
We adapt the UGRNN~\citep{collins2017} to the convolutional setting.
The update equations are as follows:
\begin{equation}\label{convrnn:eq:ugrnn}
\begin{split}
& c_t^{\ell} = \tanh(W_c^{\ell}*x_t^{\ell} + U_c^{\ell}*s_{t-1}^{\ell} + b_c^{\ell}) \\
& g_t^{\ell} = \sigma(W_g^{\ell}*x_t^{\ell} + U_g^{\ell}*s_{t-1}^{\ell} + b_g^{\ell} + 1) \\
& s_t^{\ell} = g_t^{\ell} \circ s_{t-1}^{\ell} + (1-g_t^{\ell}) \circ c_t^{\ell} \\
& h_t^{\ell} = s_t^{\ell}.
\end{split}
\end{equation}

For the TensorFlow implementation of this circuit, see the \texttt{ConvUGRNNCell()} class in the \texttt{utils.cells.py} file on our Github repository.

\subsubsection{IntersectionRNN}
\label{convrnn:sss:methods-intersectionrnn}
We adapt the IntersectionRNN~\citep{collins2017} to the convolutional setting.
The update equations are as follows:
\begin{equation}\label{convrnn:eq:intersectionrnn}
\begin{split}
& m_t^{\ell} = \tanh(W_m^{\ell}*x_t^{\ell} + U_m^{\ell}*s_{t-1}^{\ell} + b_m^{\ell}) \\
& n_t^{\ell} = \text{relu}(W_n^{\ell}*x_t^{\ell} + U_n^{\ell}*s_{t-1}^{\ell} + b_n^{\ell}) \\
& p_t^{\ell} = \sigma(W_p^{\ell}*x_t^{\ell} + U_p^{\ell}*s_{t-1}^{\ell} + b_p^{\ell} + 1) \\
& y_t^{\ell} = \sigma(W_y^{\ell}*x_t^{\ell} + U_y^{\ell}*s_{t-1}^{\ell} + b_y^{\ell} + 1) \\
& s_t^{\ell} = p_t^{\ell} \circ s_{t-1}^{\ell} + (1-p_t^{\ell}) \circ m_t^{\ell} \\
& h_t^{\ell} = y_t^{\ell} \circ x_t^{\ell} + (1-y_t^{\ell}) \circ n_t^{\ell}.
\end{split}
\end{equation}

For the TensorFlow implementation of this circuit, see the \texttt{ConvIntersectionRNNCell()} class in the \texttt{utils.cells.py} file on our Github repository.

\subsubsection{Reciprocal Gated Circuit (RGC)}
\label{convrnn:sss:methods-rgc}
Here we provide the explicit update equations for the Reciprocal Gated Circuit~\citep{nayebi2018task}, diagrammed in Figure~\ref{convrnn:fig2}a (bottom right).
The update equation for the output of the circuit, $h_t^{\ell}$, is given by a gating of both the input $x^{\ell}_t$ and prior output $h^{\ell}_{t-1}$:
\begin{equation} \label{convrnn:eq:rgc1}
\begin{split}
& a_{t}^{\ell} = (1 - \sigma(W^{\ell}_{sh} * s^{\ell}_{t-1})) \circ x^{\ell}_t + (1 - \sigma(W^{\ell}_{hh}*h^{\ell}_{t-1})) \circ h^{\ell}_{t-1}\\
& h_t^{\ell} = \text{elu}\left(a^{\ell}_t\right).
\end{split}
\end{equation}

The update equation for the memory $s^{\ell}_t$ is given by a gating of the input $x^{\ell}_t$ and the prior state $s^{\ell}_{t-1}$:
\begin{equation} \label{convrnn:eq:rgc2}
\begin{split}
& \tilde{s}^{\ell}_{t} = (1 - \sigma(W^{\ell}_{hs} * h^{\ell}_{t-1})) \circ x^{\ell}_t + (1 - \sigma(W^{\ell}_{ss}*s^{\ell}_{t-1})) \circ s^{\ell}_{t-1}\\
& s^{\ell}_t = \text{elu}(\tilde{s}^{\ell}_t).
\end{split}
\end{equation}

For the TensorFlow implementation of this circuit, see the \texttt{ReciprocalGateCell()} class in the \texttt{utils.cells.py} file on our Github repository.

\subsection{ConvRNN Searches}
\label{convrnn:ss:methods-hyperopt}
We employed a form of Bayesian optimization, a Tree-structured Parzen Estimator (TPE), to search the space of continuous and categorical hyperparameters~\citep{Bergstra2011}. 
This algorithm constructs a generative model of $P[score\mid configuration]$ by updating a prior from a maintained history $H$ of hyperparameter configuration-loss pairs. 
The fitness function that is optimized over models is the expected improvement, where a given configuration $c$ is meant to optimize $EI(c) = \int_{x < t}P[x\mid c, H]$. 
This choice of Bayesian optimization algorithm models $P[c\mid x]$ via a Gaussian mixture, and restricts us to tree-structured configuration spaces.

Models were trained synchronously 100 models at a time using the HyperOpt package~\citep{Bergstra2015}, which implements the above Bayesian optimization.
Each model was trained on its own Tensor Processing Unit (TPUv2), and during the search, ConvRNN models were trained by stochastic gradient descent on 128 pixel ImageNet for efficiency. 
The top performing ConvRNN models were then fully trained out on 224 pixel ImageNet.

\subsubsection{LSTM search}
\label{convrnn:sss:lstm-search}
The search for better LSTM architectures involved searching over training hyperparameters and common structural variants of the LSTM to better adapt this local structure to deep convolutional networks, using hundreds of second generation Google Tensor Processing Units (TPUv2s).
We searched over learning hyperparameters (e.g. gradient clip values, learning rate) as well as structural hyperparameters (e.g. gate convolution filter sizes, channel depth, whether or not to use peephole connections, etc.).

Specifically, we implanted LSTMs into convolutional layers 3, 4, and 5, of the 6-layer BaseNet described in Section~\ref{convrnn:ss:methods-cnn}.
At each of these layers, the parameters of the LSTM circuit (defined in Section~\ref{convrnn:sss:methods-lstm}) were allowed to vary per layer, as follows:
\begin{itemize}
\item The discrete number of convolutional channels was chosen from $\{64, 128, 256\}$.
\item The discrete choice of convolutional filter sizes were chosen from $\{1,4\}$.
\item The binary choice of whether or not to use layer normalization.
\item The strength of the L2 regularization of all LSTM parameters in that layer $\in [10^{-7}, 10^{-3}]$, sampled log-uniformly.
\item The scale of the He-style initialization~\citep{He2015} of the convolutional filter weights $\in [0.25, 2]$, sampled uniformly.
\item The value of the constant initialization of the biases $\in [-2, 2]$, sampled uniformly.
\item The forget gate bias $f_b^{\ell} \in [0, 6]$, sampled uniformly (defined in \eqref{convrnn:eq:lstm-update}).
\item The binary choice of whether or not to use peephole connections (as defined in \eqref{convrnn:eq:lstm-peepholes}).
\end{itemize}

Outside of the LSTM circuit at each layer, we additionally searched over the following parameters as well:
\begin{itemize}
\item The number of discrete timesteps the model is unrolled $\in [12, 26]$, sampled uniformly in consecutive groups of size 2.
\item The timestep at each the image presentation is ``turned off'' and replaced with a mean gray stimulus $\in [8, 12]$, sampled uniformly in consecutive groups of 2.
\item The discrete choice of batch size used for the training the entire model $\in \{64, 128, 256\}$.
\item The learning rate for training the entire model $\in [10^{-3}, 10^{-1}]$, sampled log-uniformly.
\item The binary choice of whether or not to use Nesterov momentum~\citep{Nesterov}.
\item The gradient clipping value $\in [0.3, 3]$, sampled log-uniformly.
\item The scale of the He-style initialization~\citep{He2015} of the convolutional filter weights of the feedforward base model $\in [0.25, 2]$, sampled uniformly.
\item The strength of the L2 regularization of the feedforward base model parameters $\in [10^{-7}, 10^{-3}]$, sampled log-uniformly.
\end{itemize}
Each search point is a sampled value from the above described search space and trained for 1 epoch on ImageNet, in order to sample as many models as much as possible with the computational resources available.
More than 1600 models were sampled in total, and we trained out the top ones and the median performing one after 1 epoch were trained out fully on 224 pixel ImageNet.
The median model from this search attained the best top-1 validation accuracy on ImageNet, which is the resultant ``LSTM Opt'' model in Figure~\ref{convrnn:supp:minextend} and otherwise referred to as ``Shallow LSTM''.
The configuration of chosen hyperparameters for this model can be found in the \texttt{configs.lstm\_shallow.npz} file on our Github repository.

\subsubsection{Reciprocal Gated Circuit (RGC) search}
From the Reciprocal Gated Circuit equations in \eqref{convrnn:eq:rgc1} and \eqref{convrnn:eq:rgc2}, there are a variety of possibilities for how $h_{t-1}^{\ell}, x_t^{\ell}, s_t^{\ell}$, and $h_t^{\ell}$ can be connected to one another (schematized in Figure~\ref{convrnn:fig2}b).

Mathematically, the search in Figure~\ref{convrnn:fig2}b can be formalized in terms of the following update equations. 
First, we define our input sets and building block functions:
\begin{equation*}
\begin{split}
& {minin} = \{h_{t-1}^{\ell-1}, x_t^{\ell}, s_{t-1}^{\ell}, h_{t-1}^{\ell}\}\\
& {minin}_a = minin \cup \{s_t^{\ell}\}\\
& {minin}_b = minin \cup \{h_t^{\ell}\}\\
& S_a \subseteq {minin}_a\\
& S_b \subseteq {minin}_b\\
& \text{Affine}(x) \in \{+, 1\times 1\text{ conv}, K\times K\text{ conv}, K\times K\text{ depth-separable conv}\}\\
& K \in \{3,\ldots, 7\}\\
& 
\end{split}
\end{equation*}
With those in hand, we have the following update equations:
\begin{equation*}
\begin{split}
& \tau_a = v_1^{\tau} + v_2^{\tau}\sigma(\text{Affine}(S_a))\\
& \tau_b = v_1^{\tau} + v_2^{\tau}\sigma(\text{Affine}(S_b))\\
& gate_a = v_1^{g} + v_2^{g}\sigma(\text{Affine}(S_a))\\
& gate_b = v_1^{g} + v_2^{g}\sigma(\text{Affine}(S_b))\\
& a_t^{\ell} = \{gate_a\}\cdot in_t^{\ell} + \{\tau_a\}\cdot h_{t-1}^{\ell}\\
& h_t^{\ell} = f(a_t^{\ell})\\
& b_t^{\ell} = \{gate_b\}\cdot in_t^{\ell} + \{\tau_b\}\cdot s_{t-1}^{\ell}\\
& s_t^{\ell} = f(b_t^{\ell})\\
& f \in \{\text{elu}, \tanh, \sigma\}.
\end{split}
\end{equation*}
For clarity, the following matrix summarizes the connectivity possibilities (with ? denoting the possibility of a connection), schematized in Figure~\ref{convrnn:fig2}b:
\renewcommand{\kbldelim}{(}
\renewcommand{\kbrdelim}{)}
\[
\kbordermatrix{
    & h_{t-1}^{\ell - 1} & x_{t}^{\ell} & s_{t-1}^{\ell} & s_{t}^{\ell} & h_{t-1}^{\ell} & h_t^{\ell} \\
    h_{t-1}^{\ell - 1} & 0 & 1 & 0 & ? & 0 & ? \\
    x_{t}^{\ell} & 0 & 0 & 0 & ? & 0 & ? \\
    s_{t-1}^{\ell} & 0 & 0 & 0 & ? & 0 & ? \\
    s_{t}^{\ell} & 0 & 0 & 0 & 0 & 0 & ? \\
    h_{t-1}^{\ell} & 0 & 0 & 0 & ? & 0 & ? \\
    h_t^{\ell} & 0 & 0 & 0 & ? & 0 & 0 \\
  }
\]

Each search point is a sampled value from the above described search space and trained for five epochs on ImageNet, in order to sample as many models as much as possible with the computational resources available.
Around 6000 models were sampled in total over the course of the search.
The top and median models from this search were then fully trained out on 224 pixel ImageNet with a batch size of 64 (which was maximum that we could fit into TPU memory).
Moreover, as explicated in the table in Section~\ref{convrnn:sss:methods-training}, the ResNet models were also trained using this same batch size, with the standard ResNet learning rate of 0.1 for a batch size of 256 linearly rescaled to accomodate, to ensure fair comparison between these two model classes.
The median model from this search attained the best top-1 validation accuracy on ImageNet of all models selected to be trained out fully on ImageNet from the search, producing the resultant ``RGC Median'' model in Figure~\ref{convrnn:fig2}c (note that this designation also includes the long-range feedback connections).
The configuration of chosen hyperparameters for this model can be found in the \texttt{configs.median\_rgcell\_cfg.py} file on our Github repository.
The ``RGC Random'' model is from the random phase of this search (400th sampled model, since models sampled earlier than that failed to train out fully on ImageNet).

\subsection{Decoders}
\label{convrnn:ss:methods-framework-decoders}
In addition to choice of ConvRNN circuit, we consider particular choices of ``light-weight'' (in terms of parameter count) decoding strategy that determines the final object category of that image.
By construction, the model will output category logit probabilities at each timestep, given by the softmax function $\text{softmax}(z;\beta) = \dfrac{e^{\beta z_i}}{\sum_{j=1}^C e^{\beta z_j}}$, where $C=1000$ is the number of ImageNet categories.
This will then be passed to a decoding function which can take one of several forms:
\begin{enumerate}
    \item \textbf{Default:} Use the logits at the last timestep and discard the remaining, with $\beta=1$.
    
    \item \textbf{Threshold Decoder:} Select the logits from the first timepoint at which the maximum logit value at that timepoint crosses a fixed threshold (set to 0.9), with $\beta=1$.
    
    \item \textbf{Max Confidence Decoder:} For the most confident category, find the timepoint at which that confidence peaks, and return the logits at that timepoint, where $\beta$ is a trainable scalar parameter initialized to 1.
\end{enumerate}
``RGC Median'' therefore refers to the model trained using the default decoder, but when using the other two decoders with the ``RGC Median'' model, we append it to the name (as is done in Figures~\ref{convrnn:supp:fig4},~\ref{convrnn:supp:decoder-ostfig}, and~\ref{convrnn:supp:fig3}a).
The TensorFlow implementations of these decoders can be found in the \texttt{utils.decoders.py} file on our Github repository.

\subsection{Model prediction of neural responses}
\label{convrnn:ss:methods-neural}
\subsubsection{Neural data}
\label{convrnn:sss:methods-neural-data}
Neural responses came from three multi-unit arrays per primate (rhesus macques): one implanted in V4, one in posterior IT (pIT), and one in central and anterior IT (cIT/aIT)~\citep{Majaj2015}.
Each image was presented approximately 50 times, using rapid visual stimulus presentation (RSVP).
Each stimulus was presented for 100\emph{ms}, followed by a mean gray stimulus interleaved between images.
Each trial lasted 260\emph{ms}.
The image set consisted of 5120 images based on 64 object categories.
These objects belonged to 8 high-level categories (tables, planes, fruits, faces, chairs, cars, boats, animals), each of which consisted of 8 unique objects.
Each image consisted of a 2D projection of a 3D model added to a random background.
The pose, size, and $x$- and $y$-position of the object was varied across the image set, whereby 2 levels of variation were used (corresponding to medium and high variation~\citep{Majaj2015}).
Multi-unit responses to these images were binned in 10ms windows, averaged across trials of the same image, and normalized to the average response to a blank image.
This produced a set of 5120 images $\times$ 256 units $\times$ 25 timebins responses, which were the targets for our model features to predict.
There were 88 units from V4, 88 units from pIT, and 80 units from cIT/aIT.

\subsubsection{Fitting procedure}
\label{convrnn:sss:methods-neural-fitting}
\textit{Generating train/test split.} The 5120 images were split 75\%-25\% within each object category into a training set and a held-out testing set.
All images were presented to the models for 10 time steps (corresponding to 100\emph{ms}), followed by a mean gray stimulus for the remaining 15 time steps, to match the image presentation to the primates.
The images are matched to the procedure when used to validate the models on ImageNet, namely they are bilinearly resized to $224\times 224$ and normalized by the ImageNet mean ($[0.485, 0.456, 0.406]$) and standard deviation ($[0.229, 0.224, 0.225]$), applied per channel.

\textit{Model layer determination.} We stipulated that units from each multi-unit array must be fit by features from a single model layer.
To determine which one, we fit the features from the relevant feedforward BaseNet (either the 6-layer BaseNet or 11-layer BaseNet) to unit's time-averaged response, and counted how many units had minimal loss for a given model layer, schematized in Step 2 of Figure~\ref{convrnn:fig0}.
This yielded a mapping from the V4 array to model layer 3 of the 6-layer BaseNet and model layers 5 \& 6 of the 11-layer BaseNet, pIT mapping to model layer 4 of the 6-layer BaseNet and model layers 7 \& 8 of the 11-layer BaseNet, and cIT/aIT mapping to layer 5 of the 6-layer BaseNet and model layers 9 \& 10 of the 11-layer BaseNet. 

\textit{Mapping transform from models to neural responses.} Model features from each image (i.e. the activations of units in a given model layer) were linearly fit to the neural responses by stochastic gradient descent with a standard L2 loss using a spatially factored mapping~\citep{Klindt2017}, where each of the 256 units was fit independently.
This spatially factored mapping is defined as follows:
Given a model feature $f^{\ell} \in \mathbb{R}^{x, y, c}$ from layer $\ell$, where $x$ and $y$ are the number of units in the spatial extent and $c$ is the number of channels, we fit a spatial mask $w_{\text{space}} \in \mathbb{R}^{x, y}$ and a channel mask $w_{\text{channels}} \in \mathbb{R}^{c}$ for each neuron $n$ to predict the ground-truth neuron's response $r_{i,n,t}$ at image $i$ and timebin $t$.
The predicted response can be written as:
\begin{equation}\label{convrnn:eq:klindt}
\hat{r}_{i,n,t; w} = \sum_{i=1}^{x}\sum_{j=1}^{y}\sum_{k=1}^{c} w_{\text{space}}[i, j] w_{\text{channels}}[k] f^{\ell}[i, j, k].
\end{equation}
This mapping is implemented in the \texttt{factored\_fc()} function of the \texttt{utils.cell\_utils.py} file on our Github repository.

\textit{Loss function.} After these layers were determined, model features were then fit to the entire set of 25 timebins for each unit using a shared linear model: that is, a single set of regression coefficients was used for all timebins, as schematized in Step 3 of Figure~\ref{convrnn:fig0}.
The loss for this fitting was the average L2 loss across training images and 25 timebins for each unit, given by
\begin{equation}\label{convrnn:neural-fit-l2-time}
\mathcal{L}(\hat{r}_{i,n,t; w}, r_{i,n,t}) = \frac{1}{|\mathcal{B}|}\sum_{t=6}^{25}\sum_{i\in \mathcal{B}}\sum_{n=1}^{256}\left(\hat{r}_{i,n,t; w} - r_{i,n,t}\right)^2.
\end{equation}
Note that $t$ indexes model timesteps, which correspond to 10\emph{ms} timebins, so $t=6$ refers to the 60-70\emph{ms} timebin, $t=7$ refers to the 70-80\emph{ms} timebin, and so forth.

We trained the temporally-fixed parameters $w = [w_{\text{space}}; w_{\text{channels}}]$ of the mapping using the Adam optimizer~\citep{kingma2014adam} with a learning rate of $1\times 10^{-4}$ and a training batch size $|\mathcal{B}| = 64$ images.
Additionally, we used a dropout~\citep{Srivastava2014} level of 0.5 on the model features, prior to the mapping, as further regularization.

\subsubsection{Metrics}
\label{convrnn:sss:methods-neural-metrics}
To estimate a noise ceiling for each neuron's response at each timebin, we computed the Spearman-Brown corrected split-half reliability $\rho_n$ of neuron $n$, averaged across 900 bootstrap iterations of split-half trials.

Let ``Neural Predictivity'' (used in Figure~\ref{convrnn:supp:fig2}) refer to
\begin{equation}
\text{Corr}(\hat{r}^{\text{test}}, r^{\text{test}}_n),
\end{equation}
namely the Pearson correlation across test set images of the model's response $\hat{r}^{\text{test}}$ to the of any neuron $n$'s response $r^{\text{test}}_n$ at a given timebin (or time-averaged).

The ``Neural Predictivity (Noise Corrected)'' (used in Figure~\ref{convrnn:neuralfig} and Figure~\ref{convrnn:supp:fig3}) for neuron $n$ is given by
\begin{equation}
\frac{\text{Corr}(\hat{r}^{\text{test}}, r^{\text{test}}_n)}{\sqrt{\rho_n}}.
\end{equation}

\subsection{Inter-animal consistency}
\label{convrnn:ss:methods-interanimal}
We provide the definition and justification of the inter-animal consistency metric mentioned in Figure~\ref{convrnn:supp:fig3}b.
Suppose we have neural responses from two primates $A$ and $B$.
Let $t_i^p$ be the vector of true responses (either at a given timebin or averaged across a set of timebins) of primate $p \in \{A,B\}$ on stimulus set $i \in \{\text{train}, \text{test}\}$.
Of course, we only receive noisy observations of $t_i^p$, so let $s_{j,i}^p$ be the $j$-th set of $n$ trials of $t_i^p$.
Finally, let $M(x)_i$ be the predictions of a mapping $M$ (e.g. PLS) when trained on input $x$ and tested on stimulus set $i$.
For example, $M\left(t^p_{\text{train}}\right)_{\text{test}}$ is the prediction of the mapping $M$ on the test stimulus trained on the true neural responses from primate $p$ on the train stimulus, and correspondingly, $M\left(s^p_{1, \text{train}}\right)_{\text{test}}$ is the prediction of the mapping $M$ on the test stimulus trained on the (trial-average) of noisy sample 1 on the train stimulus from primate $p$.

With these definitions in hand, the inter-animal mapping consistency from one primate $A$ to another primate $B$ corresponds to the following true quantity to be estimated:
\begin{equation}\label{convrnn:interancontrue}
\text{Corr}\left(M\left(t^A_{\text{train}}\right)_{\text{test}}, t^B_{\text{test}}\right),
\end{equation}
where $\text{Corr}$ is the Pearson correlation across test stimuli.
In what follows, we argue that this true quantity can be approximated with the following ratio of measurable quantities where we divide the noisy trial observations into two sets of equal samples:
\begin{equation}\label{convrnn:interancon}
\text{Corr}\left(M\left(t^A_{\text{train}}\right)_{\text{test}}, t^B_{\text{test}}\right) \sim \dfrac{\text{Corr}\left(M\left(s^A_{1,\text{train}}\right)_{\text{test}}, s^B_{2,\text{test}}\right)}{\sqrt{\text{Corr}\left(M\left(s^A_{1,\text{train}}\right)_{\text{test}}, M\left(s^A_{2,\text{train}}\right)_{\text{test}}\right) \times \text{Corr}\left(s^B_{1,\text{test}}, s^B_{2,\text{test}}\right)}}.
\end{equation}
In words, the inter-animal consistency corresponds to the predictivity of the mapping on the test set stimuli from primate $A$ to $B$ on two different (averaged) halves of noisy trials, corrected by the square root of the mapping reliability on primate $A$'s test stimuli responses on two different halves of noisy trials and the internal consistency of primate $B$.

We justify the approximation in \eqref{convrnn:interancon} by gradually eliminating the true quantities by their measurable estimates, starting from the original quantity in \eqref{convrnn:interancontrue}.
First, we make the approximation that
\begin{equation}\label{convrnn:step1}
\text{Corr}\left(M\left(t^A_{\text{train}}\right)_{\text{test}}, s^B_{2,\text{test}}\right) \sim \text{Corr}\left(M\left(t^A_{\text{train}}\right)_{\text{test}}, t^B_{\text{test}}\right) \times \text{Corr}\left(t^B_{\text{test}}, s^B_{2,\text{test}}\right).
\end{equation}
by transitivity of positive correlations (which is reasonable assumption when the number of stimuli is large).
Next, by normality assumptions in the structure of the noisy estimates and since the number of trials ($n$) between the two sets is the same, we have that
\begin{equation}\label{convrnn:step2}
\text{Corr}\left(s^B_{1,\text{test}}, s^B_{2,\text{test}}\right) \sim \text{Corr}\left(t^B_{\text{test}}, s^B_{2,\text{test}}\right)^2.
\end{equation}
Namely, the correlation between the average of two sets of noisy observations of $n$ trials each is approximately the square of the correlation between the true value and average of one set of $n$ noisy trials.
Therefore, from \eqref{convrnn:step1} and \eqref{convrnn:step2} it follows that
\begin{equation}\label{convrnn:lemma1}
\text{Corr}\left(M\left(t^A_{\text{train}}\right)_{\text{test}}, t^B_{\text{test}}\right) \sim \dfrac{\text{Corr}\left(M\left(t^A_{\text{train}}\right)_{\text{test}}, s^B_{2,\text{test}}\right)}{\sqrt{\text{Corr}\left(s^B_{1,\text{test}}, s^B_{2,\text{test}}\right)}}.
\end{equation}

We have gotten rid of $t^B_{\text{test}}$, but we still need to get rid of the $M\left(t^A_{\text{train}}\right)_{\text{test}}$ term.
We apply the same two steps by analogy though these approximations may not always be true (though are true for additive Gaussian noise):
\begin{equation*}
\begin{split}
& \text{Corr}\left(M\left(s^A_{1,\text{train}}\right)_{\text{test}}, s^B_{2,\text{test}}\right) \sim \text{Corr}\left(s^B_{2,\text{test}}, M\left(t^A_{\text{train}}\right)_{\text{test}}\right) \times \text{Corr}\left(M\left(t^A_{\text{train}}\right)_{\text{test}}, M\left(s^A_{1,\text{train}}\right)_{\text{test}}\right)\\
& \text{Corr}\left(M\left(s^A_{1,\text{train}}\right)_{\text{test}}, M\left(s^A_{2,\text{train}}\right)_{\text{test}}\right) \sim \text{Corr}\left(M\left(s^A_{1,\text{train}}\right)_{\text{test}}, M\left(t^A_{\text{train}}\right)_{\text{test}}\right)^2,
\end{split}
\end{equation*}
which taken together implies
\begin{equation}\label{convrnn:lemma2}
\text{Corr}\left(M\left(t^A_{\text{train}}\right)_{\text{test}}, s^B_{2,\text{test}}\right) \sim \dfrac{\text{Corr}\left(M\left(s^A_{1,\text{train}}\right)_{\text{test}}, s^B_{2,\text{test}}\right)}{\sqrt{\text{Corr}\left(M\left(s^A_{1,\text{train}}\right)_{\text{test}}, M\left(s^A_{2,\text{train}}\right)_{\text{test}}\right)}}.
\end{equation}
Equations \eqref{convrnn:lemma1} and \eqref{convrnn:lemma2} together imply the final estimated quantity given in \eqref{convrnn:interancon}.

\subsection{Object solution times (OSTs)}
\label{convrnn:ss:methods-ost}
\subsubsection{Generating model OSTs}
\label{convrnn:sss:methods-ost-model}
Here we describe how we defined object solution times from both feedforward models and ConvRNNs.
As depicted in Figure~\ref{convrnn:ostfig}a, this is a multi-stage process that involves first identifying the most ``IT-preferred'' layers of each model.

\textit{Determining ``IT-preferred'' model layers.}
These are identified by a standard~\citep{yamins_ventralneural, kar2019evidence} linear mapping using 25 component partial least squares regression (PLS), from model layer units to \emph{time-averaged} IT (namely, pIT/cIT/aIT) responses from the neural data described in Section~\ref{convrnn:sss:methods-neural-data}, and corroborates the results obtained by the same procedure described in Section~\ref{convrnn:sss:methods-neural-fitting}.
We use this neural data as it has both V4 and IT responses, and demonstrates a disjoint set of layers between the preferred V4 model layers and preferred IT layers.

\textit{Mapping model timepoints to IT timepoints.}
Once these ``IT-preferred'' model layers are identified, we then map these model timepoints to 10\emph{ms} timebins as in the IT data.
For ConvRNNs with intrinsic temporal dynamics, this mapping is one-to-one, we simply concatenate the model layers at each timepoint to construct an entire IT pseudopopulation, and each timepoint of the ConvRNN corresponds to a 10\emph{ms} timebin between 70-260\emph{ms}.
For feedforward models, we map each ``IT-preferred'' layer to a 10\emph{ms} timebin between 70-260\emph{ms}.
If the number of ``IT-preferred'' layers for a feedforward model matches the total number of timebins (19), then there is only one admissible mapping, corresponding to the ``uniform'' mapping, whereby the earliest (in the feedforward hierarchy) layer is matched to the earliest 10\emph{ms} timebin of 70\emph{ms}, and so forth.
On the other hand, if the number of ``IT-preferred'' layers is strictly less than the total number of timebins, then we additionally consider a ``graded'' mapping that picks a random sample of units from one layer to the next so that the number of feedforward layers exactly matches the total number of timebins.

\textit{Obtaining model $d^{'}$ values.}
Once a timepoint mapping is selected, we compute the model object solution time (OST) in the same manner as the OST is computed for IT~\citep{kar2019evidence}.
Specifically, we train an SVM ($C = 5\times 10^{4}$) separately for each model timepoint after it has been dimension reduced through PCA (with 1000 components) to solve the ten-way categorization task for each image.
The ten categories are apple, bear, bird, car, chair, dog, elephant, person, plane, and zebra.
1000 images constitute the training set of the SVM (100 images per category) and 320 images are randomly chosen to be in the test set.
We perform 20 trials each of 10 train/test splits to get errorbars, where each image is in the test set at least once.
The model $d^{'}$ for that image is computed in the same manner as previously done for the ground truth IT response $d^{'}$ (see Kar et al. 2019~\cite{kar2019evidence} for details), only being computed from the SVM when it has been in the test set and is bounded between -5 and 5.
Since this dataset consists of 1320 grayscale images presented centrally to behaving primates for 100\emph{ms}, there are therefore 1320 $d^{'}$ values (one for each image) for any given model, constituting its ``I1'' vector~\citep{rajalingham2018large}.

\textit{Correlating model OST with IT OST.}
The OST of the model therefore is the first model timepoint in which the $d^{'}$ reaches the recorded primate $d^{'}$ for that image, as was previously done to compute the ground truth IT OST~\citep{kar2019evidence}.
Using the Levenberg–Marquardt algorithm, we further linearly interpolate between 10\emph{ms} bins to determine the precise millisecond that the response surpassed the primate's behavioral output for that image (as was done analogously with the IT population's OST).
Finally, we compare the model OST to the IT OST via a Spearman correlation across the common set of images solved by \emph{both} the model and IT.

\subsubsection{Relating the linear mapping to neural responses with the OST behavioral metric}
\label{convrnn:sss:methods-ost-neural}
The IT population OST was computed from primarily anterior IT (aIT) responses~\citep{kar2019evidence}.
Therefore, to isolate the interaction a linear mapping of model features to neural responses (as we do in neural response prediction described in Section~\ref{convrnn:ss:methods-neural}) might have compared to directly computing the OST from the original model features, we turned to neural data collected from 486 aIT units on 1100 greyscale images.

For each model, we train a linear mapping on this dataset, with 550 images used for training the mapping and 550 images are held-out for the test set.
We observe similar conclusions as with the original neural data in Section~\ref{convrnn:ss:methods-neural} for both the temporally-fixed linear mapping in Figure~\ref{convrnn:supp:neuralfig} (in the ``aIT'' panel), and with a temporally-varying PLS mapping in Figure~\ref{convrnn:supp:fig3} (``aIT'' in panel (a) as well as the data used in panel (b)), all from layer 10 of the 11-layer BaseNet/ConvRNNs.

\begin{figure}[tb]
  \centering
  \includegraphics[width=1.0\columnwidth]{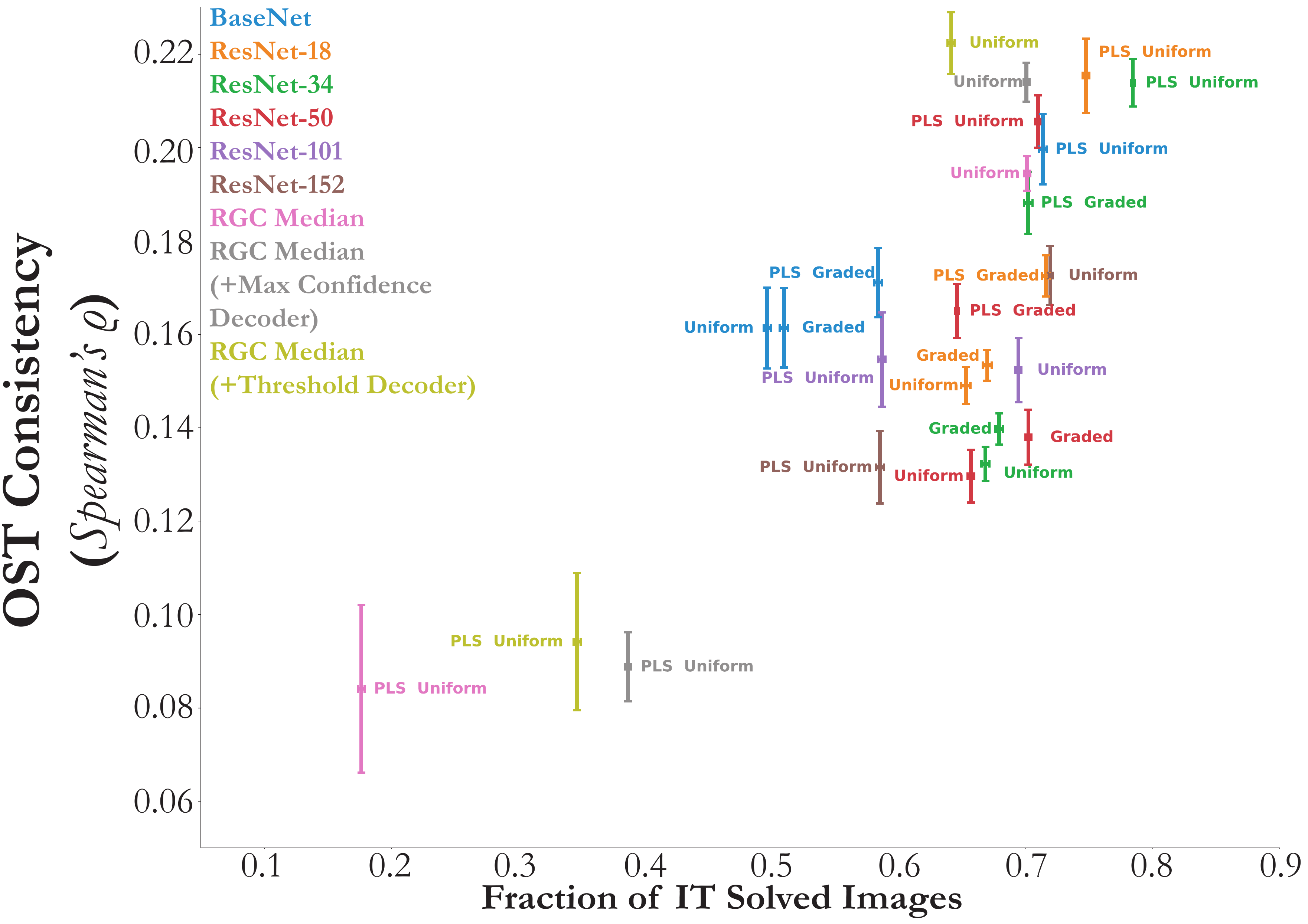}
 \caption[Behaviorally harmful effect of dimensionality reduction due to linear transform]{\textbf{Behaviorally harmful effect of dimensionality reduction due to linear transform.}
 Mean and s.e.m. are computed across train/test splits ($N=10$) when that image (of 1320 images) was a test-set image, with the Spearman correlation computed with the IT solution times across the imageset mutually solved by the given model and IT.
 As can be seen, a temporally-graded mapping directly from the model features of feedforward models always attains OST consistency at least that of the uniform one (``Graded'' vs. ``Uniform'' comparison). 
 We additionally train a 100 component PLS regression to IT responses at each defined model timepoint, where the responses are to a \emph{different} set of images than used to evaluate the OST metric.
 This procedure, detailed in Section~\ref{convrnn:sss:methods-ost-neural}, results in an image-computable model on which the OST metric is evaluated on and corresponds to ``PLS'' prepended to the name of each point on this plot, for any given model and associated temporal mapping.
 As can be seen, ``PLS Uniform'' for the BaseNet and ResNet-34 match the OST consistency of the RGC Median ConvRNNs from their original model features. However, ``PLS Uniform'' for the ConvRNNs and ResNet-101 and ResNet-152 have a significant decrease in OST consistency compared to when evaluated on their original model features, indicating the behaviorally harmful effect of dimensionality reduction due to PLS.
 }
 \label{convrnn:supp:fig4}
\end{figure}

\begin{figure}
  \centering
  \includegraphics[width=1.0\columnwidth]{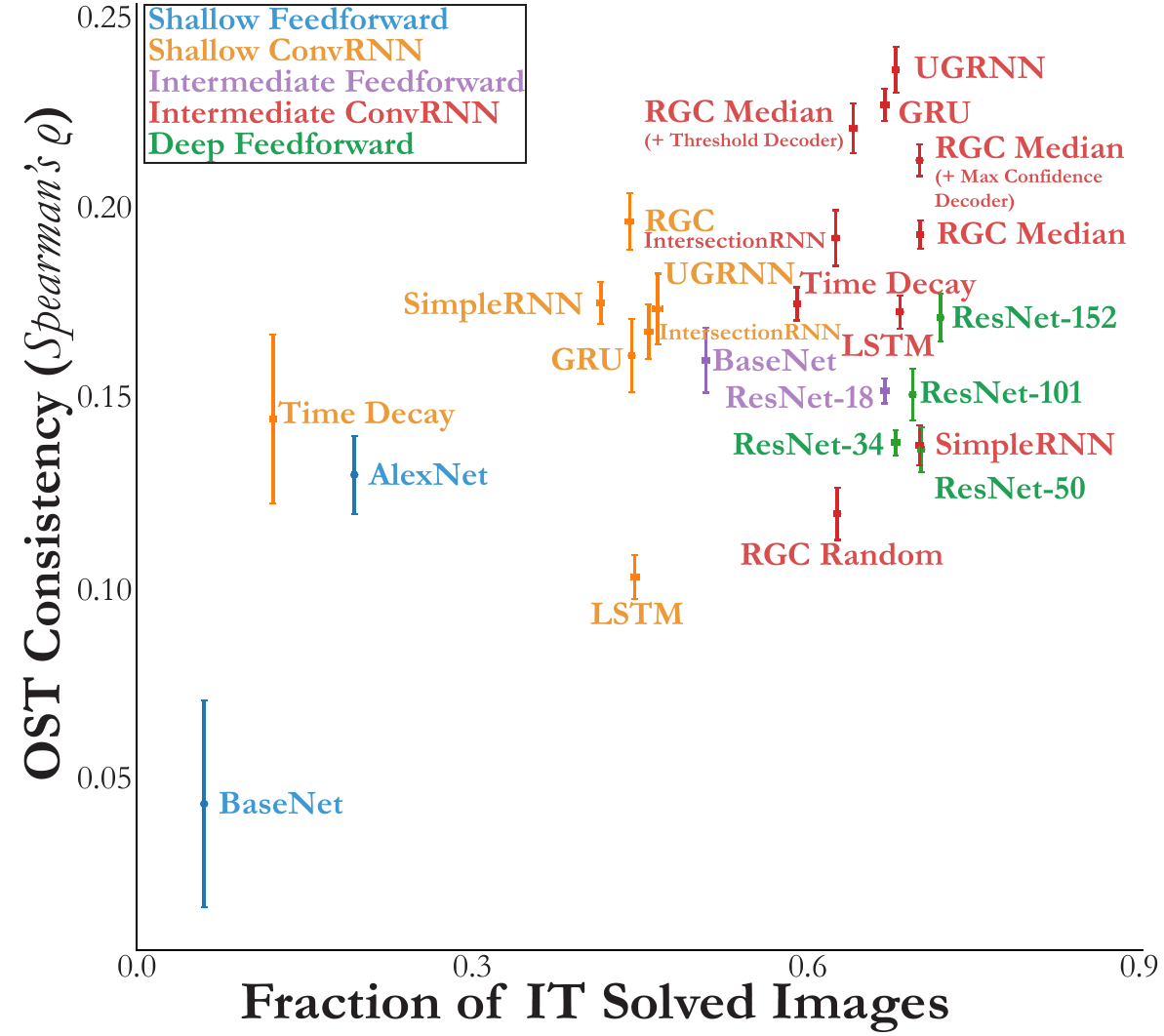}
  \caption[Decoding strategy can impact IT OST]{\textbf{Decoding strategy can impact IT OST.}
  Same as Figure~\ref{convrnn:ostfig}d, but we additionally embed decoders to the Reciprocal Gated Circuits (RGC), see definitions in Section~\ref{convrnn:ss:methods-framework-decoders}.
  Mean and s.e.m. are computed across train/test splits ($N=10$) when that image (of 1320 images) was a test-set image, with the Spearman correlation computed with the IT object solution times (analogously computed from the IT population responses) across the imageset solved by both the given model and IT, constituting the ``Fraction of IT Solved Images'' on the $x$-axis.
  We start with either a shallow base feedforward model consisting of 5 convolutional layers and 1 layer of readout (``BaseNet'' in blue) as well as an intermediate-depth variant with 10 feedforward layers and 1 layer of readout (``BaseNet'' in green), detailed in Section~\ref{convrnn:sss:methods-basenet}.
  From these base feedforward models, we embed recurrent circuits, resulting in either ``Shallow ConvRNNs'' or ``Intermediate ConvRNNs'', respectively.
}
 \label{convrnn:supp:decoder-ostfig}
\end{figure}

\begin{figure}[tb]
  \centering
  \includegraphics[width=1.0\columnwidth]{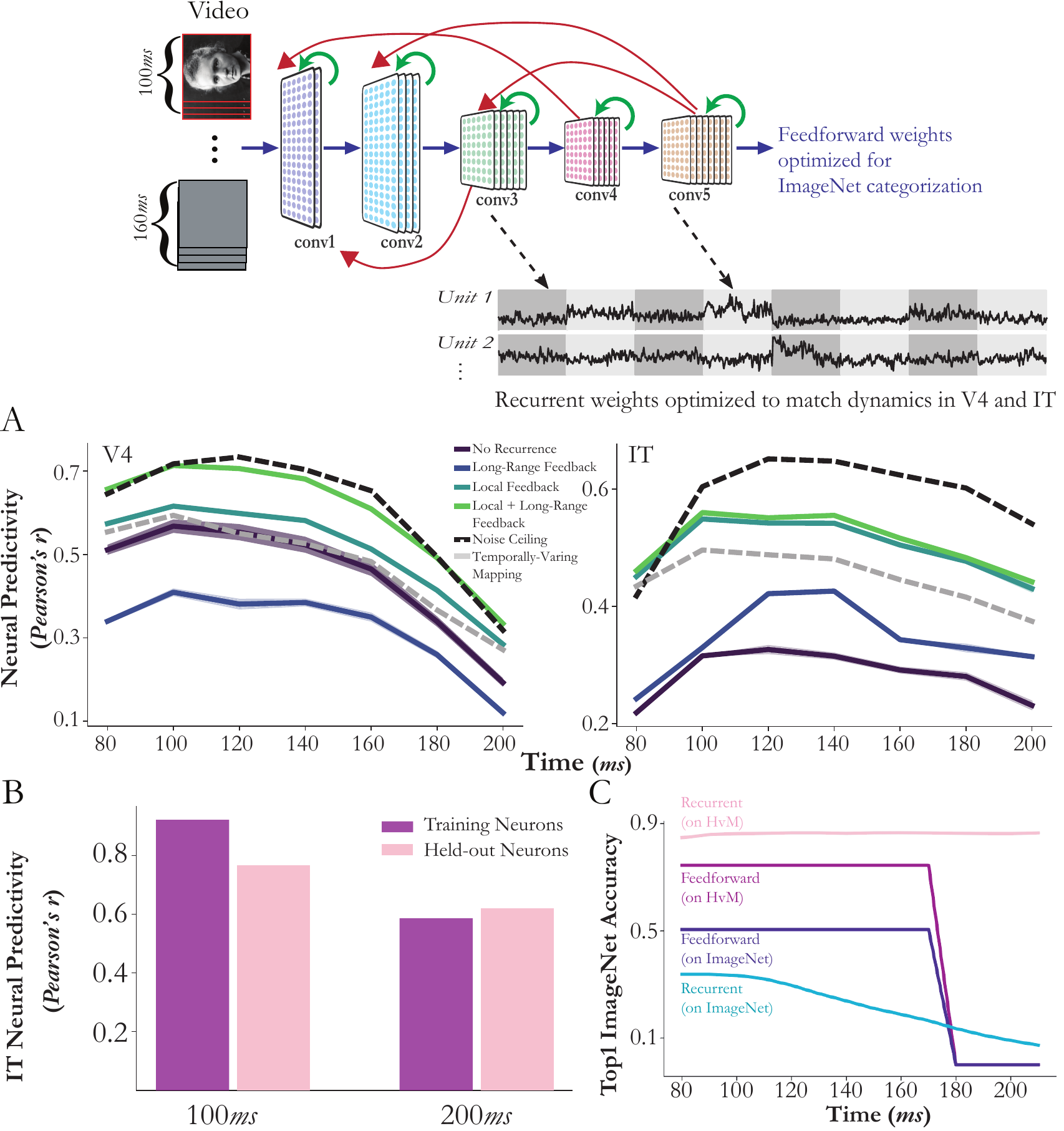}
 \caption[Subtle overfitting occurs when directly fitting dynamics to neural data]{\textbf{(a) Both local recurrence and global feedback are needed to best fit neural data.} Among a wide range of architectures with different local recurrent motifs and global feedback patterns, the best architecture was one with both gated local recurrence and a global feedback. 
 Local recurrent circuits were particularly useful for improving fits to IT neurons ($N=168$), whereas both local recurrence and global feedback were critical for improving fits to V4 neurons ($N=88$).
Except for ``temporally-varying mapping'', fixed model-unit-to-neuron linear mappings were fixed across all time bins, constraining trajectories to be produced by actual dynamics of the network. In contrast, ``temporally-varying mapping'' indicates an independent PLS regression for each time bin. 
 The fact that models with local recurrence and global feedback are better than ``temporally-varying mapping'' suggests that some nonlinear dynamics at earlier layers contributed meaningfully to network fits. S.e.m. across four splits of held-out test images. 
 \textbf{(b) Held-out neural predictivity.} At both 100\emph{ms} and 200\emph{ms}, this direct fitting procedure to the dynamics generalizes to neurons held-out (right bars) in the fitting procedure, a stronger test of generalization than held-out images depicted in the left bars.
 \textbf{(c) Underfitting to the task.} However, a subtle overfitting to the neural image distribution occurs, whereby the task-optimized network whose dynamics are trained on the V4 and IT neural dynamics no longer transfers to ImageNet.
}
 \label{convrnn:supp:fig2}
\end{figure}

\begin{figure}[tb]
  \centering
  \includegraphics[width=1.0\columnwidth]{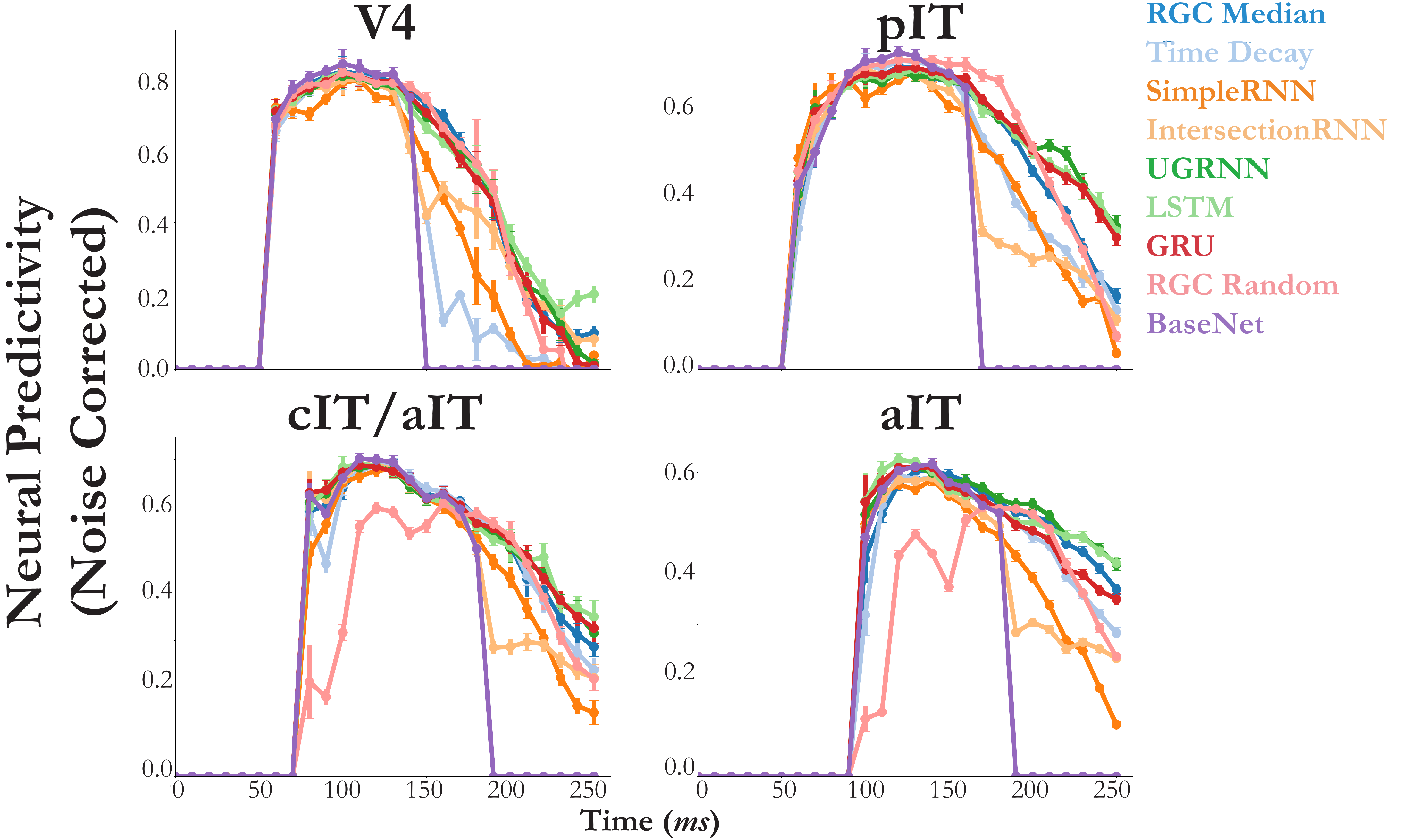}
  \caption[Intermediate ConvRNN circuits are differentiated by primate ventral stream neural dynamics]{\textbf{Intermediate ConvRNN circuits are differentiated by primate ventral stream neural dynamics.}
  Fitting model features of ConvRNNs with a temporally-fixed linear mapping to neural dynamics approaches the noise ceiling of these responses in most cases.
  The $y$-axis indicates the median across neurons of the explained variance between predictions and ground-truth responses on held-out images.
  Error bars indicates the s.e.m across neurons ($N=88$ for V4, $N=88$ for pIT, $N=80$ for cIT/aIT, and $N=486$ for aIT).
  Note that ``aIT'' refers to a separate neural dataset from primarily anterior IT neurons, detailed in Section~\ref{convrnn:sss:methods-ost-neural}.
  The onset time of the response is the first timepoint the area-preferred model layer (see Section~\ref{convrnn:sss:methods-neural-fitting} for details) of the base feedforward model (``BaseNet''), which all these circuits share, receives its input.
  As can be seen, the feedforward BaseNet model (purple) is incapable of generating a response beyond the feedforward pass, and certain types of ConvRNN circuits added to the feedforward model are less predictive than others.
}
 \label{convrnn:supp:neuralfig}
\end{figure}

\begin{figure}[tb]
  \centering
  \includegraphics[width=1.0\columnwidth]{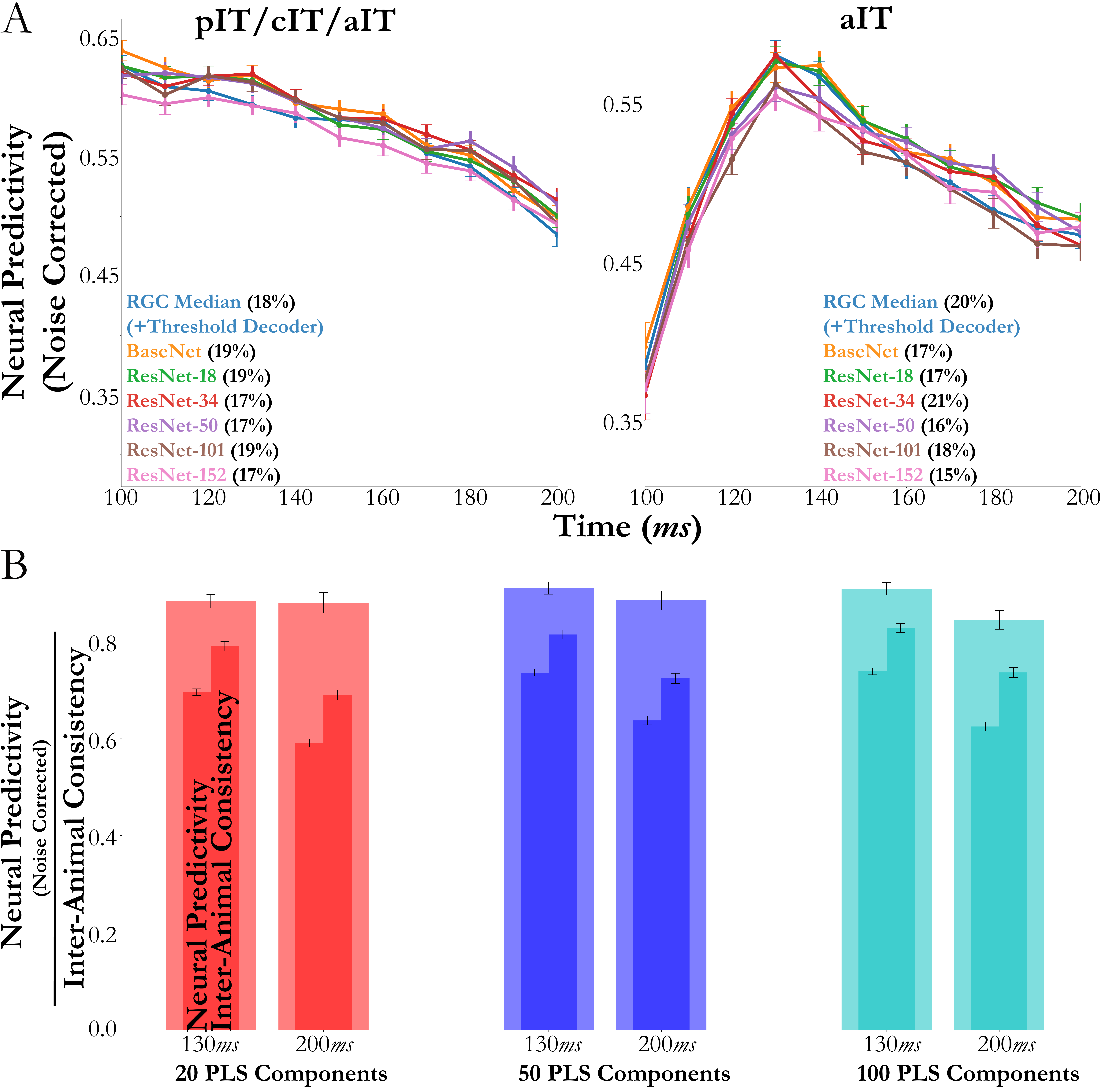}
 \caption[Neural dynamics are explained close to inter-animal consistency]{\textbf{(a) Increasing feedforward depth does \emph{not} account for drop in median explained variance from early to late timepoints.} 
 We observe a similar drop in median explained variance from 130-140\emph{ms} to 200-210\emph{ms}, between the intermediate-depth ConvRNN and deeper feedforward models, where we fix each model's training image size and batch size to be able compare across depths.
 To compare these two models, we subselect for high reliability neurons (above 0.3 split-half consistency) and use a temporally-varying mapping (PLS 25 components).
Note that the temporally-varying mapping implies providing the feedforward models with a constant input stream (unlike the primates and ConvRNNs, which are given a 100\emph{ms} presentation) in order for them to produce a (constant) response at every timepoints.
 We plot the median and s.e.m. predictivity in both panels per timebin ($N=108,113,117,123,118,118,116,115,108,99,86$ neurons for each timebin in the ``pIT/cIT/aIT'' panel, and  $N=247,313,378,441,437,411,397,391,392,384,380$ neurons for each timebin in the ``aIT'' panel).
 \textbf{(b) Drop in explained variance may be exhibited in inter-animal consistency.} Using the neural data described in Section~\ref{convrnn:sss:methods-ost-neural}, we see a similar inter-animal consistency (metric detailed in Section~\ref{convrnn:ss:methods-interanimal}) at 130-140\emph{ms} and 200-210\emph{ms}, as we do with the 11-layer BaseNet.
 Median and s.e.m. across aIT neurons ($N=441$ at 130-140\emph{ms} and $N=380$ at 200-210\emph{ms}) from the dataset described in Section~\ref{convrnn:sss:methods-ost-neural}.
 }
 \label{convrnn:supp:fig3}
\end{figure}

With these observations, we then proceeded to evaluate the effect of the linear mapping on OST correlations in Figure~\ref{convrnn:supp:fig4}.
Crucially, in this setting, we train a 100 component PLS mapping on the 526 images for which an IT $d^{'}$ is \emph{not} defined, in order to ensure that the images from Section~\ref{convrnn:sss:methods-ost-model} that the OST correlation is evaluated on are not the same images the PLS mapping was trained with.


\section*{Extended Data}
\begin{table}[ht]
    \centering
    \begin{tabular}{ | l | c | r | }
        \hline
      \textbf{Model} & \textbf{Visual Area} & \textbf{Wilcoxon test $p$-value} \\ \hline
      Time Decay & V4 & $< 0.001$ \\ \hline
      IntersectionRNN & V4 & $< 0.001$ \\ \hline
      LSTM & V4 & $< 0.001$ \\ \hline
      UGRNN & V4 & $< 0.001$ \\ \hline
      GRU & V4 & $< 0.001$ \\ \hline
      SimpleRNN & V4 & $< 0.001$ \\ \hline
      RGC Random & V4 & $< 0.001$ \\ \hline
      RGC Median & V4 & $< 0.01$ \\ \hline\hline
      Time Decay & pIT & $0.022$ \\ \hline
      IntersectionRNN & pIT & $< 0.001$ \\ \hline
      LSTM & pIT & $< 0.001$ \\ \hline
      UGRNN & pIT & $< 0.001$ \\ \hline
      GRU & pIT & $< 0.001$ \\ \hline
      SimpleRNN & pIT & $< 0.001$ \\ \hline
      RGC Random & pIT & $0.31$ \\ \hline
      RGC Median & pIT & $< 0.001$ \\ \hline\hline
      Time Decay & aIT & $< 0.001$ \\ \hline
      IntersectionRNN & aIT & $< 0.001$ \\ \hline
      LSTM & aIT & $0.47$ \\ \hline
      UGRNN & aIT & $0.09$ \\ \hline
      GRU & aIT & $0.16$ \\ \hline
      SimpleRNN & aIT & $< 0.001$ \\ \hline
      RGC Random & aIT & $< 0.001$ \\ \hline
      RGC Median & aIT & $< 0.01$ \\ \hline
      \end{tabular}
    \caption[Wilcoxon test (with Bonferroni correction) $p$-values for comparing each intermediate-depth ConvRNN's neural predictivity at the ``early'' timepoints (Figure~\ref{convrnn:neuralfig}) to the (11-layer) BaseNet]{Wilcoxon test (with Bonferroni correction) $p$-values for comparing each intermediate-depth ConvRNN's neural predictivity at the ``early'' timepoints (Figure~\ref{convrnn:neuralfig}) to the (11-layer) BaseNet.}
    \label{convrnn:tab:pval-basenet-early}
\end{table}

\section*{Supplementary Figures}

\chapter{Goal-Driven Models of Mouse Visual Cortex}
\label{ch:mouse}
\section{Chapter Abstract}
Studies of the mouse visual system have revealed a variety of visual brain areas in a roughly hierarchical arrangement, together with a multitude of behavioral capacities, ranging from stimulus-reward associations, to goal-directed navigation, and object-centric discriminations. However, an overall understanding of the mouse's visual cortex organization, and how this organization supports visual behaviors, remains unknown. Here, we take a computational approach to help address these questions. We first demonstrate that shallow hierarchical neural network architectures, with comparatively low-resolution retinal inputs, achieve a substantially better match to neuronal response patterns in mouse visual cortex than deeper, higher-resolution networks that have been successful in modeling primate visual cortex. We then combine this shallow, low-resolution network structure with a task-agnostic, self-supervised objective function based on the concept of contrastive embedding. These self-supervised networks yield highly quantitatively accurate matches to mouse visual responses, significantly surpassing that of prior supervised neural network models, and approaching the inter-animal consistency level of the data itself. We further find that these shallow, self-supervised networks transfer to a wide variety of visual tasks without the need for supervised, task-specific labels. Taken together, our results suggest a picture of the mouse visual system as a cortical network that uses its limited resources to support a task-general representation.

\section{Introduction}
\label{sec:introduction}
In systems neuroscience, the mouse has become an indispensable model organism, allowing unprecedented genetic and experimental control at the level of cell-type specificity in individual circuits~\citep{Huberman2011}.
Beyond fine-grained control, studies of mouse visual behavior have revealed a multitude of abilities, ranging from stimulus-reward associations, to goal-directed navigation, and object-centric discriminations.
These behaviors suggest that the mouse visual system is capable of supporting higher-order functions, and prior physiological studies provide evidence that higher visual cortical areas might subserve such behaviors~\citep{Glickfeld2017}.
A natural question, therefore, is what these populations of neurons code for during visually-guided behaviors.
Formal computational models are needed to test these hypotheses: if optimizing a model for a certain task leads to accurate predictions of neural responses, then that task may provide a unified, normative account for why those population responses occur in the brain.

Deep convolutional neural networks (CNNs) are a class of models that have had immense success as predictive models of the human and non-human primate ventral visual stream~\cite[e.g.,][]{yamins_ventralneural, khaligh2014deep, Gueclue2015, Cichy2016, cadena2019deep, bashivan2019neural}.
In contrast with the strong correspondence between task-optimized CNNs and the primate visual system, these CNNs are poor predictors of neural responses in mouse visual cortex~\citep{Cadena2019}.

Three fundamental problems, each grounded in the goal-driven modeling approach~\citep{Yamins2016}, confront these primate ventral stream models as potential models of the mouse visual system.
Firstly, these models are too deep to be plausible models of the mouse visual system, since mouse visual cortex is known to be more parallel and much shallower than primate visual cortex~\citep{harris2019hierarchical, Siegle2021, Felleman1991}.
Secondly, they are trained in a supervised manner on ImageNet~\citep{Schrimpf2018, Conwell2020}, which is an image set containing over one million images belonging to one thousand, mostly human-relevant, semantic categories~\citep{Deng2009}.
While such a dataset is an important technical tool for machine learning, it is highly implausible as a biological model particularly for rodents, who do not receive such category labels over development.
Finally, mice are known to have lower visual acuity than that of primates~\citep{Prusky2000, Kiorpes2019}, suggesting that the resolution of the inputs to mouse models should be lower than that of the inputs to primate models.
Given these three differences between the visual system of primates and of mice, one cannot simply use current supervised primate ventral stream models as models of the mouse visual system.

The failure of these current models may therefore be tied to a failure in the application of the principles of goal-driven modeling to mouse vision, having to do with a mismatch between the model's architecture and task and those of the system being investigated.
We addressed these three differences between the primate and mouse visual system by training shallower CNN architectures in an \emph{unsupervised} manner using lower-resolution images.
First, we noticed that AlexNet \citep{Krizhevsky2012}, which had the shallowest hierarchical architecture, provided strong correspondence to neural responses in mouse visual cortex.
However, the deepest layers of AlexNet did not correspond well in neural predictivity to any mouse visual area, suggesting that even this architecture is too deep to be a completely physically matched model of the system.
Therefore, we developed a class of novel shallower architectures with multiple parallel streams (``StreamNets'') based on the AlexNet architecture.
The parallel streams mimic the intermediate and higher visual areas identified in mice, informed by empirical work on the mouse visual hierarchy~\citep{harris2019hierarchical, Siegle2021}.
These StreamNets were able to achieve neural predictive performance competitive with that of AlexNet, while also maintaining a match between each model layer and a mouse visual area.

We then addressed the strong supervision signals used in the standard ImageNet categorization task by turning to a spectrum of unsupervised objectives including sparse autoencoding~\citep{Olshausen1996}, image-rotation prediction~\citep{Gidaris2018}, and contrastive embedding objectives~\citep{Wu2018,Chen2020simclr,Chen2020mocov2,Chen2020Siam}, as well as supervised tasks with less category labels (CIFAR-10) or ethologically relevant labels that might be available to the mouse~\citep[e.g., depth information provided by whiskers;][]{quist2014modeling,huet2016simulations,zhuang2017toward}, all while operating on lower-resolution images.

Finally, we found that lowering the resolution of the inputs during model training led to improved correspondence with the neural responses across model architectures, including current deep CNNs (VGG16 and ResNet-18) used in prior comparisons to mouse visual data~\citep{Cadena2019, Shi2019, DeVries2020, Conwell2020}.
Thus, strong constraints even at the level of input transformations improve model correspondence to the mouse visual system, although there remains a small gap between these models and the inter-animal consistency ceiling.

Overall, shallow architectures (our StreamNet variants and AlexNet) trained on unsupervised contrastive objectives using lower-resolution inputs yielded the best match to neural response patterns in mouse visual cortex, substantially improving the matches achieved by any of the supervised models we considered and approached the inter-animal consistency ceiling.
Moreover, we show that these contrastive objectives attain comparable performance as supervised models on a variety of downstream visual tasks without the need for category labels.

Taken together, our best models of the mouse visual system suggest that it is a shallow, general-purpose system operating on comparatively low-resolution inputs.
These identified factors therefore provide interpretable insight into the confluence of constraints that may have given rise to the system in the first place, suggesting that these factors were crucially important given the ecological niche in which the mouse is situated, and the resource limitations to which it is subject.


\section{Determining the animal-to-animal mapping transform}
\label{sec:results-upper}
\begin{figure}[htbp]
    \centering
    \includegraphics[width=\columnwidth]{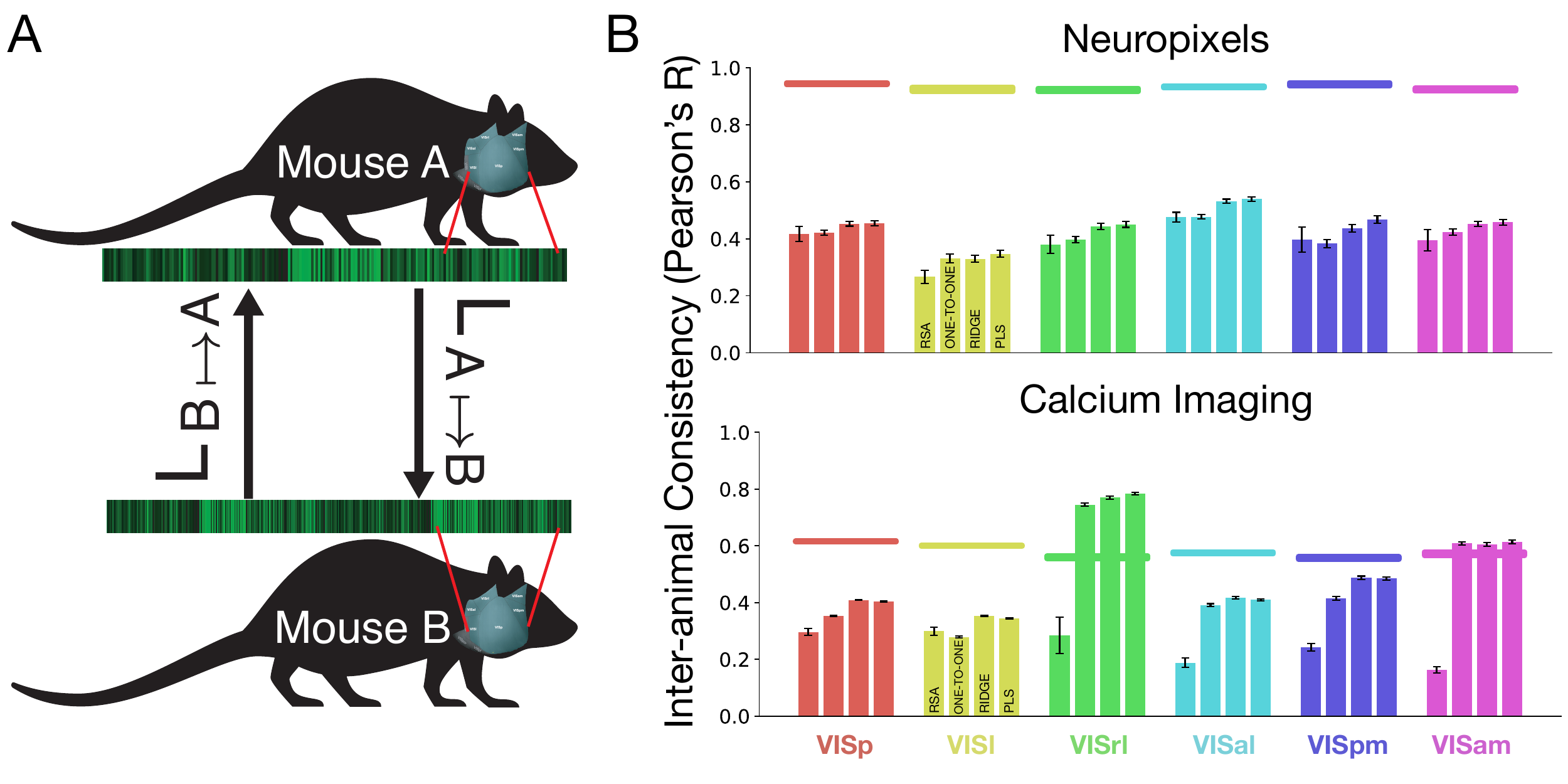}
    \caption[Evaluating the inter-animal consistency of the neural data]{\textbf{Evaluating the inter-animal consistency of the neural data.}
    \textbf{A}. Calcium imaging and Neuropixels data were collected by the Allen Institute for six mouse visual areas: VISp, VISl, VISal, VISrl, VISam, VISpm.
    We assessed the neural data for their internal consistency (split-half reliability) and their inter-animal consistency, which tells us how well one animal corresponds to a pseudo-population of pooled source animals.
    Obtaining these metrics further allows us to determine how well \emph{any} model can be expected to match the neural data, whereby each animal's visual responses are mapped onto other animal's visual responses.
    \textbf{B}. Inter-animal consistency was computed using different linear maps, showing that PLS regression provides the highest consistency.
    Horizontal bars at the top are the median and s.e.m. of the internal consistencies of the neurons in each visual area.
    Refer to Table~\ref{tab:neural-data} for $N$ units per visual area.
    }
    \label{fig:interanimal-cons}
\end{figure}

How should we map a neural network to mouse visual responses?
What firing patterns of mouse visual areas are common across multiple animals, and thus worthy of computational explanation?
A natural approach would be to map neural network features to mouse neural responses in the same manner that different animals can be mapped to each other.
Specifically, we aimed to identify the best performing class of similarity transforms needed to map the firing patterns of one animal's neural population to that of another (inter-animal consistency; Figure~\ref{fig:interanimal-cons}A).
We took inspiration from methods that have proven useful in modeling primate and human visual, auditory, and motor cortex~\citep{Yamins2016, kell2018task, michaels2020goal, nayebi2022}.
As with other cortical areas, this transform class likely cannot be so strict as to require fixed neuron-to-neuron mappings between cells.
However, the transform class for each visual area also cannot be so loose as to allow an unconstrained nonlinear mapping, since the model already yields an image-computable nonlinear response.

We explored a variety of linear mapping transform classes (fit with different constraints) between the population responses for each mouse visual area, as illustrated in Figure~\ref{fig:interanimal-cons}.
The mouse visual responses to natural scenes were collected previously using both two-photon calcium imaging and Neuropixels by the Allen Institute~\citep{DeVries2020, Siegle2021}.
For all methods, the corresponding mapping was trained on $50\%$ of all the natural scene images, and evaluated on the remaining held-out set of images (Figure~\ref{fig:interanimal-cons}B, see supplement for more details).
We also included representational similarity analyses \citep[RSA,][]{Kriegeskorte2008} as a baseline measure of population-wide similarity across animals, corresponding to no selection of individual units, unlike the other mapping transforms.
For the strictest mapping transform (One-to-One), each target unit was mapped to the single most correlated unit in the source animal. 
Overall, the One-to-One mapping tended to yield the lowest inter-animal consistency among the maps considered.
However, Ridge regression (L2-regularized) and PLS (Partial Least Squares) regression were more effective at the inter-animal mapping, yielding the most consistent fits across visual areas, with PLS regression providing the highest inter-animal consistency.
This result implies that an appropriate transform between visual areas in different mice (at least in these datasets) is a linear transform that incorporates a substantial proportion of source animal units to map to each unit in the target animal.
We therefore use this \emph{same} transform class by which to evaluate candidate models.

We further noticed a large difference between the inter-animal consistency obtained via RSA and the consistencies achieved by any of the other mapping transforms for the responses in VISrl of the calcium imaging dataset (green in Figure~\ref{fig:interanimal-cons}B).
However, this difference was not observed for responses in VISrl in the Neuropixels dataset.
This discrepancy suggested that there was a high degree of population-level heterogeneity in the responses collected from the calcium imaging dataset, which may be attributed to the fact that the two-photon FOV for VISrl spanned the boundary between the visual and somatosensory cortex, as originally noted by~\cite{DeVries2020}.
We therefore excluded it from further analyses, following~\cite{siegle2020reconciling}, who systematically compared these two datasets.
Thus, this analysis provided insight into the experiments from which the data were collected.

\section{Three key factors of quantitatively accurate goal-driven models of mouse visual cortex}
\label{sec:results}

We considered three primary ingredients that, when combined, yielded quantitatively accurate goal-driven models of mouse visual cortex: architecture (analogous to the wiring diagram), task (analogous to the visual behavior), and input resolution at which the system operates.
Having established a consistent similarity transform class between animals across visual areas (PLS regression), we proceeded to map artificial neural network responses which varied in these three factors to mouse neural response patterns under this transform class.
We delved into each factor individually before adjoining them, leading to the overall conclusion that the mouse visual system is most consistent with a low-resolution, shallow, and general-purpose visual system.
These models approached $90\%$ of the inter-animal consistency, significantly improving over the prior high-resolution, deep, and task-specific models (VGG16) which attained only $56.27\%$ of this ceiling.

\subsection{Architecture: Shallow architectures better predict mouse visual responses than deep architectures}
\label{sec:objin-shallow}
\begin{figure}[htbp]
    \centering
    \includegraphics[width=0.95\columnwidth]{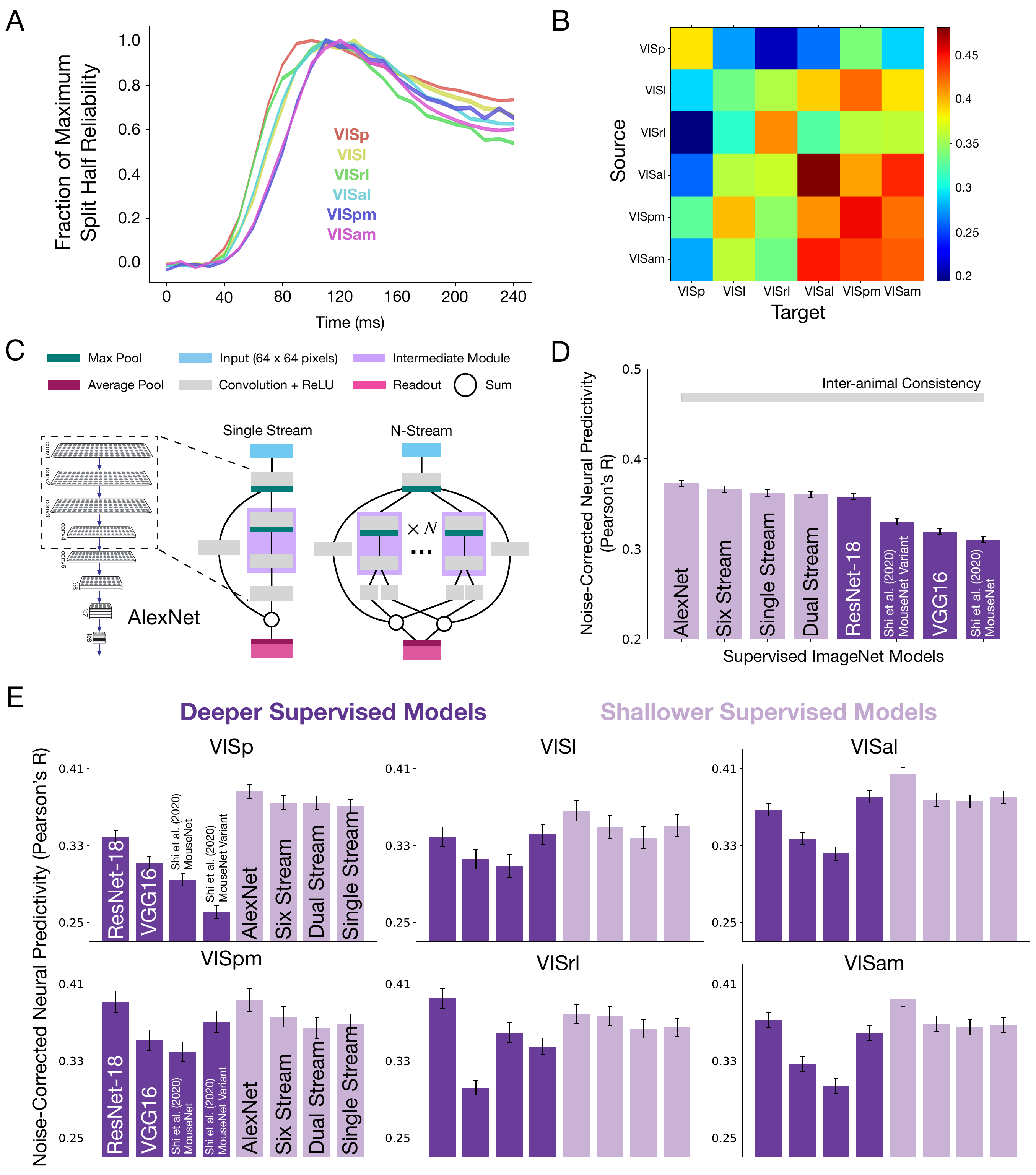}
    \caption[Hierarchically shallow models achieve competitive neural predictivity performance (Neuropixels dataset)]{\textbf{Hierarchically shallow models achieve competitive neural predictivity performance (Neuropixels dataset).}
    \textbf{A}. 
    Fraction of maximum split-half reliability for each visual area as a function of time computed from the Neuropixels dataset.
    \textbf{B}. Each entry in the matrix denotes the median neural predictivity across units for the pairwise mapping performed between a source and a target visual area (and between animals, as in Figure~\ref{fig:interanimal-cons}B) using the PLS map.
    \textbf{C}. We found that the first four convolutional layers of AlexNet best corresponded to all the mouse visual areas (see also Figure~\ref{fig:main-loss-fct-vary-alexnet}B).
    These convolutional layers were used as the basis for our StreamNet architecture variants.
    \textbf{D}. Noise-corrected neural predictivity (aggregated across all units across all visual areas) of typically primate visual stream models (i.e., AlexNet, VGG16, ResNet-18), our StreamNet variants, and the MouseNet of \cite{Shi2020} (along with our variant of it), all trained in a supervised manner on ImageNet.
    \textbf{E}. AlexNet and our StreamNet variants (light purple) provide neural predictivity on the Neuropixels dataset that is better or at least as good as those of deeper architectures (dark purple).
    Refer to Table~\ref{tab:neural-data} for $N$ units per visual area.
    See Figure~\ref{fig:supp-shallow-calcium} for neural predictivity on the calcium imaging dataset.
    }
    \label{fig:main-calc-imagenet64}
\end{figure}

The mouse visual system has a shallow hierarchy, in contrast to the primate ventral visual stream \citep{harris2019hierarchical, Siegle2021, Felleman1991}.
We further corroborated this observation by examining the internal consistencies (i.e., split-half reliability) of the neurons in each visual area from the Neuropixels dataset at each $10$-ms time bin, shown in the left panel of Figure~\ref{fig:main-calc-imagenet64}A.
The peak internal consistencies occurred in quick succession from $100$-$130$ ms, starting from VISp (hierarchically the lowest visual area), suggesting an overall three to four level architecture.
We additionally verified this observation from the functional data, under the mapping transform (PLS) that optimally matched animals to each other in Figure~\ref{fig:interanimal-cons}.
Specifically, we mapped each visual area to every other visual area between animals (Figure~\ref{fig:main-calc-imagenet64}B). 
We found generally that each visual area was predicted well by the corresponding visual area in the other animals, but that VISl, VISal, VISpm, and VISam were similarly predicted by each other.
This is suggestive of three functionally distinct groupings of visual areas.

We found that the neural response predictions of a standard deep CNN model (VGG16), used in prior comparisons to mouse visual areas~\citep{Cadena2019, Shi2019, DeVries2020}, were quite far from the inter-animal consistency ($56.27\%$).
Retraining this model with images of resolution closer to the visual acuity of mice ($64 \times 64$ pixels) improved the model's neural predictivity, reaching $67.7\%$ of the inter-animal consistency.
We dive deeper into the image resolution issue in Section \ref{sec:objin-acuity}.

We also reasoned that the substantial gap with the inter-animal consistency was partly due to the mismatch between the shallow hierarchy of the mouse visual system and the deep hierarchy of the model.
Work by \cite{Shi2020} investigated the construction of a parallel pathway model based on information provided by large-scale tract tracing data, though this model was neither task-optimized nor compared to neural responses.
We trained this network on ($64 \times 64$ pixels) ImageNet categorization, and conducted a hyperparameter sweep to identify the learning parameters that yielded the best performance on the task.
We also trained a variant of this network (denoted ``\cite{Shi2020} MouseNet Variant'', Figures~\ref{fig:main-calc-imagenet64}D and~\ref{fig:main-calc-imagenet64}E; see Methods),
and found that this yielded an approximately $2\%$ improvement in ImageNet categorization performance over the original model with its best hyperparameters. 
The neural predictivity of the original MouseNet and this variant were comparable on both datasets (see Figure~\ref{fig:main-calc-imagenet64}D for neural predictivity on the Neuropixels dataset and Figure~\ref{fig:supp-shallow-calcium} for neural predictivity on the calcium imaging dataset).

\begin{figure}[htbp]
    \centering
    \includegraphics[width=0.9\columnwidth]{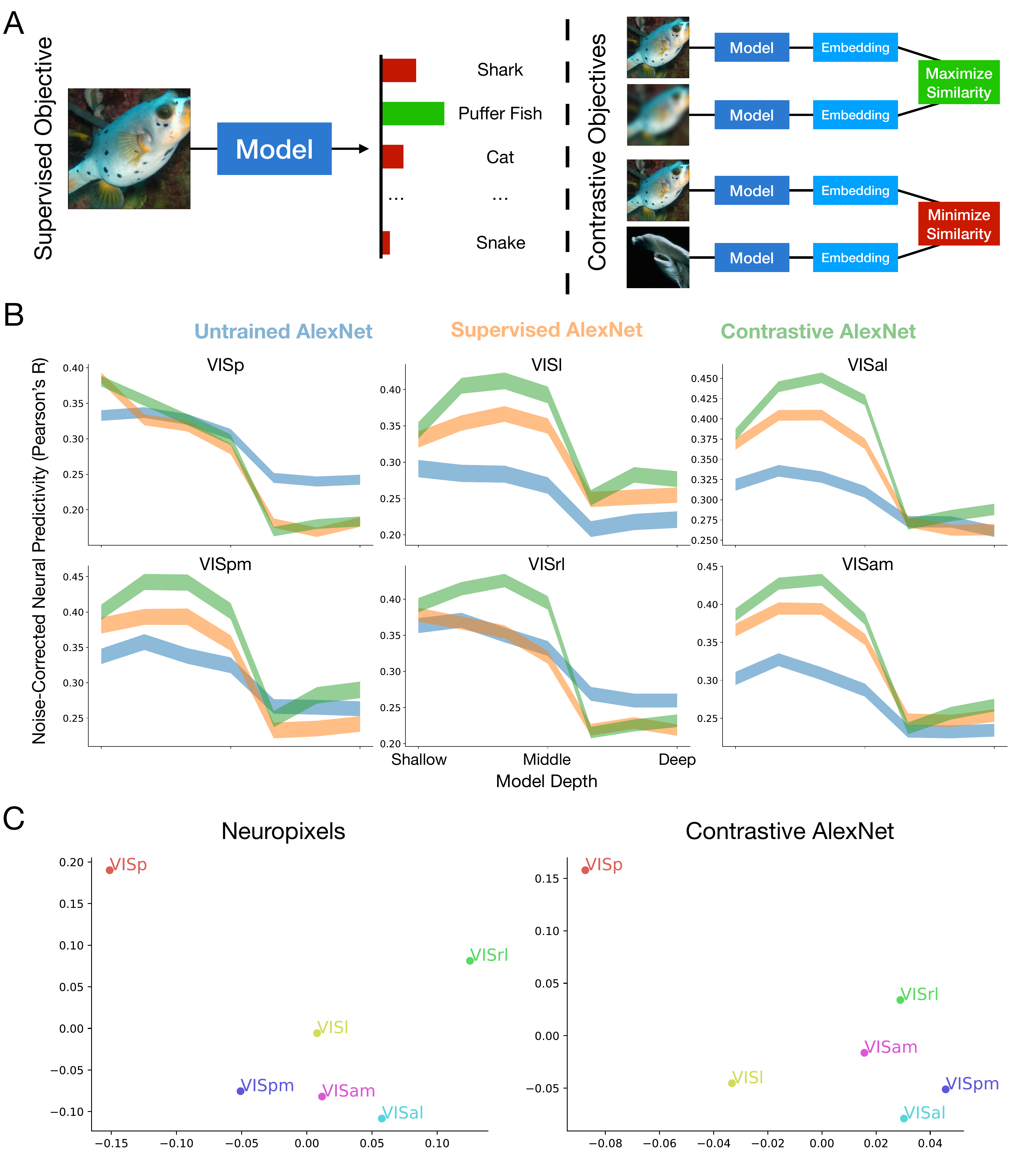}
    \caption[Unsupervised models better predict the neural responses in mouse visual cortex (Neuropixels dataset)]{\textbf{Unsupervised models better predict the neural responses in mouse visual cortex (Neuropixels dataset).}
    \textbf{A}. Models can be trained in either a supervised or an unsupervised contrastive manner.
    In supervised training (left), an image is used as input for a model and the model's prediction (bars) is compared with the labels.
    In unsupervised contrastive training (right), models are trained so that embeddings of augmentations of an image are more similar to each other (upper two rows) than to the embeddings of another image (lower two rows).
    \textbf{B}. Neural predictivity, using PLS regression, on the Neuropixels dataset across AlexNet architectures trained in two different ways (supervised [orange] and unsupervised [green]).
    We observe that the first four convolutional layers provide the best fits to the neural data while the latter three layers are not very predictive for any visual area, suggesting that an even shallower architecture may be suitable.
    This is further corroboration for our architectural decision in Figure~\ref{fig:main-calc-imagenet64}B.
    See Figure~\ref{fig:supp-loss-fct-vary-alexnet} for neural predictivity on the calcium imaging dataset.
    \textbf{C}. Multidimensional scaling (MDS) analysis of the median correlations across neurons for each area as predictors for each target area (left), as derived from Figure~\ref{fig:main-calc-imagenet64}B, and the same applied across the model layers (right) of Contrastive AlexNet (green lines in panel B).
    }
    \label{fig:main-loss-fct-vary-alexnet}
\end{figure}

For each visual area, the maximum neural predictivity of all of these models was worse than that of AlexNet (trained on $64 \times 64$ pixels images), which was the best (and shallowest) model among these architectures (Figures~\ref{fig:main-calc-imagenet64}D and \ref{fig:main-calc-imagenet64}E).
By examining neural predictivity of the best performing model (AlexNet) as a function of model layer, we found that peak neural predictivity did not occur past the fourth convolutional layer (Figure~\ref{fig:main-loss-fct-vary-alexnet}B; orange), suggesting that an even shallower network architecture might be more appropriate.
This result motivated the development of an architecture that is shallower than AlexNet (which we call ``StreamNet''), that is more physically matched to the known shallower hierarchy of the mouse visual system and known dense feedforward skip connectivity (schematic shown in Figure~\ref{fig:main-calc-imagenet64}C).

Our StreamNet model was based on the AlexNet architecture, up to the model layer of maximum predictivity across all visual areas, but allowed for potentially multiple parallel pathways with three levels, consistent with the hierarchy shown in Figure~\ref{fig:main-calc-imagenet64}A.
This yielded an architecture of four convolutional layers, where the first module consisted of one convolutional layer and the intermediate module consisted of two convolutional layers.
The final module has two ``areas'' in parallel (where each area consists of one convolutional layer).
We set the number of parallel streams, $N$, to be one (as a control; denoted ``single-stream''), two (to mimic a potential ventral/dorsal stream distinction; denoted ``dual-stream''), and six (to more closely match the known number of intermediate visual areas: VISl, VISli, VISal, VISrl, VISpl, and VISpm; denoted ``six-stream'').
Figure~\ref{fig:main-calc-imagenet64}C shows a schematic of our StreamNet architecture.
We found that our StreamNet model variants were always more predictive of the neural data across all the visual areas than the MouseNet of \cite{Shi2020}, but attain comparable predictivity to AlexNet (Figures~\ref{fig:main-calc-imagenet64}C and \ref{fig:main-calc-imagenet64}D).
The lack of improvement over AlexNet suggests that a temporally-extended, recurrent version of a shallower network (e.g., StreamNet), might be a functional reason why the added layers of AlexNet offer improved neural predictivity over the current StreamNets.
We will revisit this point in the Discussion.

\subsection{Task: Unsupervised, contrastive objectives, instead of supervised categorization, improves predictions of mouse visual responses}
Training neural network models on 1000-way ImageNet categorization is useful for obtaining visual representations that are well-matched to those of the primate visual system~\citep{Schrimpf2018, Zhuang2021} and seems to result in the best \emph{supervised} models for rodent visual cortex~\citep{Conwell2020}.
However, it is unclear that rodents can perform well on large-scale object recognition tasks when trained \citep[attaining approximately $70\%$ on a two-alternative forced-choice object classification task,][]{Froudarakis2020}, such as those where there are hundreds of labels.
Furthermore, the categories of the ImageNet dataset are rather human-centric and therefore not entirely ethologically relevant for rodents to begin with.

We therefore considered unsupervised losses instead, as these may provide more general goals for the models beyond the specifics of (human-centric) object categorization.
Advances in computer vision have yielded algorithms that are powerful unsupervised visual representation learners, and models trained in those ways are quantitatively accurate models of the primate ventral visual stream~\citep{Zhuang2021}.
Reducing the image size during task training based on rodent visual acuity, which we show in Section~\ref{sec:objin-acuity} to be important to provide good neural predictivity when controlling for task and architecture, further set a constraint on the type of unsupervised algorithms we considered.
Specifically, algorithms that involved crops were unlikely candidates, as the resultant crop would be too small to be effective or too small to have non-trivial features downstream of the network due to the architecture~\citep[e.g., relative location prediction or contrastive predictive coding (CPC) for static images,][]{Doersch2015, Oord2018}.
We instead considered objective functions that use image statistics from the \emph{entire} image.
As control models, we used relatively less powerful unsupervised algorithms including the sparse autoencoder \citep{Olshausen1996}, depth-map prediction, and image-rotation prediction \citep[RotNet,][]{Gidaris2018}.
Advances in unsupervised learning have shown that training models on contrastive objective functions yields representations that can support strong performance on downstream object categorization tasks.
Thus, the remaining four algorithms we used were from the family of contrastive objective functions: instance recognition~\citep[IR,][]{Wu2018}, simple framework for contrastive learning~\citep[SimCLR,][]{Chen2020simclr}, momentum contrast~\citep[MoCov2,][]{Chen2020mocov2}, and simple siamese representation learning~\citep[SimSiam,][]{Chen2020Siam}.

At a high-level, the goal of these contrastive objectives is to learn a representational space where embeddings of augmentations for one image (i.e., embeddings for two transformations of the \emph{same} image) are more ``similar'' to each other than to embeddings of other images (schematized in Figure~\ref{fig:main-loss-fct-vary-alexnet}A).
We found that a model trained with these contrastive objectives resulted in higher neural predictivity across all the visual areas than a model trained on supervised object categorization, for the best model architecture class in Section~\ref{sec:objin-shallow} (i.e., AlexNet) (Figure~\ref{fig:main-loss-fct-vary-alexnet}B).
We systematically explored the space of architecture and objective function combinations in Section~\ref{sec:omnibus}, and found that this observation held more generally.
Finally, a multi-dimensional scaling analysis performed on the predictivity of each visual area across model layers (in the case of ``Contrastive AlexNet'') and across visual areas (in the case of the Neuropixels data; see Figure~\ref{fig:main-calc-imagenet64}A, right) demonstrated a similar partitioning into three functional groups, with the largest functional differences existing between VISp and the remaining areas (Figure~\ref{fig:main-loss-fct-vary-alexnet}C).

\subsection{Data stream: Task-optimization on images of lower resolution improves predictions of mouse visual responses}
\label{sec:objin-acuity}
\begin{figure}[htbp]
    \centering
    \includegraphics[width=0.9\columnwidth]{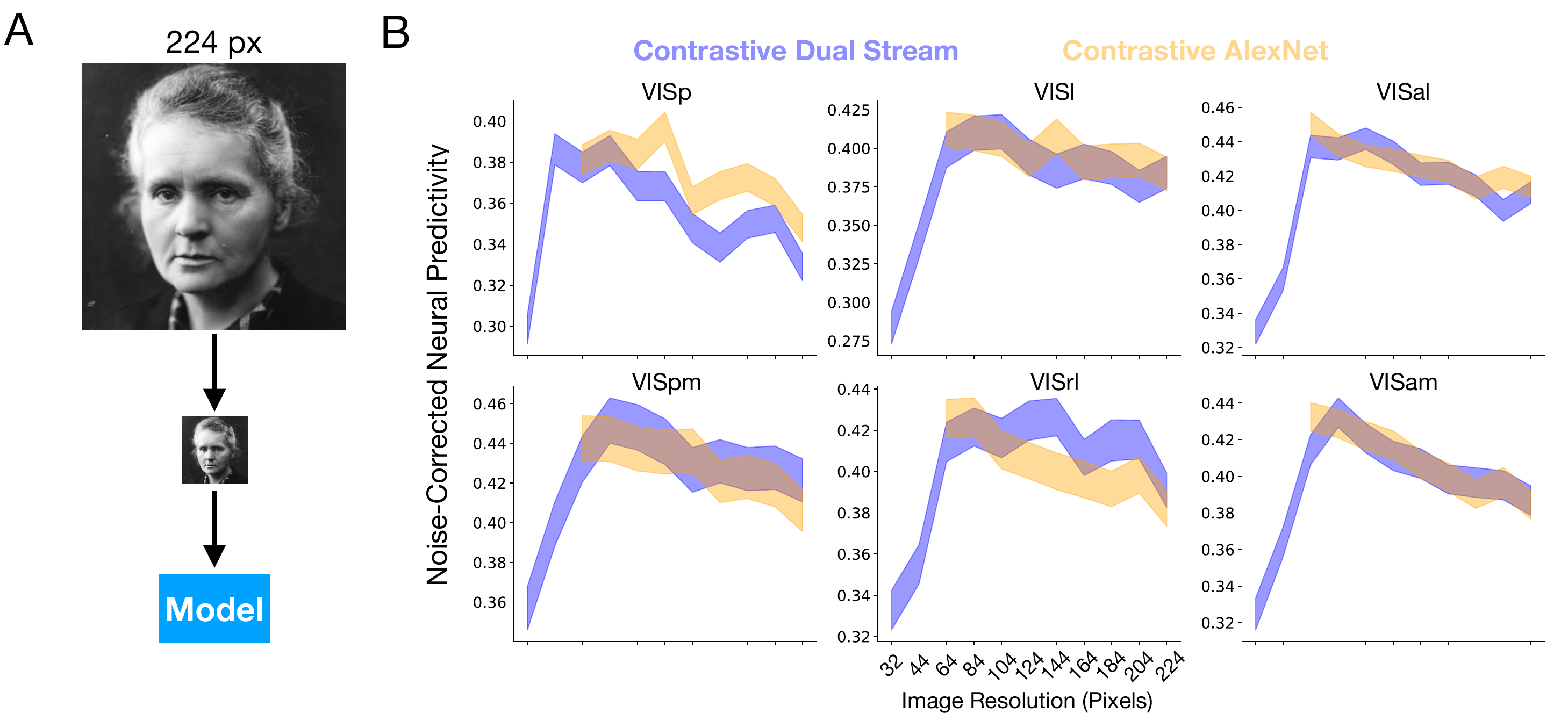}
    \caption[Lower image resolution during model training improves task-optimized neural predictivity (Neuropixels dataset)]{\textbf{Lower image resolution during model training improves task-optimized neural predictivity (Neuropixels dataset).} 
    \textbf{A}. Models with ``lower visual acuity'' were trained using lower-resolution ImageNet images.
    Each image was downsampled from $224 \times 224$ pixels, which is the size typically used to train primate ventral stream models, to various image sizes.
    \textbf{B}. We trained our dual stream variant (blue) and AlexNet (orange) on instance recognition using various image sizes ranging from $32 \times 32$ pixels to $224 \times 224$ pixels and computed their neural predictivity performance for each mouse visual area.
    Training models on resolutions lower than $224 \times 224$ pixels generally led to improved correspondence with the neural responses for both models.
    The median and s.e.m. across neurons in each visual area is reported.
    Refer to Table~\ref{tab:neural-data} for $N$ units per visual area.
    See Figure~\ref{fig:supp-acuity} for neural predictivity on the calcium imaging dataset.
    }
    \label{fig:main-acuity}
\end{figure}

The visual acuity of mice is known to be lower than the visual acuity of primates~\citep{Prusky2000, Kiorpes2019}.
We briefly mentioned previously that task-optimization with images of lower-resolution is important in building models of the mouse visual system.
Here we delved deeper and investigated how neural predictivity performances of two (shallower) architectures varied as a function of the image resolution at which models were trained.
A schematic of this is shown in Figure~\ref{fig:main-acuity}A.
We trained our dual stream variant in an unsupervised manner (instance recognition) using image resolutions that varied from $32 \times 32$ pixels to $224 \times 224$ pixels.
Similarly, we trained AlexNet on instance recognition using image resolutions that varied from $64 \times 64$ pixels to $224 \times 224$ pixels.
$64 \times 64$ pixels was the minimum image size for AlexNet due to its additional max-pooling layer.
In both cases, $224 \times 224$ pixels is the image resolution that is typically used to train neural network models of the primate ventral visual stream.

Training models using resolutions lower than what is used for primate models indeed improves neural predictivity across all visual areas, shown in Figure~\ref{fig:main-acuity}B.
Although the input resolution of $64 \times 64$ pixels may not be optimal for different architectures, it is the resolution that we used to train all the models.
This was motivated by the observation that the upper bound on mouse visual acuity is 0.5 cycles $/$ degree \citep{Prusky2000}, corresponding to 2 pixels $/$ cycle $\times$ 0.5 cycles $/$ degree $=$ 1 pixel $/$ degree, so that a simplified correction for the appropriate visual acuity would correspond to $64\times 64$ pixels, which was also used in \cite{Shi2019} and in the MouseNet of \cite{Shi2020}.
A more thorough investigation into the appropriate image transformations, however, may be needed.

Overall, our observations suggest that a change in the task via a simple change in the image statistics (i.e., data stream) is crucial to obtain an appropriate model of mouse visual encoding.
This further suggests that mouse visual encoding is the result of task-optimization at a lower ``visual acuity'' than what is typically used for primate ventral stream models.

\subsection{Putting it all together: Shallow architectures trained on contrastive objectives with low-resolution inputs best capture neural responses throughout mouse visual cortex}
\label{sec:omnibus}

\begin{figure}[htbp]
    \centering
    \includegraphics[width=\columnwidth]{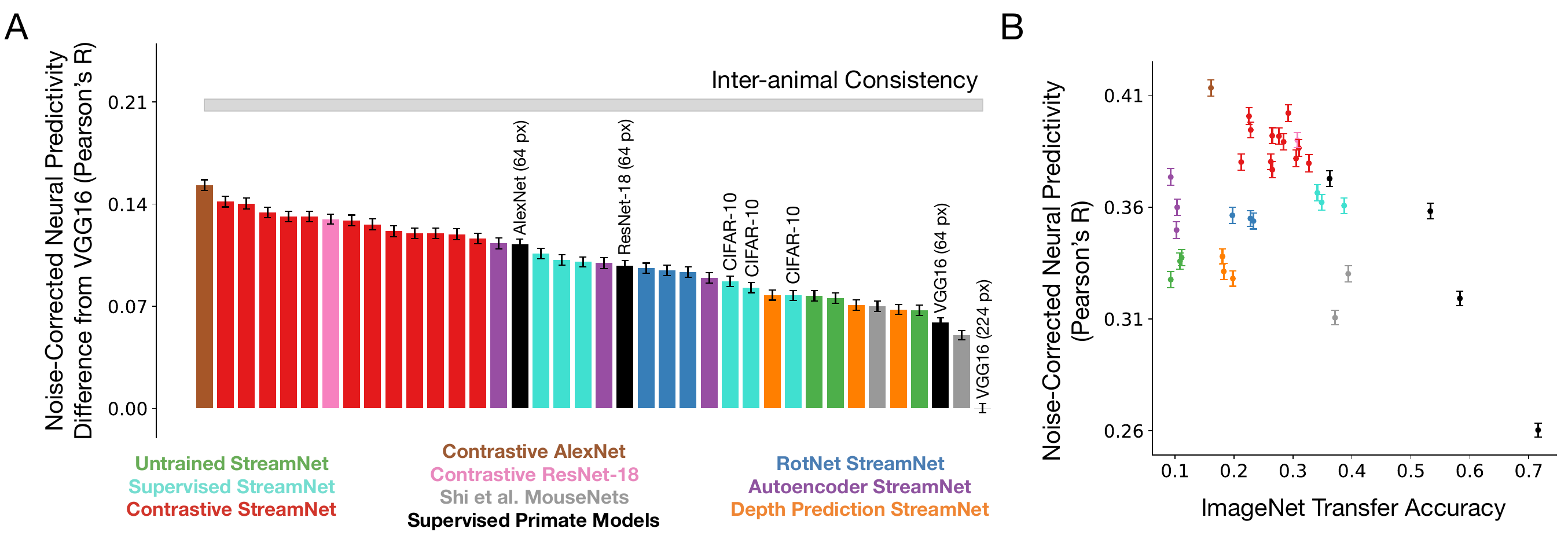}
    \caption[Shallow architectures trained with contrastive objective functions yield the best matches to the neural data (Neuropixels dataset)]{\textbf{Shallow architectures trained with contrastive objective functions yield the best matches to the neural data (Neuropixels dataset).}
    \textbf{A}. The median and s.e.m. neural predictivity, using PLS regression, across units in all mouse visual areas.
    $N=1731$ units in total.
    Red denotes our StreamNet models trained on contrastive objective functions, blue denotes our StreamNet models trained on RotNet, turquoise denotes our StreamNet models trained in a supervised manner on ImageNet and on CIFAR-10, green denotes untrained models (random weights), orange denotes our StreamNet models trained depth prediction, purple denotes our StreamNet models trained on autoencoding, brown denotes contrastive AlexNet, pink denotes contrastive ResNet-18 (both trained on instance recognition), black denotes the remaining ImageNet supervised models (primate ventral stream models), and grey denotes the MouseNet of \cite{Shi2020} and our variant of this architecture.
    Actual neural predictivity performance can be found in Table~\ref{tab:imagenet-top1-transfer}.
    \textbf{B}. Each model's performance on ImageNet is plotted against its median neural predictivity across all units from each visual area.
    All ImageNet performance numbers can be found in Table~\ref{tab:imagenet-top1-transfer}.
    Color scheme as in \textbf{A}.
    See Figure~\ref{fig:supp-loss-fct-calcium} for neural predictivity on the calcium imaging dataset.}
    \label{fig:main-lossfct}
\end{figure}



Here we combined all three ingredients, varying the architecture and task at the visual acuity of the rodent.
We found that AlexNet and our unsupervised StreamNet model variants outperformed all the other models (Figure~\ref{fig:main-lossfct}A).
Furthermore, when those models were trained with contrastive objectives, they had the highest neural predictivity, as shown by the red and brown bars on the left of Figure~\ref{fig:main-lossfct}A, attaining close to 90\% of the inter-animal consistency ceiling.
However, there was no clear separation in neural predictivity among the different contrastive objectives.
Among the unsupervised algorithms, contrastive objectives had the highest $64 \times 64$ pixels ImageNet transfer performance (red vs. blue/orange/purple in Figures~\ref{fig:main-lossfct}B and~\ref{fig:supp-loss-fct-rsa}B), indicating that powerful unsupervised loss functions are crucial for explaining the variance in the neural responses.

Higher ImageNet categorization performance also did not correspond to higher neural predictivity, in contrast to findings in models of the primate ventral visual stream \citep{yamins_ventralneural, Schrimpf2018}.
Specifically, deeper, purely supervised models that attain greater than $40\%$ accuracy had, on average, the least match to the neural data (black dots in Figures~\ref{fig:main-lossfct}B and~\ref{fig:supp-loss-fct-rsa}B).
Moreover, object recognition tasks with less categories \cite[e.g., $10$ categories in CIFAR-10,][]{Krizhevsky2009} did not improve neural predictivity for the \emph{same} architecture trained on ImageNet (turquoise bars in Figure~\ref{fig:main-lossfct}A).

As a positive control, we optimized ResNet-18 on a contrastive objective function (pink in Figure~\ref{fig:main-lossfct}) and found that although changing the objective function improved neural predictivity for ResNet-18 over its supervised counterpart, it was still worse than the shallower AlexNet trained using a contrastive objective (compare pink and brown points in Figures~\ref{fig:main-lossfct}A and~\ref{fig:main-lossfct}B).
This indicates that having an appropriately shallow architecture contributes to neural predictivity, but even with a less physically realistic deep architecture such as ResNet-18, you can greatly improve neural predictivity with a contrastive embedding loss function.
These findings are consistent with the idea that appropriate combinations of objective functions and architectures are necessary to build quantitatively accurate models of neural systems, with the objective function providing a strong constraint especially when coupled with a shallow architecture \citep{Yamins2016}.

\section{Discussion}
\label{sec:discussion}
In this work, we showed that shallow architectures trained with contrastive embedding methods operating on lower-resolution images most accurately predict image-evoked neural responses across visual areas in mice, surpassing the predictive power of supervised methods.
Deep CNNs supervised on ImageNet categorization, which have been previously evaluated as models of mouse visual cortex and are quantitatively the best models of the \emph{primate} ventral stream, were comparatively poor predictors of neural responses throughout mouse visual cortex.
Our best models approached the computed inter-animal consistency of all measured units on both neural datasets.
Additionally assessing the transfer performance of our models on a variety of visual tasks (categorization, object position localization, size estimation, pose estimation, and texture discrimination), we found specifically that unsupervised, contrastive objectives generally enabled comparable performance across all tasks as ImageNet categorization losses (Figure~\ref{fig:task_transfer_performance}), but without the need for using category labels to achieve this task-general representation.
While the primate ventral visual stream is well-modeled by a deep hierarchical system and category learning, mouse as a model visual system has not had such a coherent account heretofore. 
Here we propose that, in contrast to the primate ventral stream, mouse vision is best modeled as a shallow, low-resolution system, one that can achieve task-general representations through self-learning.

Taken together with recent work done in primates~\citep{Zhuang2021}, the results indicate that contrastive objectives appear to best explain responses across both rodent and primate species, suggesting that these objectives may be part of a species-general toolkit.
Unlike the situation in the primate ventral visual stream, however, contrastive objectives \emph{surpassed} the neural predictivity of their supervised counterparts, as increased categorization performance lead to overall \emph{worse} correspondence to mouse visual areas.
We observe that the advantage of these contrastive objectives is that they provide representations that are generally improved over those obtained by supervised methods in order to enable a diverse range of visual behaviors.
We additionally found that neural networks of larger sizes, either measured by network parameters (analogous to the number of synapses) or network units (analogous to the number of neurons), had comparatively lower neural predictivity (Figure~\ref{fig:supp-network-size}).
These results suggest that the mouse visual cortex is a light-weight, shallow, low-resolution, and general-purpose visual system in contrast to the deep, high-resolution, and more task-specific visual system in primates.
Our improved models of the mouse visual system therefore provide a different view of its goals and constraints than that provided by (comparatively) high-resolution, deep feedforward categorization models.
Furthermore, the generic nature of these unsupervised contrastive objective functions suggests the intriguing possibility that they might be used by other sensory systems, such as in barrel cortex or the olfactory system.

These results, coupled with the fact that larger, deeper networks (which are relatively better models of primate ventral visual responses than shallow networks) are among the worst models of mouse visual cortex, demonstrates a kind of ``double dissociation'' between the mouse-like architectures and tasks and the primate-like architectures and tasks. 
Thus, the failure of the ``blind application'' of deep networks to capture mouse data well -- and the subsequent success of our more structurally-and-functionally tuned approach -- illustrates not a weakness of the goal-driven neural network approach, but instead a strength of this methodology's ability to be appropriately sensitive to salient biological differences.

Overall, we have made progress in modeling the mouse visual system in three core ways: the choice of architecture class, objective function, and the data stream.

On the architectural front, we explored a range of shallow, deep, and novel multi-stream models, finding that shallower models are a more reasonable starting point for building more accurate mouse vision models.
Specifically, our focus in this work was on feedforward models, but there are many feedback connections from higher visual areas to lower visual areas~\citep{harris2019hierarchical}, which appear to play a functional role in the data as VISpm/am are lower than VISal in the data's own functional hierarchy (Figure~\ref{fig:main-loss-fct-vary-alexnet}C; left panel) despite being more reliable later in time (Figure~\ref{fig:main-calc-imagenet64}A; left panel).
Incorporating these architectural motifs into our models and training these models using dynamic inputs may be useful for modeling temporal dynamics in mouse visual cortex, as has been recently done in primates~\citep{nayebi2018task, Kubilius2019, nayebi2022}.

We also demonstrated that unsupervised, contrastive embedding functions are critical goals for a system to accurately match responses in the mouse visual system.
Thus, towards incorporating recurrent connections in the architecture, we would also like to probe the functionality of these feedback connections in scenarios with temporally-varying, dynamic inputs.
Concurrent work of \cite{Bakhtiari2021} used the unsupervised predictive objective of CPC~\citep{Oord2018} to model neural responses of mouse visual cortex to natural movies.
Given that our best performing unsupervised methods obtained good representations on static images by way of contrastive learning, it would be interesting to explore a larger spectrum of more object-centric unsupervised signals operating on dynamic inputs, such as in the context of forward prediction~\citep[e.g.,][]{Mrowca2018, Haber2018, Lingelbach2020}.

Moreover, we found that constraining the input data so that they are closer to those received by the mouse visual system, was important for improved correspondence -- specifically, resizing the images to be smaller during training as a proxy for low-pass filtering.
We believe that future work could investigate other appropriate low-pass filters and ethologically relevant pixel-level transformations to apply to the original image or video stream.
These additional types of input transformations will likely also constrain the types of unsupervised objective functions that can be effectively deployed in temporally-varying contexts, as it did in our case for static images.

Finally, our inter-animal consistency measurements make a clear recommendation for the type of future neural datasets that are likely to be helpful in more sharply differentiating future candidate models. 
In Figure~\ref{fig:interan-im-ss}A, we observed that when fitting linear maps between animals to assess inter-animal consistency, fitting values are significantly higher in training than on the evaluation (test) set, indicating that the number of stimuli is not large enough to prevent overfitting when identifying source animal neuron(s) to match any given target neuron.
Furthermore, as a function of the number of stimuli, the test set inter-animal consistencies steadily increases (see Figure~\ref{fig:interan-im-ss}B), and likely would continue to increase substantially if the dataset had more stimuli.
Thus, while much focus in methods has been on increasing the number of neurons contained in a given dataset~\citep{steinmetz2021neuropixels}, our analysis indicates that the main limiting factor in model identification is \emph{number of stimuli} in the dataset, rather than the number of neurons.
In our view, future experiments should preferentially focus more resources on increasing stimulus count.
Doing so would likely raise the inter-animal consistency, in turn providing substantially more dynamic range for separating models in terms of their ability to match the data, and thereby increasing the likelihood that more specific conclusions about precisely which circuit structure(s)~\citep{collins2017,Bergstra2015} and which specific (combinations of) objectives \citep[e.g.,][]{Wu2018, Chen2020Siam, Chen2020simclr} best describe mouse visual cortex.

\section{Methods}
\label{sec:methods}
\subsection{Neural Response Datasets}
\label{ss:methods-neural-data}
We used the Allen Brain Observatory Visual Coding dataset~\citep{DeVries2020, Siegle2021} collected using both two-photon calcium imaging and Neuropixels from areas VISp (V1), VISl (LM), VISal (AL), VISrl (RL), VISam (AM), and VISpm (PM) in mouse visual cortex.
We focused on the natural scene stimuli, consisting of $118$ images, each presented $50$ times (i.e., $50$ trials per image).

We list the number of units and specimens for each dataset in Table~\ref{tab:neural-data}, after units are selected, according to the following procedure:
For the calcium imaging data, we used a similar unit selection criterion as in~\cite{Conwell2020}, where we sub-selected units that attain a Spearman-Brown corrected split-half consistency of at least $0.3$ (averaged across $100$ bootstrapped trials), and whose peak responses to their preferred images are not significantly modulated by the mouse’s running speed during stimulus presentation ($p > 0.05$).

For the Neuropixels dataset, we separately averaged, for each specimen and each visual area, the temporal response (at the level of $10$-ms bins up to $250$ ms) on the largest contiguous time interval when the median (across the population of units in that specimen) split-half consistency reached at least $0.3$.
This procedure helps to select the most internally-consistent units in their temporally-averaged response, and accounts for the fact that different specimens have different time courses along which their population response becomes reliable.

Finally, after subselecting units according to the above criteria for both datasets, we only keep specimens that have at least the $75$th percentile number of units among all specimens for that given visual area.
This final step helped to ensure we have enough internally-consistent units per specimen for the inter-animal consistency estimation (derived in Section~\ref{ss:methods-interanimal}).

\begin{table}[ht]
    \centering
    \begin{tabular}{| c || c | c | c | } \hline
      \textbf{Dataset Type} & \textbf{Visual Area} & \textbf{Total Units} & \textbf{Total Specimens} \\ \hline\hline
      \multirow{6}{*}{Calcium Imaging} & VISp & 7080 & 29 \\ \cline{2-4}
      & VISl & 4393 & 24 \\ \cline{2-4}
      & VISal & 2064 & 9 \\ \cline{2-4}
      & VISrl & 1116 & 8 \\ \cline{2-4}
      & VISam & 847 & 9 \\ \cline{2-4}
      & VISpm & 1844 & 19 \\ \hline\hline
      \multirow{6}{*}{Neuropixels} & VISp & 442 & 8 \\ \cline{2-4}
      & VISl & 162 & 6 \\ \cline{2-4}
      & VISal & 396 & 6 \\ \cline{2-4}
      & VISrl & 299 & 7 \\ \cline{2-4}
      & VISam & 257 & 7 \\ \cline{2-4}
      & VISpm & 175 & 5 \\ \hline
      \end{tabular}
    \caption[Descriptive statistics of the neural datasets]{\textbf{Descriptive statistics of the neural datasets.} Total number of units and specimens for each visual area for the calcium imaging and Neuropixels datasets.}
    \label{tab:neural-data}
\end{table}

\subsection{Noise Corrected Neural Predictivity}
\subsubsection{Linear Regression}
\label{sss:methods-linreg}
When we perform neural fits, we choose a random $50\%$ set of natural scene images ($59$ images in total) to train the regression, and the remaining $50\%$ to use as a test set ($59$ images in total), across ten train-test splits total.
For Ridge regression, we use an $\alpha=1$, following the \texttt{sklearn.linear\_model} convention.
PLS regression was performed with $25$ components, as in prior work \citep[e.g.,][]{yamins_ventralneural, Schrimpf2018}.
When we perform regression with the One-to-One mapping, as in Figure~\ref{fig:interanimal-cons}B, we identify the top correlated (via Pearson correlation on the training images) unit in the source population for each target unit.
Once that source unit has been identified, we then fix it for that particular train-test split, evaluated on the remaining $50\%$ of images.

Motivated by the justification given in Section~\ref{ss:methods-interanimal} for the noise correction in the inter-animal consistency, the noise correction of the model to neural response regression is a special case of the quantity defined in Section~\ref{sss:methods-interanimal-multiple}, where now the source animal is replaced by model features, separately fit to each target animal (from the set of available animals $\mathcal{A}$).
Let $L$ be the set of model layers, let $r^{\ell}$ be the set of model responses at model layer $\ell \in L$, $M$ be the mapping, and let $\mathrm{s}$ be the trial-averaged pseudo-population response.

\begin{equation*}\label{modelcon}
\begin{split}
\max_{\ell\in L}\median\bigoplus_{\animalB \in \mathcal{A}} \left\langle\dfrac{\corr\left(M\left(r^{\ell}_{\text{train}};\sfBtrain\right)_{\test}, \ssBtest\right)}{\sqrt{\widetilde{\corr}\left(M\left(r^{\ell}_{\text{train}};\sfBtrain\right)_{\test}, M\left(r^{\ell}_{\text{train}};\ssBtrain\right)_{\test}\right) \times \widetilde{\corr}\left(\sfBtest, \ssBtest\right)}}\right\rangle,
\end{split}
\end{equation*}
where the average is taken over $100$ bootstrapped split-half trials, $\oplus$ denotes concatenation of units across animals $\animalB \in \mathcal{A}$ followed by the median value across units, and $\corr(\cdot,\cdot)$ denotes the Pearson correlation of the two quantities.
$\widetilde{\corr}(\cdot, \cdot)$ denotes the Spearman-Brown corrected value of the original quantity (see Section~\ref{sss:methods-interanimal-spearman-brown}).

Prior to obtaining the model features of the stimuli for linear regression, we preprocessed each stimulus using the image transforms used on the validation set during model training, resizing the shortest edge of the stimulus in both cases to $64$ pixels, preserving the aspect ratio of the input stimulus.
Specifically, for models trained using the ImageNet dataset, we first resized the shortest edge of the stimulus to $256$ pixels, center-cropped the image to $224 \times 224$ pixels, and finally resized the stimulus to $64 \times 64$ pixels.
For models trained using the CIFAR-10 dataset, this resizing yielded a $64 \times 81$ pixels stimulus.

\subsubsection{Representational Similarity Analysis (RSA)}
\label{sss:methods-rsa}
In line with prior work~\citep{Shi2019, Conwell2020}, we also used representational similarity analysis~\citep[RSA,][]{Kriegeskorte2008} to compare models to neural responses, as well as to compare animals to each other. 
Specifically, we compared (via Pearson correlation) only the upper-right triangles of the representational dissimilarity matrices (RDMs), excluding the diagonals to avoid illusory effects~\citep{Ritchie2017}.


For each visual area and a given model, we defined the predictivity of the model for that area to be the maximum RSA score across model layers after the suitable noise correction is applied, which is defined as follows.
Let $r^{\ell}$ be the model responses at model layer $\ell$ and let $\mathrm{s}$ be the trial-averaged pseudo-population response (i.e., responses aggregated across specimens).
The metric used here is a specific instance of Equation~\eqref{rsainteranconid}, where the single source animal $\animalA$ is the trial-wise, deterministic model features (which have a mapping consistency of 1 as a result) and a single target animal $\animalB$, which is the pseudo-population response:
\begin{equation}\label{rsaidmodelconcat}
\begin{split}
& \max_{\ell\in L}\left\langle\dfrac{\rsa\left(r^{\ell}, \mathrm{s}_{2}\right)}{\sqrt{\spbrsa\left(\mathrm{s}_{1}, \mathrm{s}_{2}\right)}}\right\rangle, \\
& \spbrsa\left(\mathrm{s}_{1}, \mathrm{s}_{2}\right) \coloneqq \frac{2\rsa\left(\mathrm{s}_{1}, \mathrm{s}_{2}\right)}{1 + \rsa\left(\mathrm{s}_{1}, \mathrm{s}_{2}\right)},
\end{split}
\end{equation}
where $L$ is the set of model layers, $\{\mathrm{s}_i\}_{i=1}^2$ are the animal's responses for two halves of the trials (and averaged across the trials dimension), the average is computed over 100 bootstrapped split-half trials, and $\spbrsa\left(\mathrm{s}_{1}, \mathrm{s}_{2}\right)$ denotes Spearman-Brown correction applied to the internal consistency quantity, $\rsa\left(\mathrm{s}_{1}, \mathrm{s}_{2}\right)$, defined in Section~\ref{sss:methods-interanimal-spearman-brown}.

If the fits are performed separately for each animal, then $\animalB$ corresponds to each animal among those for a given visual area (defined by the set $\mathcal{A}$), and we compute the median across animals $\animalB \in \mathcal{A}$:
\begin{equation}\label{rsaidmodelperanimal}
\max_{\ell\in L} \median_{\animalB \in \mathcal{A}} \left\langle\dfrac{\rsa\left(r^{\ell}, \ssBid \right)}{\sqrt{\spbrsa\left(\sfBid, \ssBid\right)}}\right\rangle.
\end{equation}
Similar to the above, Spearman-Brown correction is applied to the internal consistency quantity, $\rsa\left(\sfBid, \ssBid\right)$.


\subsection{Inter-Animal Consistency Derivation}
\label{ss:methods-interanimal}
\subsubsection{Single Animal Pair}
\label{sss:methods-interanimal-pair}
Suppose we have neural responses from two animals $\animalA$ and $\animalB$.
Let $\mathrm{t}_i^p$ be the vector of true responses (either at a given time bin or averaged across a set of time bins) of animal $p \in \mathcal{A} = \{\animalA,\animalB,\dots\}$ on stimulus set $i \in \{\train, \test\}$.
Of course, we only receive noisy observations of $\mathrm{t}_i^p$, so let $\mathrm{s}_{j,i}^p$ be the $j$th set of $n$ trials of $\mathrm{t}_i^p$.
Finally, let $M(x;y)_i$ be the predictions of a mapping $M$ (e.g., PLS) when trained on input $x$ to match output $y$ and tested on stimulus set $i$.
For example, $M\left(\trueA;\trueB\right)_{\test}$ is the prediction of mapping $M$ on the test set stimuli trained to match the true neural responses of animal $\animalB$ given, as input, the true neural responses of animal $\animalA$ on the train set stimuli.
Similarly, $M\left(\sfAtrain;\sfBtrain\right)_{\test}$ is the prediction of mapping $M$ on the test set stimuli trained to match the trial-average of noisy sample 1 on the train set stimuli of animal $\animalB$ given, as input, the trial-average of noisy sample 1 on the train set stimuli of animal $\animalA$.

With these definitions in hand, the inter-animal mapping consistency from animal $\animalA$ to animal $\animalB$ corresponds to the following true quantity to be estimated:
\begin{equation}\label{interancontrue}
\corr\left(M\left(\trueA;\trueB\right)_{\test}, \trueBtest\right),
\end{equation}
where $\corr(\cdot, \cdot)$ is the Pearson correlation across a stimulus set.
In what follows, we will argue that Equation~\eqref{interancontrue} can be approximated with the following ratio of measurable quantities, where we split in half and average the noisy trial observations, indexed by 1 and by 2:
\begin{equation}\label{interancon}
\begin{split}
&\corr\left(M\left(\trueA;\trueB\right)_{\test}, \trueBtest\right) \\
& \sim \dfrac{\corr\left(M\left(\sfAtrain;\sfBtrain\right)_{\test}, \ssBtest\right)}{\sqrt{\corr\left(M\left(\sfAtrain;\sfBtrain\right)_{\test}, M\left(\ssAtrain;\ssBtrain\right)_{\test}\right) \times \corr\left(\sfBtest, \ssBtest\right)}}
\end{split}.
\end{equation}
In words, the inter-animal consistency (i.e., the quantity on the left side of Equation~\eqref{interancon}) corresponds to the predictivity of the mapping on the test set stimuli from animal $\animalA$ to animal $\animalB$ on two different (averaged) halves of noisy trials (i.e., the numerator on the right side of Equation~\eqref{interancon}), corrected by the square root of the mapping reliability on animal $\animalA$'s responses to the test set stimuli on two different halves of noisy trials multiplied by the internal consistency of animal $\animalB$.

We justify the approximation in Equation~\eqref{interancon} by gradually replacing the true quantities ($\mathrm{t}$) by their measurable estimates ($\mathrm{s}$), starting from the original quantity in Equation~\eqref{interancontrue}.
First, we make the approximation that:
\begin{equation}\label{step1}
\corr\left(M\left(\trueA;\trueB\right)_{\test}, \ssBtest\right) \sim \corr\left(M\left(\trueA;\trueB\right)_{\test}, \trueBtest\right) \times \corr\left(\trueBtest, \ssBtest\right),
\end{equation}
by the transitivity of positive correlations (which is a reasonable assumption when the number of stimuli is large).
Next, by transitivity and normality assumptions in the structure of the noisy estimates and since the number of trials ($n$) between the two sets is the same, we have that:
\begin{align}\label{step2}
\corr\left(\sfBtest, \ssBtest\right) &\sim \corr\left(\sfBtest, \trueBtest\right) \times \corr\left(\trueBtest, \ssBtest\right) \nonumber \\
&\sim \corr\left(\trueBtest, \ssBtest\right)^2.
\end{align}
In words, Equation~\eqref{step2} states that the correlation between the average of two sets of noisy observations of $n$ trials each is approximately the square of the correlation between the true value and average of one set of $n$ noisy trials.
Therefore, combining Equations~\eqref{step1} and \eqref{step2}, it follows that:
\begin{equation}\label{lemma1}
\corr\left(M\left(\trueA;\trueB\right)_{\test}, \trueBtest\right) \sim \dfrac{\corr\left(M\left(\trueA;\trueB\right)_{\test}, \ssBtest\right)}{\sqrt{\corr\left(\sfBtest, \ssBtest\right)}}.
\end{equation}

From the right side of Equation~\eqref{lemma1}, we can see that we have removed $\trueBtest$, but we still need to remove the $M\left(\trueA;\trueB\right)_{\test}$ term, as this term still contains unmeasurable (i.e., true) quantities.
We apply the same two steps, described above, by analogy, though these approximations may not always be true (they are, however, true for Gaussian noise):
\begin{equation*}
\begin{split}
\corr\left(M\left(\sfAtrain;\sfBtrain\right)_{\test}, \ssBtest\right) & \sim \corr\left(\ssBtest, M\left(\trueA;\trueB\right)_{\test}\right) \\
& \times \corr\left(M\left(\trueA;\trueB\right)_{\test}, M\left(\sfAtrain;\sfBtrain\right)_{\test}\right)
\end{split}
\end{equation*}
\begin{equation*}
\begin{split}
& \corr\left(M\left(\sfAtrain;\sfBtrain\right)_{\test}, M\left(\ssAtrain;\ssBtrain\right)_{\test}\right) \\
& \sim \corr\left(M\left(\sfAtrain;\sfBtrain\right)_{\test}, M\left(\trueA;\trueB\right)_{\test}\right)^2,
\end{split}
\end{equation*}
which taken together implies the following:
\begin{equation}\label{lemma2}
\corr\left(M\left(\trueA;\trueB\right)_{\test}, \ssBtest\right) \sim \dfrac{\corr\left(M\left(\sfAtrain;\sfBtrain\right)_{\test}, \ssBtest\right)}{\sqrt{\corr\left(M\left(\sfAtrain;\sfBtrain\right)_{\test}, M\left(\ssAtrain;\ssBtrain\right)_{\test}\right)}}.
\end{equation}
Equations \eqref{lemma1} and \eqref{lemma2} together imply the final estimated quantity given in Equation~\eqref{interancon}.

\subsubsection{Multiple Animals}
\label{sss:methods-interanimal-multiple}
For multiple animals, we consider the average of the true quantity for each target in $\animalB$ in Equation~\eqref{interancontrue} across source animals $\animalA$ in the ordered pair $(\animalA,\animalB)$ of animals $\animalA$ and $\animalB$:
\begin{equation*}\label{multipleinterancon}
\begin{split}
&\left\langle \corr\left(M\left(\trueA;\trueB\right)_{\test}, \trueBtest\right)\right\rangle_{\animalA \in \mathcal{A}: (\animalA,\animalB)\in \mathcal{A}\times\mathcal{A}} \\
& \sim \left\langle\dfrac{\corr\left(M\left(\sfAtrain;\sfBtrain\right)_{\test}, \ssBtest\right)}{\sqrt{\widetilde{\corr}\left(M\left(\sfAtrain;\sfBtrain\right)_{\test}, M\left(\ssAtrain;\ssBtrain\right)_{\test}\right) \times \widetilde{\corr}\left(\sfBtest, \ssBtest\right)}}\right\rangle_{\animalA \in \mathcal{A}: (\animalA,\animalB)\in \mathcal{A}\times\mathcal{A}}.
\end{split}
\end{equation*}
We also bootstrap across trials, and have multiple train/test splits, in which case the average on the right hand side of the equation includes averages across these as well.

Note that each neuron in our analysis will have this single average value associated with it when \emph{it} was a target animal ($\animalB$), averaged over source animals/subsampled source neurons, bootstrapped trials, and train/test splits.
This yields a vector of these average values, which we can take median and standard error of the mean (s.e.m.) over, as we do with standard explained variance metrics.

\subsubsection{RSA}
\label{sss:methods-interanimal-rsa}
We can extend the above derivations to other commonly used metrics for comparing representations that involve correlation.
Since $\rsa(x,y) := \corr(\rdm(x), \rdm(y))$, then the corresponding quantity in Equation~\eqref{interancon} analogously (by transitivity of positive correlations) becomes:
\begin{equation}\label{rsainterancon}
\begin{split}
& \left\langle\rsa\left(M\left(\trueA;\trueB\right)_{\test}, \trueBtest\right)\right\rangle_{\animalA \in \mathcal{A}: (\animalA,\animalB)\in \mathcal{A}\times\mathcal{A}} \\
& \sim \left\langle\dfrac{\rsa\left(M\left(\sfAtrain;\sfBtrain\right)_{\test}, \ssBtest\right)}{\sqrt{\widetilde{\rsa}\left(M\left(\sfAtrain;\sfBtrain\right)_{\test}, M\left(\ssAtrain;\ssBtrain\right)_{\test}\right) \times \widetilde{\rsa}\left(\sfBtest, \ssBtest\right)}}\right\rangle_{\animalA \in \mathcal{A}: (\animalA,\animalB)\in \mathcal{A}\times\mathcal{A}}.
\end{split}
\end{equation}

Note that in this case, each \emph{animal} (rather than neuron) in our analysis will have this single average value associated with it when \emph{it} was a target animal ($\animalB$) (since RSA is computed over images and neurons), where the average is over source animals/subsampled source neurons, bootstrapped trials, and train/test splits.
This yields a vector of these average values, which we can take median and s.e.m. over, across animals $\animalB \in \mathcal{A}$.

For RSA, we can use the identity mapping (since RSA is computed over neurons as well, the number of neurons between source and target animal can be different to compare them with the identity mapping). 
As parameters are not fit, we can choose $\train = \test$, so that Equation~\eqref{rsainterancon} becomes:
\begin{equation}\label{rsainteranconid}
\left\langle\rsa\left(\trueAid,\trueBid\right)\right\rangle_{\animalA \in \mathcal{A}: (\animalA,\animalB)\in \mathcal{A}\times\mathcal{A}} \sim \left\langle\dfrac{\rsa\left(\sfAid, \ssBid\right)}{\sqrt{\widetilde{\rsa}\left(\sfAid, \ssAid\right) \times \widetilde{\rsa}\left(\sfBid, \ssBid\right)}}\right\rangle_{\animalA \in \mathcal{A}: (\animalA,\animalB)\in \mathcal{A}\times\mathcal{A}}.
\end{equation}

\subsubsection{Pooled Source Animal}
\label{sss:methods-interanimal-holdouts}
Often times, we may not have enough neurons per animal to ensure that the estimated inter-animal consistency in our data closely matches the ``true'' inter-animal consistency.
In order to address this issue, we holdout one animal at a time and compare it to the pseudo-population aggregated across units from the remaining animals, as opposed to computing the consistencies in a pairwise fashion.
Thus, $\animalB$ is still the target heldout animal as in the pairwise case, but now the average over $\animalA$ is over a sole ``pooled'' source animal constructed from the pseudo-population of the remaining animals.

\subsubsection{Spearman-Brown Correction}
\label{sss:methods-interanimal-spearman-brown}
The Spearman-Brown correction can be applied to each of the terms in the denominator individually, as they are each correlations of observations from half the trials of the \emph{same} underlying process to itself (unlike the numerator). Namely,
\begin{equation*}
\widetilde{\corr}\left(X,Y\right) \coloneqq \frac{2\corr\left(X,Y\right)}{1 + \corr\left(X,Y\right)}.
\end{equation*}
Analogously, since $\rsa(X,Y) := \corr(\rdm(x), \rdm(y))$, then we define
\begin{align*}
\widetilde{\rsa}\left(X,Y\right) &\coloneqq \widetilde{\corr}(\rdm(x), \rdm(y)) \\
    &= \frac{2\rsa\left(X,Y\right)}{1 + \rsa\left(X,Y\right)}.
\end{align*}


\subsection{StreamNet Architecture Variants}
\label{ss:methods-architectures}
We developed shallow, multiple-streamed architectures for mouse visual cortex, shown in Figure~\ref{fig:main-lossfct}A.
There are three main modules in our architecture: shallow, intermediate, and deep.
The shallow and deep modules each consist of one convolutional layer and the intermediate module consists of a block of two convolutional layers.
Thus, the longest length of the computational graph, excluding the readout module, is four (i.e., $1 + 2 + 1$).
Depending on the number of parallel streams in the model, the intermediate module would contain multiple branches (in parallel), each receiving input from the shallow module.
The outputs of the intermediate modules are then passed through one convolutional operation (deep module).
Finally, the outputs of each parallel branch would be summed together, concatenated across the channels dimension, and used as input for the readout module.
Table~\ref{tab:mouse_model_definition} describes the parameters of three model variants, each containing one ($N=1$), two ($N=2$), or six ($N=6$) parallel branches.

\begin{table}[ht]
\begin{center}
\begin{tabular}{| 
>{\centering\arraybackslash}m{0.15\columnwidth} || >{\centering\arraybackslash}m{0.15\columnwidth} | >{\centering\arraybackslash}m{0.18\columnwidth} | >{\centering\arraybackslash}m{0.18\columnwidth} | >{\centering\arraybackslash}m{0.18\columnwidth} |}
\hline
    \textbf{Module Name} 
        & \textbf{Output Size} 
        & \textbf{Single} ($N=1$) 
        & \textbf{Dual} ($N=2$) 
        & \textbf{Six} ($N=6$)
        \\ \hline\hline
    Input 
        & $64 \times 64$ 
        & N/A 
        & N/A 
        & N/A
        \\ \hline
    Shallow 
        & $7 \times 7$ 
        & $(64, 11, 4, 2)$ 
        & $(64, 11, 4, 2)$ 
        & $(64, 11, 4, 2)$
        \\ \hline
    Intermediate 
        & $3 \times 3$
        & $\begin{bmatrix} (192,5,1,2) \\ (384,3,1,1) \end{bmatrix}$ 
        & $\begin{bmatrix} (192,5,1,2) \\ (384,3,1,1) \end{bmatrix} \times 2$ 
        & $\begin{bmatrix} (192,5,1,2) \\ (384,3,1,1) \end{bmatrix} \times 6$
        \\ \hline
    Deep 
        & $3 \times 3$ 
        & If inputs are from intermediate: $(256,3,1,1)$, otherwise: $(256,3,2,0)$ 
        & If inputs are from intermediate: $(256,3,1,1)$, otherwise: $(256,3,2,0)$ 
        & If inputs are from intermediate: $(256,3,1,1)$, otherwise: $(256,3,2,0)$
        \\ \hline
\end{tabular}
\caption[Neural network parameters and output sizes for the convolutional layers of our StreamNet model variants containing one, two, and six parallel branches in the intermediate module]{\textbf{Neural network parameters and output sizes for the convolutional layers of our StreamNet model variants containing one, two, and six parallel branches in the intermediate module.}  One convolutional layer is denoted by a tuple: (number of filters, filter size, stride, padding).  A block of convolutional layers is denoted by a list of tuples, where each tuple in the list corresponds to a single convolutional layer. When a list of tuples is followed by ``$\times N$", this means that the convolutional parameters for each of the $N$ parallel branches are the same.}
\label{tab:mouse_model_definition}
\end{center}
\end{table}

\subsection{Neural Network Training Objectives}
\label{ss:methods-task}
In this section, we briefly describe the supervised and unsupervised objectives that were used to train our models.

\subsubsection{Supervised Training Objective}
The loss function $\L$ used in supervised training is the cross-entropy loss, defined as follows:
\begin{equation}
\label{eq:cross_entropy_loss}
    \L(\bX; \btheta) = - \frac{1}{N} \sum_{i=1}^N \log \left( \frac{\exp(\bX_i[\bc_i])}{\sum_{j=0}^{C-1} \exp(\bX_i[j])} \right),
\end{equation}
where $N$ is the batch size, $C$ is the number of categories for the dataset, $\bX \in \R^{N \times C}$ are the model outputs (i.e., logits) for the $N$ images, $\bX_i \in \R^C$ are the logits for the $i$th image, $\bc_i \in [0, C-1]$ is the category index of the $i$th image (zero-indexed), and $\btheta$ are the model parameters.
Equation~\eqref{eq:cross_entropy_loss} was minimized using stochastic gradient descent (SGD) with momentum~\citep{Bottou2010}.

\paragraph{ImageNet {\normalfont\citep{Deng2009}}} 
This dataset contains approximately $1.3$ million images in the train set and 50,000 images in the validation set.  
Each image was previously labeled into $C=1000$ distinct categories.

\paragraph{CIFAR-10 {\normalfont\citep{Krizhevsky2009}}} 
This dataset contains 50,000 images in the train set and 10,000 images in the validation set.  
Each image was previously labeled into $C=10$ distinct categories.

\subsubsection{Unsupervised Training Objectives}
\paragraph{Sparse Autoencoder {\normalfont\citep{Olshausen1996}}}
The goal of this objective is to reconstruct an image from a sparse image embedding.
In order to generate an image reconstruction, we used a mirrored version of each of our StreamNet variants.
Concretely, the loss function was defined as follows:
\begin{equation}
    \L(\bx; \btheta) = \frac{1}{2 \cdot 64^2} \lVert f(\bx) - \bx \rVert_2^2 + \frac{\lambda}{128} \lVert \bv \rVert_1,
\end{equation}
where $\bv \in \R^{128}$ is the image embedding, $f$ is the (mirrored) model, $f(\bx)$ is the image reconstruction, $\bx$ is a $64 \times 64$ pixels image, $\lambda$ is the regularization coefficient, and $\btheta$ are the model parameters.

Our single-, dual-, and six-stream variants were trained using a batch size of $256$ for $100$ epochs using SGD with momentum of $0.9$ and weight decay of $0.0005$.
The initial learning rate was set to $0.01$ for the single- and dual-stream variants and was set to $0.001$ for the six-stream variant.
The learning rate was decayed by a factor of $10$ at epochs $30$, $60$, and $90$.
For all the StreamNet variants, the embedding dimension was set to $128$ and the regularization coefficient was set to $0.0005$.

\paragraph{Depth Prediction {\normalfont\citep{Zhang2017}}}
The goal of this objective is to predict the depth map of an image.
We used a synthetically generated dataset of images known as PBRNet \citep{Zhang2017}.
It contains approximately 50,0000 images and their associated depth maps.
Similar to the loss function used in the sparse autoencoder objective, we used a mean-squared loss to train the models.
The output (i.e., depth map prediction) was generated using a mirrored version of each of our StreamNet variants.
In order to generate the depth map, we appended one final convolutional layer onto the output of the mirrored architecture in order to downsample the three image channels to one image channel.
During training, random crops of size $224 \times 224$ pixels were applied to the image and depth map (which were both subsequently resized to $64 \times 64$ pixels).
In addition, both the image and depth map were flipped horizontally with probability $0.5$.
Finally, prior to the application of the loss function, each depth map was normalized such that the mean and standard deviation across pixels were zero and one respectively.

Each of our single-, dual-, and six-stream variants were trained using a batch size of $256$ for $50$ epochs using SGD with momentum of $0.9$, and weight decay of $0.0001$.
The initial learning rate was set to $10^{-4}$ and was decayed by a factor of $10$ at epochs $15$, $30$, and $45$.

\paragraph{RotNet {\normalfont\citep{Gidaris2018}}} 
The goal of this objective is to predict the rotation of an image.  
Each image of the ImageNet dataset was rotated four ways ($0^\circ$, $90^\circ$, $180^\circ$, $270^\circ$) and the four rotation angles were used as ``pseudo-labels" or ``categories".
The cross-entropy loss was used with these pseudo-labels as the training objective (i.e., Equation~\eqref{eq:cross_entropy_loss} with $C=4$).

Our single-, dual-, and six-stream variants were trained using a batch size of $192$ (which is effectively a batch size of $192 \times 4 = 768$ due to the four rotations for each image) for $50$ epochs using SGD with nesterov momentum of $0.9$, and weight decay of $0.0005$.
An initial learning rate of $0.01$ was decayed by a factor of $10$ at epochs $15$, $30$, and $45$.

\paragraph{Instance Recognition {\normalfont\citep{Wu2018}}} The goal of this objective is to be able to differentiate between embeddings of augmentations of one image from embeddings of augmentations of other images.
Thus, this objective function is an instance of the class of contrastive objective functions.

A random image augmentation is first performed on each image of the ImageNet dataset (random resized cropping, random grayscale, color jitter, and random horizontal flip).
Let $\bx$ be an image augmentation, and $f(\cdot)$ be the model backbone composed with a one-layer linear multi-layer perceptron (MLP) of size $128$.
The image is then embedded onto a $128$-dimensional unit-sphere as follows:
\begin{align*}
    \bz &= f(\bx) / \lVert f(\bx) \rVert_2, \qquad \bz \in \R^{128}.
\end{align*}
Throughout model training, a memory bank containing embeddings for each image in the train set is maintained (i.e., the size of the memory bank is the same as the size of the train set).
The embedding $\bz$ will be ``compared" to a subsample of these embeddings.
Concretely, the loss function $\L$ for one image $\bx$ is defined as follows:
\begin{align}
\label{eq:instance_discrimination_loss}
    h(\bu) &= \frac{\exp(\bu \cdot \bz / \tau) / Z}{\exp(\bu \cdot \bz / \tau) / Z + (m/N)}, \nonumber \\
    \L(\bx; \btheta) &= - \log h(\bv) - \sum_{j=1}^m \log \left( 1 - h(\bv_j) \right),
\end{align}
where $\bv \in \R^{128}$ is the embedding for image $\bx$ that is currently stored in the memory bank, $N$ is the size of the memory bank, $m=4096$ is the number of ``negative" samples used, $\{\bv_j\}_{j=1}^m$ are the negative embeddings sampled from the memory bank uniformly, $Z$ is some normalization constant, $\tau=0.07$ is a temperature hyperparameter, and $\btheta$ are the parameters of $f$.
From Equation~\eqref{eq:instance_discrimination_loss}, we see that we want to maximize $h(\bv)$, which corresponds to maximizing the similarity between $\bv$ and $\bz$ (recall that $\bz$ is the embedding for $\bx$ obtained using $f$).
We can also see that we want to maximize $1 - h(\bv_j)$ (or minimize $h(\bv_j)$).
This would correspond to minimizing the similarity between $\bv_j$ and $\bz$ (recall that $\bv_j$ are the negative embeddings).

After each iteration of training, the embeddings for the current batch are used to update the memory bank (at their corresponding positions in the memory bank) via a momentum update.
Concretely, for image $\bx$, its embedding in the memory bank $\bv$ is updated using its current embedding $\bz$ as follows:
\begin{align*}
    \bv &\leftarrow \lambda \bv + (1 - \lambda) \bz, \\
    \bv &\leftarrow \bv / \lVert \bv \rVert_2,
\end{align*}
where $\lambda=0.5$ is the momentum coefficient.
The second operation on $\bv$ is used to project $\bv$ back onto the $128$-dimensional unit sphere.

Our single-, dual-, and six-stream variants were trained using a batch size of $256$ for $200$ epochs using SGD with momentum of $0.9$, and weight decay of $0.0005$.
An initial learning rate of $0.03$ was decayed by a factor of $10$ at epochs $120$ and $160$.

\paragraph{SimSiam {\normalfont\citep{Chen2020Siam}}} The goal of this objective is to maximize the similarity between the embeddings of two augmentations of the same image.
Thus, SimSiam is another instance of the class of contrastive objective functions.

Two random image augmentations (e.g., random resized crop, random horizontal flip, color jitter, random grayscale, and random Gaussian blur) are first generated for each image in the ImageNet dataset.
Let $\bx_1$ and $\bx_2$ be the two augmentations of the same image, $f(\cdot)$ be the model backbone, $g(\cdot)$ be a three-layer non-linear MLP, and $h(\cdot)$ be a two-layer non-linear MLP.
The three-layer MLP has hidden dimensions of $2048$, $2048$, and $2048$.
The two-layer MLP has hidden dimensions of $512$ and $2048$ respectively.
Let $\btheta$ be the parameters for $f$, $g$, and $h$.
The loss function $\L$ for one image $\bx$ of a batch is defined as follows (recall that $\bx_1$ and $\bx_2$ are two augmentations of one image):
\begin{align}
\label{eq:simsiam_loss}
    \bp_1 = h \circ g \circ f(\bx_1), \qquad \bp_2 &= h \circ g \circ f(\bx_2), \qquad \bz_1 = g \circ f(\bx_1), \qquad \bz_2 = g \circ f(\bx_2), \nonumber \\
    \L(\bx_1, \bx_2; \btheta) &= -\frac{1}{2} \left( \frac{\bz_1 \cdot \bp_2}{\lVert \bz_1 \rVert_2 \lVert \bp_2 \rVert_2} + \frac{\bz_2 \cdot \bp_1}{\lVert \bz_2 \rVert_2 \lVert \bp_1 \rVert_2} \right),
\end{align}
where $\bz_1, \bz_2, \bp_1, \bp_2 \in \R^{2048}$.
Note that $\bz_1$ and $\bz_2$ are treated as constants in this loss function (i.e., the gradients are not back-propagated through $\bz_1$ and $\bz_2$).
This ``stop-gradient" method was key to the success of this objective function.

Our single-, dual-, and six-stream variants were trained using a batch size of $512$ for $100$ epochs using SGD with momentum of $0.9$, and weight decay of $0.0001$.
An initial learning rate of $0.1$ was used, and the learning rate was decayed to $0.0$ using a cosine schedule (with no warm-up).

\paragraph{MoCov2 {\normalfont\citep{He2020, Chen2020mocov2}}} The goal of this objective is to be able to distinguish augmentations of one image (i.e., by labeling them as ``positive") from augmentations of other images (i.e., by labeling them as ``negative").
Intuitively, embeddings of different augmentations of the same image should be more ``similar" to each other than to embeddings of augmentations of other images.
Thus, this algorithm is another instance of the class of contrastive objective functions and is similar conceptually to instance recognition.

Two image augmentations are first generated for each image in the ImageNet dataset by applying random resized cropping, color jitter, random grayscale, random Gaussian blur, and random horizontal flips.
Let $\bx_1$ and $\bx_2$ be the two augmentations for one image.
Let $f_q(\cdot)$ be a query encoder, which is a model backbone composed with a two-layer non-linear MLP of dimensions $2048$ and $128$ respectively and let $f_k(\cdot)$ be a key encoder, which has the same architecture as $f_q$.
$\bx_1$ is encoded by $f_q$ and $\bx_2$ is encoded by $f_k$ as follows:
\begin{align*}
    \bv = f_q(\bx_1), \qquad \bk_0 = f_k(\bx_2), \qquad \bv, \bk_0 \in \R^{128}.
\end{align*}
During each iteration of training, a dictionary of size $K$ of image embeddings obtained from previous iterations is maintained (i.e., the dimensions of the dictionary are $K \times 128$).
The image embeddings in this dictionary are used as ``negative" samples.
The loss function $\L$ for one image of a batch is defined as follows:
\begin{equation}
\label{eq:moco_loss}
    \L(\bx_1, \bx_2; \btheta_q) = - \log \frac{\exp(\bv \cdot \bk_{0}/ \tau)}{\sum_{i=0}^K \exp(\bv \cdot \bk_i / \tau)},
\end{equation}
where $\btheta_q$ are the parameters of $f_q$, $\tau=0.2$ is a temperature hyperparameter, $K={65536}$ is the number of ``negative" samples, and $\{\bk_i\}_{i=1}^{K}$ are the embeddings of the negative samples (i.e., the augmentations for other images which are encoded using $f_k$, and are stored in the dictionary).
From Equation~\eqref{eq:moco_loss}, we see that we want to maximize $\bv \cdot \bk_0$, which corresponds to maximizing the similarity between the embeddings of the two augmentations of an image.

After each iteration of training, the dictionary of negative samples is enqueued with the embeddings from the most recent iteration, while embeddings that have been in the dictionary for the longest are dequeued.
Finally, the parameters $\btheta_k$ of $f_k$ are updated via a momentum update, as follows:
\begin{align*}
    \btheta_k \leftarrow \lambda \btheta_k + (1 - \lambda) \btheta_q,
\end{align*}
where $\lambda = 0.999$ is the momentum coefficient.
Note that only $\btheta_q$ are updated with back-propagation.

Our single-, dual-, and six-stream variants were trained using a batch size of $512$ for $200$ epochs using SGD with momentum of $0.9$, and weight decay of $0.0005$.
An initial learning rate of $0.06$ was used, and the learning rate was decayed to $0.0$ using a cosine schedule (with no warm-up).

\paragraph{SimCLR {\normalfont\citep{Chen2020simclr}}} The goal of this objective is conceptually similar to that of MoCov2, where the embeddings of augmentations of one image should be distinguishable from the embeddings of augmentations of other images.
Thus, SimCLR is another instance of the class of contrastive objective functions.

Similar to other contrastive objective functions, two image augmentations are first generated for each image in the ImageNet dataset (by using random cropping, random horizontal flips, random color jittering, random grayscaling and random Gaussian blurring).  Let $f(\cdot)$ be the model backbone composed with a two-layer non-linear MLP of dimensions $2048$ and $128$ respectively.  The two image augmentations are first embedded into a $128$-dimensional space and normalized:
\begin{align*}
    \bz_1 &= f(\bx_1) / \lVert f(\bx_1) \rVert_2, \qquad \bz_2 = f(\bx_2) / \lVert f(\bx_2) \rVert_2, \qquad \bz_1, \bz_2 \in \R^{128}.
\end{align*}
The loss function $\L$ for a single pair of augmentations of an image is defined as follows:
\begin{equation}
    \label{eq:simclr_loss}
    \L(\bx_1, \bx_2; \btheta) = -\log \frac{\exp(\bz_1 \cdot \bz_2 / \tau)}{\sum_{i=1}^{2N} \indicator [i \neq 1] \exp(\bz_1 \cdot \bz_i / \tau)},
\end{equation}
where $\tau = 0.1$ is a temperature hyperparameter, $N$ is the batch size, $\indicator [i \neq 1 ]$ is equal to $1$ if $i \neq 1$ and $0$ otherwise, and $\btheta$ are the parameters of $f$.  The loss defined in Equation~\eqref{eq:simclr_loss} is computed for every pair of images in the batch (including their augmentations) and subsequently averaged.  

Our single-, dual-, and six-stream variants were trained using a batch size of $4096$ for $200$ epochs using layer-wise adaptive rate scaling \citep[LARS,][]{you2017large} with momentum of $0.9$, and weight decay of $10^{-6}$.
An initial learning rate of $4.8$ was used and decayed to $0.0$ using a cosine schedule.
A linear warm-up of $10$ epochs was used for the learning rate with warm-up ratio of $0.0001$.

\subsection{Top-1 Validation Set Performance}
\label{ss:top1-perf}
\subsubsection{Performance of primate models on $224 \times 224$ pixels and $64 \times 64$ pixels ImageNet}
Here we report the top-1 validation set accuracy of models trained in a supervised manner on $64 \times 64$ pixels and $224 \times 224$ pixels ImageNet.

\begin{center}
\begin{tabular}{| c | c | c | c |} \hline
\textbf{Architecture} & \textbf{Image Size} & \textbf{Objective Function} & \textbf{Top-1 Accuracy} \\ \hline\hline
\multirow{2}{*}{AlexNet} & $224 \times 224$ & \multirow{6}{*}{Supervised (ImageNet)} & 56.52\% \\ \cline{2-2}\cline{4-4}
 & $64 \times 64$ &  & 36.22\% \\ \cline{1-2}\cline{4-4}
\multirow{2}{*}{VGG16} & $224 \times 224$ &  & 71.59\% \\ \cline{2-2}\cline{4-4}
 & $64 \times 64$ &  & 58.32\% \\ \cline{1-2}\cline{4-4}
\multirow{2}{*}{ResNet-18} & $224 \times 224$ &  & 69.76\% \\ \cline{2-2}\cline{4-4}
 & $64 \times 64$ &  & 53.31\% \\ \hline
\end{tabular}
\end{center}

\subsubsection{Performance of StreamNet Variants on $64 \times 64$ pixels CIFAR-10 and $64 \times 64$ pixels ImageNet}
Here we report the top-1 validation set accuracy of our model variants trained in a supervised manner on $64 \times 64$ pixels CIFAR-10 and ImageNet.

\begin{center}
\begin{tabular}{| c | c | c | c | c |} \hline
\textbf{Architecture} & \textbf{Dataset} & \textbf{Objective Function} & \textbf{Top-1 Accuracy} \\ \hline\hline
\multirow{2}{*}{Single Stream} & CIFAR-10 & \multirow{6}{*}{Supervised} & 76.52\% \\ \cline{2-2}\cline{4-4}
 & ImageNet &  & 34.87\% \\ \cline{1-2}\cline{4-4}
\multirow{2}{*}{Dual Stream} & CIFAR-10 &  & 81.13\% \\ \cline{2-2}\cline{4-4}
 & ImageNet &  & 38.68\% \\ \cline{1-2}\cline{4-4}
\multirow{2}{*}{Six Stream} & CIFAR-10 &  & 78.73\% \\ \cline{2-2}\cline{4-4}
 & ImageNet &  & 34.15\% \\ \hline
\end{tabular}
\end{center}

\subsubsection{Transfer Performance of StreamNet Variants on $64 \times 64$ pixels ImageNet Under Linear Evaluation for Models Trained with Unsupervised Objectives}
In this subsection, we report the top-1 ImageNet validation set performance under linear evaluation for models trained with unsupervised objectives.
After training each model on a unsupervised objective, the model backbone weights are then held fixed and a linear readout head is trained on top of the fixed model backbone.
In the case where the objective function is ``untrained'', model parameters were randomly initialized and held fixed while the linear readout head was trained.
The image augmentations used during transfer learning were random cropping and random horizontal flipping.
The linear readout for every unsupervised model was trained with the cross-entropy loss function (i.e., Equation~\eqref{eq:cross_entropy_loss} with $C=1000$) for $100$ epochs, which was minimized using SGD with momentum of $0.9$, and weight decay of $10^{-9}$.
The initial learning rate was set to $0.1$ and reduced by a factor of $10$ at epochs $30$, $60$, and $90$.

\begin{table}
    \centering
\begin{tabular}{| c | c | c | c | c |} \hline
\multirow{2}{*}{\textbf{Architecture}} & \multirow{2}{*}{\textbf{Objective Function}} & \textbf{ImageNet Transfer} & \textbf{Neural Predictivity} \\
 & & \textbf{Top-1 Accuracy} & \textbf{Neuropixels; Calcium Imaging} \\\hline\hline
\multirow{8}{*}{Single Stream} & Untrained & 9.28\% &  32.76\%; 28.65\%\\ \cline{2-4}
 & Supervised & 34.87\% & 36.21\%; 29.73\%\\ \cline{2-4}
 & Autoencoder & 10.37\% & 35.99\%; 28.69\%\\ \cline{2-4}
 & Depth Prediction & 18.04\% & 33.79\%; 27.54\%\\ \cline{2-4}
 & RotNet & 19.72\% & 35.63\%; 29.27\%\\ \cline{2-4}
 & Instance Recognition & 21.22\% & 38.01\%; 30.88\%\\ \cline{2-4}
 & SimSiam & 26.48\% & 39.19\%; 30.48\%\\ \cline{2-4}
 & MoCov2 & 27.63\% & 39.17\%; 30.30\%\\ \cline{2-4}
 & SimCLR & 22.84\% & 39.45\%; 29.50\%\\ \hline\hline
\multirow{8}{*}{Dual Stream} & Untrained & 10.85\% & 33.58\%; 29.24\%\\ \cline{2-4} 
 & Supervised & 38.68\% & 36.07\%; 29.43\%\\ \cline{2-4}
 & Autoencoder & 10.26\% & 34.97\%; 28.74\%\\ \cline{2-4}
 & Depth Prediction & 19.81\% & 32.81\%; 27.20\%\\ \cline{2-4}
 & RotNet & 23.29\% & 35.37\%; 29.15\%\\ \cline{2-4}
 & Instance Recognition & 22.55\% & 40.07\%; 30.64\%\\ \cline{2-4}
 & SimSiam & 29.21\% & 40.20\%; 30.60\%\\ \cline{2-4}
 & MoCov2 & 31.00\% & 38.64\%; 30.33\%\\ \cline{2-4}
 & SimCLR & 26.25\% & 38.03\%; 29.08\%\\ \hline\hline
\multirow{8}{*}{Six Stream} & Untrained & 11.12\% & 33.74\%; 29.26\%\\ \cline{2-4} 
 & Supervised & 34.15\% &  36.64\%; 29.79\%\\ \cline{2-4}
 & Autoencoder & 9.27\% & 37.34\%; 31.12\%\\ \cline{2-4}
 & Depth Prediction & 18.27\% & 33.12\%; 27.63\%\\ \cline{2-4}
 & RotNet & 22.78\% & 35.49\%; 28.97\%\\ \cline{2-4}
 & Instance Recognition & 26.49\% & 37.67\%; 31.18\%\\ \cline{2-4}
 & SimSiam & 30.52\% & 38.17\%; 30.46\%\\ \cline{2-4}
 & MoCov2 & 32.70\% & 37.96\%; 30.44\%\\ \cline{2-4}
 & SimCLR & 28.42\% & 38.92\%; 29.19\%\\ \hline\hline
AlexNet & Supervised & 36.22\% & 37.28\%; 30.34\%\\ \hline
AlexNet & Instance Recognition & 16.09\% & 41.33\%; 31.60\%\\ \hline\hline
ResNet-18 & Supervised & 53.31\% & 35.82\%; 28.93\%\\ \hline
ResNet-18 & Instance Recognition & 30.75\% & 38.99\%; 30.11\%\\ \hline\hline
VGG16 & Supervised & 58.32\% & 31.92\%; 27.09\%\\ \hline
VGG16 ($224$ px) & Supervised & 71.59\% & 26.03\%; 20.40\%\\ \hline\hline
MouseNet of & \multirow{2}{*}{Supervised} & \multirow{2}{*}{37.14\%} & \multirow{2}{*}{31.05\%; 25.89\%} \\
\cite{Shi2020} & & & \\ \hline
MouseNet & \multirow{2}{*}{Supervised} & \multirow{2}{*}{39.37\%} & \multirow{2}{*}{33.02\%; 26.53\%} \\
Variant & & & \\ \hline
\end{tabular}
    \caption[ImageNet top-1 validation set accuracy via linear transfer or via supervised training and neural predictivity for each model]{\textbf{ImageNet top-1 validation set accuracy via linear transfer or via supervised training and neural predictivity for each model.}
    We summarize here the top-1 accuracy for each unsupervised and supervised model on ImageNet as well as their noise-corrected neural predictivity obtained via the PLS map (aggregated across all visual areas).
    These values are plotted in Figures~\ref{fig:main-lossfct}C and ~\ref{fig:supp-loss-fct-calcium}.
    Unless otherwise stated, each model is trained and validated on $64 \times 64$ pixels images.
    }
    \label{tab:imagenet-top1-transfer}
\end{table}

\subsection{Parameter and Unit Counts for Each Model}
Table~\ref{tab:network-size} summarizes the total number of trainable parameters and the total number of units for each model.
The number of trainable parameters reported excludes those specific to the loss function itself (i.e., the embedding or classification head).
The total number of units for each model was defined as the total number of features used in the neural response fitting procedure for each model layer, summed across all model layers used for neural response fitting.

\begin{table}
    \centering
\begin{tabular}{|c|c|c|c|}
\hline
\textbf{Architecture}           & \textbf{Objective Function}    & \textbf{Parameter Count} & \textbf{Unit Count} \\ \hline\hline
\multirow{10}{*}{Single Stream} & Untrained & 2029632 & 8896 \\ \cline{2-4} 
    & Supervised (CIFAR-10) & 2029632         & 11712      \\ \cline{2-4} 
    & Supervised (ImageNet) & 2029632         & 8896       \\ \cline{2-4} 
    & Autoencoder           & 2029632         & 8896       \\ \cline{2-4} 
    & Depth Prediction      & 2029632         & 8896       \\ \cline{2-4} 
    & RotNet                & 2029632         & 8896       \\ \cline{2-4} 
    & Instance Recognition  & 2029632         & 8896       \\ \cline{2-4} 
    & SimSiam               & 2029632         & 8896       \\ \cline{2-4} 
    & MoCov2                & 2029632         & 8896       \\ \cline{2-4} 
    & SimCLR                & 2029632         & 8896       \\ \hline\hline
\multirow{10}{*}{Dual Stream}   & Untrained & 5806848 & 14656      \\ \cline{2-4} 
    & Supervised (CIFAR-10) & 5806848         & 19392      \\ \cline{2-4} 
    & Supervised (ImageNet) & 5806848         & 14656      \\ \cline{2-4} 
    & Autoencoder           & 5806848         & 14656      \\ \cline{2-4} 
    & Depth Prediction      & 5806848         & 14656      \\ \cline{2-4} 
    & RotNet                & 5806848         & 14656      \\ \cline{2-4} 
    & Instance Recognition  & 5806848         & 14656      \\ \cline{2-4} 
    & SimSiam               & 5806848         & 14656      \\ \cline{2-4} 
    & MoCov2                & 5806848         & 14656      \\ \cline{2-4} 
    & SimCLR                & 5806848         & 14656      \\ \hline\hline
\multirow{10}{*}{Six Stream}    & Untrained & 16780800 & 28480      \\ \cline{2-4} 
    & Supervised (CIFAR-10) & 16780800        & 37824      \\ \cline{2-4} 
    & Supervised (ImageNet) & 16780800        & 28480      \\ \cline{2-4} 
    & Autoencoder           & 16780800        & 28480      \\ \cline{2-4} 
    & Depth Prediction      & 16780800        & 28480      \\ \cline{2-4} 
    & RotNet                & 16780800        & 28480      \\ \cline{2-4} 
    & Instance Recognition  & 16780800        & 28480      \\ \cline{2-4} 
    & SimSiam               & 16780800        & 28480      \\ \cline{2-4} 
    & MoCov2                & 16780800        & 28480      \\ \cline{2-4} 
    & SimCLR                & 16780800        & 28480      \\ \hline\hline
AlexNet & Supervised (ImageNet) & 57003840  & 19072      \\ \hline
AlexNet ($224$ px) & Supervised (ImageNet) & 57003840        & 204672     \\ \hline
AlexNet & Instance Recognition  & 57022528   & 19072      \\ \hline\hline
ResNet-18 & Supervised (ImageNet) & 11176512 & 143872     \\ \hline
ResNet-18 & Instance Recognition  & 11176512 & 143872     \\ \hline\hline
VGG16 & Supervised (ImageNet) & 134260544 & 133120     \\ \hline
VGG16 ($224$ px) & Supervised (ImageNet) & 134260544       & 1538560    \\ \hline\hline
MouseNet of & \multirow{2}{*}{Supervised} & \multirow{2}{*}{5974858} & \multirow{2}{*}{823296} \\
\cite{Shi2020} & & & \\ \hline
MouseNet & \multirow{2}{*}{Supervised} & \multirow{2}{*}{5974858} & \multirow{2}{*}{823296} \\
Variant & & & \\ \hline
\end{tabular}
    \caption[Parameter and unit counts for each model]{\textbf{Parameter and unit counts for each model.}
    Each model is summarized by its total number of trainable parameters (parameter count) and the total number of features used in neural predictions (unit count), excluding those specific to the loss function itself.
    Unless otherwise stated, each model is trained on $64 \times 64$ pixels images.}
    \label{tab:network-size}
\end{table}

\subsection{Evaluating Model Performance on Downstream Visual Tasks}
To evaluate transfer performance on downstream visual tasks, we used the activations from the outputs of the shallow, intermediate, and deep modules of our StreamNet variants.
We also included the average-pooling layer in all the variants (the model layer prior to the fully-connected readout layer).
The dimensionality of the activations was then reduced to $1000$ dimensions using principal components analysis (PCA), if the number of features exceeded $1000$.
PCA was not used if the number of features was less than or equal to $1000$.
A linear readout on these features was then used to perform five transfer visual tasks.

For the first four object-centric visual tasks (object categorization, pose estimation, position estimation, and size estimation), we used a stimulus set that was used previously in the evaluation of neural network models of the primate visual system \citep{Schrimpf2018, Rajalingham2018, Zhuang2021}.
The stimulus set consists of objects in various poses (object rotations about the $x$, $y$, and $z$ axes), positions (vertical and horizontal coordinates of the object), and sizes, each from eight categories.
We then performed five-fold cross-validation on the training split of the low variation image subset \citep[``Var0'' and ``Var3'', defined in][]{Majaj2015} consisting of $3200$ images, and computed the performance (metrics defined below) on the test split of the high variation set (``Var6'') consisting of $1280$ images.
Ten different category-balanced train-test splits were randomly selected, and the performance of the best model layer (averaged across train-test splits) was reported for each model.
All images were resized to $64\times 64$ pixels prior to fitting, to account for the visual acuity adjustment.
The final non-object-centric task was texture recognition, using the Describable Textures Dataset \citep{Cimpoi2014}.

\paragraph{Object Categorization}
We fit a linear support vector classifier to each model layer activations that were transformed via PCA.
The regularization parameter,
\begin{multline}\label{svm-c-val}
C \in [10^{-8}, 5 \times 10^{-8}, 10^{-7}, 5 \times 10^{-7}, 10^{-6}, 5 \times 10^{-6}, 10^{-5}, 5 \times 10^{-5}, \\ 10^{-4}, 5 \times 10^{-4}, 10^{-3}, 5 \times 10^{-3}, 10^{-2}, 5\times 10^{-2}, 10^{-1}, 5\times 10^{-1}, \\ 1, 5, 10^2, 5\times 10^2, 10^3, 5 \times 10^3, 10^4, 5\times 10^4, \\ 10^5, 5 \times 10^5, 10^6, 5 \times 10^6, 10^7, 5 \times 10^7, 10^8, 5 \times 10^8], 
\end{multline}
was chosen by five-fold cross validation.
The categories are Animals, Boats, Cars, Chairs, Faces, Fruits, Planes, and Tables.
We reported the classification accuracy average across the ten train-test splits.

\paragraph{Position Estimation}
We predicted both the vertical and the horizontal locations of the object center in the image.
We used Ridge regression where the regularization parameter was selected from:
\begin{equation}\label{ridge-alpha-val}
\alpha = 1 / C,
\end{equation}
where $C$ was selected from the list defined in \eqref{svm-c-val}.
For each network, we reported the correlation averaged across both locations for the best model layer.

\paragraph{Pose Estimation}
This task was similar to the position prediction task except that the prediction target were the $z$-axis (vertical axis) and the $y$-axis (horizontal axis) rotations, both of which ranged between $-90$ degrees and $90$ degrees. 
The $(0, 0, 0)$ angle was defined in a per-category basis and was chosen to make the $(0, 0, 0)$ angle ``semantically'' consistent across different categories.
We refer the reader to \cite{Hong2016} for more details.
We used Ridge regression with $\alpha$ chosen from the range in \eqref{ridge-alpha-val}.

\paragraph{Size Estimation}
The prediction target was the three-dimensional object scale, which was used to generate the image in the rendering process. 
This target varied between $0.625$ to $1.6$, which was a relative measure to a fixed canonical size of $1$. 
When objects were at the canonical size, they occluded around $40\%$ of the image on the longest axis.
We used Ridge regression with $\alpha$ chosen from the range in \eqref{ridge-alpha-val}.

\paragraph{Texture Recognition}
We trained linear readouts of the model layers on texture recognition using the Describable Textures Dataset~\citep{Cimpoi2014}, which consists of $5640$ images organized according to $47$ categories, with $120$ images per category.
We used ten category-balanced train-test splits, provided by their benchmark.
Each split consists of $3760$ train-set images and $1880$ test-set images.
A linear support vector classifier was then fit with $C$ chosen in the range \eqref{svm-c-val}.
We reported the classification accuracy average across the ten train-test splits.

\section*{Data Availability}
The calcium imaging and Neuropixels datasets were both obtained from the Allen Brain Atlas using the AllenSDK package (\url{https://allensdk.readthedocs.io/en/latest/index.html}).

\section{Supplemental Figures}

\begin{figure}[htbp]
    \centering
    \includegraphics[width=\columnwidth]{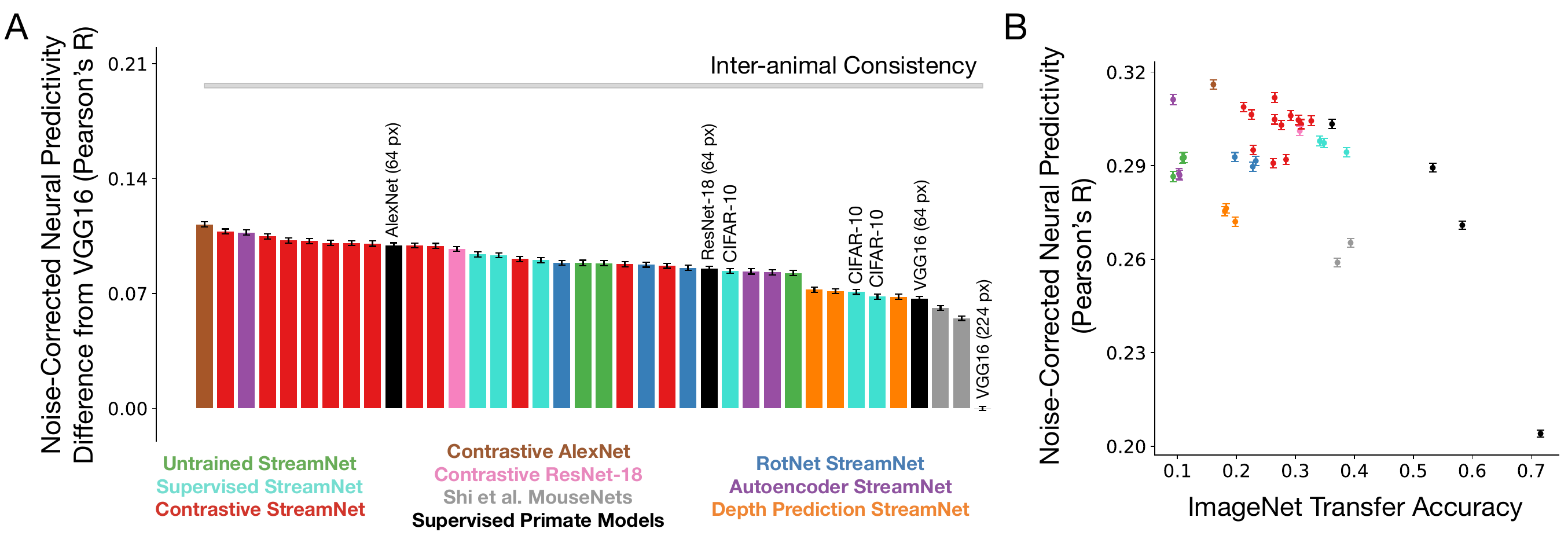}
    \caption[Shallow architectures trained with contrastive objective functions yield the best matches to the neural data (calcium imaging dataset)]{\textbf{Shallow architectures trained with contrastive objective functions yield the best matches to the neural data (calcium imaging dataset).}
    As in Figure~\ref{fig:main-lossfct}, but for the calcium imaging dataset.
    \textbf{A}. The median and s.e.m. neural predictivity, using PLS regression, across neurons in all mouse visual areas except VISrl.
    $N=16228$ units in total (VISrl is excluded, as mentioned in Section~\ref{sec:results-upper}).
    Red denotes our StreamNet models trained on contrastive objective functions, blue denotes our StreamNet models trained on RotNet, turquoise denotes our StreamNet models trained in a supervised manner on ImageNet and on CIFAR-10, green denotes untrained models (random weights), orange denotes our StreamNet models trained depth prediction, purple denotes our StreamNet models trained on autoencoding, brown denotes contrastive AlexNet, pink denotes contrastive ResNet-18 (both trained on instance recognition), black denotes the remaining ImageNet supervised models (primate ventral stream models), and grey denotes the MouseNet of \cite{Shi2020} and our variant of this architecture.
    Actual neural predictivity performance can be found in Table~\ref{tab:imagenet-top1-transfer}.
    \textbf{B}. Each model's performance on ImageNet is plotted against its median neural predictivity across all units from each visual area.
    All ImageNet performance numbers can be found in Table~\ref{tab:imagenet-top1-transfer}.
    Color scheme as in \textbf{A}.
    }
\label{fig:supp-loss-fct-calcium}
\end{figure}

\begin{figure}[htbp]
    \centering
    \includegraphics[width=\columnwidth]{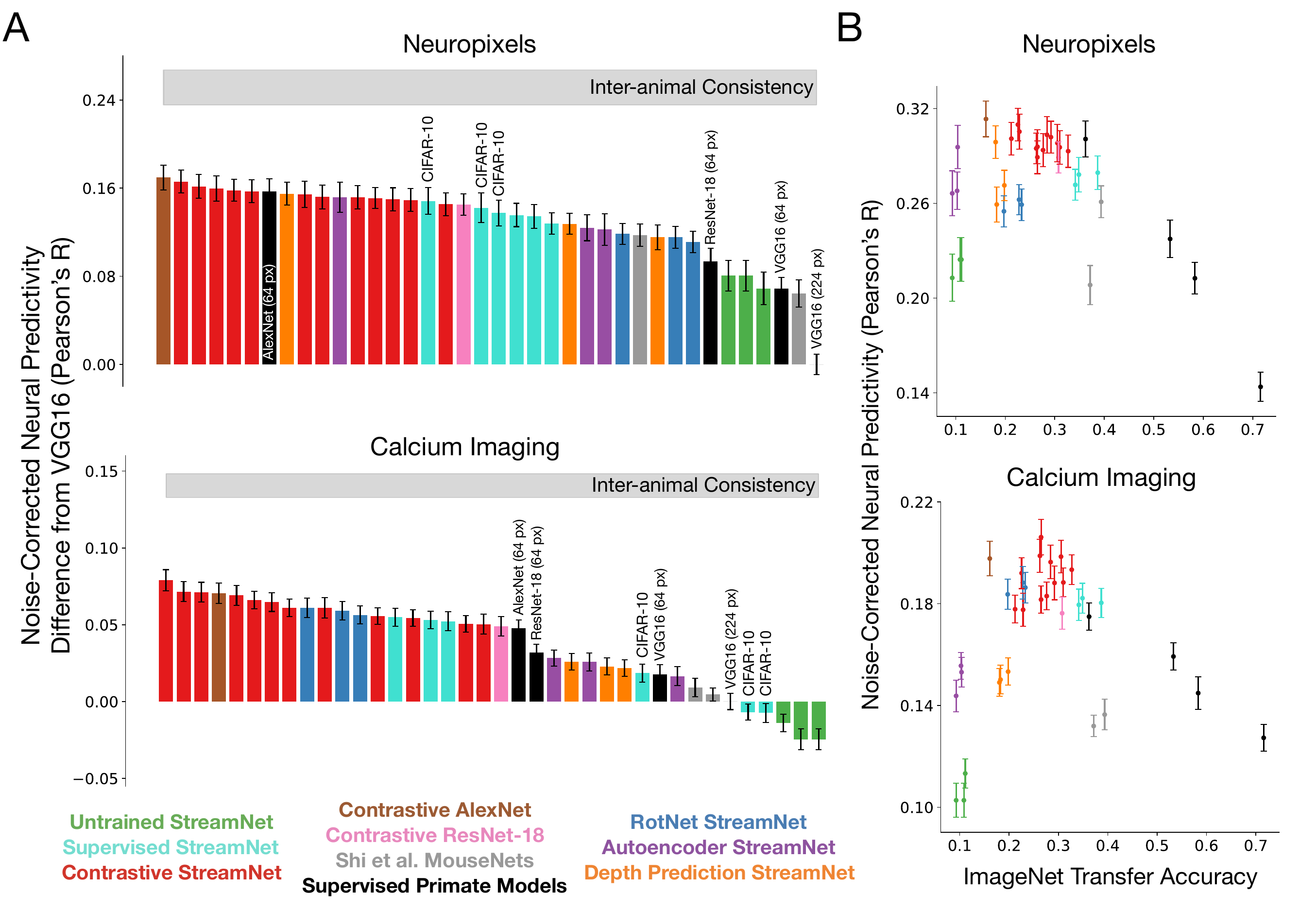}
    \caption[Shallow architectures trained with contrastive objective functions yield the best matches to the neural data (RSA)]{\textbf{Shallow architectures trained with contrastive objective functions yield the best matches to the neural data (RSA).}
    \textbf{A}. The median and s.e.m. noise-corrected neural predictivity, using RSA, across $N=39$ and $N=90$ animals for the Neuropixels and calcium imaging dataset respectively (across all visual areas, with VISrl excluded for the calcium imaging dataset, as mentioned in Section~\ref{sec:results-upper}).
    Red denotes our StreamNet models trained on contrastive objective functions, blue denotes our StreamNet models trained on RotNet, turquoise denotes our StreamNet models trained in a supervised manner on ImageNet and on CIFAR-10, green denotes untrained models (random weights), orange denotes our StreamNet models trained depth prediction, purple denotes our StreamNet models trained on autoencoding, brown denotes contrastive AlexNet, pink denotes contrastive ResNet-18 (both trained on instance recognition), black denotes the remaining ImageNet supervised models (primate ventral stream models), and grey denotes the MouseNet of \cite{Shi2020} and our variant of this architecture.
    \textbf{B}. We plot each model's performance on ImageNet against its median neural predictivity, using RSA, across visual areas.
    All ImageNet performance numbers can be found in Table~\ref{tab:imagenet-top1-transfer}.
    Color scheme as in \textbf{A}.}
\label{fig:supp-loss-fct-rsa}
\end{figure}

\begin{figure}[htbp]
    \centering
    \includegraphics[width=\columnwidth]{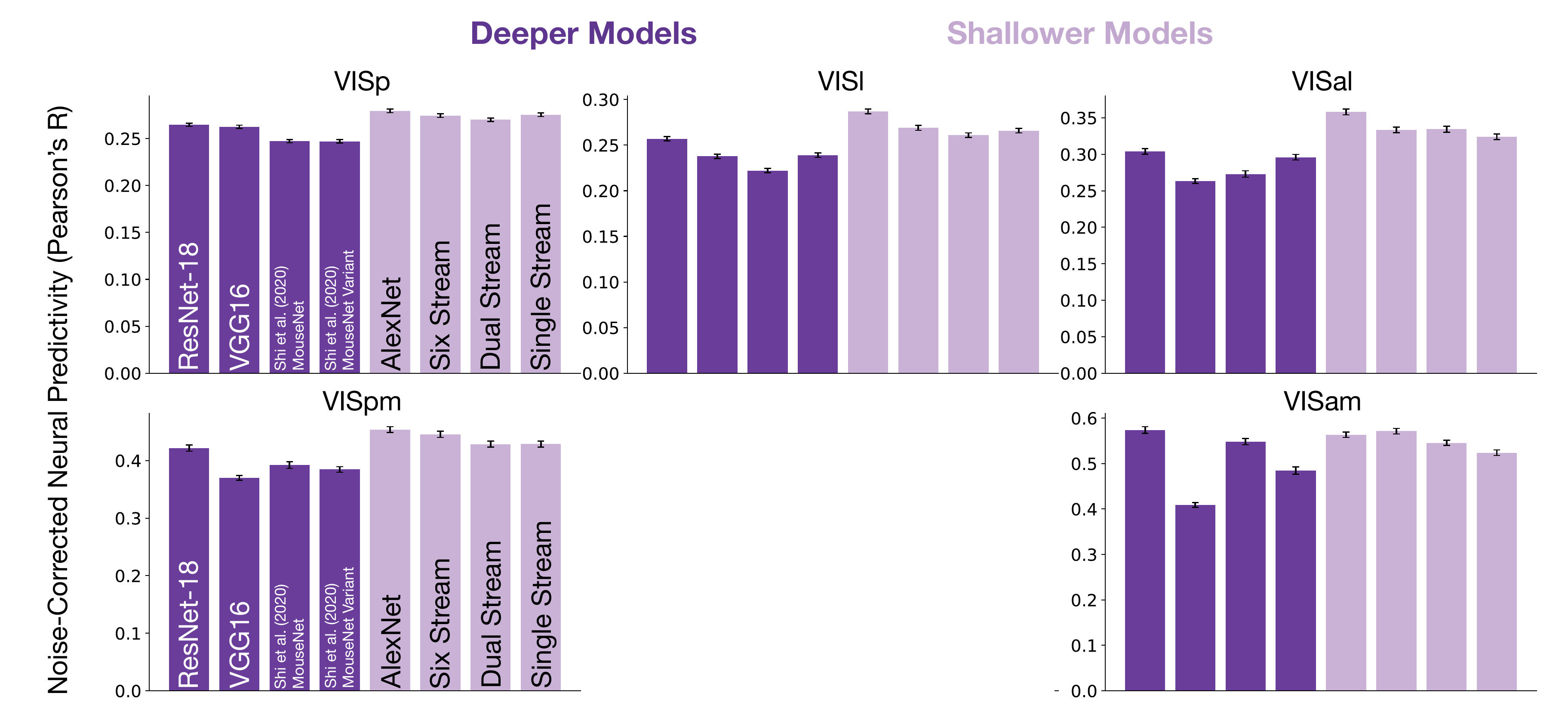}
    \caption[Hierarchically shallow models achieve competitive neural predictivity performance (calcium imaging dataset)]{\textbf{Hierarchically shallow models achieve competitive neural predictivity performance (calcium imaging dataset).}
    As in Figure~\ref{fig:main-calc-imagenet64}C, AlexNet and our StreamNet variants (light purple) were trained in a supervised manner on ImageNet and provide neural predictivity on the calcium imaging dataset that is better or at least as good as those of deeper architectures (dark purple).
    Refer to Table~\ref{tab:neural-data} for $N$ units per visual area.
    As mentioned in Section~\ref{sec:results-upper}, visual area VISrl was removed from the calcium imaging neural predictivity results.
    }
\label{fig:supp-shallow-calcium}
\end{figure}

\begin{figure}[htbp]
    \centering
    \includegraphics[width=\columnwidth]{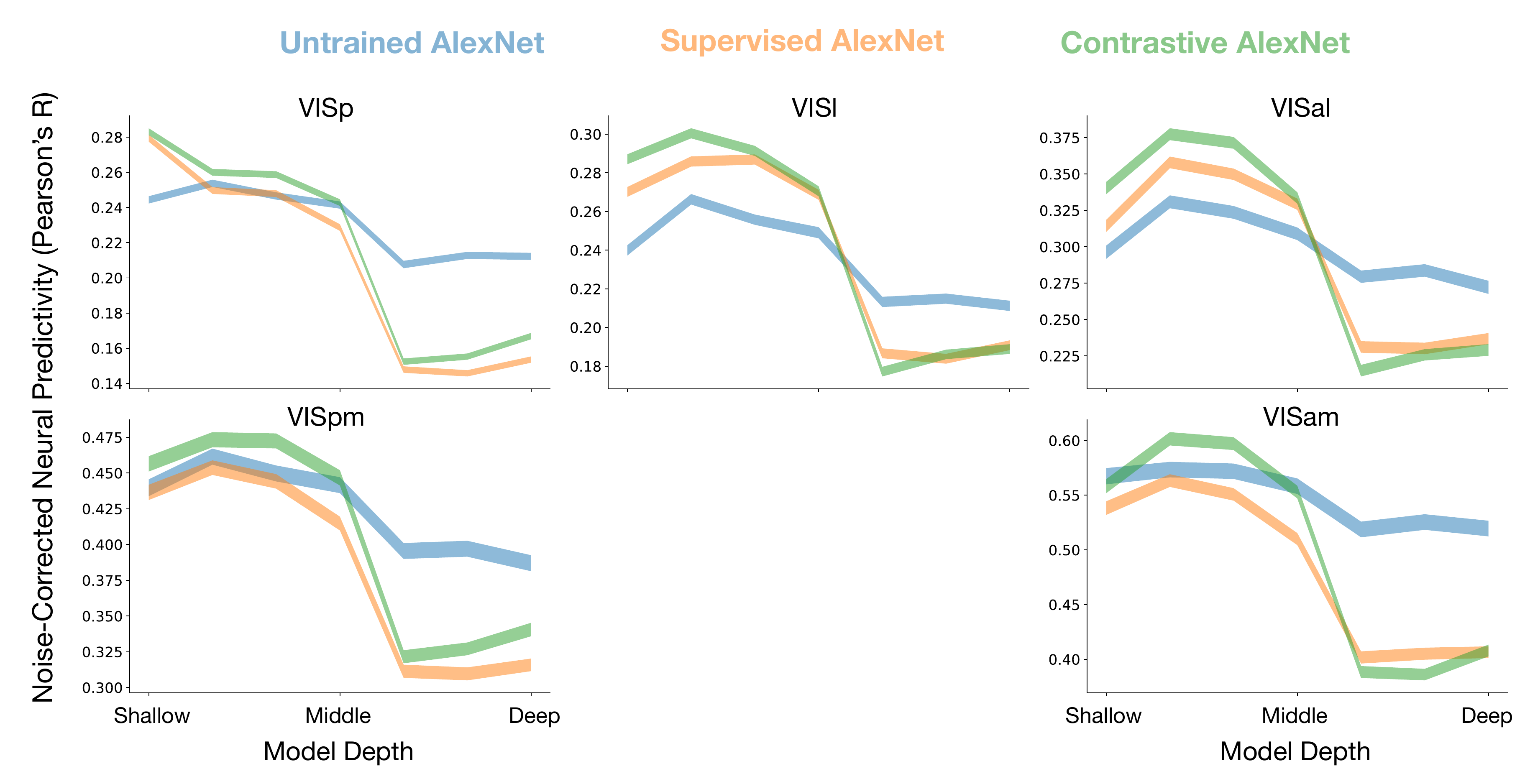}
    \caption[Unsupervised models better predict the neural responses in mouse visual cortex (calcium imaging dataset)]{\textbf{Unsupervised models better predict the neural responses in mouse visual cortex (calcium imaging dataset).}
    As in Figure~\ref{fig:main-loss-fct-vary-alexnet}, AlexNet was either untrained (blue), trained in a supervised manner (orange) or trained in an unsupervised manner (green).
    We observe that the first four convolutional layers provide the best fits to the neural responses for all the visual areas while the latter three layers are not very predictive for any visual area.
    This suggests that an even shallower architecture may be suitable and further corroborates our architectural decision in Figure~\ref{fig:main-calc-imagenet64}B.
    As mentioned in Section~\ref{sec:results-upper}, visual area VISrl was removed from the calcium imaging neural predictivity results.
    }
\label{fig:supp-loss-fct-vary-alexnet}
\end{figure}

\begin{figure}[htbp]
    \centering
    \includegraphics[width=\columnwidth]{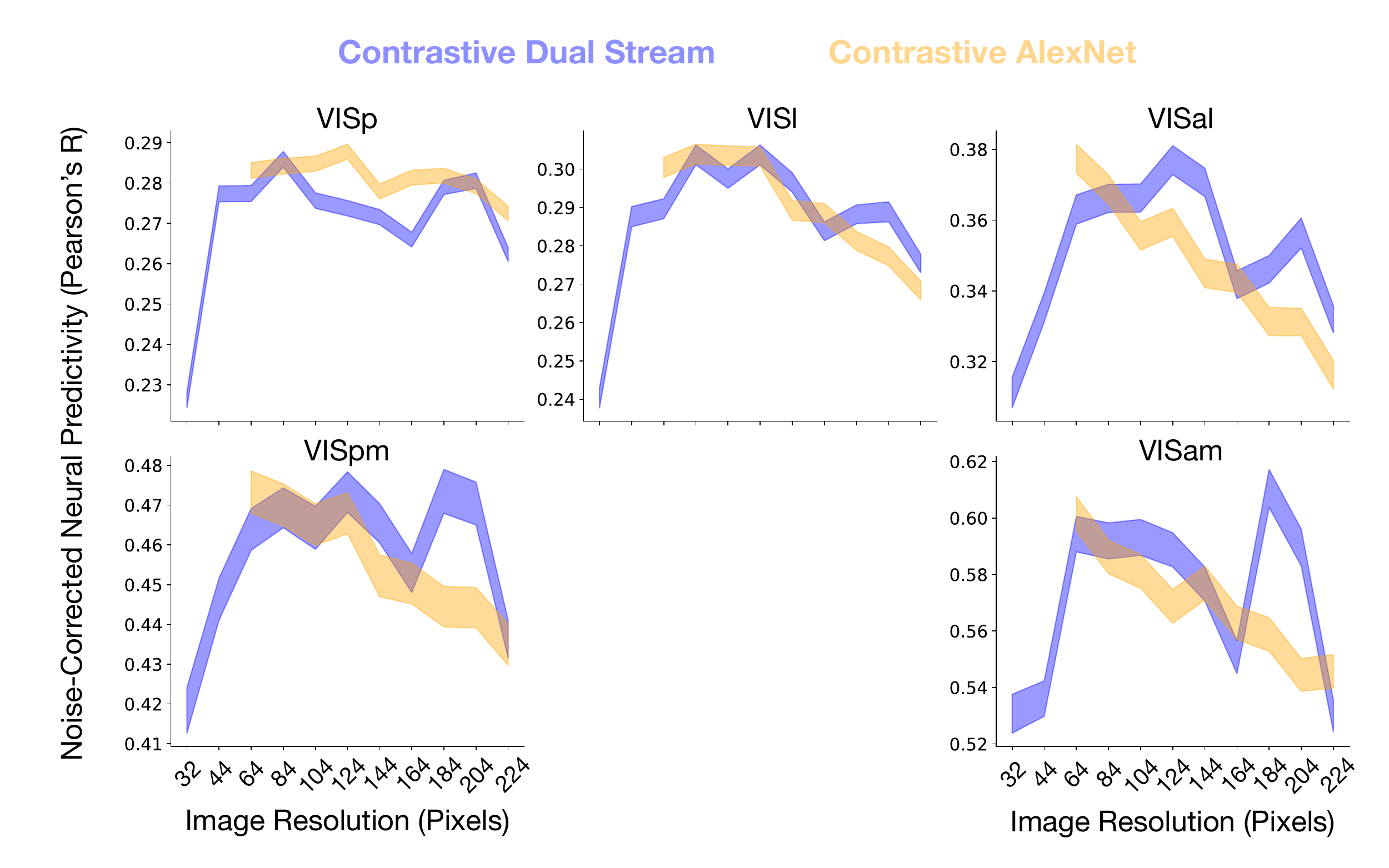}
    \caption[Lower image resolution during model training improves task-optimized neural predictivity (calcium imaging dataset)]{\textbf{Lower image resolution during model training improves task-optimized neural predictivity (calcium imaging dataset)}
    As in Figure~\ref{fig:main-acuity}, models with ``lower visual acuity'' were trained using lower resolution ImageNet images.
    Each image was downsampled from $224 \times 224$ pixels, the image size typically used to train primate ventral stream models, to various image sizes (image sizes on horizontal axis).
    Our dual stream variant (blue) and AlexNet (orange) were trained using various image sizes on instance recognition and their neural predictivity performances were computed for each mouse visual area.
    Training models on resolutions lower than $224 \times 224$ pixels generally led to improved correspondence with the neural responses for both models.
    The median and s.e.m. across neurons in each visual area is reported.
    As mentioned in Section~\ref{sec:results-upper}, visual area VISrl was removed from the calcium imaging neural predictivity results.
    Refer to Table~\ref{tab:neural-data} for $N$ units per visual area.
    }
\label{fig:supp-acuity}
\end{figure}

\begin{figure}[htbp]
    \centering
    \includegraphics[width=\columnwidth]{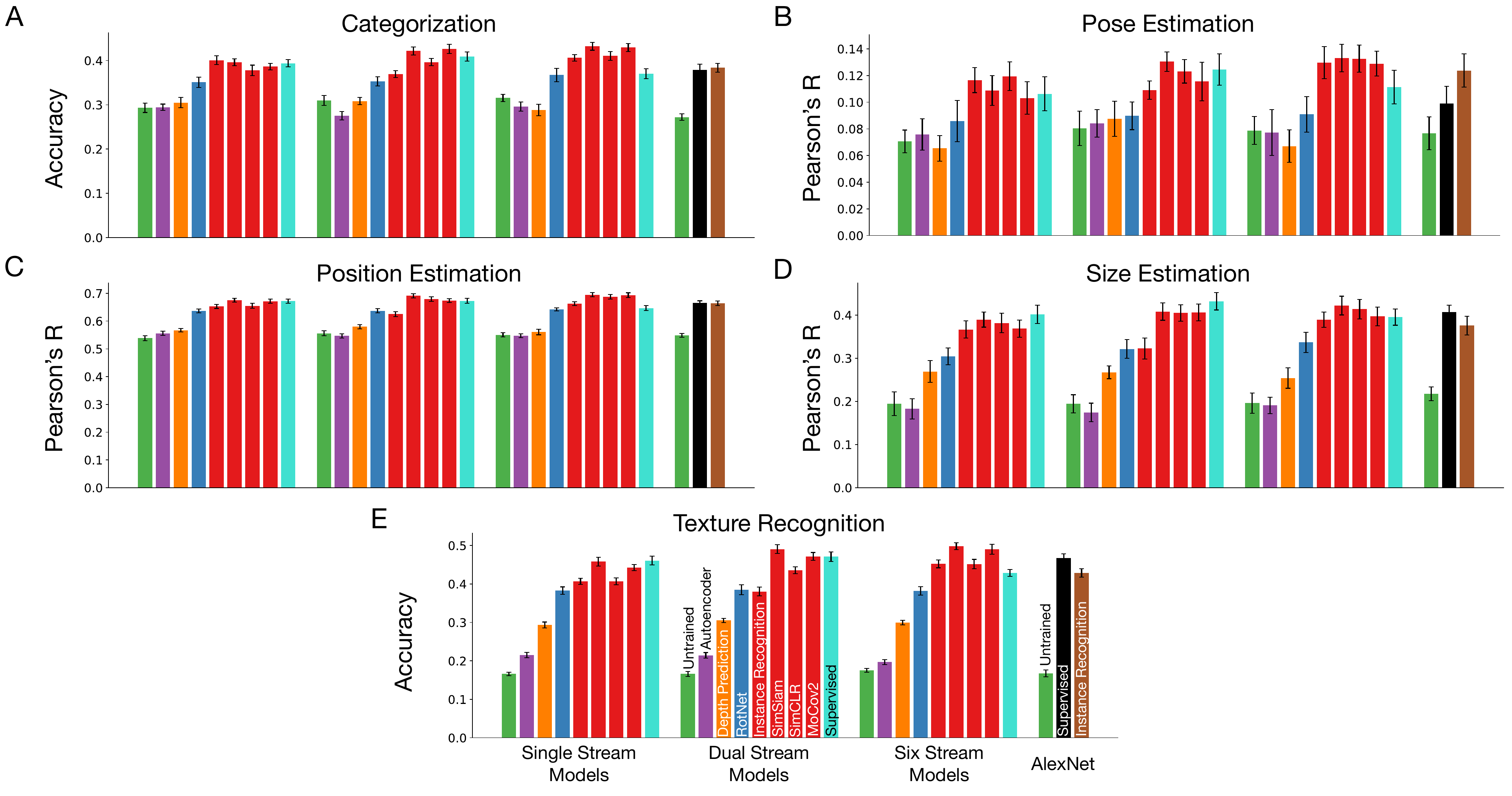}
    \caption[Evaluating visual representations of model variants learned in an unsupervised manner on object-centric and non-object-centric visual tasks]{\textbf{Evaluating visual representations of model variants learned in an unsupervised manner on object-centric and non-object-centric visual tasks.} Red denotes our StreamNet variants trained on contrastive objective functions.
    Blue denotes our models trained on rotation prediction (RotNet), orange denotes our depth prediction models, purple denotes autoencoding models, green are the untrained model.
    The three bars on the right are AlexNet architectures that are, respectively, untrained (green), supervised (black), and contrastive (brown).
    The average performance and its standard deviation (mean and s.t.d.) across $10$ train-test image splits is reported for each transfer task.
    \textbf{A.} Maximum linear transfer performance across model layers on the categorization of objects that are highly varied in terms of their rotation, sizes, and positions in the image.
    \textbf{B.} Object pose estimation accuracy.
    \textbf{C.} Object position estimation accuracy.
    \textbf{D.} Object size estimation accuracy.
    \textbf{E.} Maximum linear transfer performance on $47$-way texture classification (a non-object-centric task).
    }
    \label{fig:task_transfer_performance}
\end{figure}

\begin{figure}[htbp]
    \centering
    \includegraphics[width=\columnwidth]{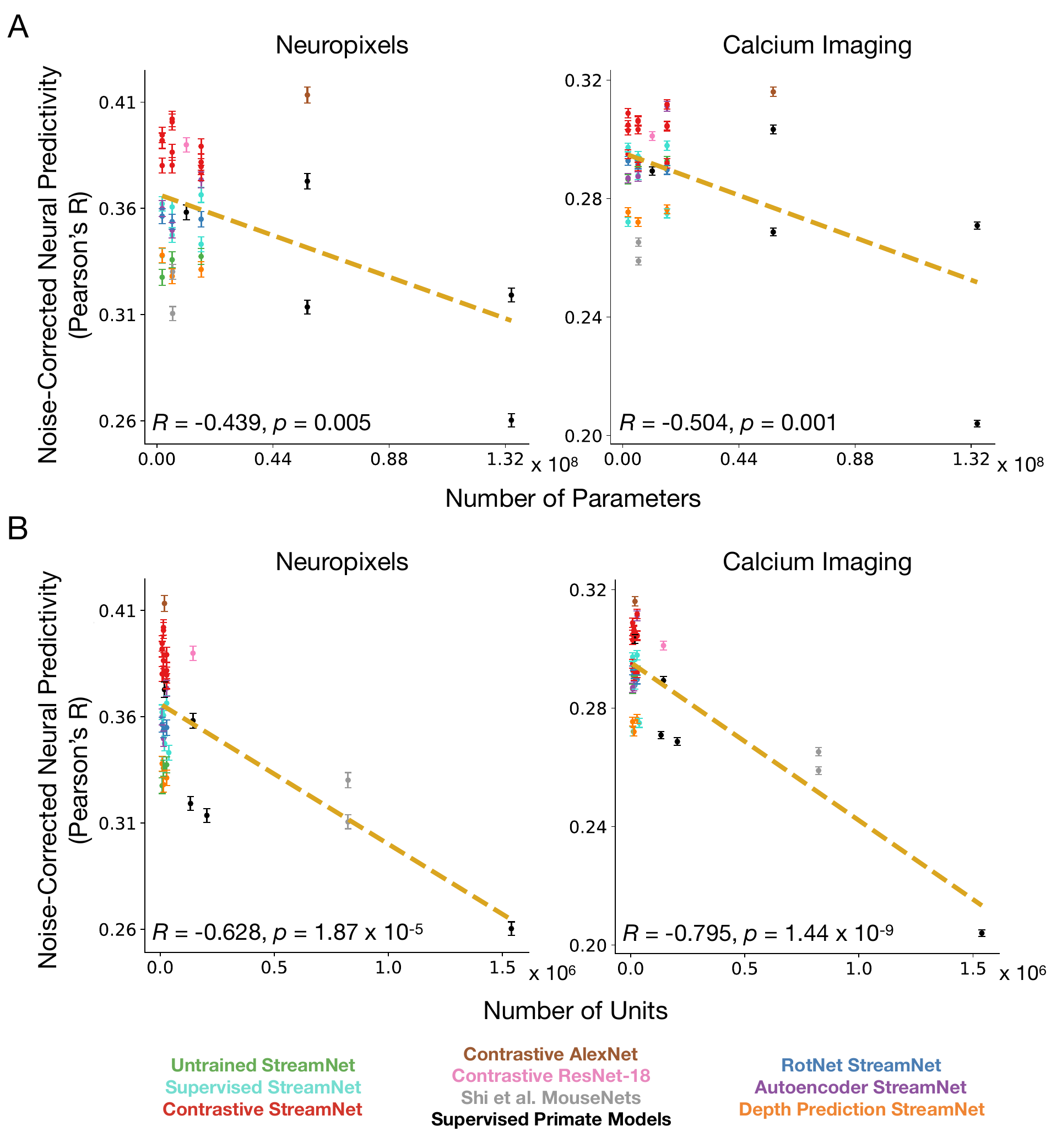}
    \caption[Increasing neural network size can decrease the model's neural predictivity of responses in mouse visual areas]{\textbf{Increasing neural network size can decrease the model's neural predictivity of responses in mouse visual areas.}
    \textbf{A}. Each model's neural predictivity is plotted as a function of its architecture size in terms of number of parameters, for both Neuropixels and calcium imaging datasets.
    \textbf{B}. Each model's neural predictivity is plotted as a function of its architecture size in terms of number of units, for both Neuropixels and calcium imaging datasets.
    The median and s.e.m. neural predictivity across neurons for each model is reported in all panels.
    Refer to Table~\ref{tab:imagenet-top1-transfer} for the neural predictivity values of each model and to Table~\ref{tab:network-size} for the parameter and unit counts of each model.
    Two-sided $p$-value obtained via Wald test with $t$-distribution of the test statistic; $N = 39$ models.
    }
\label{fig:supp-network-size}
\end{figure}

\begin{figure}[htbp]
    \centering
    \includegraphics[width=\columnwidth]{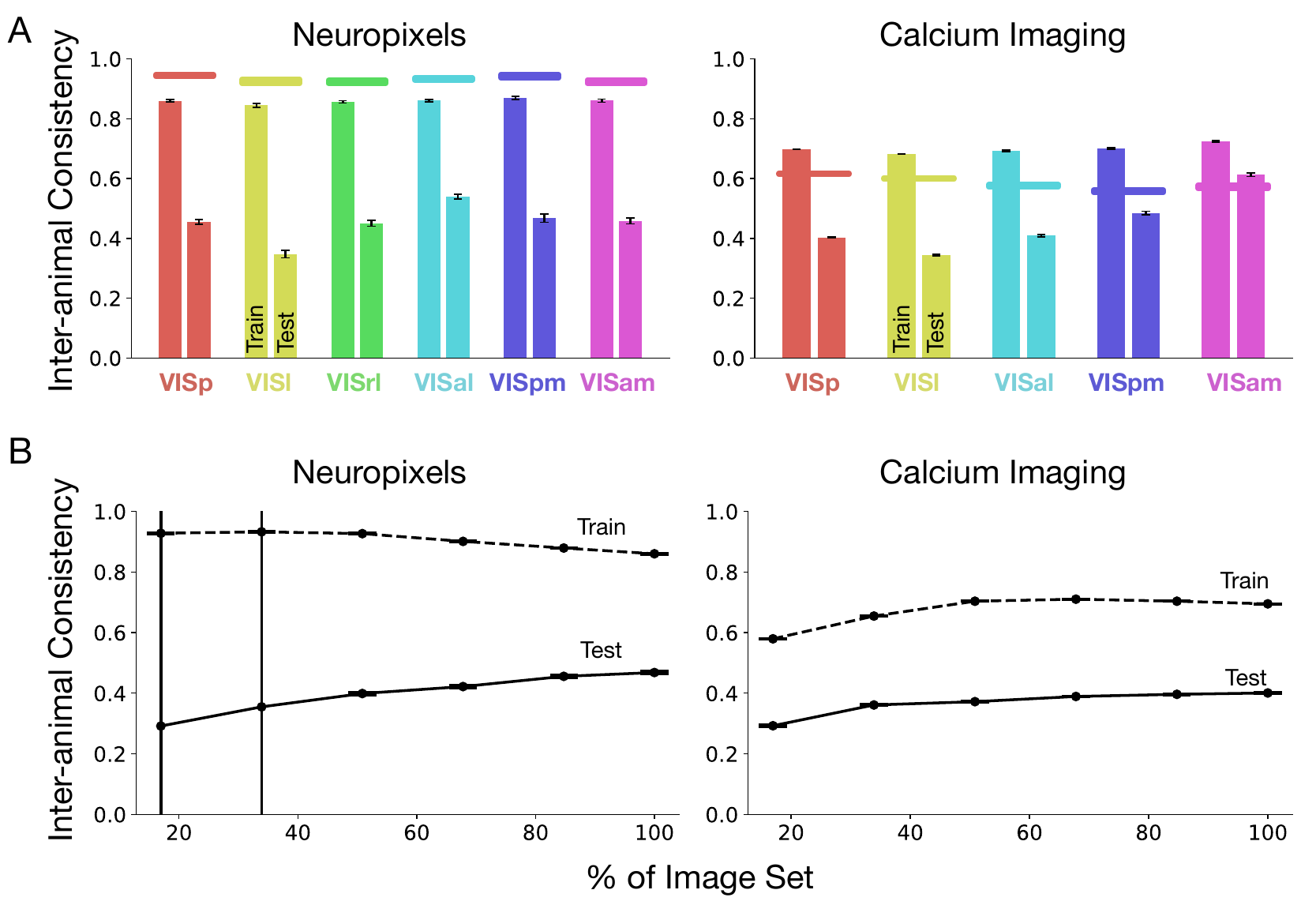}
    \caption[Inter-animal consistency can increase with more stimuli]{\textbf{Inter-animal consistency can increase with more stimuli.}
    \textbf{A}. Inter-animal consistency under PLS regression evaluated on the train set (left bars for each visual area) and test set (right bars for each visual area), for both Neuropixels and calcium imaging datasets. The horizontal lines are the internal consistency (split half reliability).
    \textbf{B}. Inter-animal consistency under PLS regression on the train set (dotted lines) and test set (straight lines), aggregated across visual areas. Each dot corresponds to the inter-animal consistency evaluated across 10 train-test splits, where each split is a sample of the natural scene image set corresponding to the percentage (x-axis). Note that VISrl is excluded for calcium imaging, as explained in the text.
    The median and s.e.m. across neurons is reported for both panels.
    Refer to Table~\ref{tab:neural-data} for $N$ units per visual area.
    }
\label{fig:interan-im-ss}
\end{figure}

\chapter{Heterogeneity in Rodent Medial Entorhinal Cortex}
\label{ch:mec}
\section{Chapter Abstract}
Medial entorhinal cortex (MEC) supports a wide range of navigational and memory related behaviors.
Well-known experimental results have revealed specialized cell types in MEC --- e.g. grid, border, and head-direction cells --- whose highly stereotypical response profiles are suggestive of the role they might play in supporting MEC functionality. 
However, the majority of MEC neurons do not exhibit stereotypical firing patterns.  
How should the response profiles of these more ``heterogeneous'' cells be described, 
and how do they contribute to behavior?
In this work, we took a computational approach to addressing these questions. 
We first performed a statistical analysis that shows that heterogeneous MEC cells are just as reliable in their response patterns as the more stereotypical cell types, suggesting that they have a coherent functional role.
Next, we evaluated a spectrum of candidate models in terms of their ability to describe the response profiles of both stereotypical and heterogeneous MEC cells.   
We found that recently developed task-optimized neural network models are substantially better than traditional grid cell-centric models at matching most MEC neuronal response profiles --- including those of grid cells themselves --- despite not being explicitly trained for this purpose.
Specific choices of network architecture (such as gated nonlinearities and an explicit intermediate place cell representation) have an important effect on the ability of the model to generalize to novel scenarios, with the best of these models closely approaching the noise ceiling of the data itself.
We then performed \emph{in silico} experiments on this model to address questions involving the relative functional relevance of various cell types, finding that heterogeneous cells are likely to be just as involved in downstream functional outcomes (such as path integration) as grid and border cells.
Finally, inspired by recent data showing that, going beyond their spatial response selectivity, MEC cells are also responsive to non-spatial rewards, we introduce a new MEC model that performs reward-modulated path integration.  
We find that this unified model matches neural recordings across all variable-reward conditions.
Taken together, our results point toward a conceptually principled goal-driven modeling approach for moving future experimental and computational efforts beyond overly-simplistic single-cell stereotypes.

\section{Introduction}
\label{mec:sec:intro}
From exploring new areas, planning shortcuts, and returning to remembered locations, the ability to self-localize within an environment subserves a range of navigational behaviors that are essential for survival.
The hippocampus (HPC) and medial entorhinal cortex (MEC) are known to contain cells that encode the position of an animal by displaying firing fields with strikingly regular response patterns, influenced by self-motion~\citep{o1971hippocampus,hafting2005microstructure,kropff2015speed,solstad2008representation, sargolini2006conjunctive}.
For example, MEC grid cells possess characteristically symmetric and periodic tuning curves, yielding hexagonal arrays of neural activity over physical space.
Additionally, border cells that fire maximally near environmental boundaries and head direction cells that fire only when an animal faces a particular direction are among the other MEC cell types with interpretable tuning curves.

However, a large fraction of the MEC population have unconventional and heterogeneous tuning to navigational variables and are less obviously well-described in terms of simple, stereotypical tuning patterns~\citep{hinman2016multiple,hardcastle2017multiplexed}.
How can we characterize what these other, more heterogeneous, populations of cells do?
Are they critical to the abilities of the animal in real-world memory tasks?
And if so, how is their role distinct from the more stereotypical grid-like cells? 
Here, we take a quantitative computational modeling approach to answering these questions.  

To start with, we address the question of whether there is a phenomenon to model in the first place, and how one would measure model accuracy quantitatively. 
To this end, we identify the \emph{similarity transform} between neural populations in different animals --- that is, a mapping which takes neuronal population vectors in a ``source'' animal and maps it to corresponding neuronal population vectors in a ``target'' animal.
To the extent that robust similarity transforms can be found that match up MEC population responses across multiple animals measured in multiple experimental conditions, then a reliable pattern of neural response behaviors (across conditions) has been identified.  
This identification strategy is well-defined even when there is no known \emph{a priori} taxonomy of functional response types.
We find here that with the right mapping class (mostly Ridge-like linear regression), the inter-animal consistency of MEC neuronal responses is in absolute terms very high ($> 0.8$), and in relative terms just as high as for heterogeneous cells as for more stereotypical grid cells. 

Using this same similarity transform, we then evaluate the ability of each of multiple computational models to explain response variance of MEC neurons, treating each candidate model as a potential ``source animal'' and measuring how well it maps to each target real animal. 
We look to the recent literature to identify potential candidate models. 
Over the past several decades, collaborations between experimental and computational neuroscientists have led to the formulation of dynamical-systems models of grid cell formation, and helped illustrate possible functional roles for grid cells in supporting path integration-based hippocampal place cells~\citep{skaggs1992information,zhang1996representation,fuhs2006spin,burak2009accurate}.
These powerful models make a number of non-obvious predictions about MEC neural properties, some of which have been confirmed in subsequent experimental work~\citep{ocko2018emergent,campbell2018principles}.
Despite their success, such models are limited in their explanatory scope, hand-designed to capture the properties of one stereotypical cell-type class (e.g. grid or border cells) at a time, or combinations of several cell types via multiple dedicated type-specific modules~\citep{couey2013recurrent,yoon2013specific}.

A potential solution to this problem arises out of recent work creating learned neural networks that achieve path integration~\citep{cueva2018emergence,banino2018vector}.
Intriguingly, these models have been found to contain internal units that resemble grid cells, suggesting that such stereotypical cells embody a computational solution to path integration that naturally arises from satisfying an end-to-end functional constraint.
Recent work has demonstrated that the underlying mathematical reason for this fact is due to pattern forming dynamics under a nonnegativity constraint~\citep{sorscher2019unified,sorscher2020unified}.
Intriguingly, in addition to having units that resemble stereotypical grid or border cells, these learned neural networks also naturally possess a wide variety of other less easily described unit types, raising the possibility that these artificial ``heterogeneous'' units might be somehow resemble the actual heterogeneous cells making up the majority of real MEC populations.
 
Motivated by these ideas, we generate a wide variety of candidate model networks
by varying architectural structure and end-to-end optimization objectives, each expressing a different hypothesis for MEC circuit structure and function. 
Architecturally, we formulate variants based on using different types of nonlinearities and different types of local recurrent memory circuits (e.g. RNNs~\citep{elman1990}, UGRNNs~\citep{collins2017}, GRUs~\citep{Cho2014}, and LSTMs~\citep{Hochreiter1997}).
From a task point of view, we test both simple dimensionality reduction~\citep{stachenfeld2014design, dordek2016extracting} as well as place cell mediated vector path-integration~\citep{banino2018vector} and direct position estimation~\citep{cueva2018emergence}.

Our core result is that there is substantial variation in the models' abilities to explain MEC responses, especially those of the heterogeneous cells, as a function of a model architecture and task.  Some models, such as the classic ``Grid Cell'' model based on low-rank decomposition of place cell fields, do a reasonable job explaining grid cell responses but are quite poor at explaining most other neurons.  Task-optimized learned models typically do better, especially those optimized for place cell-mediated vector path-integration.  The best model -- with memory-gated rectified nonlinearities -- essentially \emph{solves} the neurons, capturing nearly 100\% of the noise ceiling of the data.  This is of substantial interest, given that the nonlinear components of this model are not directly optimized to match neural data (just to solve the task), and given that a series of strong control models capture much less of the MEC neural variability.  
This same model also best generalizes to a variety of novel experimental conditions, and has the best match to the empirical data on a grid score distribution metric. 

With this predictive model in hand, we then begin to address our second core question: what is the functional role of heterogeneous neurons? We generate several results suggesting that heterogeneous neurons are important for path integration, including cell-type specific virtual knockout experiments, in which we compare performance degradation when deleting grid and/or border cells as compared to heterogeneous units. Overall, we find that models are quite robust to knockouts, and differences between heterogeneous and stereotypical cell knockouts are very small, suggesting that stereotypical cell-types may not be especially more functionally important than units with less easily-characterized response profiles. 

Building on the above results, we extend models to encompass MEC cell responses as a function of reward as well as spatial position, introducing a simple modeling paradigm that performs reward-modulated foraging in the context of the path-integration task.  We find that this unified task-optimized model matches neural responses across all reward and spatial conditions, and that a reward-response mechanism at an intermediate point on the explore-exploit continuum best explains neural responses. 
Taken together, our results suggest how specific processes of biological performance optimization may have directly shaped the neural mechanisms in MEC as a whole, and provides a path for enlarging the study of MEC beyond overly-restrictive response stereotypes.

\section{Reliability of heterogeneous cell response profiles}
\label{mec:sec:interanimal}

\begin{figure}[t]
    \centering
    \includegraphics[width=1.0\columnwidth]{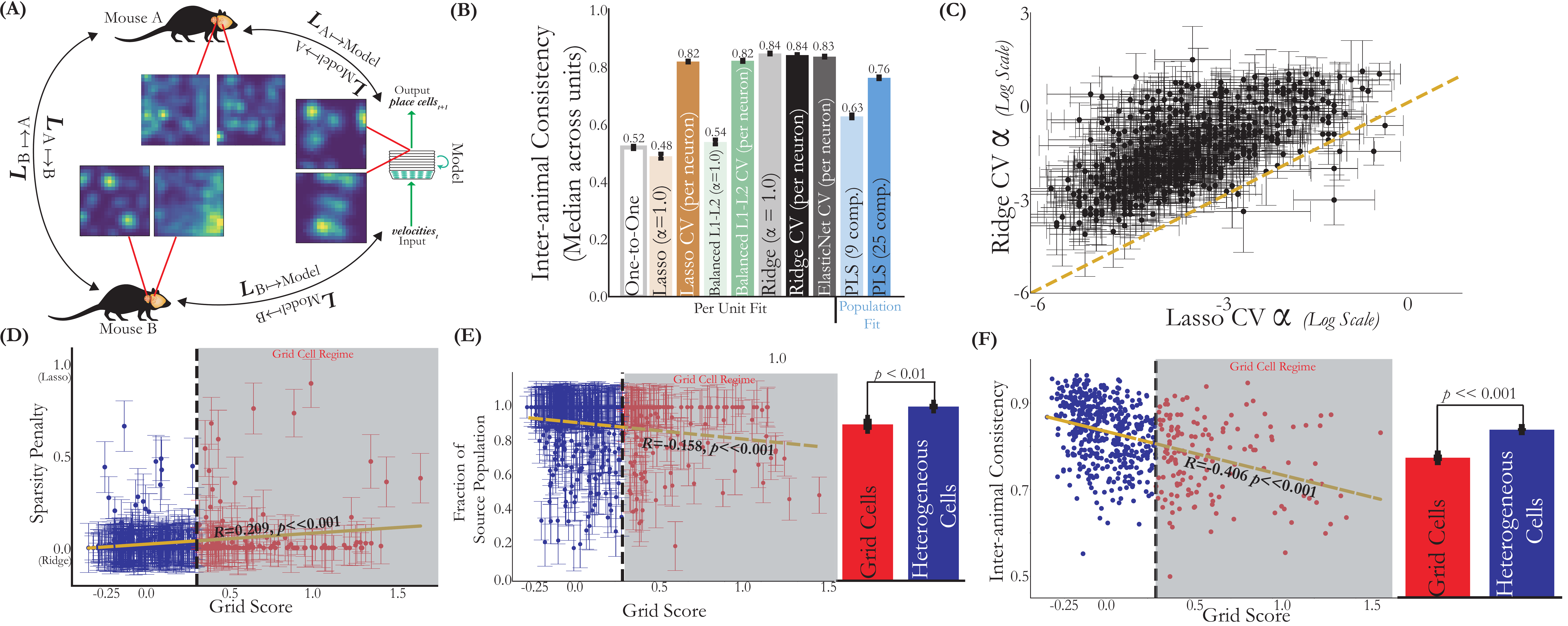}
    \caption[Heterogeneous cells are just as reliable as grid cells across 2D environments and task conditions]{\textbf{Heterogeneous cells are just as reliable as grid cells across 2D environments and task conditions.} {\textbf{(A)} We establish inter-animal consistency levels by mapping animals to each, using inter-animal transforms $L$ of at-most-linear functional complexity. Computational models are then mapped to real animal's MEC data using the same transform class. \textbf{(B)} Inter-animal consistency levels as assessed with mapping transform classes of different complexity. Median and s.e.m. across 620 cells.
    \textbf{(C)} The alpha value per neuron as chosen by Ridge CV plotted against the alpha value chosen by Lasso CV, on a log scale. Median and s.e.m. across ten train-test splits. The unity line is in yellow. \textbf{(D)} Per unit sparsity penalty under ElasticNet CV (grey bar in (B)) as a function of grid score. Median and s.e.m. across ten train-test splits. \textbf{(E)} The fraction of source units assigned nonzero weight per target cell plotted against its grid score (median and s.e.m. across ten train-test splits). Red denotes grid cells with grid score $> 0.3$ and blue denotes heterogeneous cells with grid score $\le$ 0.3. The bar plot on the right is the mean and s.e.m. of this median quantity for the identified grid and heterogeneous cell populations. The $p$-value is obtained from an independent $t$-test across neurons between populations. \textbf{(F)} Inter-animal consistency of each unit plotted against its grid score. The bar plot on the right is the mean and s.e.m. of this quantity for the identified grid and heterogeneous cell populations. The $p$-value is obtained from an independent $t$-test across neurons between populations.}}
    \label{mec:fig:interanimal-cons}
\end{figure}

What firing patterns of MEC cell populations are common across multiple animals, and thus worthy of computational explanation?
This question is comparatively straightforward for stereotypical MEC cells, because the very presence of these stereotypical features (e.g. hexagonal grids of a given orientation and spatial frequency) allows the definition and measurement of observables that arise reliably across trials and animals (e.g. the grid score distribution).
But how can this be done generally for populations in which one does not have a prior characterization of what each cell encodes?
We take inspiration from methods that have proven useful in modeling visual, auditory, and motor cortex~\citep{yamins2016using, kell2018task, michaels2020goal}.
Specifically, we aim to identify the \emph{narrowest} class of similarity transforms needed to map the firing patterns of one animal's MEC population to that of another (Figure~\ref{mec:fig:interanimal-cons}a).
As with other cortical areas, this transform class likely cannot be so strict as to require fixed neuron-to-neuron mappings between MEC cells, since even within the same animal at different times, MEC and HPC populations can undergo remapping that shift cell responses across the population~\citep{farhoodi2020estimating,low2020dynamic}.
However, the transform class for MEC also cannot be so loose as to allow a completely unconstrained linear mapping, since the highly structured response patterns of stereotypical MEC cell types such as grid cells may not be guaranteed to be preserved under arbitrary linear transforms.

To identify this transform class, we utilize data collected from electrophysiology in 12 awake behaving mice ($n=620$ cells) performing open foraging for randomly scattered crushed cereal in a 100${cm}^2$ 2D arena~\citep{mallory2021mouse}.
We explore a variety of mapping transform classes between the population rate maps of these neurons (Figure~\ref{mec:fig:interanimal-cons}b).
As a baseline we evaluate a strict one-to-one mapping transform, in which each target unit is mapped to the single most correlated unit in the source animal. 
We also evaluate more powerful linear transforms, including Lasso, Ridge, and ElasticNet regression, as well as population-level Partial Least Squares (PLS) regression.
For all methods with fittable parameters, mapping fit is performed on a random 20\% of spatial position bins, and evaluated on the remaining 80\% of position bins (see supplement for more details). 
 
The strict one-to-one mapping yielded low inter-animal consistency among the maps considered, capturing around 50\% of the target neural response variability.  
Linear regression with strong sparseness priors, such as Lasso (L1 penalty) and balanced Lasso-Ridge (equal L1 and L2 penalty) regression, also proved to be too strict when evaluated with fixed regularization level $\alpha=1$, yielding hardly any improvement over the one-to-one mapping.  
However, pure Ridge (L2 penalty) regression at this regularization level was highly effective, recovering nearly all target variability for most neurons.
Cross-validating the L1 regularization constant on a per-cell basis improved the fits, at the cost of requiring substantially looser regularization than for L2 (Figure~\ref{mec:fig:interanimal-cons}c). 
Under an ElasticNet mapping in which both sparsity penalty and regularization constants were chosen with cross validation, most cells were generally still best explained by an essentially Ridge-like transform with no sparsity penalty (Figure~\ref{mec:fig:interanimal-cons}d). However, different target cells required different numbers of source units to achieve effective mapping (Figure~\ref{mec:fig:interanimal-cons}e).
As expected, cells with more stereotypically grid-like response patterns on average chose a higher sparsity penalty (slight positive slope in Figure~\ref{mec:fig:interanimal-cons}d) and required fewer source cells to capture (slight negative slope in Figure~\ref{mec:fig:interanimal-cons}e) as compared to heterogeneous cells, though this effect is weak. 
Critically, heterogeneous cells did not have lower inter-animal consistency than grid cells (Figure~\ref{mec:fig:interanimal-cons}f). 

These results show that the heterogeneous non-stereotyped cell populations are reasonably similar across animals -- at least up to (mostly Ridge-regularized) linear transform -- establishing that there is a reliable target pattern to study in the first place.

\section{Task-optimized models of MEC spatial response variation}
\label{mec:sec:task}

\begin{figure}[t]
    \centering
    \includegraphics[width=1.0\columnwidth]{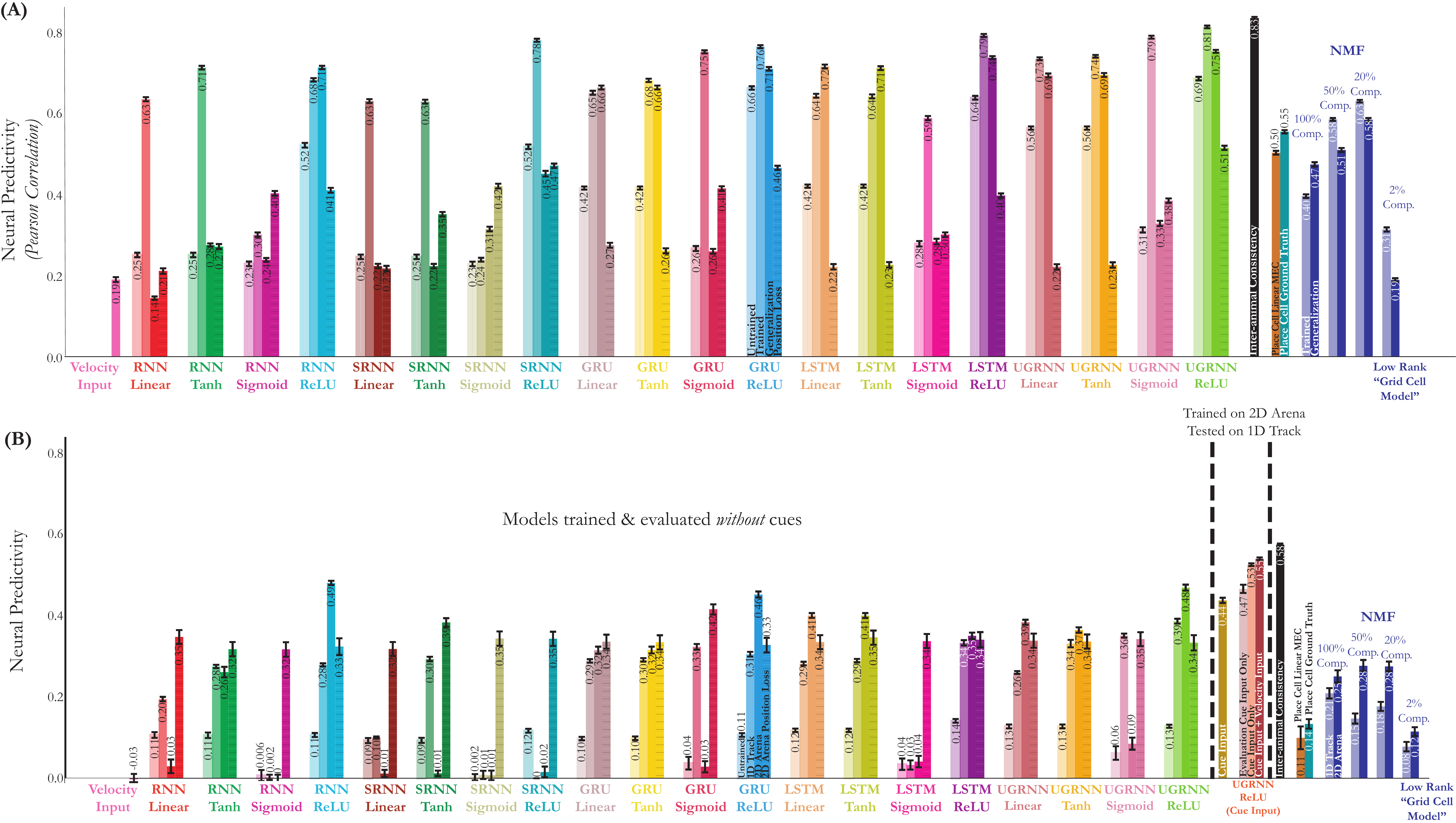}
    \caption[Task-optimized navigational models best predict the entire MEC population]{\textbf{Task-optimized navigational models best predict the entire MEC population.} {\textbf{(A)} Neural predictivity of the model MEC units to the real MEC responses of 12 animals in a 100${cm}^2$ 2D open field under the ElasticNet CV regression transform (grey bar in Figure~\ref{mec:fig:interanimal-cons}b). Median and s.e.m. across 620 total units. \textbf{(B)} Same as (A), but now the models are evaluated against responses of MEC units while the animal is traversing a 400$cm$ 1D track. ``1D Track'' refers to models trained to path integrate on this same 1D track. ``2D Arena'' refers to models trained to path integrate on the 2.2$m^2$ arena and evaluated on the 1D track. Median and s.e.m. across 2861 units from 8 different animals.}}
    \label{mec:fig:neural-pred}
\end{figure}

\textbf{Evaluating a spectrum of candidate models.}
We evaluated models of several basic types. First, we considered learnable neural networks, all of which accept a stream of two-dimensional velocity input vectors, have a single layer of hidden neurons identified as the putative MEC population, and which are optimized to perform some form of path integration readout on simulated motion paths (\cite{erdem2012goal}) in a fixed-sized arena. 
The networks varied according to the nature of their local recurrent cell structure (ungated RNN, UGRNN, GRU, or LSTM), activation functions (Linear, Tanh, Sigmoid, or ReLU), and output objective function (explicitly constructing a set of simulated place field neurons~\citep{banino2018vector} or directly performing two-dimensional position integration~\citep{cueva2018emergence}).
Following \cite{banino2018vector}, all networks consisted of three nonlinear layers, though for the ungated RNN we included its original two layer version (following \cite{cueva2018emergence}) and a three layer version (``SRNN'').
We ensured that comparisons were fair by equalizing the size of the hidden layer across models, and included an architecture-only control with untrained filter weights.

We also implemented a class of models describing MEC activations as Non-negative Matrix Factorization (NMF) operating on simulated place cells.
The lowest-rank version acts a ``Grid Cell'' model, inspired by recent work positing MEC as a low-dimensional embedding of hippocampal place fields \citep{stachenfeld2014design, dordek2016extracting}, and further validated by the mathematical theory developed in \cite{sorscher2019unified, sorscher2020unified} to explain the emergence of grid cells.
Higher-rank NMF models titrate between the low-rank Grid Cell model and a full-rank control that measures how well MEC cells can be explained as a linear projection of their putative place cell outputs. 

For evaluation, we map each model's proposed MEC activations to empirically measured firing patterns of the real neurons, using the same type of cross-validated ElasticNet regression transform used to measure inter-animal similarity as in Section~\ref{mec:sec:interanimal}.
The predicted neural responses under the mapping are then compared to real target neural responses on a neuron-by-neuron basis, and the median of accuracy of these predictions taken over target neurons.
In addition to evaluating model-data match when constructed on arenas of the same size in which the neural data were collected (100${cm}^2$), we also evaluated models trained on larger arenas (2.2$m^2$) but tested on the 100${cm}^2$ arena.

\textbf{Neural prediction results.} The results of our model evaluation, shown in Figure~\ref{mec:fig:neural-pred}a, support several inferences:
\begin{itemize}
 \item Different models are substantially different in their ability to predict neural responses. MEC electrophysiology data collected during 2D open-field foraging is thus a strong model target that effectively separates candidate models from each other. 
 
 \item Task-optimized models are reliably better than their untrained controls, across all architecture types and objective functions. 

 \item Rectification is substantially better than Linear, Tanh, or Sigmoid activation, especially for promoting generalization to new arena sizes.  
 
 \item Under the rectification nonlinearity, gated circuit architectures (UGRNN, GRU, LSTM) improve model fits compared to the simple ungated alternatives (RNN and SRNN). 

 \item The explicit place cell construction task leads to substantially better fits than the direct two-dimensional path integration task (``Position Loss''), even though both were reported in the literature to create grid cell-like units. 
 Enforcing a place field representation at the output of the network thus appears to be an important constraint in order to recapitulate responses in MEC.
 
 \item The Grid Cell (low-rank NMF) model is a poor fit to the MEC population overall. 
 
 \item Full-rank NMF and ``Place Cell Linear MEC'' output-based controls capture approximately half the explainable variance of MEC neurons, significantly more than the simple velocity linear input control, but substantially less than any of the input-driven ReLU networks.
 
 \item The best model is the UGRNN ReLU trained with the place cell loss (``UGRNN-ReLU-Place Cell'').
 This model captures nearly 100\% of the explainable variance of the MEC neural population response when trained for place cell construction on the arenas of the same size as that on which neural data was collected. 
\end{itemize}

\textbf{Generalization to novel experimental conditions.} If a model is truly correct, then once trained, it should capture neural responses in any new tested condition.  
We tested generalization in two ways. 
First, as shown in the second from the right bars of Figure~\ref{mec:fig:neural-pred}a for each model architecture class, we performed a 2D arena size generalization test by constructing models on one arena size, testing against neural data on another.  
We found this generalization test gives essentially the same rank-order comparison as in the same-arena-size test, with one striking exception: the RNN Tanh model performs well within arena size, but fails to generalize. 
The UGRNN-ReLU-Place Cell model again performs the best, capturing 90\% explained variance of the neural responses on the novel arena size. 

Second, we also evaluated the same 2D-pretrained models by running them on a 1D track, comparing models to neural data collected from mice in a 1D virtual reality setup, using the same ElasticNet mapping procedure as in the 2D comparisons (see supplement for more experimental details). 
We found (Figure~\ref{mec:fig:neural-pred}b) that 2D arena trained model results had some similar rank order as for the original 2D results (0.43 Spearman rank correlation), with the UGRNN-ReLU-Place Cell model trained in the 2D arena (and then evaluated on the 1D track) achieving the best match (82\%) when trained for place-field construction.
We hypothesized that the larger gap between the best model and the inter-animal noise ceiling, as compared to the 2D case, was due to the fact that during the 1D experimental data collection, mice were also presented with visual cues. 
MEC neurons are known to respond to visual input, but as none of our evaluated models had a visual front-end, they could not respond accordingly.  
To test this hypothesis we trained the UGRNN-ReLU-Place Cell model in 2D, but with phantom visual cue locations added as input while performing the path integration task.
We then evaluated this model against 1D neural prediction (with the real visual cue locations as input), and found that including cues rescued model performance back to essentially the noise ceiling, as in 2D (Figure~\ref{mec:fig:neural-pred}b, orange bars). 

\begin{figure}[t]
    \centering
    \includegraphics[width=1.0\columnwidth]{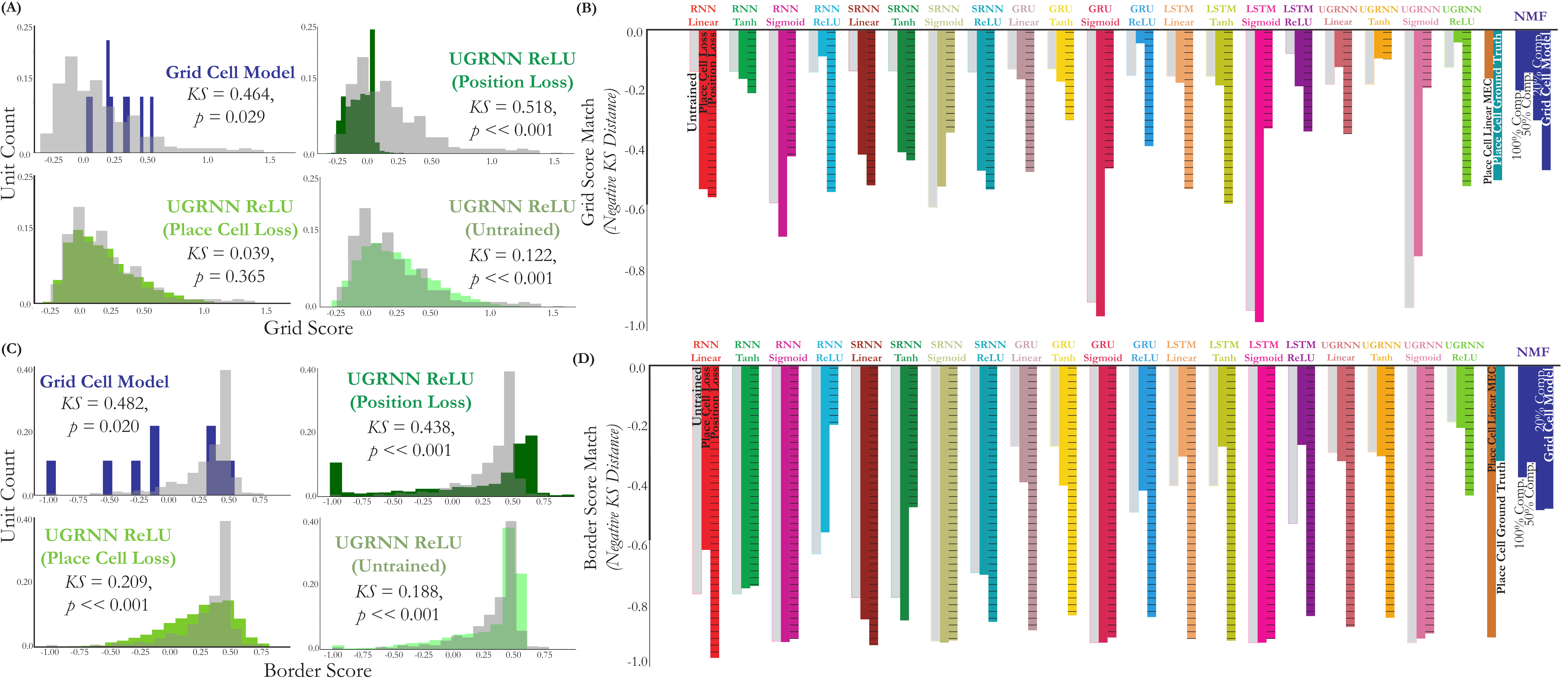}
    \caption[Relationships to grid and border score distribution]{\textbf{Relationships to grid and border score distribution.} {\textbf{(A,C)} Example distributions of grid and border scores from selected models against the ground truth distribution in the neural data (grey). Negative Kolmogorov-Smirnov distance between the distribution of grid scores of the model units to those of the 620 units in the data. The models have all been trained in a different 2.2$m^2$ arena, but are evaluated on a 100${cm}^2$ environment that the animals are in. \textbf{(B, D)} Quantification across all models.}}
    \label{mec:fig:grid}
\end{figure}

\textbf{Assessing grid and border score distribution match.}
The \emph{grid score} is a metric of how stereotypically grid-like the response pattern of a given unit is (see supplementary material for specific definitions), with ``grid cell'' typically defined as having a grid score of greater than 0.3.
Similarly, the border score is a measure of how responsive a given unit is in the presence of environmental boundaries, with a ``border cell'' typically defined as having a border score greater than 0.5~\citep{solstad2008representation}.
The ground-truth distribution of grid and border scores for cells in an unbiased experimental population (gray bars in each subpanel of Figure~\ref{mec:fig:grid}a,c) characterize the extent to which real MEC populations are non-stereotypical in their responses.  
To further assess model accuracy, we also compared the distribution of grid and border scores within each model to that of the real data, using the (negative) Kolmogorov-Smirnov (KS) distance of the empirical and model grid score distributions as a quantitative metric.
This metric is both stronger and weaker than the mapping accuracy metric used above --- stronger in the sense that, since there is no parameter fitting in the metric, to fit it well a model has to have the correct distribution in its raw feature output; and weaker in that it only assesses cells on one component of their profile (i.e. ``grid-ness'' or ``border-ness'').   
We observed that the low-rank Grid Cell model has poor fit on this metric, essentially because it contains \emph{too many} grid cells (see Figure~\ref{mec:fig:grid}a, upper left). 
In contrast, the same model that achieves best fits on the neural-fit metric (UGRNN-ReLU-Place Cell) also achieves the best on the grid cell distribution match metric, and is the only architecture that cannot be distinguished from the ground truth at the $KS<0.05$ significance level.   
Across model architectures, KS-distance was generally in line with neural regression fit metrics (see Figure~\ref{mec:fig:grid}b,d and Figure~\ref{mec:suppfig:npvsgs}). 
One key difference between the metrics, however, is that for most architectures, the untrained filters provided \emph{better} grid score distribution matches as models with task-trained parameters --- except for the that best matched architecture, UGRNN ReLU, where the trained place cell construction model is better than its untrained counterpart (and similar to it for the border score) --- and models trained for direct path integration were especially poor on this metric.
These results suggest the two metrics in Figs. \ref{mec:fig:neural-pred} and \ref{mec:fig:grid} are complementary which aspects of model correctness they address, and together help zero in on the most effective models overall.

\section{Predicting the functional relevance of heterogeneous cells}

\begin{figure}[t]
    \centering
    \includegraphics[width=1.0\columnwidth]{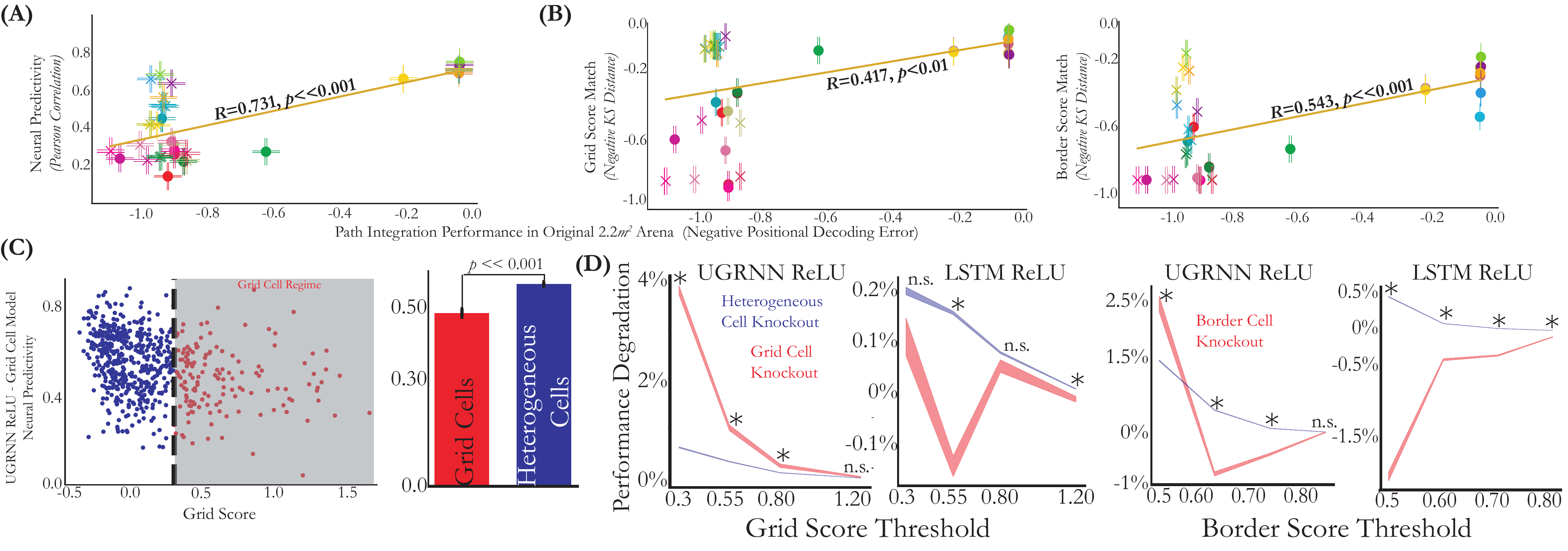}
    \caption[Heterogeneous cells are relevant to navigation]{\textbf{Heterogeneous cells are relevant to navigation.} {\textbf{(A)} Neural predictivity (median and s.e.m. across 620 units) versus path integration performance (mean and s.e.m. across 20,000 episodes), measured by negative positional decoding error, of the neural network models. The models are either untrained (``X'') or trained with the place cell loss (``O'') in the 2.2$m^2$ arena. Positional decoding error is measured by taking the top 3 most active place cell outputs in each model, evaluated on the 100${cm}^2$ arena. \textbf{(B)} Same as (A), but with grid score (\emph{Left}) and border score (\emph{Right}) distribution instead of neural predictivity. \textbf{(C)} (\emph{Left}) Per unit neural predictivity difference between the UGRNN-ReLU-Place Cell and Grid Cell models trained on the 2.2$m^2$ arena and evaluated on the 100${cm}^2$ arena, plotted against that unit's grid score. (\emph{Right}) Quantification of this difference aggregated across the grid cell and heterogeneous cell populations, respectively (mean and s.e.m). The $p$-value is obtained from an independent $t$-test across neurons between populations. \textbf{(D)} For the UGRNN-ReLU-Place Cell and LSTM-ReLU-Place Cell models, we measure the normalized performance degradation as evaluated on the 100${cm}^2$ arena, relative to the full model trained on the 2.2$m^2$ arena with the place cell loss. We identify units in this trained model with grid and border scores of varying thresholds and knockout the same number of heterogeneous cells (randomly sampled 100 times), to yield the ``heterogeneous cell knockout'' (blue). The $x$-axis denotes the threshold used for the score. Mean and s.e.m. over evaluation episodes. * denotes a $p$-value $ < 0.01$ obtained from an independent $t$-test between the performance degradations of the two knockouts at a given grid or border score threshold.}}
    \label{mec:fig:task}
\end{figure}

\textbf{Correlating task performance and neural predictivity.}  To begin to address the question of the functional relevance of heterogeneous cells, we first looked at the overall correlation between model task performance and neural fit (Figure~\ref{mec:fig:task}a).  
Though the correlation is imperfect, the most task performant models (e.g. UGRNN- and LSTM-ReLU-Place Cell) achieve the best matches to neural fit suggesting that improved task performance may be causally related to the ability capture neuron response patterns across the population.
In contrast, there is a comparatively weaker relationship between grid and border score distribution match and model performance (Figure~\ref{mec:fig:task}b).
While the most task performant models achieve the best match here, models with a gating architecture and ReLU are strong matches to these metrics with untrained filters, illustrating the importance in matching cell properties other than ``grid-ness'' or ``border-ness'' in predicting task performance. 

\textbf{Relative predictivity gain for heterogeneous cells.} The above result is put into greater perspective by comparing neural predictivity differential between a task-trained UGRNN-ReLU-Place Cell model and the Grid Cell model, on a per-neuron basis, as a function of grid score (Figure~\ref{mec:fig:task}c). The task-trained model has improved neural predictivity relative to the Grid Cell model for grid cells (grid score $> 0.3$) and heterogeneous cells, but the improvements on the latter (as well as border cells, Figure~\ref{mec:suppfig:borderdiff}) are larger than for grid cells. 
This again suggests that the heterogeneous cells are playing a substantial role in allowing the trained model to achieve improved performance.

\textbf{Virtual knockout experiment.} 
To address the question most directly, we performed a cell-type-targeted virtual knockout comparison experiment. 
In doing this, we used the UGRNN-ReLU-Place Cell and LSTM-ReLU-Place Cell models, the two best models emerging from the previous section with essentially similar neural predictivity across multiple metrics. 
We identified the units in the trained model with high grid score ($> 0.3$) or border score ($> 0.5$), and gradually increased the threshold, while measuring task generalization performance (see supplement for details of this knockout procedure).  We similarly ablated matched numbers of heterogeneous units (grid score $\le 0.3$ or border score $\le 0.5$ or both, see Figure~\ref{mec:suppfig:bgknockout} for the latter), again measuring model performance.

The main result of this experiment (Figure~\ref{mec:fig:task}d) is that all networks are highly robust to knockouts, experiencing only at most 1-4\% performance degradation relative to the full model even when substantial fractions of units are knocked out (corresponding to 21-24\% of a single layer's units for grid score $> 0.3$ and 16-20\% of these units for border score $>0.5$). 
At stricter grid and border cell thresholds, the heterogeneous knockout is similarly injurious to the cell-type specific knockouts, across the two model architectures, suggesting that highly-stereotypical ``classical'' border and grid score cells are not more essential to the path integration behavior than heterogeneous cells.
At low thresholds (when counting many relatively heterogeneous cells as ``grid'' or ``border'' cells), the two model architectures give divergent predictions, with the UGRNN model showing a small but significant effect of grid cells relative heterogeneous cells, and the LSTM model showing the opposite.
It would be of substantial future interest to confirm or reject either of these models' predictions with a real targeted knockout experiment \emph{in vivo}.

\section{Modeling reward-driven modulation in MEC}
Recent work in both rodents and humans has uncovered that MEC and HPC neurons represent not only literal space, but also capture spatialized layouts in a more abstract sense in modalities other than spatial position~\citep{constantinescu2016organizing, aronov2017mapping}.
It has also been seen~\citep{butler2019remembered, boccara2019entorhinal} that non-spatial rewards can influence the shape of MEC response profiles, restructuring them to incorporate the location of the learned reward. 
These intriguing phenomena represent a natural direction for modeling, but are not captured by any of the neural network models described in Section~\ref{mec:sec:task}, as they do not have reward-modulated inputs. (Note that while \cite{banino2018vector} used a pretrained LSTM-Tanh-Place Cell network as a front-end on which reinforcement-based navigation tasks are evaluated downstream, reward state is not input to their network or otherwise propagated back into the MEC-like layers of their model, and thus cannot address the neural modeling question raised here.)

\begin{figure}[t]
    \centering
    \includegraphics[width=0.95\columnwidth]{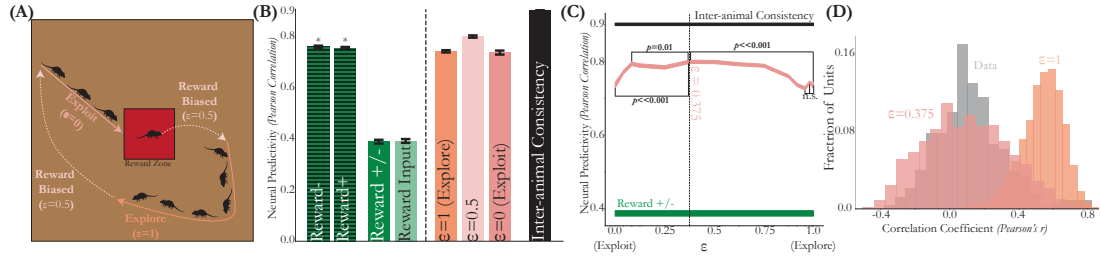}
    \caption[Reward biased path integration captures remapping of responses in the presence of a reward]{\textbf{Reward biased path integration captures remapping of responses in the presence of a reward.} {\textbf{(A)} Schematic of the three conditions. ``Explore'' is the original set of random bout trajectories used to train the agent as before. ``Exploit'' involves the agent navigating directly to the reward zone (red box) within a fixed number of timesteps per episode (7), and spending the remaining 13 timesteps only path integrating in the reward zone. ``Reward Biased'' refers to training with each of the above two trajectories on $0 < \varepsilon < 1$ of the training episodes.
    \textbf{(B)} Neural predictivity of the UGRNN-ReLU-Place Cell model, evaluated on the 150${cm}^2$ arena. 
    Median and s.e.m. of 598 cells from 7 rats.
    Green bars denote the model evaluated with random walk velocity inputs.
    $*$ denotes comparison of the model to each condition \emph{separately}.
    Coral bars (to the right of the vertical line) denote the model evaluated with random walk and reward biased velocity input modulation.
    \textbf{(C)} Neural predictivity of the reward biased UGRNN-ReLU-Place Cell model as a function of $\varepsilon$. 
    Median and s.e.m. of 598 cells from 7 rats.
    The $p$-value is obtained from an independent $t$-test across neurons between selected model pairs in brackets.
    \textbf{(D)} Histogram of the correlation coefficient of unit rate maps between the random foraging and velocity-input modulated reward condition in the $\varepsilon=0.375$ and $\varepsilon=1$ UGRNN-ReLU-Place Cell models.
    The same metric applied to the data is shown in grey.}
    }
    \label{mec:fig:reward}
\end{figure}

We thus sought to build new network models that respond to the existence of an extrinsic reward in a behaviorally-meaningful fashion. 
One natural approach would be to build a full reinforcement-learning (RL) agent in which reward depends on environmental features (e.g.total number of food units encountered during a forage path). A model optimized end-to-end to forage under these conditions could be successful in creating a representation that matches MEC reward-response modulation. 
However, we took a more direct approach: simply change the behavior of the agent in response to reward in the way RL would be expected to do if successful, by modifying the foraging trajectories used during training and testing to no longer be pure random walks. 

Specifically, we trained the UGRNN-ReLU-Place Cell model with a variety of foraging policies titrating between pure exploration and exploitation.
Paths could either purely exploit a known reward resource by directly moving to the reward location when it is present (Figure~\ref{mec:fig:reward}a, $\varepsilon=0$ condition), purely explore with a random walk as in the original unmodulated model ($\varepsilon=1$), or take an intermediate policy ($0 < \varepsilon < 1$). 
In creating these scenarios, we sought to roughly mimic experimental observations showing animals often take rapid and direct paths to the reward zone~\citep{butler2019remembered} 
(see supplement for details of implementation). 
Throughout, the actual training task of the network remained the same as above --- place cell construction --- but the network now would have to be tolerant to reward-modulated input velocity changes while maintaining positional knowledge.

We then compared these networks to neural data from \citep{butler2019remembered} for animals in conditions both with reward (\textbf{reward+}) and without it (\textbf{reward-}), in which the animal navigates to a 20${cm}^2$ reward zone within a 150${cm}^2$ arena to receive 0.5-1 units of cereal. We first performed inter-animal consistency checks, using the same method as we did with the purely spatial-condition data in Section 2 above, finding that inter-animal consistency for neural responses is high both within reward condition (e.g. just spatial modulation in \textbf{reward+} and \textbf{reward-} separately), and across both spatial and reward-modulated conditions overall (Figure~\ref{mec:suppfig:interan}).

As a baseline, we then evaluated the ability of the original (non-reward-modulated) UGRNN-ReLU-Place Cell model to match neural data. Consistent with results reported in the previous sections, this model achieved high neural predictivity in \textbf{reward+} and \textbf{reward-} separately (Figure~\ref{mec:fig:reward}b, hatched bars).  However, as expected, this non-modulated model was ineffective at predicting response patterns across both reward conditions (Figure~\ref{mec:fig:reward}b, \textbf{reward+/-}). 
Moreover, simply augmenting the network input to receive an additional binary reward state (Figure~\ref{mec:fig:reward}b,``Reward Input''), but without using that input to specifically modulate input velocities or output behavior, did not substantially improve neural predictivity, showing the need for nontrivial integration of reward-modulated state.

We then evaluated the reward-modulated networks, generating network outputs for comparison with the \textbf{reward+} and \textbf{reward-} conditions by modulating velocity inputs to match each condition during testing (see supplement for details). 
We found that the pure-explore ($\varepsilon=1$) model is substantially better at matching neural responses across reward conditions than the \textbf{reward+/-} baseline.  
This is perhaps somewhat surprising since in this comparison the underlying neural network model is identical, just evaluated with or without reward-modulated input data during testing.  
This result suggests that a substantial fraction of the reported reward-modulated effect in MEC may actually simply be input-driven, lending a new interpretation to results of \citep{butler2019remembered}. (It may be useful to note that this result represents a model-based control that was inaccessible to the authors of \citep{butler2019remembered}.) 
However, we did find that exposing the network to a mixture of exploration and exploitation behaviors during training does lead to networks with somewhat improved neural predictivity.
Results were largely robust to the specific proportion of exploration-vs-exploitation (Figure~\ref{mec:fig:reward}c), though the best model (at $\varepsilon=0.375$) model was statistically-significantly better than alternatives and had a substantially closer match ($KS=0.13$ vs. $KS=0.81$) to the data's unit-level remapping across conditions than the original $\varepsilon=1$ model (Figure~\ref{mec:fig:reward}d). 
These results are consistent with there being some nontrivial within-MEC reward-modulated responses beyond simple input modulation alone.
\section{Discussion}
We have identified a goal-driven neural network model of MEC that is quantitatively accurate across a wide variety of common experimental conditions, and that can be used to generate nontrivial insights about the underlying mechanisms and functional roles of mouse MEC. 
More generally, our results suggest that constraint-driven neural networks may provide a fruitful approach for studying navigation and memory in the MEC, and beyond. 

Our work suggests the existence not of a specialized class of heterogeneous cells that is functionally segregated from classic cell types, but rather a continuum of cells within a single unified network that naturally encompasses grid, border, and heterogeneous cells. Future research on the computational foundations of MEC/HPC may thus be well-served by putting less emphasis on identifying cleanly-stereotypical cell types (such as grid cells) or perfecting mathematically simple models of single such cell-types, and looking instead for holistic computational ideas that move beyond ``easy to visualize'' but perhaps overly-simplistic tuning-curve categories.

The nature of the understanding afforded by such a modeling approach comes from their ability to make inferences about what constraints (both structural and functional) are consistent with the data. The current work rules out a variety of simple network connectivity diagrams as inconsistent with MEC cell data, and narrows the space of functional goals MEC circuit weights might be optimized for over evolutionary timescales. Improving the results here will hopefully narrow these constraints yet further, with the ultimate goal of identifying constraints that yield the uniquely correct MEC circuit diagram and synaptic weights -- or at least, the narrowest set of such networks consistent with the inherent variability between animals in the real population.  

There are some key limitations, however, on our results.  First, we attempted to identify the simplest underlying transform that would map animals within a population to each other, using this as the basis for conclusions both about unit-type reliability and model-data comparisons.  While the philosophy of this approach may be sound \citep{cao2021explanatory1}, in practice it is possible that we could narrow the transform class further by (e.g.) enforcing that it be invariant to one more more properties of classical cell-types (e.g. periodicity). Finding algorithms to better identify sharp inter-animal transform classes will be an important topic for future work. 

Moreover, we remain unconvinced that MEC just ``is'' a UGRNN-ReLU-Place Cell network, despite a network of this architecture having explained essentially all the data we had available to challenge it. It is possible that matching all our existing data is too easy a test. Would this network generalize to more complex situations, with increased variability along key axes such as spatial structure (e.g. environments with corridors and looping interconnections), nontrivial but spatially informative cues, and rewards exhibiting complex and temporally-variable patterns? We do not think it is at all obvious that it would. 
The proper conclusion from this work is thus not that our current best model is actually correct, but rather that the fairly simplified setup typical of experiments in MEC is not sufficiently complex to falsify it. This present situation is a concrete manifestation of the ``contravariance principle'' of neural modeling \citep{cao2021explanatory2} -- the idea that working in a \emph{more} complex experimental environment might actually make it \emph{easier} to identify an actually correct model, by virtue of reducing susceptibility to spurious models that are apparently consistent with too-simple data. Future experiments should engage mice in more complex environments and behaviors.

\section{Comparisons to Neural Data}
\label{mec:neuralfit}

\subsection{Experimental Data}
\label{mec:supp:neuralfit-data}
We analyze three neural datasets in total (two from tetrode recordings of freely moving animals in 2D arenas and one from Neuropixels recordings of head-fixed mice on a 1D virtual track).
The 2D open-field foraging dataset in the 100${cm}^2$ arena came from~\cite{mallory2021mouse} across 12 mice, totalling 620 MEC cells.
The 2D open-field foraging (\textbf{reward-} condition) and reward dataset (\textbf{reward+} condition) in the 150${cm}^2$ arena came from~\cite{butler2019remembered} across 7 rats, totalling 598 MEC cells.
Note that the \emph{same} neural population was used across these two conditions, whereby in the \textbf{reward+} condition, the animal navigates to a 20${cm}^2$ reward zone within this arena to receive 0.5-1 units of cereal.
The reward zone location was fixed across sessions for each animal, but varied between animals.
Finally, the 1D VR data was used by~\cite{mallory2021mouse} across 8 mice, totalling 2861 MEC cells.
There were five landmarks (towers) total at 0, 80, 160, 240, and 320$cm$.
For further experimental details, please refer to the respective cited paper referred above.

In the 1D VR data, we identified remapping events, depicted in Figure~\ref{mec:suppfig:remapping-1d}, following the procedure used by~\cite{low2020dynamic}, by examining the trial-by-trial similarity matrices as well as the test-set (90-10\% splits) $R^2$ of $k$-means clustering (red) applied to the responses and identifying where it diverges from PCA (blue).

\subsection{Rate Map Representation}
\label{mec:supp:neuralfit-ratemap}

We bin the positions in each environment using 5 cm bins, following prior work~\citep{hardcastle2017multiplexed,butler2019remembered,low2020dynamic}.
Thus, the 100${cm}^2$ environment used 400 ($20\times 20$) bins, the 150${cm}^2$ environment used 900 ($30\times 30$) bins, and the $400 {cm}$ 1D track used 80 bins.
We then calculated each animal’s binned occupancy in seconds as well as how many times each MEC cell spiked in each bin.
Binned firing rates for each MEC cell were calculated as the ratio of the number of spikes and the time spent in that bin in seconds.
Finally, a Gaussian filter ($\sigma=1$ bin) with dimensionality matching the environment (1D or 2D) was used to smooth the rate maps.

Since the model units do not have spikes but rates, the procedure is analogous for the agent but without Gaussian smoothing.
Specifically, to generate the model rate map (from layer $g$ for the path integrator networks, defined in \eqref{mec:eq:pathint}), we evaluate the model on 100 batches consisting of 200 evaluation trajectories (see Section~\ref{mec:ss:model-training-place} for more details), with each episode being seeded to ensure the same evaluation trajectories are used across all models.
Every timestep corresponded to 20$ms$ increments of time.

See Figure~\ref{mec:suppfig:gs} for visualizations of the rate maps of both the neural data and the model.

\subsection{Mapping Transforms}
\label{mec:supp:neuralfit-transform}
\begin{figure}[htbp]
    \centering
    \includegraphics[width=1.0\columnwidth]{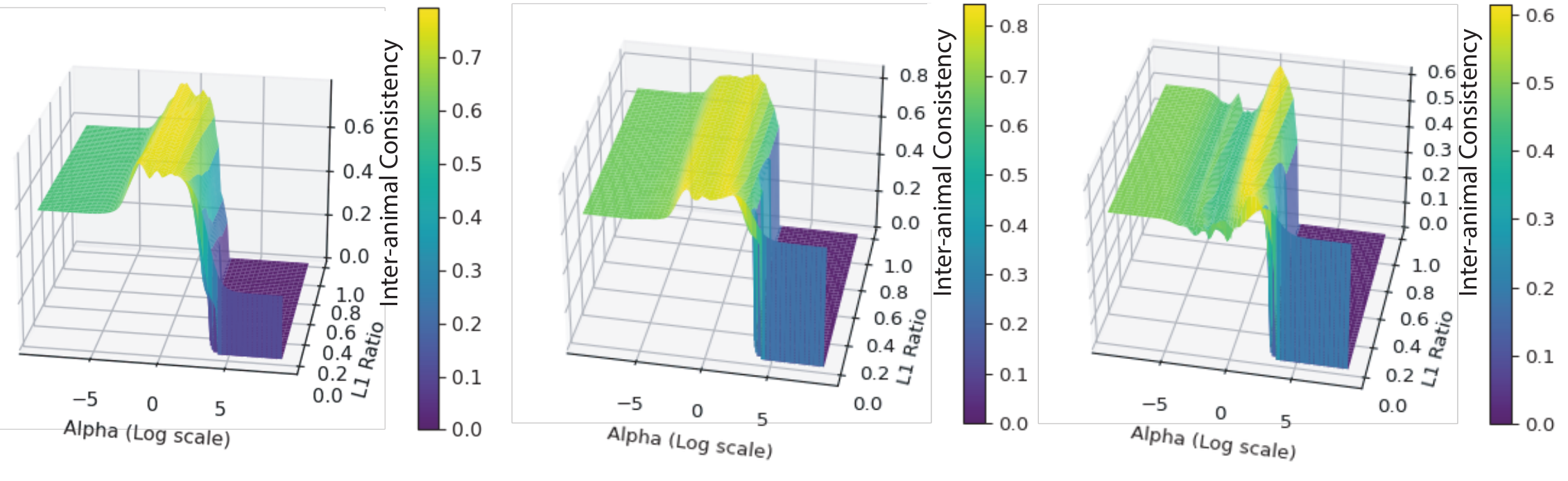}
    \caption[Inter-animal consistency as a function of $\alpha$ and L1 ratio (sparsity penalty strength)]{\textbf{Inter-animal consistency as a function of $\alpha$ and L1 ratio (sparsity penalty strength).} For $\alpha \in [10^{-9},10^{9}]$ logspaced uniformly and L1 ratio spaced uniformly $\in [0,1]$, we plot the inter-animal consistency evaluated on 80\% of position bins for a given train-test split and cell on the 100${cm}^2$ arena.}
    \label{mec:suppfig:surface}
\end{figure}

\begin{figure}[htbp]
    \centering
    \includegraphics[width=0.85\columnwidth]{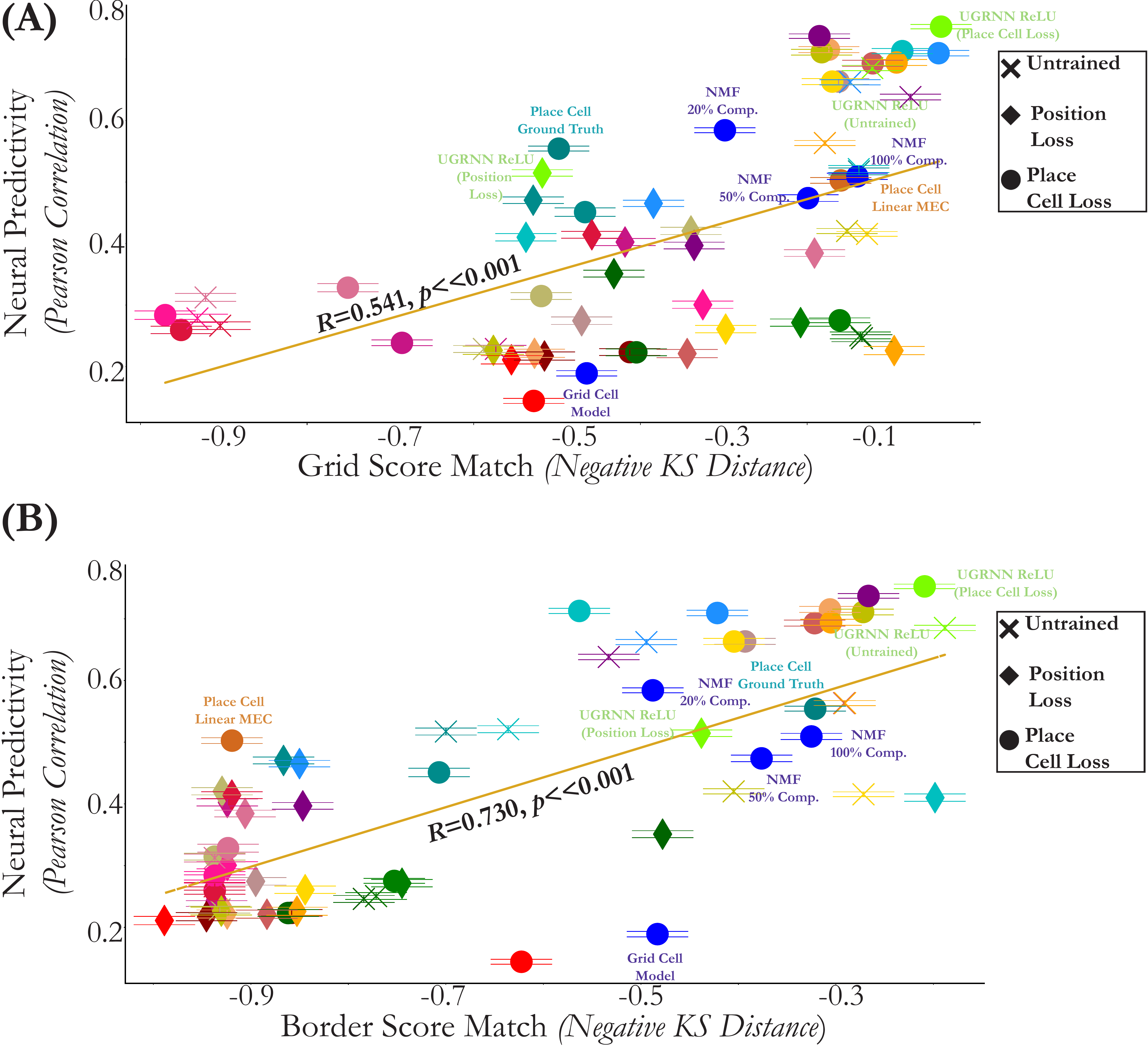}
    \caption[Neural predictivity and grid \& border score distribution match are related]{\textbf{Neural predictivity and grid \& border score distribution match are related.} Each model's neural predictivity on the 100${cm}^2$ foraging data versus grid score (top) and border score (bottom) distribution match. The neural predictivity is the median and s.e.m. across 620 cells. The models are either untrained (``X''), trained with the place cell loss (``O'') in the 2.2$m^2$ arena, or trained with the position loss (``Diamond'') in the 2.2$m^2$ arena.}
    \label{mec:suppfig:npvsgs}
\end{figure}

\begin{figure}[htbp]
    \centering
    \includegraphics[width=1.0\columnwidth]{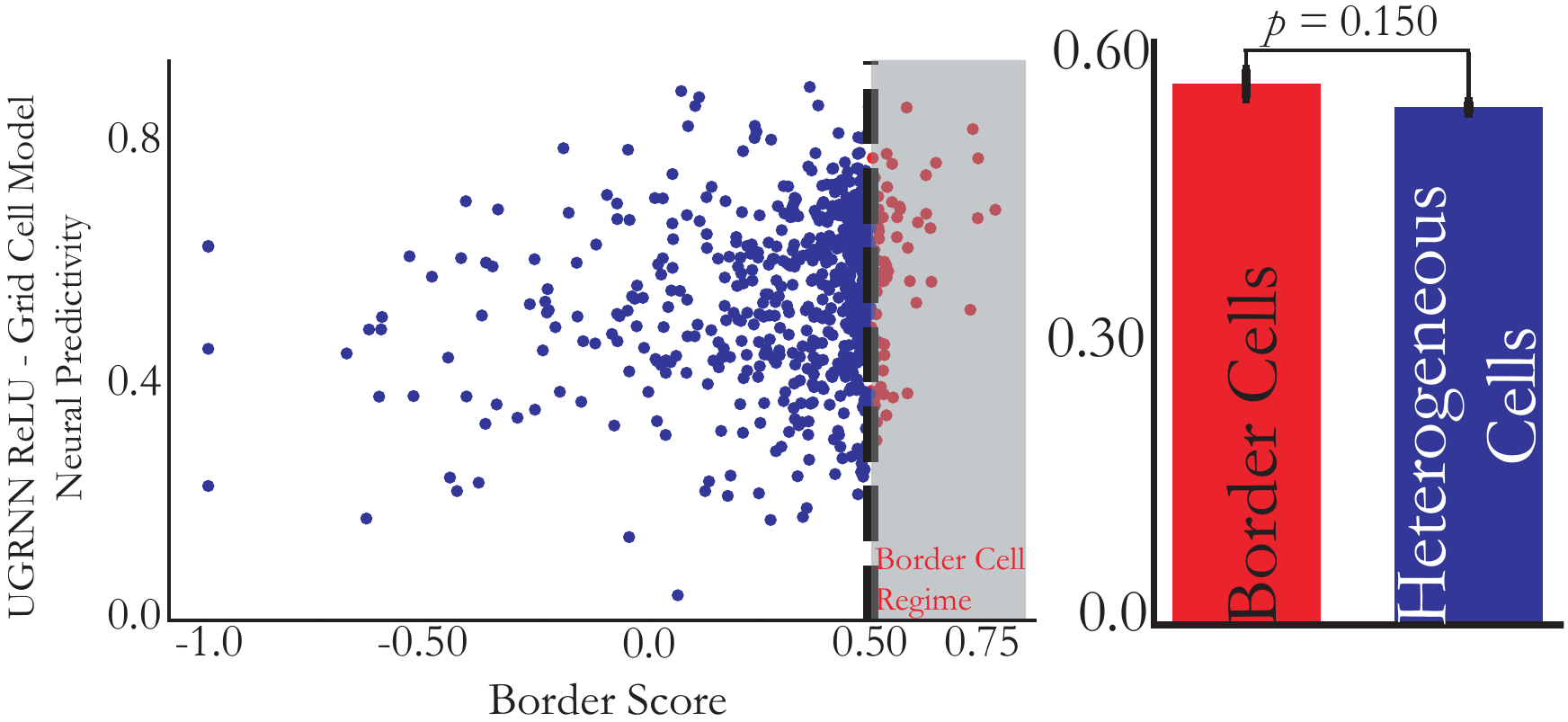}
    \caption[Difference between UGRNN-ReLU-Place Cell and Grid Cell models (Border Score)]{\textbf{Difference between UGRNN-ReLU-Place Cell and Grid Cell models (Border Score).} (\emph{Left}) Per unit neural predictivity difference between the UGRNN-ReLU-Place Cell and Grid Cell models trained on the 2.2$m^2$ arena and evaluated on the 100${cm}^2$ arena, plotted against that unit's border score. (\emph{Right}) Quantification of this difference aggregated across the border cell and heterogeneous (non-border) cell populations, respectively (mean and s.e.m).}
    \label{mec:suppfig:borderdiff}
\end{figure}

\begin{figure}[htbp]
    \centering
    \includegraphics[width=1.0\columnwidth]{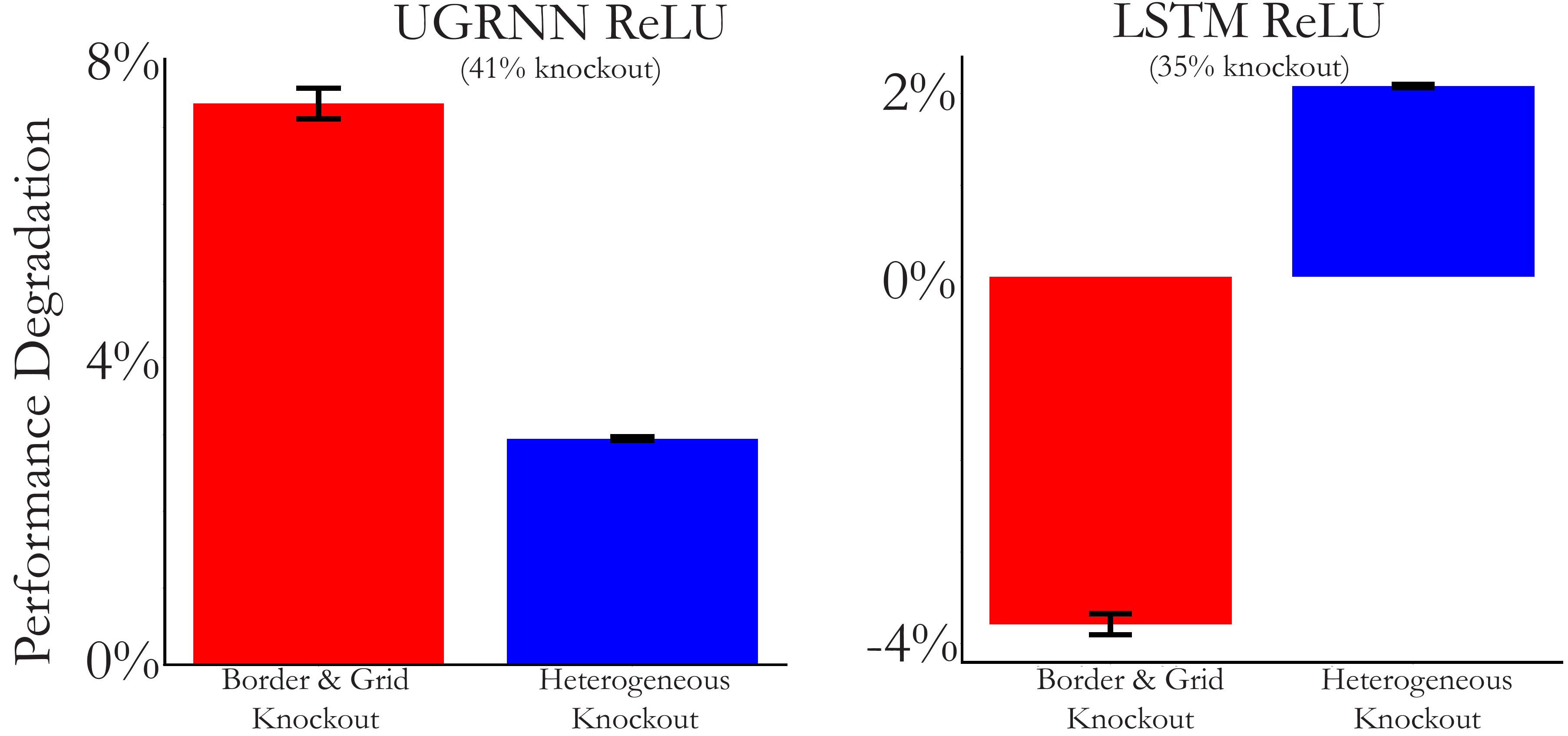}
    \caption[Combined border and grid cell knockout]{\textbf{Combined border and grid cell knockout.} For the UGRNN-ReLU-Place Cell and LSTM-ReLU-Place Cell models trained on the 2.2$m^2$ arena, we select grid cells (grid score $> 0.3$) and border cells (border score $> 0.5$) to knockout (red), and knockout the same number of heterogeneous cells (neither border nor grid cell), randomly sampled 100 times. The percentage knockout refers to the percentage of total units knocked out in the current layer, corresponding to the intermediate layer of the three layer network. Mean and s.e.m. over evaluation episodes.}
    \label{mec:suppfig:bgknockout}
\end{figure}

When we perform neural fits, we choose a random 20\% set of these position bins to train and cross-validate the regression, and the remaining 80\% to use as a test set, across ten train-test splits total.
For ElasticNet-based regression (including Lasso, Balanced Lasso-Ridge, and Ridge regression as special cases), we use the \texttt{sklearn.linear\_model} class.
When we perform cross-validation, which we do for the 100${cm}^2$ arena in 2D (Figure~\ref{mec:fig:interanimal-cons}b), the 400${cm}$ 1D track (Figure~\ref{mec:fig:neural-pred}b), and the 150${cm}^2$ for the 2D open foraging, reward, and combined across conditions datasets (Figure~\ref{mec:fig:reward}b), we search over $\alpha \in [10^{-9},10^{9}]$ logspaced uniformly and L1 ratio spaced uniformly $\in [0,1]$ with two-fold cross-validation on 20\% of the position bins per train-test split and neuron individually.
See Figure~\ref{mec:suppfig:surface} for the inter-animal consistency (2D data) across $\alpha$ and L1 ratio.
For the 1D data, we choose the parameters on the validation set that perform best averaged across the maps for a neuron.
The implementations of all of the transforms can be found here: \url{https://github.com/neuroailab/mec}.

\subsection{Grid Score and Border Score}
We calculated each cell (and model unit) grid score by taking a circular sample of the spatial rate map autocorrelation centered on the central peak and compared it to rotated versions of the same circular sample (60$^\circ$ and 120$^\circ$ versus 30$^\circ$, 90$^\circ$, and 150$^\circ$). 
The grid score~\citep{langston2010development,butler2019remembered} was defined as the mean correlation at 60$^\circ$ and 120$^\circ$ minus the mean correlation at 30$^\circ$, 90$^\circ$, and 150$^\circ$. 

The border score was computed following \cite{solstad2008representation}, by calculating $\dfrac{CM-DM}{CM+DM}$, where $CM$ is the proportion of high firing rate bins along one wall, and $DM$ is the normalized mean product of the firing rate of each bin and its distance to the nearest wall.

\begin{figure}[htbp]
    \centering
    \includegraphics[width=0.5\columnwidth]{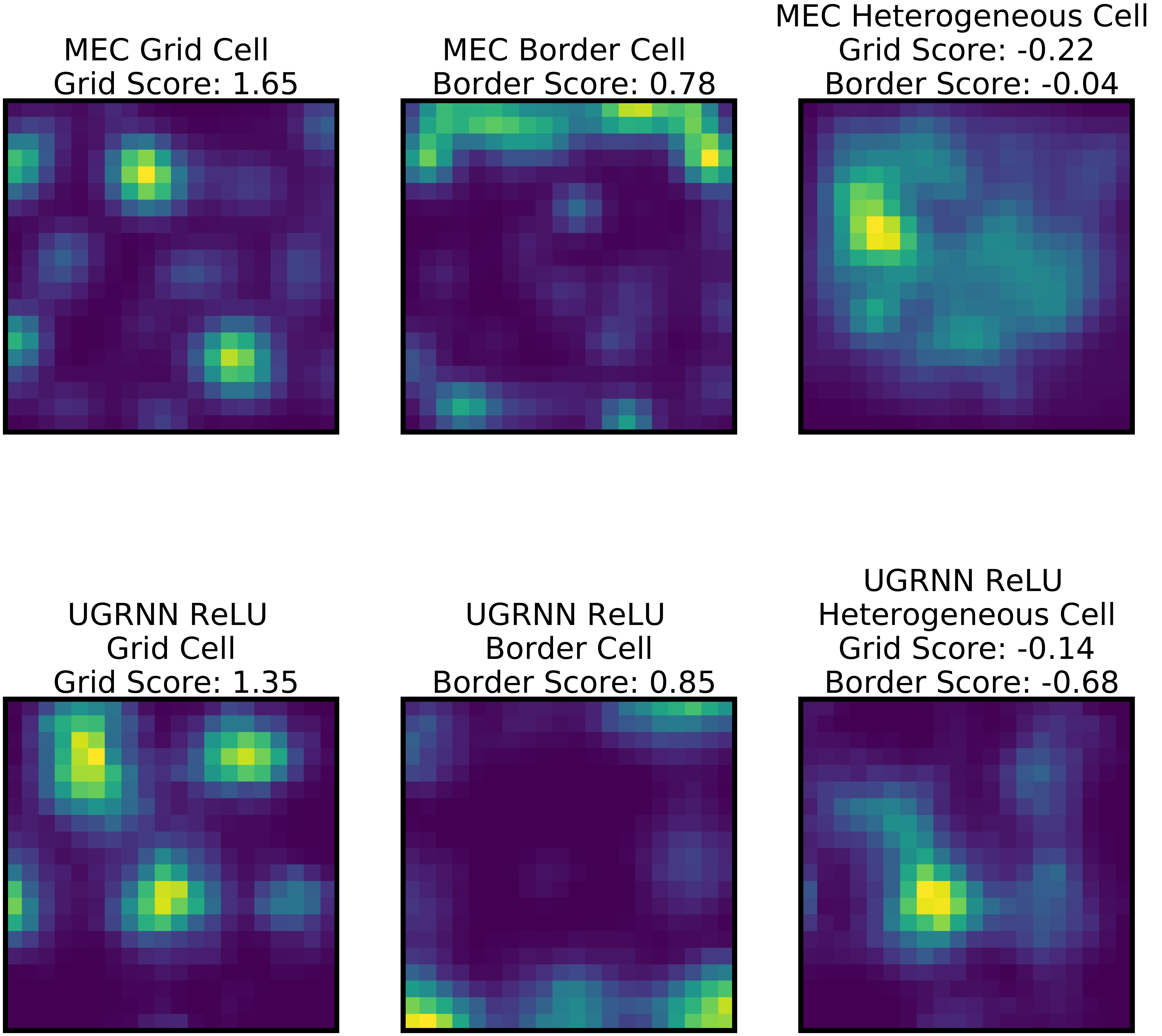}
    \caption[Example rate maps]{\textbf{Example rate maps.} Example rate maps obtained from animals foraging in the 100${cm}^2$ arena (top row) and rate maps from the UGRNN-ReLU-Place Cell trained in the 2.2$m^2$ arena and evaluated on the 100${cm}^2$ arena. We also include the grid and border scores of these example units.}
    \label{mec:suppfig:gs}
\end{figure}

\begin{figure}[htbp]
    \centering
    \includegraphics[width=0.5\columnwidth]{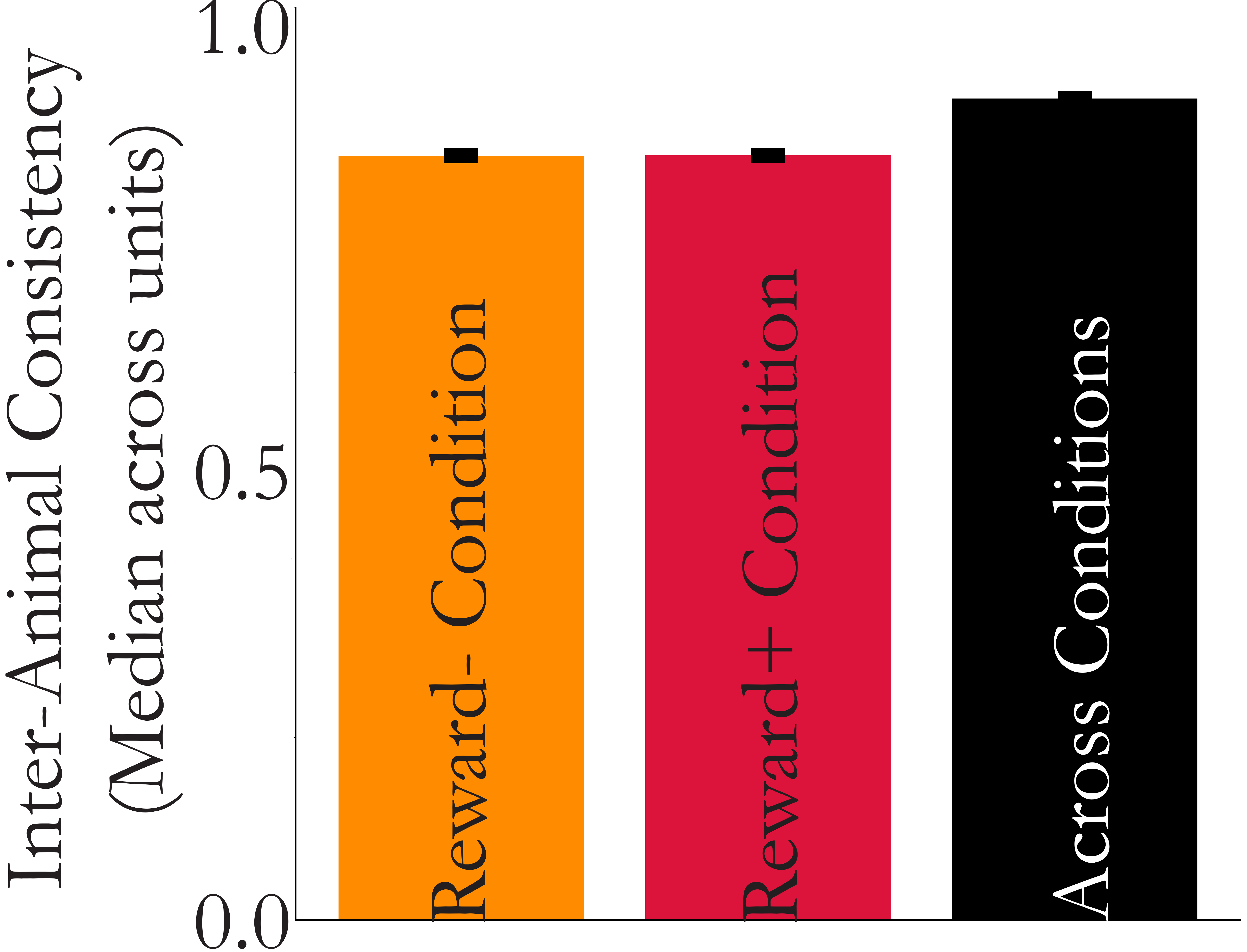}
    \caption[Inter-animal consistency is high per condition and across conditions]{\textbf{Inter-animal consistency is high per condition and across conditions.} Inter-animal consistency of neural responses (under ElasticNet CV regression) is computed per \textbf{reward-} and \textbf{reward+} condition separately, and across both spatial and reward-modulated conditions overall. Median and s.e.m. across 598 MEC cells from 7 rats in the 150${cm}^2$ arena.}
    \label{mec:suppfig:interan}
\end{figure}

The grid and border scores in Figure~\ref{mec:fig:grid} were always computed for each model at the layer of maximum neural predictivity (median across neurons). 
See Figure~\ref{mec:suppfig:gs} for visualizations of grid, border, and heterogeneous cells in both the neural data and the UGRNN-ReLU-Place Cell model.

\section{Neural Fitting Procedure and Inter-animal Consistency Definition}
\label{mec:supp:interancons}
Suppose we have neural responses from two animals $A$ and $B$.
Let $t_i^p$ be the ``true'' rate map of animal $p \in \mathcal{A} = \{A,B,...\}$ on stimulus set $i \in \{\train, \test\}$ given by positions in the rate map.
Of course, we only receive noisy observations of $t_i^p$, so let $s_{j,i}^p$ be the $j$-th set of $n$ trials of $t_i^p$.
Finally, let $M(x;y)_i$ be the predictions of a mapping $M$ (e.g. Ridge regression) when trained on input $x$ to match output $y$ and tested on stimulus set $i$.
For example, $M\left(\trueA;\trueB\right)_{\test}$ is the prediction of the mapping $M$ on the test stimulus trained to match the true neural responses from animal $B$ given input from the true neural responses from animal $A$ on the train stimulus, and correspondingly, $M\left(\sfAtrain;\sfBtrain\right)_{\test}$ is the prediction of the mapping $M$ on the test stimulus trained to match the (trial-average) of noisy sample 1 on the train stimulus from animal $B$ given inputs from the (trial-average) of noisy sample 1 on the train stimulus from animal $A$.
Finally, $r$ is the rate map constructed from the model units, which by construction are deterministic across trials.

With these definitions in hand, we now define the inter-animal consistency and the model neural predictivity.
Namely, when we have repeated trials in the same environment (as in the case of 1D data), we compute the following quantity for all units in target animal $B$:
\begin{equation}\label{mec:interanconfull}
\begin{split}
& {\text{Inter-animal Consistency}}_{\text{1D}}^B := \\
& \left\langle\dfrac{\corr\left(M\left(\sfAtrain;\sfBtrain\right)_{\test}, \ssBtest\right)}{\sqrt{\widetilde{\corr}\left(M\left(\sfAtrain;\sfBtrain\right)_{\test}, M\left(\ssAtrain;\ssBtrain\right)_{\test}\right) \times \widetilde{\corr}\left(\sfBtest, \ssBtest\right)}}\right\rangle_{A\in\mathcal{A}: (A,B)\in \mathcal{A}\times\mathcal{A}},
\end{split}
\end{equation}
\begin{equation}\label{mec:model1d}
\begin{split}
& {\text{Model Neural Predictivity}}_{\text{1D}}^B := \\
& \left\langle\dfrac{\corr\left(M\left(r_{\text{train}};\sfBtrain\right)_{\test}, \ssBtest\right)}{\sqrt{\widetilde{\corr}\left(M\left(r_{\text{train}};\sfBtrain\right)_{\test}, M\left(r_{\text{train}};\ssBtrain\right)_{\test}\right) \times \widetilde{\corr}\left(\sfBtest, \ssBtest\right)}}\right\rangle_{A\in\mathcal{A}: (A,B)\in \mathcal{A}\times\mathcal{A}},
\end{split}
\end{equation}
where the outermost average is across all source animals $A$ that regress to the current target animal $B$, ten train-test splits, and 100 bootstrapped trials, $\corr$ is Pearson correlation across test stimuli (in this case, held-out position bins), and $\widetilde{\corr}$ is Pearson correlation with Spearman-Brown correction applied to it, namely
\begin{equation*}
\widetilde{\corr}\left(X,Y\right) := \frac{2\corr\left(X,Y\right)}{1 + \corr\left(X,Y\right)}.
\end{equation*}
In 1D, we can also have \emph{multiple} maps within subsets of trials~\citep{low2020dynamic}, which we identify in Figure~\ref{mec:suppfig:remapping-1d}.
To account for this, we treat each map, $B^{m_{i}}$, which corresponds to responses of the same population in the target animal $B$ to a subset of trials, as its own target of explanation.
The average over source animals $A \in \mathcal{A}: A \ne B$ in \eqref{mec:interanconfull} now is an average over source animals and their respective maps, $A^{m_{j}}$.
Note that in the standard limit of one map per animal, this is exactly the same as the original quantity in \eqref{mec:interanconfull}.

In the absence of repeated trials in the same environment (which was the case for the 2D data) and therefore also a single map per environment per animal, the terms in the denominator of \eqref{mec:interanconfull} are trivially 1, giving us the following quantity for all units in target animal $B$:
\begin{equation}\label{mec:interanconnotrials}
\begin{split}
{\text{Inter-animal Consistency}}_{\text{2D}}^B := \left\langle\corr\left(M\left(s_{\text{train}}^{\hat{A}};s_{\text{train}}^{B}\right)_{\test}, s_{\text{test}}^B\right)\right\rangle,
\end{split}
\end{equation}
\begin{equation}\label{mec:model2d}
\begin{split}
{\text{Model Neural Predictivity}}_{\text{2D}}^B := \left\langle\corr\left(M\left(r_{\text{train}};s_{\text{train}}^{B}\right)_{\test}, s_{\text{test}}^B\right)\right\rangle,
\end{split}
\end{equation}
where now this outermost average is just across the ten train-test splits, and $\hat{A}$ is the sole ``pooled'' source animal constructed from the pseudo-population of the remaining animals distinct from the target animal $B$, namely, $A \in \mathcal{A}: A \ne B$.
As expanded on in Section~\ref{mec:ss:methods-interanimal-holdouts}, the latter use of the pooled source animal $\hat{A}$ is to ensure that we have a relatively comparable number of units in the source animal from the 2D tetrode data as in the 1D (Neuropixels) data, since otherwise there would be a small number of units in any single tetrode session.

The inter-animal consistency and model neural predictivity then is the concatenation (denoted by $\oplus$) of all of the inter-animal consistencies of all units in each animal $B \in \mathcal{A}$, over which we take median and s.e.m.:
\begin{equation}\label{mec:interancon1d}
\begin{split}
& {\text{Inter-animal Consistency}}_{\text{1D}} := \bigoplus_{B\in \mathcal{A}} \left\langle{\text{Inter-animal Consistency}}_{\text{1D}}^{B_{m_i}}\right\rangle_{\text{Maps $m_i \in B$}},\\
& {\text{Model Neural Predicitivity}}_{\text{1D}} := \bigoplus_{B\in \mathcal{A}} \left\langle{\text{Model Neural Predictivity}}_{\text{1D}}^{B_{m_i}}\right\rangle_{\text{Maps $m_i \in B$}},
\end{split}
\end{equation}
\begin{equation}\label{mec:interancon2d}
\begin{split}
& {\text{Inter-animal Consistency}}_{\text{2D}} := \bigoplus_{B\in \mathcal{A}} {\text{Inter-animal Consistency}}_{\text{2D}}^B,\\
& {\text{Model Neural Predictivity}}_{\text{2D}} := \bigoplus_{B\in \mathcal{A}} {\text{Model Neural Predictivity}}_{\text{2D}}^B,
\end{split}
\end{equation}
Thus, Figures~\ref{mec:fig:interanimal-cons}b,~\ref{mec:fig:neural-pred}a, and~\ref{mec:fig:reward}b are the median and s.e.m. of the quantities in \eqref{mec:interancon2d}.
Figure~\ref{mec:fig:neural-pred}b is the quantity in \eqref{mec:interancon1d}, which we pass through $\tanh$ to be $\in [-1,1]$, the same scale for visual comparison as \eqref{mec:interancon2d}, which we then compute median and s.e.m. over.
Note that we compute these quantities in the 2D and 1D data at \emph{each} model layer, and then report the performance at the layer of maximum median neural predictivity for each model.

In the following subsections, we give background on how the full quantity \eqref{mec:interanconfull} can be obtained from the base ``true'' quantity we want to estimate \eqref{mec:interancontrue}.
Note that these are not formal proofs (as they rely on assumptions which do not necessarily hold in all cases), but are meant to outline the motivations behind the final quantity.

\subsection{Single Animal Pair Motivation}
\label{mec:ss:derivation-pair}
The inter-animal consistency from one animal $A$ to another animal $B$ corresponds to the following ``true'' quantity to be estimated:
\begin{equation}\label{mec:interancontrue}
\corr\left(M\left(\trueA;\trueB\right)_{\test}, \trueBtest\right),
\end{equation}
where $\corr$ is the Pearson correlation across test stimuli.
In what follows, we argue that this true quantity can be approximated with the following ratio of measurable quantities where we divide the noisy trial observations into two sets of equal samples:
\begin{equation}\label{mec:interancon}
\begin{split}
&\corr\left(M\left(\trueA;\trueB\right)_{\test}, \trueBtest\right) \\
& \sim \dfrac{\corr\left(M\left(\sfAtrain;\sfBtrain\right)_{\test}, \ssBtest\right)}{\sqrt{\corr\left(M\left(\sfAtrain;\sfBtrain\right)_{\test}, M\left(\ssAtrain;\ssBtrain\right)_{\test}\right) \times \corr\left(\sfBtest, \ssBtest\right)}}.
\end{split}
\end{equation}
In words, the inter-animal consistency corresponds to the predictivity of the mapping on the test set stimuli from animal $A$ to $B$ on two different (averaged) halves of noisy trials, corrected by the square root of the mapping reliability on animal $A$'s test stimuli responses on two different halves of noisy trials and the internal consistency of animal $B$.

We justify the approximation in \eqref{mec:interancon} by gradually eliminating the true quantities by their measurable estimates, starting from the original quantity in \eqref{mec:interancontrue}.
First, we make the approximation that
\begin{equation}\label{mec:step1}
\begin{split}
& \corr\left(M\left(\trueA;\trueB\right)_{\test}, \ssBtest\right) \\
& \sim \corr\left(M\left(\trueA;\trueB\right)_{\test}, \trueBtest\right) \times \corr\left(\trueBtest, \ssBtest\right).
\end{split}
\end{equation}
by transitivity of positive correlations (which is a reasonable assumption when the number of stimuli is large).
Next, by transitivity and normality assumptions in the structure of the noisy estimates and since the number of trials ($n$) between the two sets is the same, we have that
\begin{equation}\label{mec:step2}
\begin{split}
\corr\left(\sfBtest, \ssBtest\right) & \sim \corr\left(\sfBtest, \trueBtest\right) \times \corr\left(\trueBtest, \ssBtest\right)\\
& \sim \corr\left(\trueBtest, \ssBtest\right)^2.
\end{split}
\end{equation}
Namely, the correlation between the average of two sets of noisy observations of $n$ trials each is approximately the square of the correlation between the true value and average of one set of $n$ noisy trials.
Therefore, from \eqref{mec:step1} and \eqref{mec:step2} it follows that
\begin{equation}\label{mec:lemma1}
\corr\left(M\left(\trueA;\trueB\right)_{\test}, \trueBtest\right) \sim \dfrac{\corr\left(M\left(\trueA;\trueB\right)_{\test}, \ssBtest\right)}{\sqrt{\corr\left(\sfBtest, \ssBtest\right)}}.
\end{equation}

We have gotten rid of $\trueBtest$, but we still need to get rid of the $M\left(\trueA;\trueB\right)_{\test}$ term.
We apply the same two steps by analogy though these approximations may not always be true (though are true for Gaussian noise):
\begin{equation*}
\begin{split}
\corr\left(M\left(\sfAtrain;\sfBtrain\right)_{\test}, \ssBtest\right) & \sim \corr\left(\ssBtest, M\left(\trueA;\trueB\right)_{\test}\right) \\
& \times \corr\left(M\left(\trueA;\trueB\right)_{\test}, M\left(\sfAtrain;\sfBtrain\right)_{\test}\right)
\end{split}
\end{equation*}
\begin{equation*}
\begin{split}
& \corr\left(M\left(\sfAtrain;\sfBtrain\right)_{\test}, M\left(\ssAtrain;\ssBtrain\right)_{\test}\right) \\
& \sim \corr\left(M\left(\sfAtrain;\sfBtrain\right)_{\test}, M\left(\trueA;\trueB\right)_{\test}\right)^2,
\end{split}
\end{equation*}
which taken together implies
\begin{equation}\label{mec:lemma2}
\begin{split}
& \corr\left(M\left(\trueA;\trueB\right)_{\test}, \ssBtest\right) \\
& \sim \dfrac{\corr\left(M\left(\sfAtrain;\sfBtrain\right)_{\test}, \ssBtest\right)}{\sqrt{\corr\left(M\left(\sfAtrain;\sfBtrain\right)_{\test}, M\left(\ssAtrain;\ssBtrain\right)_{\test}\right)}}.
\end{split}
\end{equation}
Equations \eqref{mec:lemma1} and \eqref{mec:lemma2} together imply the final estimated quantity given in \eqref{mec:interancon}.

\subsection{Multiple Animals}
\label{mec:ss:multiple-animals}

For multiple animals, we simply consider the average of the true quantity for each target in $B$ in \eqref{mec:interancontrue} across source animals $A$ in the ordered pair $(A,B)$ of animals $A$ and $B$:
\begin{equation*}\label{mec:multipleinterancon}
\begin{split}
&\left\langle \corr\left(M\left(\trueA;\trueB\right)_{\test}, \trueBtest\right)\right\rangle_{A\in\mathcal{A}: (A,B)\in \mathcal{A}\times\mathcal{A}} \\
& \sim \left\langle\dfrac{\corr\left(M\left(\sfAtrain;\sfBtrain\right)_{\test}, \ssBtest\right)}{\sqrt{\corr\left(M\left(\sfAtrain;\sfBtrain\right)_{\test}, M\left(\ssAtrain;\ssBtrain\right)_{\test}\right) \times \corr\left(\sfBtest, \ssBtest\right)}}\right\rangle_{A\in\mathcal{A}: (A,B)\in \mathcal{A}\times\mathcal{A}}.
\end{split}
\end{equation*}
Typically, we may bootstrap across split-half trials and have multiple train/test splits, in which case the average on the right hand side of the equation includes averages across these as well.

Note that each neuron in our analysis will have this single average value associated with it when \emph{it} was a target animal ($B$), averaged over source animals, bootstrapped split-half trials, and train/test splits.
This yields a vector of these average values, which we can take median and s.e.m. over as we do with standard explained variance metrics.

\subsection{Spearman-Brown Correction}
\label{mec:ss:spearman-brown}
The Spearman-Brown correction is to be applied to each of the terms in the denominator individually, as they are each correlations of observations from half the trials of the \emph{same} underlying process to itself (unlike the numerator).

\subsection{Pooled Source Animal}
\label{mec:ss:methods-interanimal-holdouts}
\begin{figure}[htbp]
    \centering
    \includegraphics[width=0.5\columnwidth]{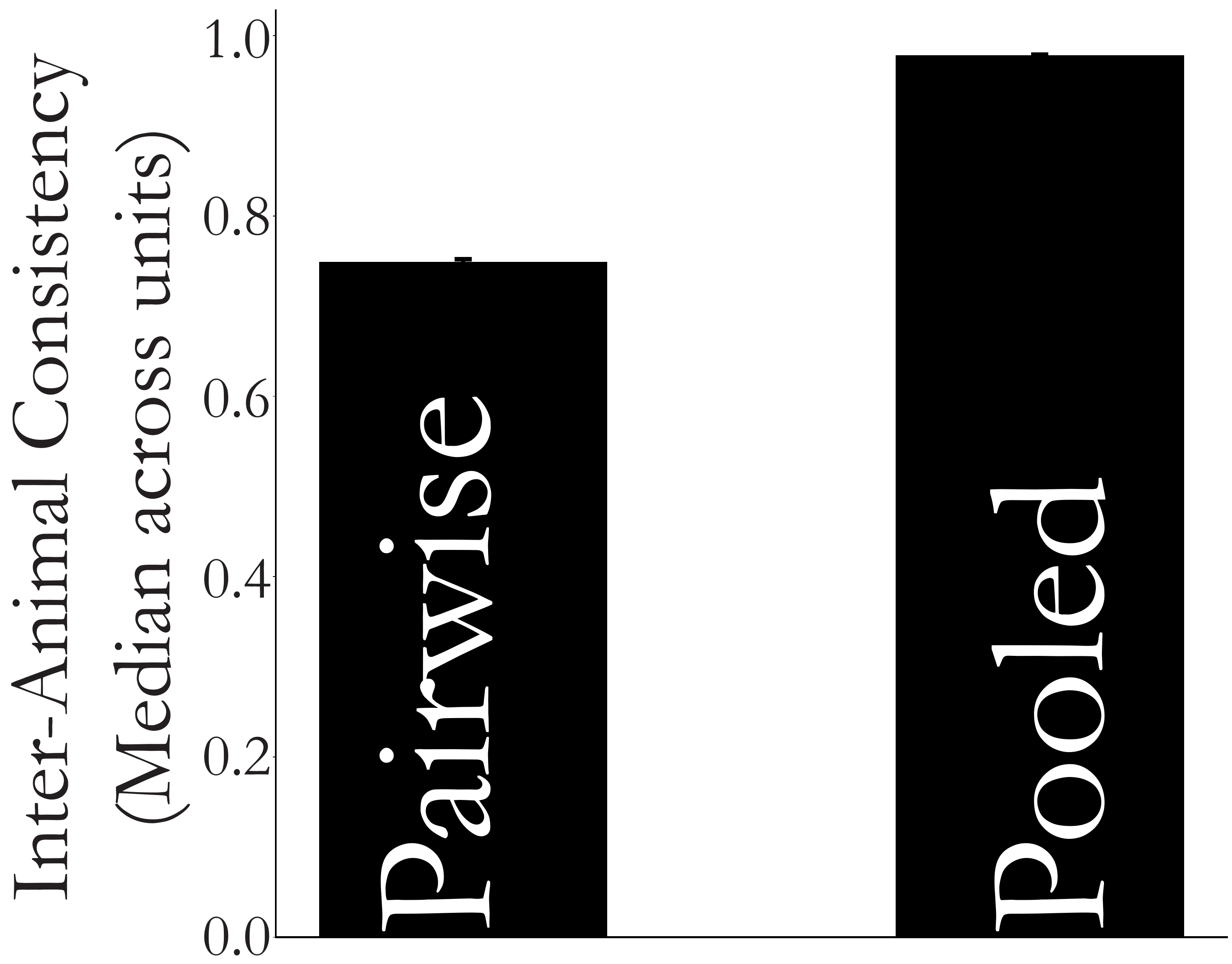}
    \caption[Pooling across source animals]{\textbf{Pooling across source animals.} Inter-animal consistency under Ridge regression ($\alpha=1$) trained with 50\% position bins (averaged across ten train-test splits) from 12 mice foraging in the 100${cm}^2$ arena. Median and s.e.m. across 620 cells. ``Pooled'' refers to computing the inter-animal consistency using a single ``pooled'' source animal from units gathered from all animals except the target animal, as described in Section~\ref{mec:ss:methods-interanimal-holdouts}. ``Pairwise'' refers to computing this quantity mapping one source animal at a time to the target animal.}
    \label{mec:suppfig:pooled}
\end{figure}

\begin{figure}[htbp]
    \centering
    \includegraphics[width=0.29\columnwidth]{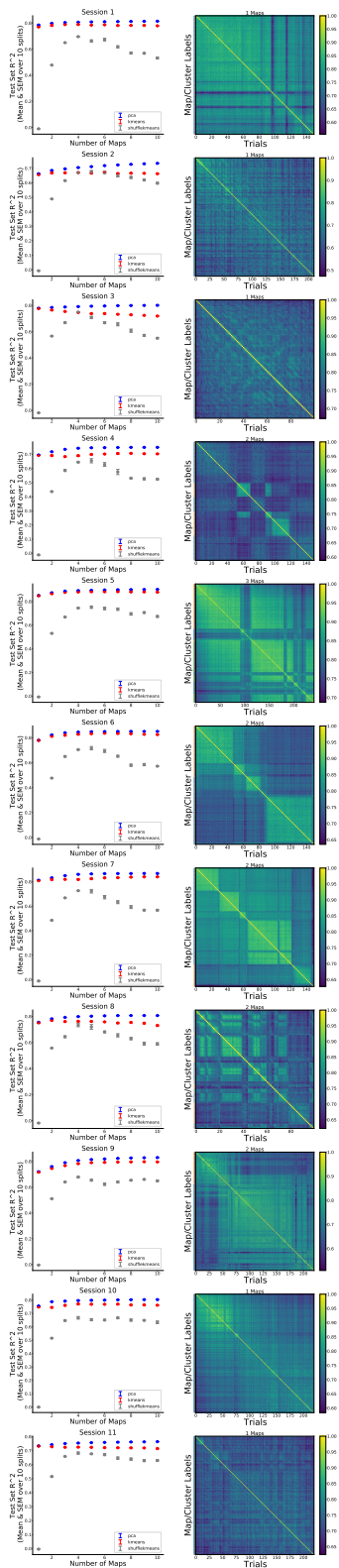}
    \caption[Remapping analysis for 1D VR data]{\textbf{Remapping analysis for 1D VR data.} For each recording session (11 total), we examine the trial-by-trial similarity matrices (right column) and the test set $R^2$ of $k$-means clustering as a function of $k$ (left column) relative to both PCA (blue) and a shuffled control (gray). The number of identified maps per session is listed at thee top of each trial-by-trial similarity matrix, along with the colored cluster labels as to the assignments for each trial.}
    \label{mec:suppfig:remapping-1d}
\end{figure}

Often times, we may not have enough neurons per animal to ensure that the estimated inter-animal consistency in our data closely matches the ``true'' inter-animal consistency.
In order to address this issue, we holdout one animal at a time and compare it to the pseudo-population aggregated across units from the remaining animals, as opposed to computing the consistencies in a pairwise fashion.
Thus, $B$ is still the target heldout animal as in the pairwise case, but now the average over $A$ is over the sole ``pooled'' source animal $\hat{A}$ constructed from the pseudo-population of the remaining animals.
We found that this pooling of the source animal units helped improve the estimated inter-animal consistency, as demonstrated in Figure~\ref{mec:suppfig:pooled}.

\section{Model Training Details}
\label{mec:supp:model-training}
All model code be found here: \url{https://github.com/neuroailab/mec}.

\subsection{Simulated Trajectories and Place Cell Representation}
\label{mec:ss:model-training-place}
Place cell receptive field centers $\vec{c}_i$, $i=1,\ldots,n_P$, distributed uniformly randomly across each environment.
This environment is the 2.2$m^2$ arena for all models except for the ``Trained'' bars in Figure~\ref{mec:fig:neural-pred}a, which corresponds to training in the 100${cm}^2$ environment, and the ``1D Track'' bar in Figure~\ref{mec:fig:neural-pred}b which corresponds to training on the 400$cm$ track.
We take $n_P = 512$ place cells in all environments and models, following~\cite{banino2018vector}.

The response of the $i$-th place cell is simulated using a difference of Gaussians tuning curve,
$p_i(x) = e^{-\|x-c_i\|^2_2/2\sigma_1^2} - e^{-\|x-c_i\|^2_2/2\sigma_2^2}$, where $x$ is the current location of the agent, $\sigma_1=0.12m$ and $\sigma_2=0.12\sqrt{2}m$.
Agent trajectories are generated using the rat motion model of~\cite{erdem2012goal}.
In 1D, during either training or evaluation we prevent the agent from making turns, in order to simulate the head-fixed condition that the mice experience.
We collect the place cell activations at $n_x$ locations as the animal explores its environment in a matrix $P \in \mathbb{R}^{n_x\times n_P}$.

\subsection{Place Cell Input Models}
While the path integrator networks are trained with the place cells as supervised \emph{outputs}, defined in Section~\ref{mec:ss:model-training-loss}, we have several controls based on the place cell representation.
The ``Place Cell Ground Truth'' model directly corresponds to the matrix $P$.

NMF corresponds to Non-negative Matrix Factorization on the matrix $P$, implemented via \texttt{sklearn.decomposition.NMF}.
As noted by~\cite{dordek2016extracting}, this corresponds to a 1-layer neural network with $n_G$ hidden units via unsupervised Hebbian learning on inputs $P$, subject to a nonnegativity constraint.
The ``Grid Cell Model'' corresponds specifically to NMF with $n_G=9$ components, following~\cite{sorscher2019unified}.

Finally, the ``Place Cell Linear MEC'' model is intended to be a neural data constrained linear alternative to NMF on the place cell matrix $P$.
This is ElasticNet CV regression trained on 20\% of the position bins in the current evaluation environment (100${cm}^2$ arena in Figure~\ref{mec:fig:neural-pred}a and $400cm$ track in Figure~\ref{mec:fig:neural-pred}b), fitted to the neurons of animals distinct from the current target neural population (namely the units of the source animal defined in Section~\ref{mec:supp:interancons}).

\subsection{Loss Functions}
\label{mec:ss:model-training-loss}
The ``Place Cell Loss'' corresponds to the loss function used by~\cite{banino2018vector}, which is the softmax cross-entropy loss between the ground truth timestep $t$ place cell targets $p^t_i$ and model outputs $\hat{p}^t_i$, given by
\begin{equation}\label{mec:eq:place-loss}
\mathcal{L}(\hat{p}, p) := -\frac{1}{T}\sum_{t=1}^T\sum_{i=1}^{N_p}p^t_i\log\hat{p}^t_i.
\end{equation}
The ``Position Loss''~\citep{cueva2018emergence} is given by
\begin{equation}\label{mec:eq:pos-loss}
\mathcal{L}(\hat{p}, p) := \frac{1}{2}\frac{1}{T}\sum_{t=1}^T\left(\left({{p}}^t_x - \widehat{{p}}^t_x\right)^2 + \left({{p}}^t_y - \widehat{{p}}^t_y\right)^2\right),
\end{equation}
where ${p}^t_x$ and ${p}^t_y$ are the \emph{Cartesian} coordinates $(x,y)$ of the agent's ground truth position at timestep $t$.

Both loss functions are averaged across the batch, where the path length in each batch for both loss functions is $T = 20$ timesteps.
Additionally, for either loss function, we apply an L2 penalty of $1\times 10^{-4}$ to the path integrator weights.

\subsection{Network Architectures and Hyperparameters}
\label{mec:ss:model-training-hp}

\subsubsection{Path Integrators}
\label{mec:sss:model-training-of}
We use Tensorflow 2.0 for these models~\citep{Abadi2016}.
The RNN path integrator network takes in 2 linear input units for $x$ and $y$ velocity, a set of recurrently connected input units, and linear readout units.
The network update equations are as follows:
\begin{equation}\label{mec:eq:pathint}
\begin{split}
& g^{t+1} = f\left(Jg^{t} + Mv^{t}\right) \\
& \hat{p}^{t+1} = Wg^{t+1},
\end{split}
\end{equation}
where $g$ is the vector of model MEC activities (4096 units total), $J$ is the matrix of recurrent weights, $M$ is the network's velocity input weights, $v$ is the agent’s 2D velocity in the arena, $f$ is the element-wise nonlinearity (or the identity function if it is ``Linear''), $\hat{p}$ is the vector of estimated place cell activities (or outputted Cartesian coordinates if the network is being trained with the ``Position Loss''), and $W$ is the place cell (or Cartesian coordinate) readout weights.

The SRNN, GRU, LSTM, and UGRNN path integrator networks had an identical task and training protocol as \eqref{mec:eq:pathint}.
The model architecture was reproduced from~\cite{banino2018vector} except that our models had 4096 units (rather than 128) in order to match the number of units of the MEC layer ($g$) of the RNN path integrator above.
Specifically, it consists of 2D velocity inputs to a recurrent circuit (SRNN, GRU, LSTM, or UGRNN) with 4096 units with nonlinearity either being Linear, Tanh, Sigmoid, or ReLU, followed by a nonlinear layer of 4096 units (which constitutes the model's activities $g$, with the nonlinearity matching that of the recurrent circuit, following~\cite{sorscher2019unified}), followed by a final readout to the estimated 512 place cell activities (or 2 Cartesian units of position if training with the ``Position Loss'').
Furthermore, the initial cell state and hidden state are initialized by computing a linear transformation of the ground truth place cells (or Cartesian positions if training with the ``Position Loss'') at time 0.
We did not employ any dropout at the $g$ layer of these networks during training.

All networks are trained with Adam~\citep{kingma2014adam} with a learning rate of $1\times 10^{-4}$, batch size of 200, and 100 training epochs consisting of 1000 batches of trajectories per epoch.

\subsubsection{Cue Input}
\label{mec:sss:model-training-cue}
MEC neurons are known to respond to visual input, but as none of our evaluated models had a visual front-end, they could not respond accordingly when evaluated on a cue-rich environment (which was the case for the 1D data). 
To test this hypothesis, we trained the UGRNN-ReLU-Place Cell model in the 2.2$m^2$ 2D arena, but with input visual cue locations concatenated with the 2D velocity input, corresponding to the ``Cue Input + Velocity Input'' model in Figure~\ref{mec:fig:neural-pred}b.
Specifically, these visual cue locations were a vector $\ell \in \mathbb{R}^5$, corresponding to 5 cues placed in fixed, arbitrary locations in the 2D arena with widths between 0.06 to 0.3$m$ on each side.
Each element $\ell_i$ of this cue input vector corresponds to the Euclidean distance of the agent at current time $t$ to the nearest boundary of the $i$-th cue (and 0 for that entry if the agent is within the boundaries of this $i$-th cue).
We also considered a UGRNN-ReLU-Place Cell network trained in 2D without any velocity input and only the cue input, corresponding to the ``Cue Input Only'' bar in Figure~\ref{mec:fig:neural-pred}b.

Finally, we evaluated these networks with the 1D cue input which matched the widths and locations of the cues in the 1D virtual track.
The neural predictivity of this 1D cue input is the ``Cue Input'' bar in Figure~\ref{mec:fig:neural-pred}b.
As a control, we also included a UGRNN-ReLU-Place Cell path integrator trained with the usual 2D velocity input but with constant 0 concatenated to the velocities in place of the active cue input (thus being functionally equivalent to the path integrators in Section~\ref{mec:sss:model-training-of}).
The network was then provided the 1D cue inputs during evaluation only, corresponding to the ``Evaluation Cue Input Only'' bar in Figure~\ref{mec:fig:neural-pred}b.

\subsubsection{Reward Biased Path Integrator}
\label{mec:sss:model-training-reward}
For the reward biased path integrator (parametrized by $\varepsilon$ in Figure~\ref{mec:fig:reward}b), we trained the UGRNN-ReLU-Place Cell in the 2.2$m^2$ arena, where on $(1-\varepsilon)$ fraction of training batches (each example in the batch corresponding to a 20 timestep path-length episode), we had the agent navigate directly within a fixed number of timesteps (7) to the center of a 20${cm}^2$ reward zone placed in an arbitrary, fixed location of the environment.
In the remaining 13 timesteps of these episodes, the agent path integrated using the motion trajectories of~\citep{erdem2012goal} but restricted to the 20${cm}^2$ reward zone.
In the other $\varepsilon$ fraction of episodes, the trajectories were unchanged from before.

As a control, we consider a ``Reward Input'' path integrator (Figure~\ref{mec:fig:reward}b,c), which does \emph{not} employ the reward biased trajectories (so $\varepsilon=1$), but instead takes an additional scalar reward signal (concatenated with the 2D velocity input) during training, indicating if it is in the reward zone or not at current timestep $t$.

\section{Performance Measure and Ablation}
\label{mec:ss:model-training-performance}
Positional decoding error for the place cell loss models is measured first by computing a predicted Cartesian position $\hat{p}^t_x, \hat{p}^t_y$, obtained by taking the top 3 most active place cell outputs at each timestep in each model and averaging them.
Finally, for each trajectory episode (of length $T=20$), we compute the following measure of path integration error,
\begin{equation}\label{mec:eq:perf}
\mathcal{E}(\hat{p}, p) := -\frac{1}{T}\sum_{t=1}^T\sqrt{\left(\left({{p}}^t_x - \widehat{{p}}^t_x\right)^2 + \left({{p}}^t_y - \widehat{{p}}^t_y\right)^2\right)},
\end{equation}
where the error is additionally averaged over the batch dimension (20,000 examples computed from 100 batches of 200 evaluation trajectory episodes each).

The performance degradation metric that is the $y$-axis of Figure~\ref{mec:fig:grid}d is given by $\left(\mathcal{E}^{curr} - \mathcal{E}^{full}\right)/\mathcal{E}^{full}$, where $\mathcal{E}^{full}$ is the performance of the trained UGRNN-ReLU-Place Cell model, and $\mathcal{E}^{curr}$ is the performance of the same model but with a subset of its population of units in the $g$ layer set to 0.
In Figure~\ref{mec:fig:task}d, for the ``grid cell knockout'', the outputs of the set of units in layer $g$ with grid score $> 0.3$ are set to 0 during evaluation, and for the ``heterogeneous cell knockout'', the same number of units in layer $g$ with grid score $\le 0.3$ are randomly set to 0 (subsampled 100 times).
Analogously, for the ``border cell knockout'', the border score was set to a threshold of 0.5, and the heterogeneous knockout in that case was border score $\le 0.5$.

\section{Supplementary Figures}

\chapter{Building Performant Biological Learning Rules}
\label{ch:lrperf}
\section{Chapter Abstract}
The neural plausibility of backpropagation has long been disputed, primarily for its use of non-local weight transport --- the biologically dubious requirement that one neuron instantaneously measure the synaptic weights of another. 
Until recently, attempts to create local learning rules that avoid weight transport have typically failed in the large-scale learning scenarios where backpropagation shines, e.g. ImageNet categorization with deep convolutional networks.
Here, we investigate a recently proposed local learning rule that yields competitive performance with backpropagation and find that it is highly sensitive to metaparameter choices, requiring laborious tuning that does not transfer across network architecture.
Our analysis indicates the underlying mathematical reason for this instability, allowing us to identify a more robust local learning rule that better transfers without metaparameter tuning.
Nonetheless, we find a performance and stability gap between this local rule and backpropagation that widens with increasing model depth.
We then investigate several non-local learning rules that relax the need for instantaneous weight transport into a more biologically-plausible ``weight estimation'' process, showing that these rules match state-of-the-art performance on deep networks and operate effectively in the presence of noisy updates.
Taken together, our results suggest two routes towards the discovery of neural implementations for credit assignment without weight symmetry: further improvement of local rules so that they perform consistently across architectures and the identification of biological implementations for non-local learning mechanisms.

\section{Introduction}
\label{lrperf:intro}

Backpropagation is the workhorse of modern deep learning and the only known learning algorithm that allows multi-layer networks to train on large-scale tasks.
However, any exact implementation of backpropagation is inherently non-local, requiring instantaneous weight transport in which backward error-propagating weights are the transpose of the forward inference weights. 
This violation of locality is biologically suspect because there are no known neural mechanisms for instantaneously coupling distant synaptic weights. 
Recent approaches such as feedback alignment \cite{lillicrap_random_2016} and weight mirror \cite{Akrout2019} have identified circuit mechanisms that seek to approximate backpropagation while circumventing the weight transport problem. 
However, these mechanisms either fail to operate at large-scale \cite{bartunov_assessing_2018} or, as we demonstrate, require complex and fragile metaparameter scheduling during learning. 
Here we present a unifying framework spanning a space of learning rules that allows for the systematic identification of robust and scalable alternatives to backpropagation. 

To motivate these rules, we replace tied weights in backpropagation with a regularization loss on untied forward and backward weights. 
The forward weights parametrize the global cost function, the backward weights specify a descent direction, and the regularization constrains the relationship between forward and backward weights. 
As the system iterates, forward and backward weights dynamically align, giving rise to a pseudogradient. 
Different regularization terms are possible within this framework. 
Critically, these regularization terms decompose into geometrically natural primitives, which can be parametrically recombined to construct a diverse space of credit assignment strategies. 
This space encompasses existing approaches (including feedback alignment and weight mirror), but also elucidates novel learning rules. 
We show that several of these new strategies are competitive with backpropagation on real-world tasks (unlike feedback alignment), without the need for complex metaparameter tuning (unlike weight mirror). 
These learning rules can thus be easily deployed across a variety of neural architectures and tasks. 
Our results demonstrate how high-dimensional error-driven learning can be robustly performed in a biologically motivated manner.

\section{Related Work}
\label{lrperf:sec:related}

Soon after \citet{rumelhart_learning_1986} published the backpropagation algorithm for training neural networks, its plausibility as a learning mechanism in the brain was contended \cite{crick_recent_1989}.  The main criticism was that backpropagation requires exact transposes to propagate errors through the network and there is no known physical mechanism for such an ``operation" in the brain. 
This is known as the \textit{weight transport problem} \cite{grossberg_competitive_1987}. Since then many credit assignment strategies have proposed circumventing the problem by introducing a distinct set of feedback weights to propagate the error backwards. Broadly speaking, these proposals fall into two groups: those that encourage symmetry between the forward and backward weights \cite{lillicrap_random_2016,nokland_direct_2016,bartunov_assessing_2018,liao_how_2016,xiao_biologically-plausible_2019,moskovitz_feedback_2018,Akrout2019}, and those that encourage preservation of information between neighboring network layers \cite{bengio_how_2014, lee_difference_2015, bartunov_assessing_2018}.

The latter approach, sometimes referred to as target propagation, encourages the backward weights to locally invert the forward computation \cite{bengio_how_2014}.  
Variants of this approach such as difference target propagation \cite{lee_difference_2015} and simplified difference target propagation \cite{bartunov_assessing_2018} differ in how they define this inversion property.  
While some of these strategies perform well on shallow networks trained on MNIST and CIFAR10, they fail to scale to deep networks trained on ImageNet \cite{bartunov_assessing_2018}.

A different class of credit assignment strategies focuses on encouraging or enforcing symmetry between the weights, rather than preserving information.
\citet{lillicrap_random_2016} introduced a strategy known as feedback alignment in which backward weights are chosen to be fixed random matrices. 
Empirically, during learning, the forward weights partially align themselves to their backward counterparts, so that the latter transmit a meaningful error signal. \citet{nokland_direct_2016} introduced a variant of feedback alignment where the error signal could be transmitted across long range connections. However, for deeper networks and more complex tasks, the performance of feedback alignment and its variants break down \cite{bartunov_assessing_2018}.

\citet{liao_how_2016} and \citet{xiao_biologically-plausible_2019} took an alternative route to relaxing the requirement for exact weight symmetry by transporting just the sign of the forward weights during learning. \citet{moskovitz_feedback_2018} combined sign-symmetry and feedback alignment with additional normalization mechanisms. These methods outperform feedback alignment on scalable tasks, but still perform far worse than backpropation. It is also not clear that instantaneous sign transport is more biologically plausible than instantaneous weight transport.

More recently, \citet{Akrout2019} introduced weight mirror (WM), a learning rule that incorporates dynamics on the backward weights to improve alignment throughout the course of training. Unlike previous methods, weight mirror achieves backpropagation level performance on ResNet-18 and ResNet-50 trained on ImageNet.

Concurrently, \citet{kunin_loss_2019} suggested training the forward and backward weights in each layer as an encoder-decoder pair, based on their proof that $L_2$-regularization induces symmetric weights for linear autoencoders.  This approach incorporates ideas from both information preservation and weight symmetry. 

A complementary line of research \cite{Xie2003Equivalence, Scellier2017Equilibrium, bengio2017STDP, guerguiev_towards_2017, whittington2017approximation,  Sacramento2018Dendritic, guerguiev_spike-based_2019} investigates how learning rules, even those that involve weight transport, could be implemented in a biologically mechanistic manner, such as using  spike-timing dependent plasticity rules and obviating the need for distinct phases of training. 
In particular, \citet{guerguiev_spike-based_2019} show that key steps in the \citet{kunin_loss_2019} regularization approach could be implemented by a spike-based mechanism for approximate weight transport.

In this work, we extend this regularization approach to formulate a more general framework of credit assignment strategies without weight symmetry, one that encompasses existing and novel learning rules.
Our core result is that the best of these strategies are substantially more robust across architectures and metaparameters than previous proposals. 

\section{Regularization Inspired Learning Rule Framework}
\label{lrperf:sec:framework}

\begin{figure}[h]
\begin{center}
\centerline{
\includegraphics[width=0.75\linewidth]{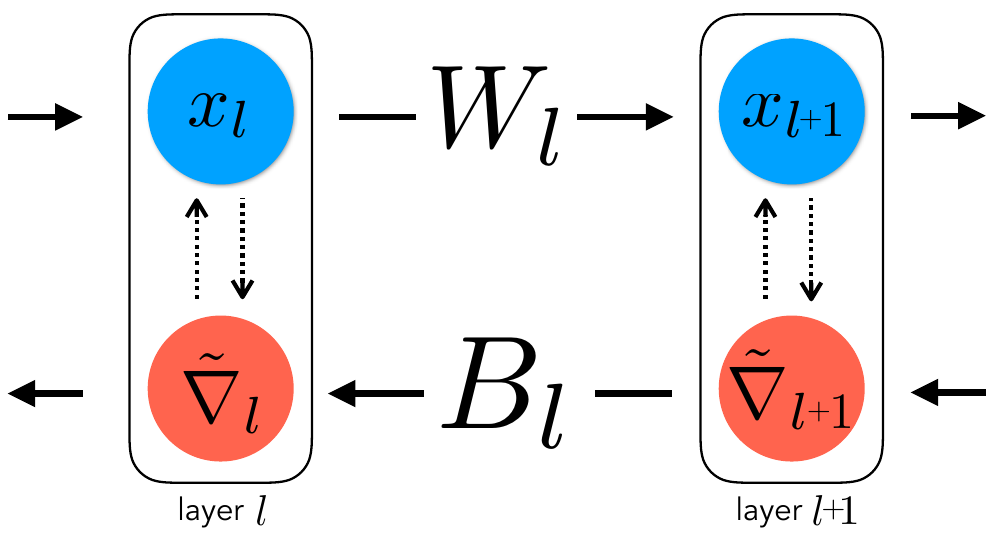}
}
\caption[Learning circuit notational diagram]{\textbf{Notational diagram.} The forward weight $W_l$ propagates the input signal $x_l$ downstream through the network.  The backward weight $B_l$ propagates the pseudogradient $\widetilde{\nabla}_{l+1}$ of $\mathcal{J}$ upstream through the network.  The regularization function $\mathcal{R}$ is constructed layer-wise from $x_l$, $x_{l+1}$, $W_l$ and $B_l$. Similar to \citet{Akrout2019}, we assume lateral pathways granting the backward weights access to $x$ and the forward weights access to $\widetilde{\nabla}$.  Biases and non-linearities are omitted from the diagram.}
\label{lrperf:fig:conceptual-framework}
\end{center}
\end{figure}

We consider the credit assignment problem for neural networks as a layer-wise regularization problem. 
We consider a network parameterized by forward weights $\theta_f$ and backward weights $\theta_b$. 
Informally, the network is trained on the sum of a global task function $\mathcal{J}$ and a layer-wise regularization function\footnote{ $\mathcal{R}$ is not regularization in the traditional sense, as it does not directly penalize the forward weights $\theta_f$ from the cost function $\mathcal{J}$.} $\mathcal{R}$:
\begin{equation*}
\mathcal{L}(\theta_f,\theta_b) = \mathcal{J}(\theta_f) + \mathcal{R}(\theta_b).
\end{equation*}
Formally, every step of training consists of two updates, one for the forward weights and one for the backward weights. 
The forward weights are updated according to the error signal on $\mathcal{J}$ propagated through the network by the backward weights, as illustrated in Figure~\ref{lrperf:fig:conceptual-framework}. 
The backward weights are updated according to gradient descent on $\mathcal{R}$. 
$$\Delta \theta_f \propto \widetilde{\nabla}J \qquad \Delta \theta_b \propto \nabla R$$
Thus, $\mathcal{R}$ is responsible for introducing dynamics on the backward weights, which in turn impacts the dynamics of the forward weights. 
The functional form of $\mathcal{R}$ gives rise to different learning rules and in particular the locality of a given learning rule depends solely on the locality of the computations involved in $\mathcal{R}$.

\subsection{Regularization Primitives}
\label{lrperf:sec:framework-primitives}

In this work, the regularization function $\mathcal{R}$ is built from a set of simple \emph{primitives} $\mathcal{P}$, which at any given layer $l$ are functions of the forward weight $W_l \in \theta_f$, backward weight $B_l \in \theta_b$, layer input $x_l$, and layer output $x_{l+1}$ as depicted in Figure~\ref{lrperf:fig:conceptual-framework}.  These primitives are biologically motivated components with strong geometric interpretations, from which more complex learning rules may be algebraically constructed.  

The primitives we use, displayed in Table~\ref{lrperf:tab:prim}, can be organized into two groups: those that involve purely local operations and those that involve at least one non-local operation.
To classify the primitives, we use the criteria for locality described in \citet{whittington2017approximation}: 
(1) \textit{Local computation}. Computations only involve synaptic weights acting on their associated inputs.
(2) \textit{Local plasticity}.  Weight modifications only depend on pre-synaptic and post-synaptic activity.
A primitive is local if it satisfies both of these constraints and non-local otherwise.  

We introduce three \textbf{local primitives}: $\mathcal{P}^{\text{decay}}$, $\mathcal{P}^{\text{amp}}$, and $\mathcal{P}^{\text{null}}$. 
The \textit{decay} primitive can be understood as a form of energy efficiency penalizing the Euclidean norm of the backward weights.  The \textit{amp} primitive promotes alignment of the layer input $x_l$ with the reconstruction $B_lx_{l+1}$. The \textit{null} primitive imposes sparsity in the layer-wise activity through a Euclidean norm penalty on the reconstruction $B_lx_{l+1}$.

We consider two \textbf{non-local primitives}: $\mathcal{P}^{\text{sparse}}$ and $\mathcal{P}^{\text{self}}$. 
The \textit{sparse} primitive promotes energy efficiency by penalizing the Euclidean norm of the activation $x_l^\intercal B_l$.  This primitive fails to meet the \textit{local computation} constraint, as $B_l$ describes the synaptic connections from the $l+1$ layer to the $l$ layer and therefore cannot operate on the input $x_l$. 

The \textit{self} primitive promotes alignment of the forward and backward weights by directly promoting their inner product.  This primitive fails the \textit{local plasticity} constraint, as its gradient necessitates that the backward weights can exactly measure the strengths of their forward counterparts.  

\begin{table}
\centering
\begin{tabular}{cccc} \toprule
    & \multicolumn{1}{c}{Local} & \multicolumn{1}{c}{$\mathcal{P}_l$} & $\nabla\mathcal{P}_l$\\ \midrule 
    & decay & $\frac{1}{2}||B_l||^2$ & $B_l$\\
    & amp & $-\mathrm{tr}(x_l^\intercal B_l x_{l+1})$ & $-x_lx_{l+1}^\intercal$ \\
    & null & $\frac{1}{2}||B_lx_{l+1}||^2$ & $B_lx_{l+1}x_{l+1}^\intercal$\\
    \midrule
    & \multicolumn{1}{c}{Non-local} & \multicolumn{1}{c}{$\mathcal{P}_l$} & $\nabla\mathcal{P}_l$\\ \midrule
    & sparse & $\frac{1}{2}||x_l^\intercal B_l||^2$ & $x_lx_l^\intercal B_l$\\
    & self & $-\mathrm{tr}(B_l W_l)$ & $-W_l^\intercal$ \\\bottomrule
\end{tabular}
\caption[Regularization primitives]{\textbf{Regularization primitives.} Mathematical expressions for local and non-local primitives and their gradients with respect to the backward weight $B_l$.
Note, both $x_l$ and $x_{l+1}$ are the post-nonlinearity rates of their respective layers.
\label{lrperf:tab:prim}}
\end{table}

\subsection{Building Learning Rules from Primitives}
\label{lrperf:sec:framework-learning_circuits}

These simple primitives can be linearly combined to encompass existing credit assignment strategies, while also elucidating natural new approaches. 

\textbf{Feedback alignment (FA)} \cite{lillicrap_random_2016} corresponds to no regularization, $\mathcal{R}_{\text{FA}} \equiv 0$, effectively fixing the backward weights at their initial random values\footnote{We explore the consequences of this interpretation analytically in Appendix~\ref{lrperf:sec:analysis-beyond_fa}.}.

The \textbf{weight mirror (WM)} \cite{Akrout2019} update, $\Delta B_l = \eta x_lx_{l+1}^\intercal - \lambda_{\text{WM}} B_l$, where $\eta$ is the learning rate and $\lambda_{\text{WM}}$ is a weight decay constant, corresponds to gradient descent on the layer-wise regularization function
$$\mathcal{R}_{\text{WM}} = \sum_{l \in \text{layers}} \alpha\mathcal{P}^{\text{amp}}_l + \beta\mathcal{P}^{\text{decay}}_l,$$
for $\alpha = 1$ and $\beta = \frac{\lambda_{\text{WM}}}{\eta}$.

If we consider primitives that are functions of the pseudogradients $\widetilde{\nabla}_{l}$ and $\widetilde{\nabla}_{l+1}$, then the \textbf{Kolen-Pollack (KP)} algorithm, originally proposed by \citet{Kolen1994backpropagation} and modified by \citet{Akrout2019}, can be understood in this framework as well.
See Appendix~\ref{lrperf:sec:kolen-pollack} for more details.

The range of primitives also allows for learning rules not yet investigated in the literature. 
In this work, we introduce several such novel learning rules, including Information Alignment (IA), Symmetric Alignment (SA), and Activation Alignment (AA).  
Each of these strategies is defined by a layer-wise regularization function composed from a linear combination of the primitives (Table~\ref{lrperf:tab:taxonomy}).  
Information Alignment is a purely local rule, but unlike feedback alignment or weight mirror, contains the additional null primitive. 
In Section~\ref{lrperf:sec:local-learning-rules}, we motivate this addition theoretically, and show empirically that it helps make IA a higher-performing and substantially more stable learning rule than previous local strategies.
SA and AA are both non-local, but as shown in Section~\ref{lrperf:sec:non-local-learning-rules} perform even more robustly than any local strategy we or others have found, and may be implementable by a type of plausible biological mechanism we call ``weight estimation.''

\begin{table}[tb]
\setlength\tabcolsep{4pt}
\hspace*{-0.25cm}
\centering
\begin{tabular}{@{}|c|l|c|c|c|c|c|@{}} \toprule
    & \multicolumn{1}{c|}{\textbf{Alignment}} & \multicolumn{1}{c|}{$\mathcal{P}^{\text{decay}}$} & \multicolumn{1}{c|}{$\mathcal{P}^{\text{amp}}$} & \multicolumn{1}{c|}{$\mathcal{P}^{\text{null}}$} & \multicolumn{1}{c|}{$\mathcal{P}^{\text{sparse}}$} & \multicolumn{1}{c|}{$\mathcal{P}^{\text{self}}$} \\ \hline
    \multirow{3}{*}{\STAB{\rotatebox[origin=c]{90}{\small Local}}}
    & Feedback & & & & & \\
    & Weight Mirror & \checkmark &  \checkmark & &  & \\
    & Information & \checkmark & \checkmark & \checkmark & & \\\midrule
    \multirow{3}{*}{\STAB{\rotatebox[origin=c]{90}{\small \shortstack{Non-\\Local}\hspace{-0.8em}}}}
    & Symmetric & \checkmark & & & & \checkmark \\
    & Activation & & \checkmark & & \checkmark & \\
    \bottomrule
\end{tabular}
\caption[Taxonomy of learning rules based on the locality and composition of their primitives]{\textbf{Taxonomy of learning rules} based on the locality and composition of their primitives. \label{lrperf:tab:taxonomy}}
\end{table}

\subsection{Evaluating Learning Rules}
\label{lrperf:sec:guidelines}
For all the learning rules, we evaluate two desirable target metrics.

\textbf{Task Performance.} 
Performance-optimized CNNs on ImageNet provide the most effective quantitative description of neural responses of cortical neurons throughout the primate ventral visual pathway~\cite{yamins2014performance, cadena2019deep}, indicating the biological relevance of task performance. 
Therefore, our first desired target will be ImageNet top-1 validation accuracy, in line with \citet{bartunov_assessing_2018}.

\textbf{Metaparameter Robustness.}
Extending the proposal of \citet{bartunov_assessing_2018}, we also consider whether a proposed learning rule's metaparameters, such as learning rate and batch size, transfer across architectures. 
Specifically, when we optimize for metaparameters on a given architecture (e.g. ResNet-18), we will fix these metaparameters and use them to train both deeper (e.g. ResNet-50) and different variants (e.g. ResNet-v2).
Therefore, our second desired target will be ImageNet top-1 validation accuracy \emph{across} models for \emph{fixed} metaparameters.

\section{Local Learning Rules}
\label{lrperf:sec:local-learning-rules}

\textbf{Instability of Weight Mirror.} \citet{Akrout2019} report that the weight mirror update rule matches the performance of backpropagation on ImageNet categorization.
The procedure described in \citet{Akrout2019} involves not just the weight mirror rule, but a number of important additional training details, including alternating learning modes and using layer-wise Gaussian input noise. 
After reimplementing this procedure in detail, and using their prescribed metaparameters for the ResNet-18 architecture, the best top-1 validation accuracy we were able to obtain was 63.5\% ($\mathcal{R}_{\text{WM}}$ in Table~\ref{lrperf:tab:hp-local}), substantially below the reported performance of 69.73\%.
To try to account for this discrepancy, we considered the possibility that the metaparameters were incorrectly set.
We thus performed a large-scale metaparameter search over the continuous $\alpha$, $\beta$, and the standard deviation $\sigma$ of the Gaussian input noise, jointly optimizing these parameters for ImageNet validation set performance using a Bayesian Tree-structured Parzen Estimator (TPE) \cite{Bergstra2011}. 
After considering 824 distinct settings (see Appendix~\ref{lrperf:sup:hp-ss-details} for further details), the optimal setting achieved a top-1 performance of 64.07\% ($\mathcal{R}^{\text{TPE}}_{\text{WM}}$ in Table~\ref{lrperf:tab:hp-local}), still substantially below the reported performance in \citet{Akrout2019}.

Considering the second metric of robustness, we found that the WM learning rule is very sensitive to metaparameter tuning. 
Specifically, when using either the metaparameters prescribed for ResNet-18 in \citet{Akrout2019} or those from our metaparameter search, directly attempting to train other network architectures failed entirely (Figure~\ref{lrperf:fig:hp-deeper}, brown line).

Why is weight mirror under-performing backpropagation on both performance and robustness metrics?
Intuition can be gained by simply considering the functional form of $\mathcal{R}_{\text{WM}}$, which can become \emph{arbitrarily} negative even for fixed values of the forward weights.
$\mathcal{R}_{\text{WM}}$ is a combination of a primitive which depends on the input ($\mathcal{P}^{\text{amp}}$) and a primitive which is independent of the input ($\mathcal{P}^{\text{decay}}$).
Because of this, the primitives of weight mirror and their gradients may operate at different scales and careful metaparameter tuning must be done to balance their effects.
This instability can be made precise by considering the dynamical system given by the symmetrized gradient flow on $\mathcal{R}_{\text{WM}}$ at a given layer $l$.

\begin{figure}[h]
\begin{subfigure}{0.32\columnwidth}
    \centering
    \includegraphics[width=\textwidth]{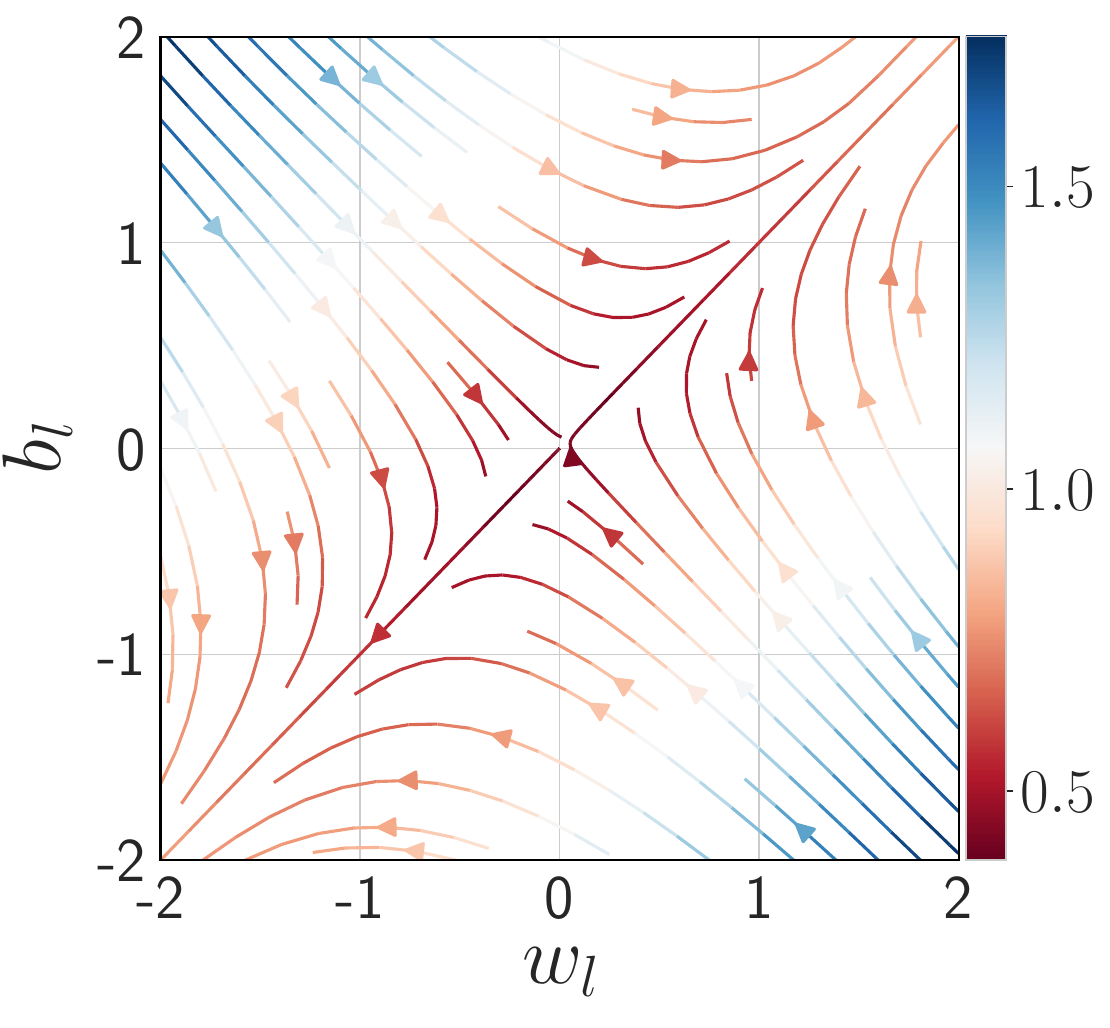}
    \caption{$\lambda_{\text{WM}} < x_l^2$}
\end{subfigure}
\begin{subfigure}{0.32\columnwidth}
    \centering
    \includegraphics[width=\textwidth]{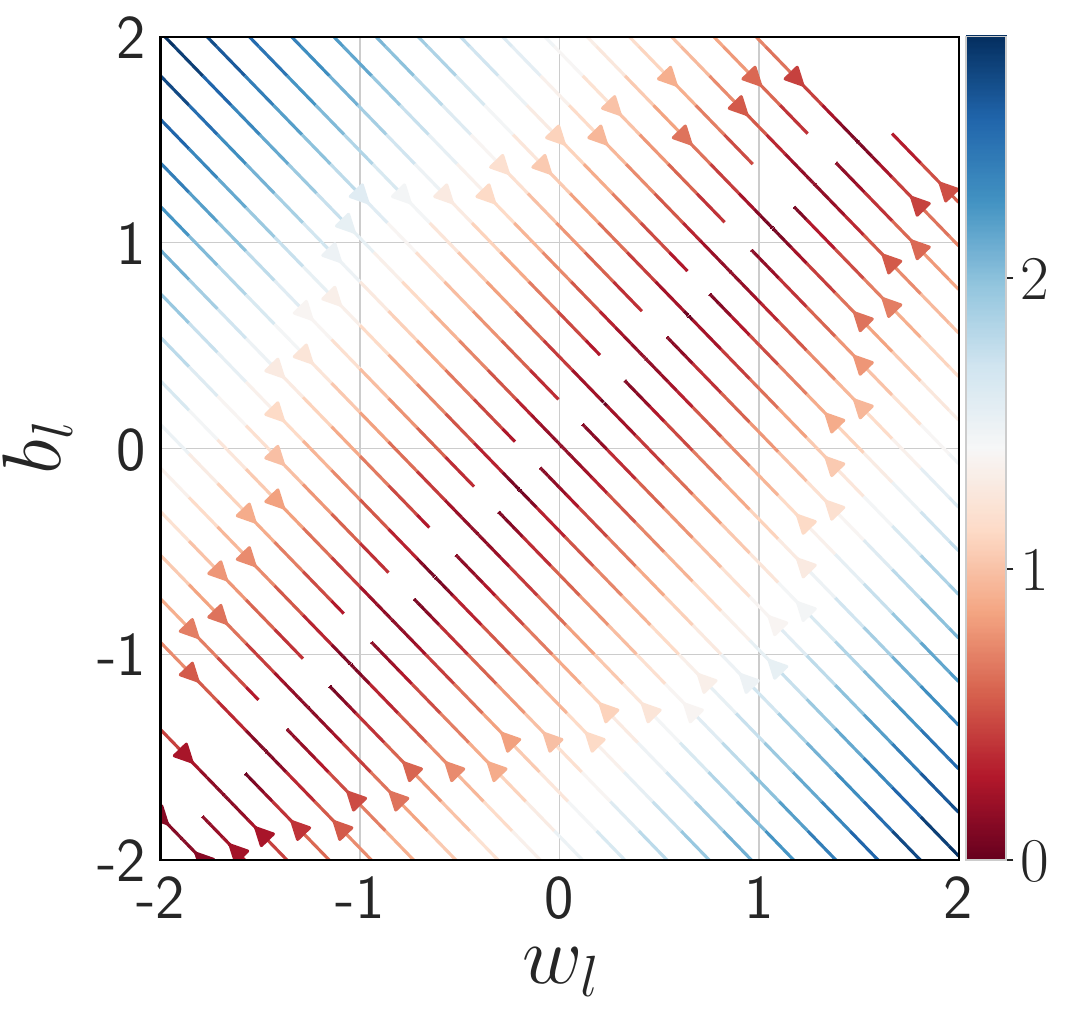}
    \caption{$\lambda_{\text{WM}} = x_l^2$}
\end{subfigure}
\begin{subfigure}{0.32\columnwidth}
    \centering
    \includegraphics[width=\textwidth]{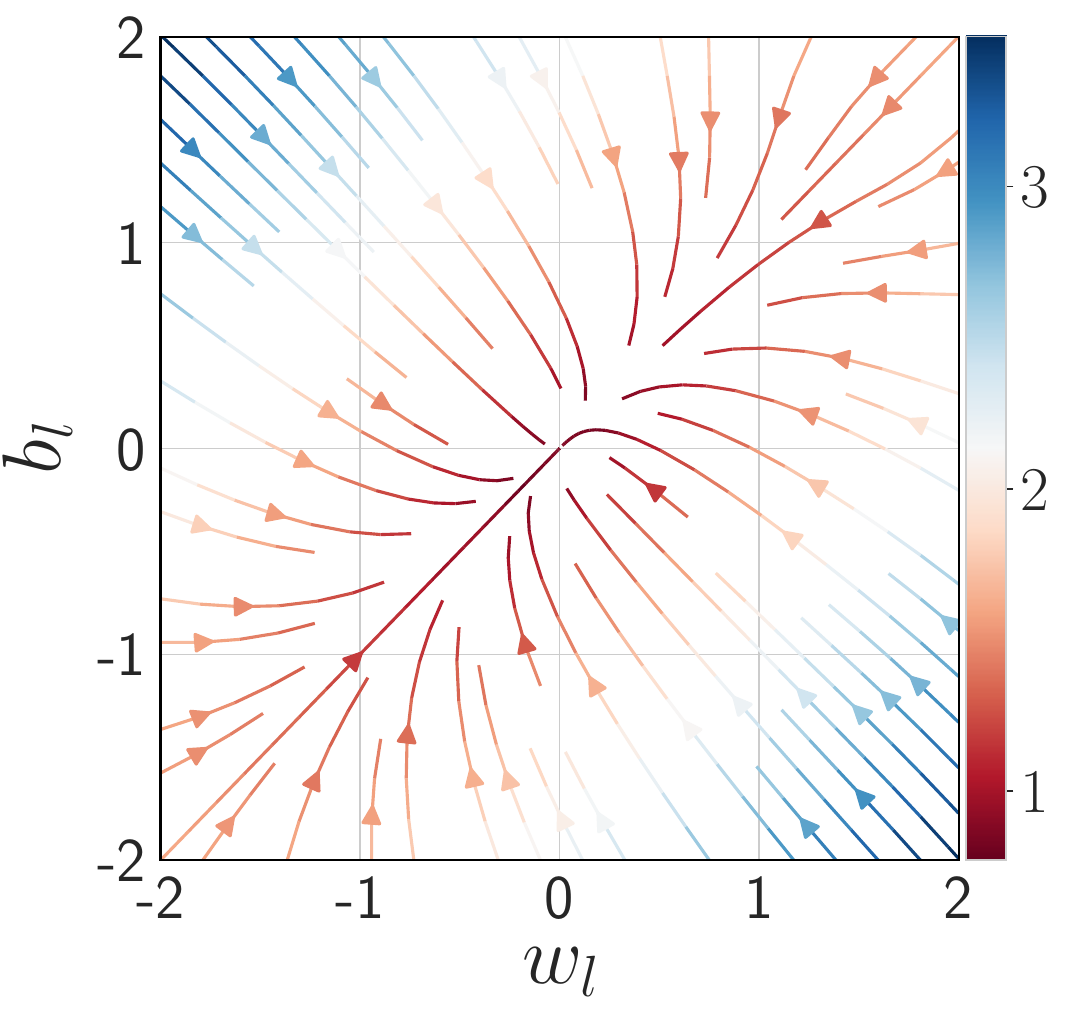}
    \caption{$\lambda_{\text{WM}} > x_l^2$}
\end{subfigure}
\begin{subfigure}{0.32\columnwidth}
    \centering
    \includegraphics[width=\textwidth]{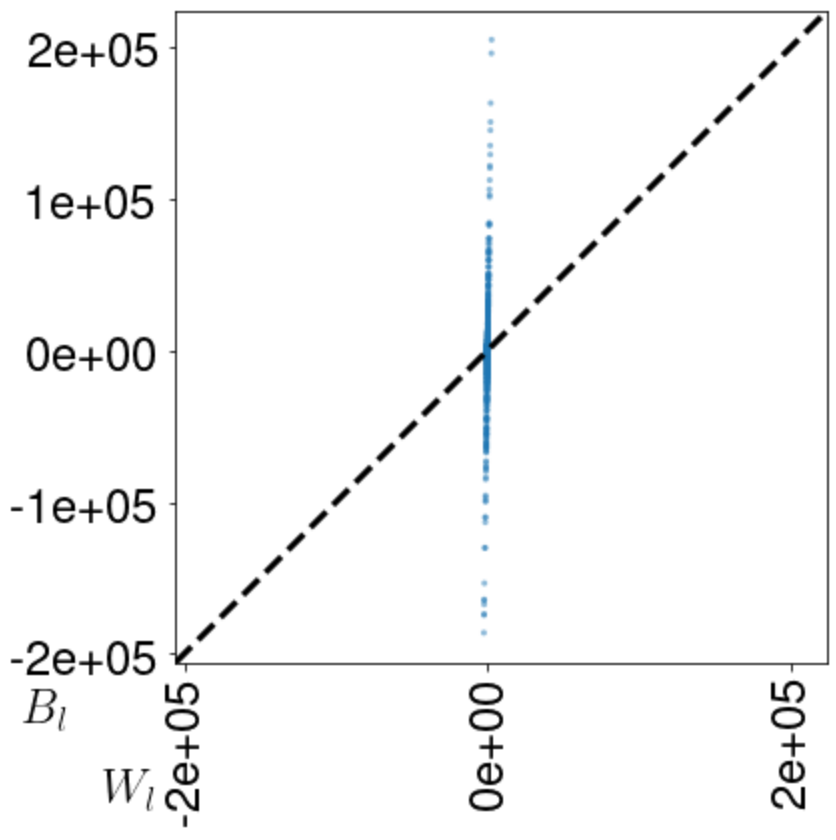}
    \caption{Conv 1}
\end{subfigure}
\begin{subfigure}{0.32\columnwidth}
    \centering
    \includegraphics[width=\textwidth]{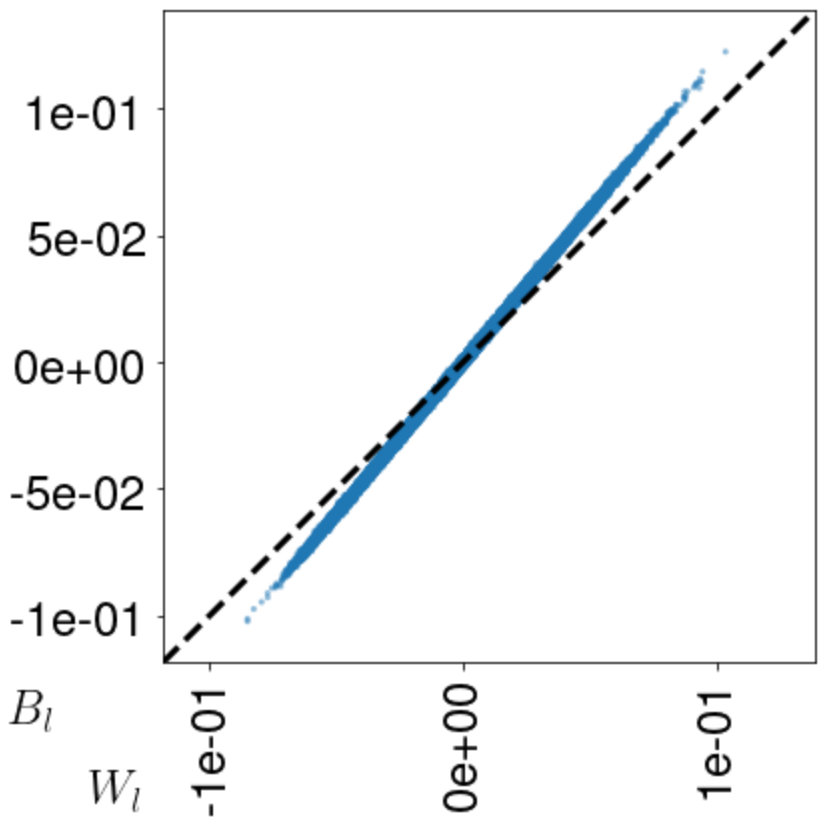}
    \caption{Conv 16}
\end{subfigure}
\begin{subfigure}{0.32\columnwidth}
    \centering
    \includegraphics[width=\textwidth]{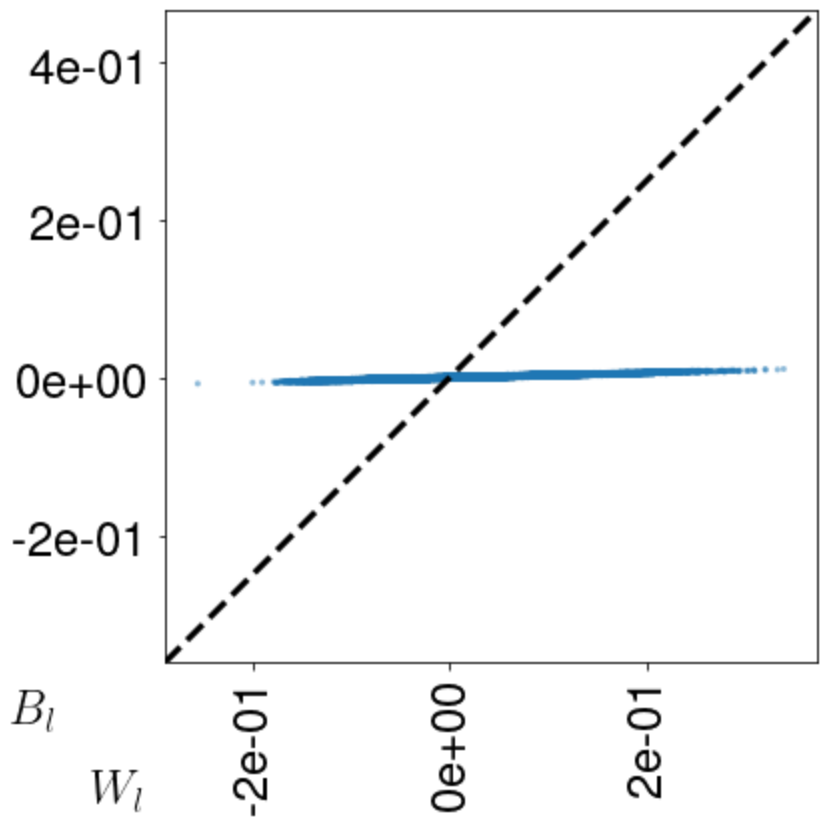}
    \caption{Dense}
\end{subfigure}
\caption[Examples of unstable dynamics]{\textbf{Unstable dynamics (a-c).}  Symmetrized gradient flow on $\mathcal{R}_{\text{WM}}$ at layer $l$ with scalar weights and $x_l = 1$.  The color and arrow indicate respectively the magnitude and direction of the flow.
\textbf{Empirical instability (d-f).} Weight scatter plots for the first convolution, an intermediate convolution and the final dense layer of a ResNet-18 model trained with weight mirror for five epochs.
Each dot represents an element in layer $l$'s weight matrix and its $(x,y)$ location corresponds to its forward and backward weight values, $(W_l^{(i,j)},B_l^{(j,i)} )$. 
The dotted diagonal line shows perfect weight symmetry, as is the case in backpropagation.
Different layers demonstrate one of the the three dynamics outlined by the gradient flow analysis in Section~\ref{lrperf:sec:local-learning-rules}: diverging, stable, and collapsing backward weights.
\label{lrperf:fig:stability}}
\end{figure}

\begin{figure}[tb]
\centering
\includegraphics[width=1.0\columnwidth]{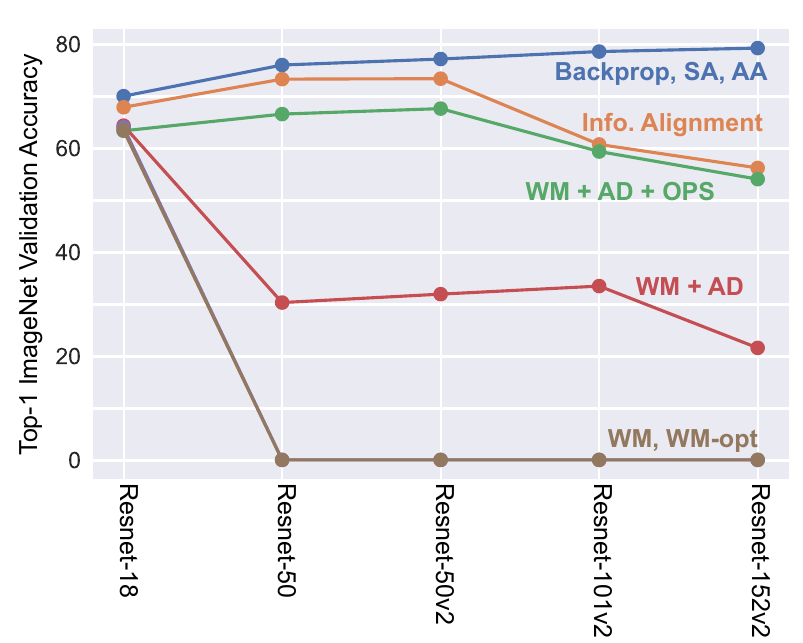}
\caption[Performance of local and non-local rules across architectures]{\textbf{Performance of local and non-local rules across architectures.} We fixed the categorical and continuous metaparameters for ResNet-18 and applied them directly to deeper and different ResNet variants (e.g. v2). A performance of 0.001 indicates the alignment loss became NaN within the first thousand steps of training. Our local rule Information Alignment (IA) consistently outperforms the other \emph{local} alternatives across architectures, despite not being optimized for these architectures. The non-local rules, Symmetric Alignment (SA) and Activation Alignment (AA), consistently perform as well as backpropagation.}
\label{lrperf:fig:hp-deeper}
\end{figure}

In the following analysis we ignore non-linearities, include weight decay on the forward weights, set $\alpha = \beta$, and consider the gradient with respect to both the forward and backward weights.  
When the weights, $w_l$ and $b_l$, and input, $x_l$, are all scalar values, the gradient flow gives rise to the dynamical system
\begin{equation} \label{lrperf:eq:dynamical_system}
\dfrac{\partial}{\partial t}\begin{bmatrix}
w_l\\
b_l
\end{bmatrix} = - A \begin{bmatrix}
w_l\\
b_l
\end{bmatrix},
\end{equation}
where $A$ is an indefinite matrix (see Appendix~\ref{lrperf:sup:stability-analysis} for details.) $A$ can be diagonally decomposed by the eigenbasis $\{u,v\}$, where $u$ spans the symmetric component and $v$ spans the skew-symmetric component of any realization of the weight vector $\begin{bmatrix}
w_l &
b_l
\end{bmatrix}^\intercal$. 
Under this basis, the dynamical system decouples into a system of ODEs governed by the eigenvalues of $A$.
The eigenvalue associated with the skew-symmetric eigenvector $v$ is strictly positive, implying that this component decays exponentially to zero.
However, for the symmetric eigenvector $u$, the sign of the corresponding eigenvalue depends on the relationship between $\lambda_{\text{WM}}$ and $x_l^2$.
When $\lambda_{\text{WM}} > x_l^2$, the eigenvalue is positive and the symmetric component decays to zero (i.e. too much regularization).
When $\lambda_{\text{WM}} < x_l^2$, the eigenvalue is negative and the symmetric component exponentially grows (i.e. too little regularization).
Only when $\lambda_{\text{WM}} = x_l^2$ is the eigenvalue zero and the symmetric component stable. These various dynamics are shown in Figure~\ref{lrperf:fig:stability}.

This analysis suggests that the sensitivity of weight mirror is not due to the misalignment of the forward and backward weights, but rather due to the stability of the symmetric component throughout training.
Empirically, we find that this is true.
In Figure~\ref{lrperf:fig:stability}, we show a scatter plot of the backward and forward weights at three layers of a ResNet-18 model trained with weight mirror.
At each layer there exists a linear relationship between the weights, suggesting that the backward weights have aligned to the forward weights up to magnitude.
Despite being initialized with similar magnitudes, at the first layer the backward weights have grown orders larger, at the last layer the backward weights have decayed orders smaller, and at only one intermediate layer were the backward weights comparable to the forward weights, implying symmetry.

This analysis also clarifies the stabilizing role of Gaussian noise, which was found to be essential to weight mirror's performance gains over feedback alignment~\cite{Akrout2019}. 
Specifically, when the layer input $x_l \sim N\left(0,\sigma^2\right)$ and $\sigma^2 = \lambda_{\text{WM}}$, then $x_l^2 \approx \lambda_{\text{WM}}$, implying the dynamical system in equation (\ref{lrperf:eq:dynamical_system}) is stable.

\textbf{Strategies for Reducing Instability.} 
Given the above analysis, can we identify further strategies for reducing instability during learning beyond the use of Gaussian noise?

\textit{Adaptive Optimization.} 
One option is to use an adaptive learning rule strategy, such as Adam~\cite{kingma2014adam}. 
An adaptive learning rate keeps an exponentially decaying moving average of past gradients, allowing for more effective optimization of the alignment regularizer even in the presence of exploding or vanishing gradients.

\textit{Local Stabilizing Operations.} 
A second option to improve stability is to consider local layer-wise operations to the backward path such as choice of non-linear activation functions, batch centering, feature centering, and feature normalization.
The use of these operations is largely inspired by Batch Normalization \cite{Ioffe2015} and Layer Normalization \cite{Ba2016} which have been observed to stabilize learning dynamics.
The primary improvement that these normalizations allow for is the further conditioning of the covariance matrix at each layer's input, building on the benefits of using Gaussian noise.
In order to keep the learning rule fully local, we use these normalizations, which unlike Batch and Layer Normalization, do not add any additional learnable parameters.

\textit{The Information Alignment (IA) Learning Rule.} 
There is a third option for improving stability that involves modifying the local learning rule itself. 

Without decay, the update given by weight mirror, $\Delta B_l = \eta x_{l}x_{l+1}^\intercal$, is Hebbian.
This update, like all purely Hebbian learning rules, is unstable and can result in the norm of $B_l$ diverging.
This can be mitigated by weight decay, as is done in \citet{Akrout2019}. 
However, an alternative strategy to dealing with the instability of a Hebbian update was given by \citet{oja_neuron} in his analysis of learning rules for linear neuron models.  
In the spirit of that analysis, assume that we can normalize the backward weight after each Hebbian update such that
$$B_l^{(t+1)} = \frac{B_l^{(t)} + \eta x_{l}x_{l+1}^\intercal}{||B_l^{(t)} + \eta x_{l}x_{l+1}^\intercal||},$$
and in particular $||B_l^{(t)}|| = 1$ at all time $t$. Then, for small learning rates $\eta$, the right side can be expanded as a power series in $\eta$, such that
$$B_l^{(t+1)} = B_l^{(t)} + \eta \left(x_{l}x_{l+1}^\intercal - B_l^{(t)} x_l^\intercal B_l^{(t)} x_{l+1}\right) + O(\eta^2).$$
Ignoring the $O(\eta^2)$ term gives the non-linear update
$$\Delta B_l = \eta \left( x_{l}x_{l+1}^\intercal - B_l x_{l}^\intercal B_l x_{l+1}\right).$$
If we assume $x_l^\intercal B_l = x_{l+1}$ and $B_l$ is a column vector rather than a matrix, then by Table~\ref{lrperf:tab:prim}, this is approximately the update given by the null primitive introduced in Section~\ref{lrperf:sec:framework-primitives}.

Thus motivated, we define \textbf{Information Alignment (IA)} as the local learning rule defined by adding a (weighted) null primitive to the other two local primitives already present in the weight mirror rule. 
That is, the layer-wise regularization function
$$\mathcal{R}_{\text{IA}} = \sum_{l \in \text{layers}}\alpha\mathcal{P}^{\text{amp}}_l +  \beta\mathcal{P}^{\text{decay}}_l + \gamma\mathcal{P}^{\text{null}}_l.$$
In the specific setting when $x_{l+1} = W_lx_l$ and $\alpha = \gamma$, then the gradient of $\mathcal{R}_{\text{IA}}$ is proportional to the gradient with respect to $B_l$ of
$\frac{1}{2}||x_l - B_lW_lx_l||^2 + \frac{\beta}{2}\left(||W_l||^2 + ||B_l||^2\right)$, a quadratically regularized linear autoencoder\footnote{In this setting, the resulting learning rule is a member of the target propagation framework introduced in Section~\ref{lrperf:sec:related}.}.
As shown in \citet{kunin_loss_2019}, all critical points of a quadratically regularized linear autoencoder attain symmetry of the encoder and decoder.

\begin{table}
    \setlength\tabcolsep{3pt}
    \centering
    \begin{tabular}{ccc}
    \toprule
    Learning Rule & Top-1 Val Accuracy & Top-5 Val Accuracy\\
    \midrule
     $\mathcal{R}_{\text{WM}}$ & 63.5\% & 85.16\% \\
    \midrule
     $\mathcal{R}_{\text{WM}}^{\text{TPE}}$ & 64.07\% & 85.47\% \\
     \midrule
     $\mathcal{R}_{\text{WM}+\text{AD}}^{\text{TPE}}$ & 64.40\% & 85.53\% \\
     \midrule
     $\mathcal{R}_{\text{WM}+\text{AD}+\text{OPS}}^{\text{TPE}}$ & 63.41\% & 84.83\% \\
     \midrule
     $\mathcal{R}_{\text{IA}}^{\text{TPE}}$ & \textbf{67.93\%} & \textbf{88.09\%} \\
    \midrule
    \text{Backprop.} & 70.06\% & 89.14\%\\
    \bottomrule
    \end{tabular}
    \caption[Performance of local learning rules with ResNet-18 on ImageNet]{\textbf{Performance of local learning rules with ResNet-18 on ImageNet.} $\mathcal{R}_{\text{WM}}$ is weight mirror as described in \citet{Akrout2019}, $\mathcal{R}_{\text{WM}}^{\text{TPE}}$ is weight mirror with learning metaparameters chosen through an optimization procedure.  $\mathcal{R}_{\text{WM}+\text{AD}}^{\text{TPE}}$ is weight mirror with an adaptive optimizer. $\mathcal{R}_{\text{WM}+\text{AD}+\text{OPS}}^{\text{TPE}}$ involves the addition of stabilizing operations to the network architecture. The best local learning rule, $\mathcal{R}_{\text{IA}}^{\text{TPE}}$, additionally involves the null primitive.  
    For details on metaparameters for each local rule, see Appendix~\ref{lrperf:sup:hp-ss-details}.}
    \label{lrperf:tab:hp-local}
\end{table}

\textbf{Empirical Results.} 
To evaluate the three strategies for stabilizing local weight updates, we performed a neural architecture search implementing all three strategies, again using TPE.  
This search optimized for Top-1 ImageNet validation performance with the ResNet-18 architecture, comprising a total of 628 distinct settings.   
We found that validation performance increased significantly, with the optimal learning rule $\mathcal{R}_{\text{IA}}^{\text{TPE}}$, attaining 67.93\% top-1 accuracy (Table \ref{lrperf:tab:hp-local}). 
More importantly, we also found that the parameter robustness of $\mathcal{R}_{\text{IA}}^{\text{TPE}}$ is dramatically improved as compared to weight mirror (Figure \ref{lrperf:fig:hp-deeper}, orange line), nearly equaling the parameter robustness of backpropagation across a variety of deeper architectures.  
Critically, this improvement was achieved not by directly optimizing for robustness across architectures, but simply by finding a parameter setting that achieved high task performance on one architecture.

To assess the importance of each strategy type in achieving this result, we also performed several ablation studies, involving neural architecture searches using only various subsets of the stabilization strategies (see Appendix~\ref{lrperf:sup:hp-wm-ad-ops-tpe-details} for details).
Using just the adaptive optimizer while otherwise optimizing the weight mirror metaparameters yielded the learning rule $\mathcal{R}_{\text{WM}+\text{AD}}^{\text{TPE}}$, while adding stabilizing layer-wise operations yielded the learning rule $\mathcal{R}_{\text{WM}+\text{AD}+\text{OPS}}^{\text{TPE}}$ (Table \ref{lrperf:tab:hp-local}).  
We found that while the top-1 performance of these ablated learning rules was not better for the ResNet-18 architecture than the weight-mirror baseline, each of the strategies did individually contribute significantly to improved parameter robustness (Figure \ref{lrperf:fig:hp-deeper}, red and green lines). 

Taken together, these results indicate that the regularization framework allows the formulation of local learning rules with substantially improved performance and, especially, metaparameter robustness characteristics. 
Moreover, these improvements are well-motivated by mathematical analysis that indicates how to target better circuit structure via improved learning stability.

\section{Non-Local Learning Rules}
\label{lrperf:sec:non-local-learning-rules}
While our best local learning rule is substantially improved as compared to previous alternatives, it still does not quite match backpropagation, either in terms of performance or metaparameter stability over widely different architectures (see the gap between blue and orange lines in Figure \ref{lrperf:fig:hp-deeper}). 
We next introduce two novel non-local learning rules that entirely eliminate this gap.

\textbf{Symmetric Alignment (SA)} is defined by the layer-wise regularization function
$$\mathcal{R}_{\text{SA}} = \sum_{l \in \text{layers}} \alpha\mathcal{P}^{\text{self}}_l + \beta\mathcal{P}^{\text{decay}}_l.$$
When $\alpha=\beta$, then the gradient of $\mathcal{R}_{\text{SA}}$ is proportional to the gradient with respect to $B_l$ of $\frac{1}{2}||W_l - B_l^\intercal ||^2$, which encourages symmetry of the weights.

\textbf{Activation Alignment (AA)} is defined by the layer-wise regularization function
$$\mathcal{R}_{\text{AA}} = \sum_{l \in \text{layers}} \alpha\mathcal{P}^{\text{amp}}_l + \beta\mathcal{P}^{\text{sparse}}_l.$$
When $x_{l+1} = W_lx_l$ and $\alpha = \beta$, then the gradient of $\mathcal{R}_{\text{AA}}$ is proportional to the gradient with respect to $B_l$ of $\frac{1}{2}||W_lx_l - B_l^\intercal x_l||^2$, which encourages alignment of the activations.

Both SA and AA give rise to dynamics that encourage the backward weights to become transposes of their forward counterparts.
When $B_l$ is the transpose of $W_l$ for all layers $l$ then the updates generated by the backward pass are the exact gradients of $\mathcal{J}$.
It follows intuitively that throughout training the pseudogradients given by these learning rules might converge to better approximations of the exact gradients of $\mathcal{J}$, leading to improved learning.
Further, in the context of the analysis in equation (\ref{lrperf:eq:dynamical_system}), the matrix $A$ associated with SA and AA is positive semi-definite, and unlike the case of weight mirror, the eigenvalue associated with the symmetric eigenvector $u$ is zero, implying stability of the symmetric component.

While weight mirror and Information Alignment introduce dynamics that implicitly encourage symmetry of the forward and backward weights, the dynamics introduced by SA and AA encourage this property explicitly.

Despite not having the desirable locality property, we show that SA and AA perform well empirically in the weight-decoupled regularization framework --- meaning that they \emph{do} relieve the need for exact weight symmetry.  
As we will discuss, this may make it possible to find plausible biophysical mechanisms by which they might be implemented.

\textbf{Parameter Robustness of Non-Local Learning Rules.}
To assess the robustness of SA and AA, we trained ResNet-18 models with standard 224-sized ImageNet images (training details can be found in Appendix~\ref{lrperf:sup:non-local-hp}).
Without any metaparameter tuning, SA and AA were able to match backpropagation in performance. 
Importantly, for SA we did not need to employ any specialized or adaptive learning schedule involving alternating modes of learning, as was required for all the local rules.
However, for AA we did find it necessary to use an adaptive optimizer when minimizing $\mathcal{R}_{\text{AA}}$, potentially due to the fact that it appears to align less exactly than SA (see Figure~\ref{lrperf:fig:full_weight_scatter}).
We trained deeper ResNet-50, 101, and 152 models \cite{He2016} with larger 299-sized ImageNet images. 
As can be seen in Table~\ref{lrperf:tab:empimnet}, both SA and AA maintain consistent performance with backpropagation despite changes in image size and increasing depth of network architecture, demonstrating their robustness as a credit assignment strategies.

\begin{table}
\centering
\begin{tabular}{cccc}
\toprule
Model & Backprop. & Symmetric & Activation \\
\midrule
ResNet-18 & 70.06\% & 69.84\% & 69.98\% \\
\midrule
ResNet-50 & 76.05\% & 76.29\% & 75.75\% \\
\midrule
ResNet-50v2 & 77.21\% & 77.18\% & 76.67\% \\
\midrule
ResNet-101v2 & 78.64\% & 78.74\% & 78.35\% \\
\midrule
ResNet-152v2 & 79.31\% & 79.15\% & 78.98\% \\
\bottomrule
\end{tabular}
\caption[Symmetric and Activation Alignment consistently match backpropagation]{\textbf{Symmetric and Activation Alignment consistently match backpropagation.}
Top-1 validation accuracies on ImageNet for each model class and non-local learning rule, compared to backpropagation.
\label{lrperf:tab:empimnet}}
\end{table}

\begin{figure}[ht]
    \centering
    \begin{subfigure}{\columnwidth}
        \centering
        \includegraphics[width=0.85\textwidth]{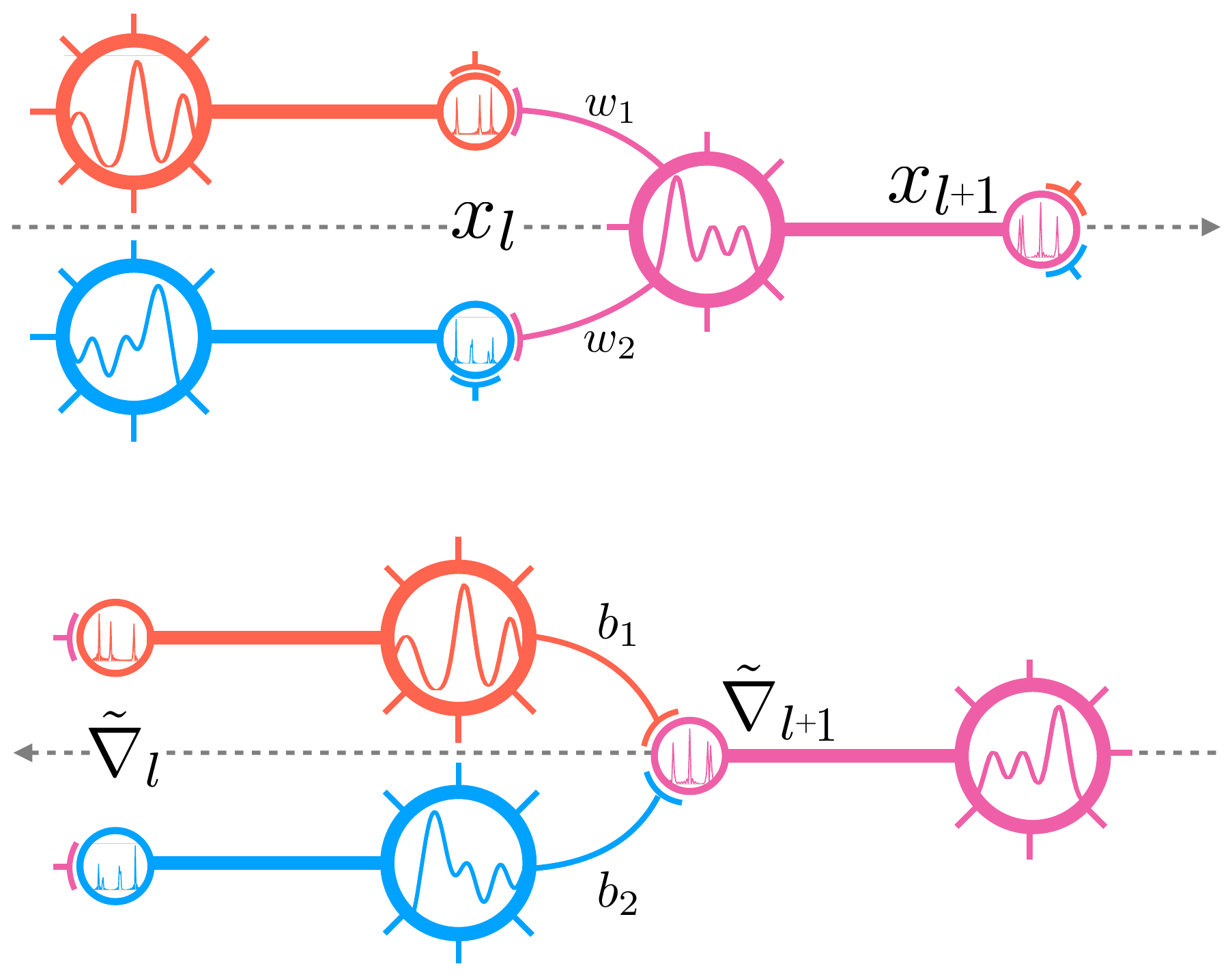}
        \caption{Neural Circuit Diagram}
    \end{subfigure}
    \begin{subfigure}{\columnwidth}
        \centering
        \includegraphics[width=0.49\textwidth]{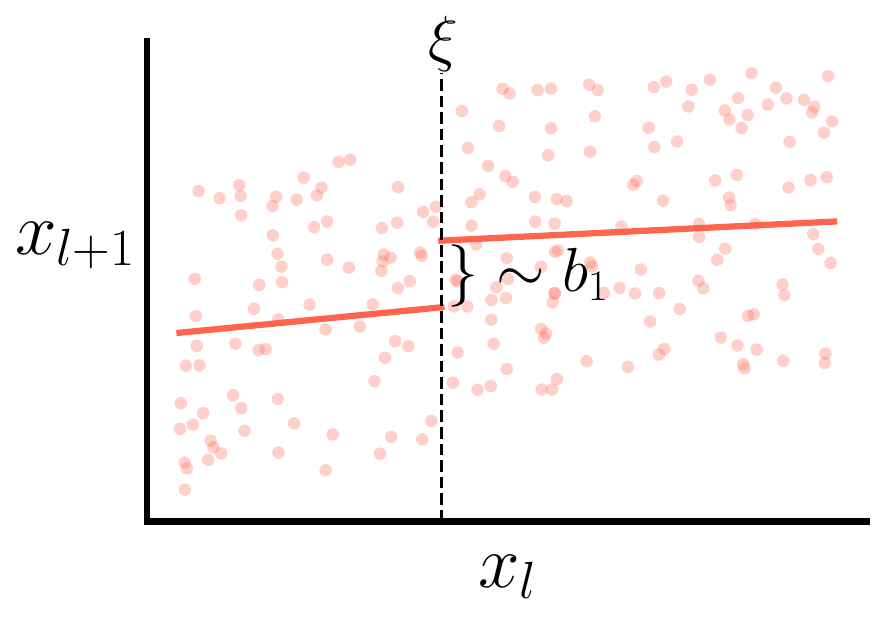}
        \includegraphics[width=0.49\textwidth]{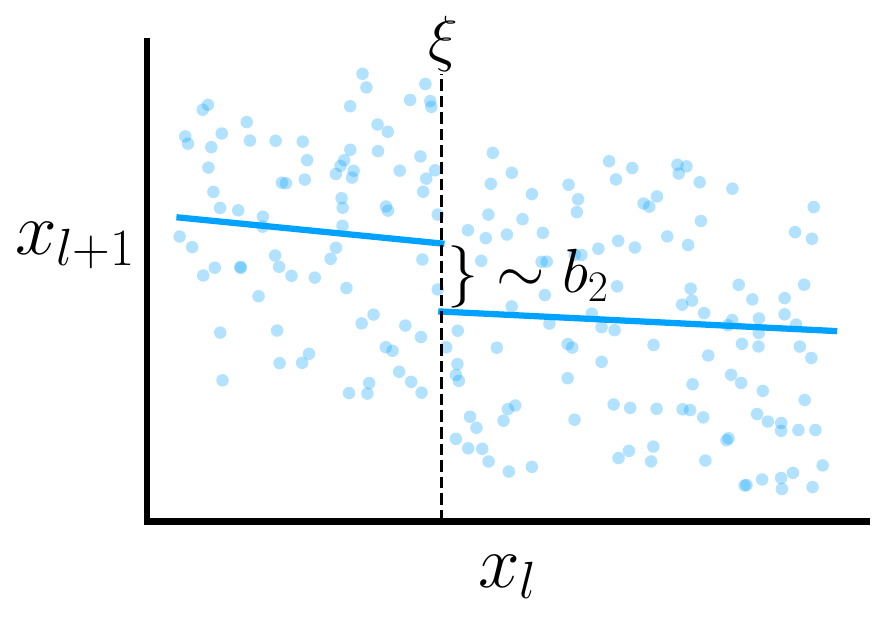}
        \caption{Regression Discontinuity Design}
    \end{subfigure}
    \caption[Weight estimation mechanisms]{\textbf{Weight estimation.} \textbf{(a)} The notational diagram, as shown in Figure~\ref{lrperf:fig:conceptual-framework}, is mapped into a neural circuit.
    The top and bottom circuits represent the forward and backward paths respectively.
    Large circles represent the soma of a neuron, thick edges the axon, small circles the axon terminal, and thin edges the dendrites.
    Dendrites corresponding to the lateral pathways between the circuits are omitted.
    \textbf{(b)} A mechanism such as regression discontinuity design, as explained by \citet{lansdell2019spiking} and \citet{guerguiev_spike-based_2019}, could be used independently at each neuron to do weight estimation by quantifying the causal effect of $x_l$ on $x_{l+1}$.}
    \label{lrperf:fig:rdd_neurons}
\end{figure}

\textbf{Weight Estimation, Neural Plausibility, and Noise Robustness.}
Though SA is non-local, it does avoid the need for instantaneous weight transport --- as is shown simply by the fact that it optimizes effectively in the framework of decoupled forward-backward weight updates, where alignment can only arise over time due to the structure of the regularization circuit rather than instantaneously by \emph{fiat} at each timepoint.
Because of this key difference, it may be possible to find plausible biological implementations for SA, operating on a principle of iterative ``weight estimation'' in place of the implausible idea of instantaneous weight transport.

By ``weight estimation'' we mean any process that can measure changes in post-synaptic activity relative to varying synaptic input, thereby providing a temporal estimate of the synaptic strengths.
Prior work has shown how noisy perturbations in the presence of spiking discontinuities \cite{lansdell2019spiking} could provide neural mechanisms for weight estimation, as depicted in Figure~\ref{lrperf:fig:rdd_neurons}.
In particular, \citet{guerguiev_spike-based_2019} present a spiking-level mechanism for estimating forward weights from noisy dendritic measurements of the implied effect of those weights on activation changes. 
This idea, borrowed from the econometrics literature, is known as regression discontinuity design \cite{imbens2008regression}. 
This is essentially a form of iterative weight estimation, and is used in \citet{guerguiev_spike-based_2019} for minimizing a term that is mathematically equivalent to $\mathcal{P}^{\text{self}}$.
\citet{guerguiev_spike-based_2019} demonstrate that this weight estimation mechanism works empirically for small-scale networks.

\begin{figure}
\centering
\begin{subfigure}{0.9\columnwidth}
    \centering
    \includegraphics[width=\textwidth]{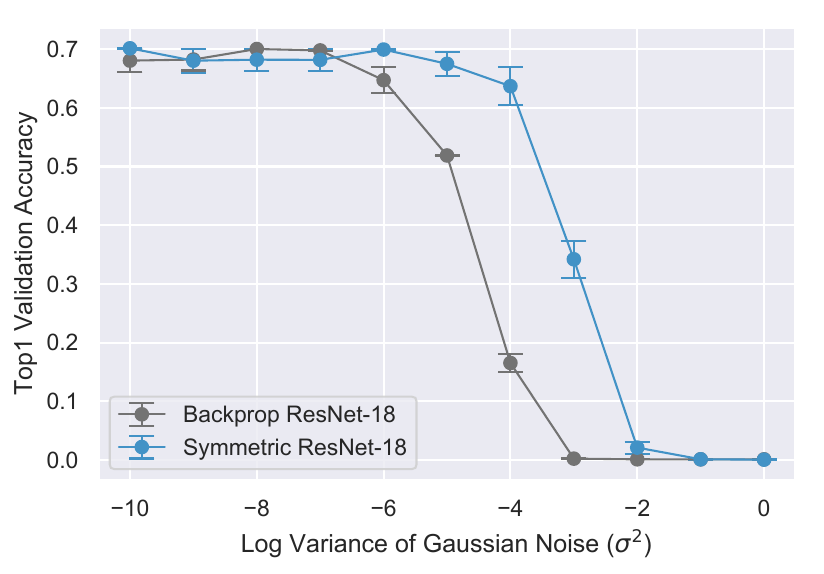}
    \caption{ResNet-18}
    \label{lrperf:fig:noise-plot-a}
\end{subfigure}
\begin{subfigure}{0.9\columnwidth}
    \centering
    \includegraphics[width=\textwidth]{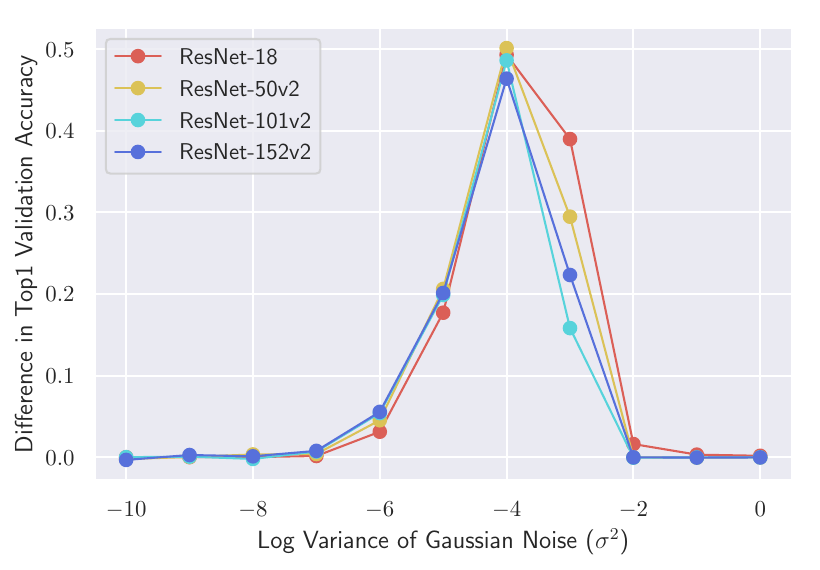}
    \caption{Deeper Models}
    \label{lrperf:fig:noise-plot-b}
\end{subfigure}
\caption[Symmetric Alignment is more robust to noisy updates than backpropagation]{\textbf{Symmetric Alignment is more robust to noisy updates than backpropagation.} \textbf{(a)} Symmetric Alignment is more robust than backpropagation to increasing levels of Gaussian noise added to its updates for ResNet-18. 
\textbf{(b)} Symmetric Alignment maintains this robustness for deeper models.
See Appendix~\ref{lrperf:sup:noisy-updates} for more details and similar experiments with Activation Alignment.}
\label{lrperf:fig:noise-plot}
\end{figure}

Our performance and robustness results above for SA can be interpreted as providing evidence that a rate-coded version of weight estimation scales effectively to training deep networks on large-scale tasks.
However, there remains a gap between what we have shown at the rate-code level and the spike level, at which the weight estimation mechanism operates.
Truly showing that weight estimation could work at scale would involve being able to train deep spiking neural networks, an unsolved problem that is beyond the scope of this work. 
One key difference between any weight estimation process at the rate-code and spike levels is that the latter will be inherently noisier due to statistical fluctuations in whatever local measurement process is invoked --- e.g. in the \citet{guerguiev_spike-based_2019} mechanism, the noise in computing the regression discontinuity.

As a proxy to better determine if our conclusions about the scalable robustness of rate-coded SA are likely to apply to spiking-level equivalents, we model this uncertainty by adding Gaussian noise to the backward updates during learning.
To the extent that rate-coded SA is robust to such noise, the more likely it is that a spiking-based implementation will have the performance and parameter robustness characteristics of the rate-coded version. 
Specifically, we modify the update rule as follows:
\begin{equation*}
\Delta \theta_b \propto \nabla \mathcal{R} +  \mathcal{N}(0,\sigma^2),\qquad \Delta \theta_f \propto \widetilde{\nabla}\mathcal{J}.
\end{equation*}
As shown in Figure~\ref{lrperf:fig:noise-plot}, the performance of SA is very robust to noisy updates for training ResNet-18.
In fact, for comparison we also train backpropagation with Gaussian noise added to its gradients,
$\Delta \theta \propto \nabla \mathcal{J} + \mathcal{N}(0,\sigma^2)$, and find that SA is substantially \emph{more} robust than backpropagation.
For deeper models, SA maintains this robustness, implying that pseudogradients generated by backward weights with noisy updates leads to more robust learning than using equivalently noisy gradients directly.

\section{Discussion}
\label{lrperf:sec:discussion}

\begin{figure}
\centering
\includegraphics[width=0.8\columnwidth]{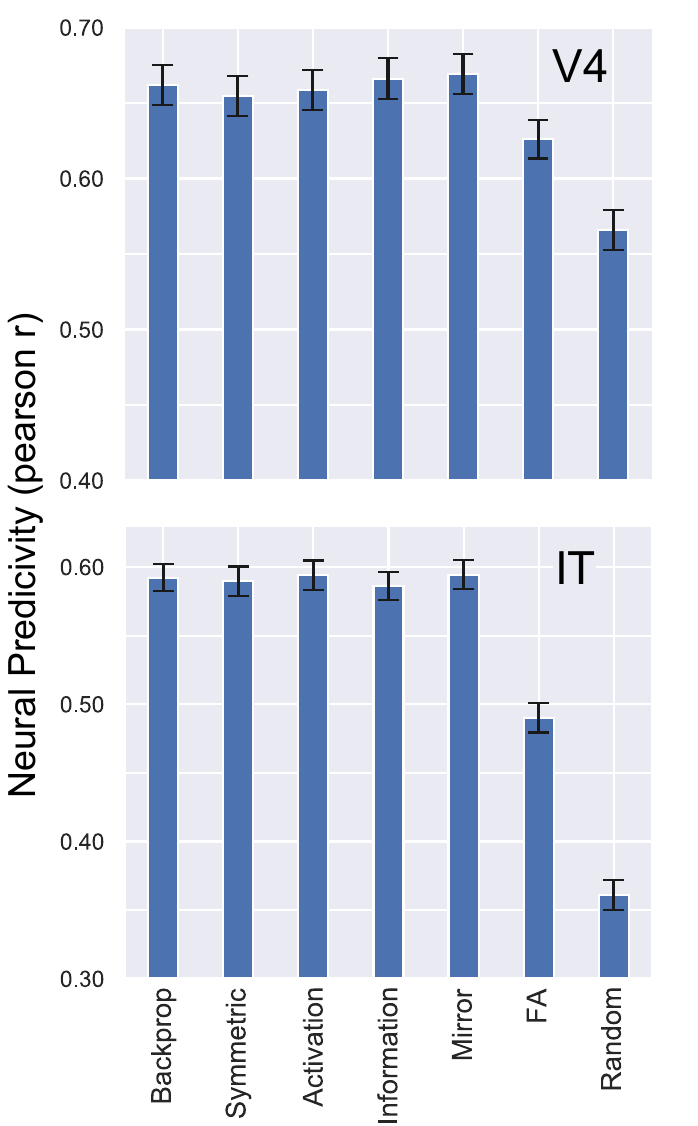}
\caption[Neural fits to temporally-averaged V4 and IT responses]{
\textbf{Neural fits to temporally-averaged V4 and IT responses.}
Neural fits to V4 (top) and IT (bottom) time-averaged responses \cite{Majaj2015}, using a 25 component PLS mapping on a ResNet-18.
The median (across neurons) Pearson correlation over images, with standard error of mean (across neurons) denoting the error bars.
``Random'' refers to a ResNet-18 architecture at initialization. 
For details, see Appendix~\ref{lrperf:sup:neural-fit}.
\label{lrperf:fig:neural-fits}}
\end{figure}

In this work, we present a unifying framework that allows for the systematic identification of robust and scalable alternatives to backpropagation.
We obtain, through large-scale searches, a local learning rule that transfers more robustly across architectures than previous local alternatives.
Nonetheless, a performance and robustness gap persists with backpropagation.
We formulate non-local learning rules that achieve competitive performance with backpropagation, requiring almost no metaparameter tuning and are robust to noisy updates.
Taken together, our findings suggest that there are two routes towards the discovery of robust, scalable, and neurally plausible credit assignment without weight symmetry.

The first route involves further improving local rules. 
We found that the local operations and regularization primitives that allow for improved approximation of non-local rules perform better and are much more stable. 
If the analyses that inspired this improvement could be refined, perhaps further stability could be obtained. 
To aid in this exploration going forward, we have written an open-source TensorFlow library\footnote{\url{https://github.com/neuroailab/neural-alignment}}, allowing others to train arbitrary network architectures and learning rules at scale, distributed across multiple GPU or TPU workers.
The second route involves the further refinement and characterization of scalable biological implementations of weight estimation mechanisms for Symmetric or Activation Alignment, as \citet{guerguiev_spike-based_2019} initiate.

Given these two routes towards neurally-plausible credit assignment without weight symmetry, how would we use neuroscience data to adjudicate between them?
It would be convenient if functional response data in a ``fully trained'' adult animal showed a signature of the underlying learning rule, without having to directly measure synaptic weights during learning.
Such data have been very effective in identifying good models of the primate ventral visual pathway~\cite{Majaj2015, yamins2014performance}. 
As an initial investigation of this idea, we compared the activation patterns generated by networks trained with each local and non-local learning rule explored here, to neural response data from several macaque visual cortical areas, using a regression procedure similar to that in \citet{yamins2014performance}.
As shown in Figure~\ref{lrperf:fig:neural-fits}, we found that, with the exception of the very poorly performing feedback alignment rule, all the reasonably effective learning rules achieve similar V4 and IT neural response predictivity, and in fact match that of the network learned via backpropagation. 
Such a result suggests the interesting possibility that the functional response signatures in an already well-learned neural representation may be relatively independent of which learning rule created them. 
Perhaps unsurprisingly, the question of identifying the operation of learning rules in an \emph{in vivo} neural circuit will likely require the deployment of more sophisticated neuroscience techniques.

\section{Code base}
In this section we describe our implementation and highlight the technical details that allow its generality for use in any architecture. 
We used TensorFlow version 1.13.1 to conduct all experiments, and adhered to its interface. All code can be found at \url{https://github.com/neuroailab/neural-alignment}.

\subsection{Layers}
The essential idea of our code base is that by implementing custom layers that match the TensorFlow API, but use custom operations for matrix multiplication (\texttt{matmul}) and two-dimensional convolutions (\texttt{conv2d}), then we can efficiently implement arbitrary feedforward networks using any credit assignment strategies with untied forward and backward weights.
Our custom \texttt{matmul} and \texttt{conv2d} operations take in a forward and backward kernel. 
They use the forward kernel for the forward pass, but use the backward kernel when propagating the gradient.
To implement this, we leverage the \texttt{@tf.custom\_gradient} decorator, which allows us to explicitly define the forward and backward passes for that op. 
Our \texttt{Layer} objects implement custom dense and convolutional layers which use the custom operations described above. 
Both layers take in the same arguments as the native TensorFlow layers and an additional argument for the learning rule.

\subsection{Alignments}
A learning rule is defined by the form of the layer-wise regularization $\mathcal{R}$ added to the model at each layer.  
The custom layers take an instance of an alignment class which when called will define its alignment specific regularization and add it to the computational graph. 

The learning rule are specializations of a parent \texttt{Alignment} object which implements a \texttt{\_\_call\_\_} method that creates the regularization function. 
The regularization function uses tensors that prevent the gradients from flowing to previous layers via \texttt{tf.stop\_gradient}, keeping the alignment loss localized to a single layer.
Implementation of the \texttt{\_\_call\_\_} method is delegated to subclasses, such that they can define their alignment specific regularization as a weighted sum of primitives, each of which is defined as a function.

\subsection{Optimizers}
The total network loss is defined as the sum of the global cost function $\mathcal{J}$ and the local alignment regularization $\mathcal{R}$.
The optimizer class provides a framework for specifying how to optimize each part of the total network loss as a function of the global step.

In the \texttt{Optimizers} folder you will find two important files:
\begin{itemize}
    \item \texttt{rate\_scheduler.py} defines a scheduler which is a function of the global step, that allows you to adapt the components of the alignment weighting based on where it is in training. If you do not pass in a scheduling function, it will by default return a constant rate.
    \item \texttt{optimizers.py} provides a class which takes in a list of optimizers, as well as a list of losses to optimize. Each loss element is optimized with the corresponding optimizer at each step in training, allowing you to have potentially different learning rate schedules for different components of the loss. Minibatching is also supported.
\end{itemize}

\section{Experimental Details}

\begin{table*}[t]
\resizebox{\textwidth}{!}{
    \begin{tabular}{@{}rrrcrrcrrcrr@{}}
    \toprule
                        & $\mathcal{R}_{\text{WM}}^{\text{TPE}}$ & $\mathcal{R}_{\text{WM}+\text{AD}}^{\text{TPE}}$ & $\mathcal{R}_{\text{WM}+\text{AD}+\text{OPS}}^{\text{TPE}}$ & $\mathcal{R}_{\text{IA}}^{\text{TPE}}$ \\ \midrule
    Alternating Minimization & True & True & True & True \\
    Delay Epochs ($\text{de}$) & 2 & 0 & 0 & 1 \\
    Train Batch Size ($|\mathcal{B}|$) & 2048 & 256 & 256 & 256 \\
    SGDM Learning Rate & 1.0 & 0.125 & 0.125 & 0.125 \\
    Alignment Optimizer & Vanilla GD & Adam & Adam & Adam \\
    Alignment Learning Rate ($\eta$) & 1.0 & 0.0053 & 0.0025 & 0.0098 \\
    $\sigma$ & 0.6905 & 0.9500 & 0.6402 & 0.8176 \\
    $\alpha/\beta$ & 15.6607 & 13.9040 & 0.1344 & 129.1123 \\
    $\beta$ & 0.0283 & $2.8109 \times 10^{-8}$ & 232.7856 & 7.9990 \\
    $\gamma$ & N/A & N/A & N/A & $3.1610 \times 10^{-6}$ \\
    Forward Path Output (FO) Bias & True & True & True & True \\
    FO ReLU & True & True & True & True \\
    FO BWMC & True & True & True & True \\
    FO FWMC & False & False & True & True \\
    FO FWL2N & False & False & False & False \\
    Backward Path Output (BO) Bias & False & False & False & True \\
    BO ReLU & False & False & True & False \\
    BO FWMC & False & False & False & True \\
    BO FWL2N & False & False & True & True \\
    Backward Path Input (BI) BWMC & True & True & True & False \\
    BI FWMC & False & False & False & False \\
    BI FWL2N & False & False & False & False \\
    \bottomrule
    \end{tabular}
}
\caption[Metaparameter settings for each of the learning rules obtained by large-scale searches]{Metaparameter settings (rows) for each of the learning rules obtained by large-scale searches (columns). Continuous values were rounded up to 4 decimal places.}
\label{lrperf:tab:metaparameters}
\end{table*}

In what follows we describe the metaparameters we used to run each of the experiments reported above, tabulated in Table~\ref{lrperf:tab:metaparameters}.
Any defaults from TensorFlow correspond to those in version 1.13.1.

\subsection{TPE search spaces}
\label{lrperf:sup:hp-ss-details}
We detail the search spaces for each of the searches performed in Section~\ref{lrperf:sec:local-learning-rules}.
For each search, we trained approximately 60 distinct settings at a time using the HyperOpt package \cite{Bergstra2011} using the ResNet-18 architecture and L2 weight decay of $\lambda = 10^{-4}$ \cite{He2016} for 45 epochs, corresponding to the point in training midway between the first and second learning rate drops. 
Each model was trained on its own Tensor Processing Unit (TPUv2-8 and TPUv3-8).

We employed a form of Bayesian optimization, a Tree-structured Parzen Estimator (TPE), to search the space of continuous and categorical metaparameters \cite{Bergstra2011}. This algorithm constructs a generative model of $P[score\mid configuration]$ by updating a prior from a maintained history $H$ of metaparameter configuration-loss pairs. The fitness function that is optimized over models is the expected improvement, where a given configuration $c$ is meant to optimize $EI(c) = \int_{x < t}P[x\mid c, H]$. This choice of Bayesian optimization algorithm models $P[c\mid x]$ via a Gaussian mixture, and restricts us to tree-structured configuration spaces.

\subsubsection{$\mathcal{R}_{\text{WM}}^{\text{TPE}}$ search space}
\label{lrperf:sup:hp-wm-tpe-details}
Below is a description of the metaparameters and their ranges for the search that gave rise to $\mathcal{R}_{\text{WM}}^{\text{TPE}}$ in Table~\ref{lrperf:tab:hp-local}.

\begin{itemize}
\item Gaussian input noise standard deviation $\sigma \in [10^{-10}, 1]$ used in the backward pass, sampled uniformly.

\item Ratio between the weighting of $\mathcal{P}^{\text{amp}}$ and $\mathcal{P}^{\text{decay}}$ given by $\alpha/\beta \in [0.1, 200]$, sampled uniformly. 

\item The weighting of $\mathcal{P}^{\text{decay}}$ given by $\beta \in [10^{-11}, 10^{7}]$, sampled log-uniformly.
\end{itemize}
We fix all other metaparameters as prescribed by \citet{Akrout2019}, namely batch centering the backward path inputs and forward path outputs in the backward pass, as well as applying a ReLU activation function and bias to the forward path but not to the backward path in the backward pass.
To keep the learning rule fully local, we do not allow for any transport during the mirroring phase of the batch normalization mean and standard deviation as \citet{Akrout2019} allow.

\subsubsection{$\mathcal{R}_{\text{WM}+\text{AD}}^{\text{TPE}}$ search space}
\label{lrperf:sup:hp-wm-ad-tpe-details}
Below is a description of the metaparameters and their ranges for the search that gave rise to $\mathcal{R}_{\text{WM}+\text{AD}}^{\text{TPE}}$ in Table~\ref{lrperf:tab:hp-local}.
\begin{itemize}
\item Train batch size $|\mathcal{B}| \in \{256, 1024, 2048, 4096\}$.  
This choice also determines the forward path Nesterov momentum learning rate on the \emph{pseudogradient} of the categorization objective $\mathcal{J}$, as it is set to be $|\mathcal{B}|/2048$, and linearly warm it up to this value for 6 epochs followed by $90\%$ decay at 30, 60, and 80 epochs, training for 100 epochs total, as prescribed by \citet{buchlovsky2019tf}.

\item Alignment learning rate $\eta \in [10^{-6}, 10^{-2}]$, sampled log-uniformly.
This parameter sets the adaptive learning rate on the Adam optimizer applied to the \emph{gradient} of the alignment loss $\mathcal{R}$, and which will be dropped synchronously by $90\%$ decay at 30, 60, and 80 epochs along with the Nesterov momentum learning rate on the \emph{pseudogradient} of the categorization objective $\mathcal{J}$.

\item Number of delay epochs $de \in \{0, 1, 2\}$ for which we delay optimization of the categorization objective $\mathcal{J}$ and solely optimize the alignment loss $\mathcal{R}$.
If $de > 0$, we use the alignment learning rate $\eta$ during this delay period and the learning rate drops are shifted by $de$ epochs; otherwise, if $de = 0$, we linearly warmup $\eta$ for 6 epochs as well.

\item Whether or not to perform alternating minimization of $\mathcal{J}$ and $\mathcal{R}$ each step, or instead simultaneously optimize these objectives in a single training step. 
\end{itemize}

The remaining metaparameters and their ranges were the same as those from Appendix \ref{lrperf:sup:hp-wm-tpe-details}.




We fix the layer-wise operations as prescribed by \citet{Akrout2019}, namely batch centering the backward path input and forward path outputs in the backward pass (\textbf{BI BWMC} and \textbf{FO BWMC}, respectively), as well as applying a ReLU activation function and bias to the forward path (\textbf{FO ReLU} and \textbf{FO Bias}, respectively) but not to the backward path in the backward pass (\textbf{BO ReLU} and \textbf{BO Bias}, respectively).

\subsubsection{$\mathcal{R}_{\text{WM}+\text{AD}+\text{OPS}}^{\text{TPE}}$ search space}
\label{lrperf:sup:hp-wm-ad-ops-tpe-details}
Below is a description of the metaparameters and their ranges for the search that gave rise to $\mathcal{R}_{\text{WM}+\text{AD}+\text{OPS}}^{\text{TPE}}$ in Table~\ref{lrperf:tab:hp-local}. 
In this search, we expand the search space described in Appendix \ref{lrperf:sup:hp-wm-ad-tpe-details} to include boolean choices over layer-wise operations performed in the \emph{backward pass}, involving either the inputs, the forward path $f_l$ (involving only the forward weights $W_l$), or the backward path $b_l$ (involving only the backward weights $B_l$):

Use of biases in the forward and backward paths:
\begin{itemize}
\item \textbf{FO Bias:} Whether or not to use biases in the forward path.

\item \textbf{BO Bias:} Whether or not to use biases in the backward path.
\end{itemize}

Use of nonlinearities in the forward and backward paths:
\begin{itemize}
\item \textbf{FO ReLU:} Whether or not to apply a ReLU to the forward path output.

\item \textbf{BO ReLU:} Whether or not to apply a ReLU to the backward path output.
\end{itemize}

Centering and normalization operations in the forward and backward paths:
\begin{itemize}
\item \textbf{FO BWMC:} Whether or not to mean center (across the \emph{batch} dimension) the forward path output $f_l = f_l - \bar{f}_l$.

\item \textbf{BI BWMC:} Whether or not to mean center (across the \emph{batch} dimension) the backward path \emph{input}. 

\item \textbf{FO FWMC:} Whether or not to mean center (across the \emph{feature} dimension) the forward path output $f_l = f_l - \hat{f}_l$.

\item \textbf{BO FWMC:} Whether or not to mean center (across the \emph{feature} dimension) the backward path output $b_l = b_l - \hat{b}_l$.

\item \textbf{FO FWL2N:} Whether or not to L2 normalize (across the feature dimension) the forward path output $f_l = (f_l - \hat{f}_l)/||f_l - \hat{f}_l||_2$. 

\item \textbf{BO FWL2N:} Whether or not to L2 normalize (across the feature dimension) the backward path output $b_l = (b_l - \hat{b}_l)/||b_l - \hat{b}_l||_2$. 
\end{itemize}

Centering and normalization operations applied to the inputs to the backward pass:
\begin{itemize}
\item \textbf{BI FWMC:} Whether or not to mean center (across the feature dimension) the backward pass input $x_l = x_l - \hat{x}_l$.

\item \textbf{BI FWL2N:} Whether or not to L2 normalize (across the feature dimension) the backward pass input $x_l = (x_l - \hat{x}_l)/||x_l - \hat{x}_l||_2$. 
\end{itemize}

The remaining metaparameters and their ranges were the same as those from Appendix \ref{lrperf:sup:hp-wm-ad-tpe-details}.







\subsubsection{$\mathcal{R}_{\text{IA}}^{\text{TPE}}$ search space}
\label{lrperf:sup:hp-ia-tpe-details}
Below is a description of the metaparameters and their ranges for the search that gave rise to $\mathcal{R}_{\text{IA}}^{\text{TPE}}$ in Table~\ref{lrperf:tab:hp-local}. In this search, we expand the search space described in Appendix \ref{lrperf:sup:hp-wm-ad-ops-tpe-details}, to now include the additional $\mathcal{P}^{\text{null}}$ primitive.

\begin{itemize}
\item The weighting of $\mathcal{P}^{\text{null}}$ given by $\gamma \in [10^{-11}, 10^{7}]$, sampled log-uniformly.
\end{itemize}

The remaining metaparameters and their ranges were the same as those from Appendix \ref{lrperf:sup:hp-wm-ad-ops-tpe-details}.

\subsection{Symmetric and Activation Alignment metaparameters}
\label{lrperf:sup:non-local-hp}
We now describe the metaparameters used to generate Table~\ref{lrperf:tab:empimnet}. 
We used a batch size of 256, forward path Nesterov with Momentum of 0.9 and a learning rate of 0.1 applied to the categorization objective $\mathcal{J}$, warmed up linearly for 5 epochs, with learning rate drops at 30, 60, and 80 epochs, trained for a total of 90 epochs, as prescribed by \citet{He2016}.

For Symmetric and Activation Alignment ($\mathcal{R}_{\text{SA}}$ and $\mathcal{R}_{\text{AA}}$), we used Adam on the alignment loss $\mathcal{R}$ with a learning rate of $0.001$, along with the following weightings for their primitives:
\begin{itemize}
\item Symmetric Alignment: $\alpha=10^{-3}, \beta=2\times 10^{-3}$
\item Activation Alignment: $\alpha=10^{-3}, \beta=2\times 10^{-3}$
\end{itemize}
We use biases in both the forward and backward paths of the backward pass, but do \emph{not} employ a ReLU nonlinearity to either path.

\subsection{Noisy updates}
\label{lrperf:sup:noisy-updates}

We describe the experimental setup and metaparameters used in Section~\ref{lrperf:sec:non-local-learning-rules} to generate Figure~\ref{lrperf:fig:noise-plot}.  

Figure~\ref{lrperf:fig:noise-plot-a} was generated by running 10 trials for each experiment configuration. 
The error bars show the standard error of the mean across trials. 

For backpropagation  we used a momentum optimizer with an initial learning rate of $0.1$, standard batch size of $256$, and learning rate drops at $30$ and $60$ epochs.

For Symmetric and Activation Alignment we used the same metaparameters as backpropagation for the categorization objective $\mathcal{J}$ and an Adam optimizer with an initial learning rate of $0.001$ and learning rate drops at $30$ and $60$ epochs for the alignment loss $\mathcal{R}$.  All other metaparameters were the same as described in Appendix \ref{lrperf:sup:non-local-hp}.

\begin{figure}[tb]
\centering
\includegraphics[width=0.48\textwidth]{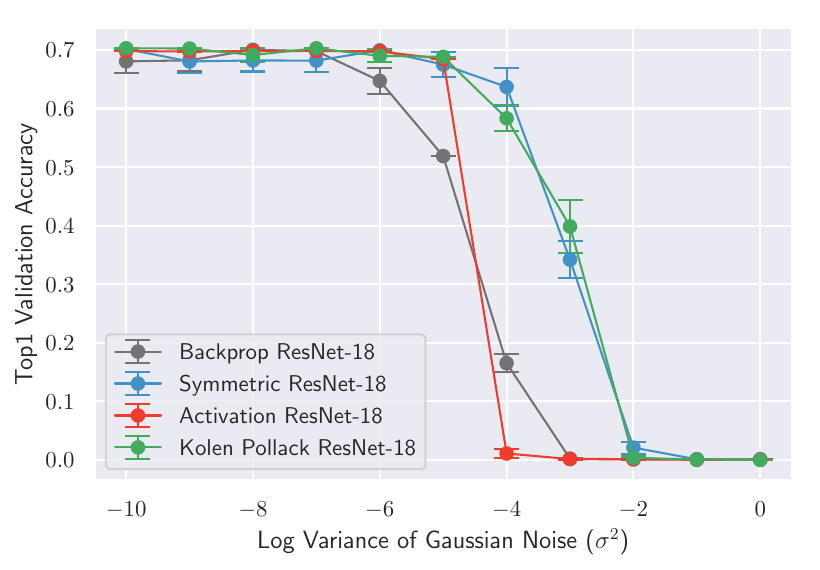}
\caption[Robustness to noisy updates]{\textbf{Noisy updates.} Symmetric Alignment, Activation Alignment, and Kolen-Pollack are \emph{more} robust to noisy updates than backpropagation for ResNet-18.}
\label{lrperf:fig:all-noise}
\end{figure}

In all experiments we added the noise to the update given by the respective optimizers and scaled by the current learning rate, that way at learning rate drops the noise scaled appropriately.  To account for the fact that the initial learning rate for the backpopagation experiments was $0.1$, while for symmetric and activation experiments it was $0.001$, we shifted the latter two curves by $10^4$ to account for the effective difference in variance.
See Figure~\ref{lrperf:fig:all-noise}.

\subsection{Metaparameter importance quantification}
\label{lrperf:sup:meta-analysis-hp}
\begin{figure}
\centering
\includegraphics[width=0.96\columnwidth]{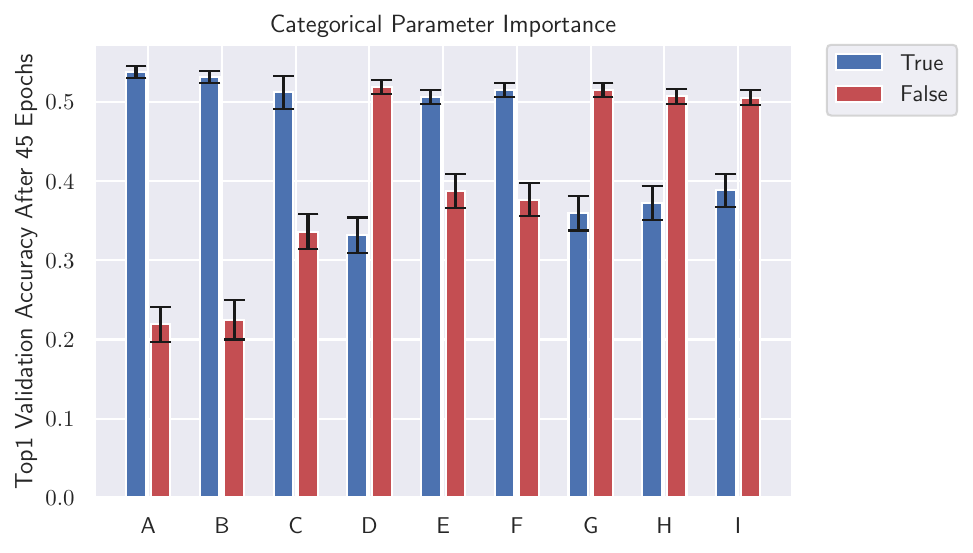}
\caption[Analysis of important categorical metaparameters of top performing local rule $\mathcal{R}_{\text{IA}}^{\text{TPE}}$]{\textbf{Analysis of important categorical metaparameters of top performing local rule $\mathcal{R}_{\text{IA}}^{\text{TPE}}$.} Mean across models, and the error bars indicate SEM across models.}
\label{lrperf:fig:hp-metaanalysis}
\end{figure}

We include here the set of discrete metaparameters that mattered the most across hundreds of models in our large-scale search, sorted by most to least important, plotted in Figure~\ref{lrperf:fig:hp-metaanalysis}. 
Specifically, these amount to choices of activation, layer-wise normalization, input normalization, and Gaussian noise in the forward and backward paths of the backward pass. 
The detailed labeling is given as follows: \textbf{A:} Whether or not to L2 normalize (across the feature dimension) the backward path outputs in the backward pass. 
\textbf{B:} Whether to use Gaussian noise in the backward pass inputs. 
\textbf{C:} Whether to solely optimize the alignment loss in the first 1-2 epochs of training. 
\textbf{D, E:} Whether or not to apply a non-linearity in the backward or forward path outputs in the backward pass, respectively. 
\textbf{F:} Whether or not to apply a bias in the forward path outputs (pre-nonlinearity). 
\textbf{I:} Same as \textbf{A}, but instead applied to the forward path outputs in the backward pass.

\subsection{Neural Fitting Procedure}
\label{lrperf:sup:neural-fit}
We fit trained model features to multi-unit array responses from \cite{Majaj2015}.
Briefly, we fit to 256 recorded sites from two monkeys. These came from three multi-unit arrays per monkey: one implanted in V4, one in posterior IT, and one in central and anterior IT.
Each image was presented approximately 50 times, using rapid visual stimulus presentation (RSVP).
Each stimulus was presented for 100 ms, followed by a mean gray background interleaved between images.
Each trial lasted 250 ms. The image set consisted of 5120 images based on 64 object categories.
Each image consisted of a 2D projection of a 3D model added to a random background.
The pose, size, and $x$- and $y$-position of the object was varied across the image set, whereby 2 levels of variation were used (corresponding to medium and high variation from \cite{Majaj2015}.)
Multi-unit responses to these images were binned in 10ms windows, averaged across trials of the same image, and normalized to the average response to a blank image.
They were then averaged 70-170 ms post-stimulus onset, producing a set of (5120 images x 256 units) responses, which were the targets for our model features to predict.
The 5120 images were split 75-25 within each object category into a training set and a held-out testing set.

\section{Visualizations}
\label{lrperf:sup:weight-scatter}

\begin{figure}
\begin{subfigure}{0.3\columnwidth}
    \centering
    \includegraphics[width=\textwidth]{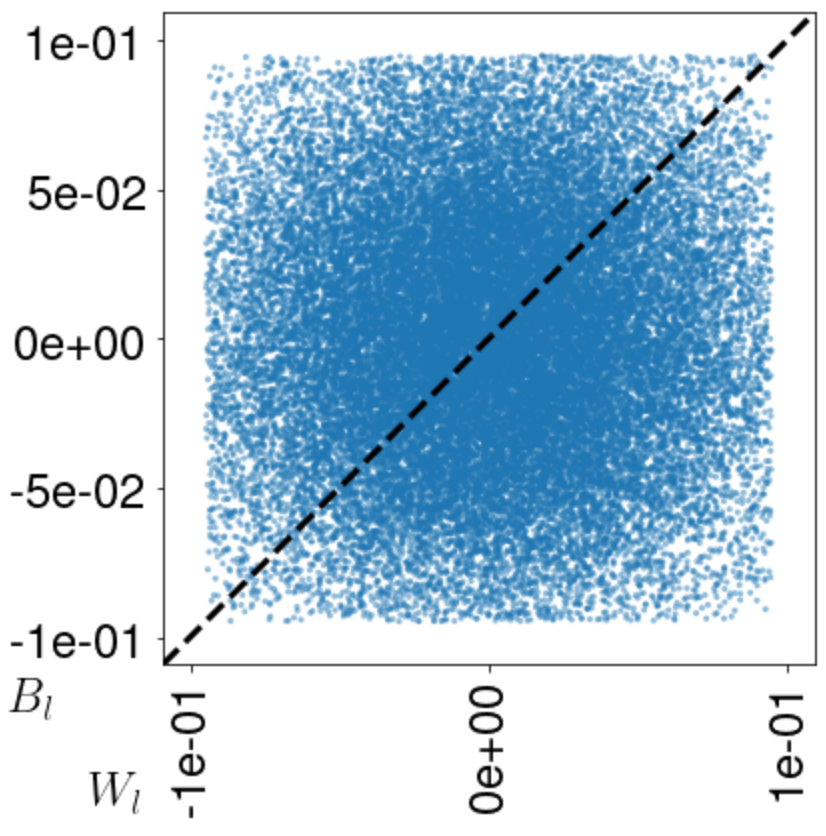}
    \caption{SA Epoch 0}
\end{subfigure}
\begin{subfigure}{0.3\columnwidth}
    \centering
    \includegraphics[width=\textwidth]{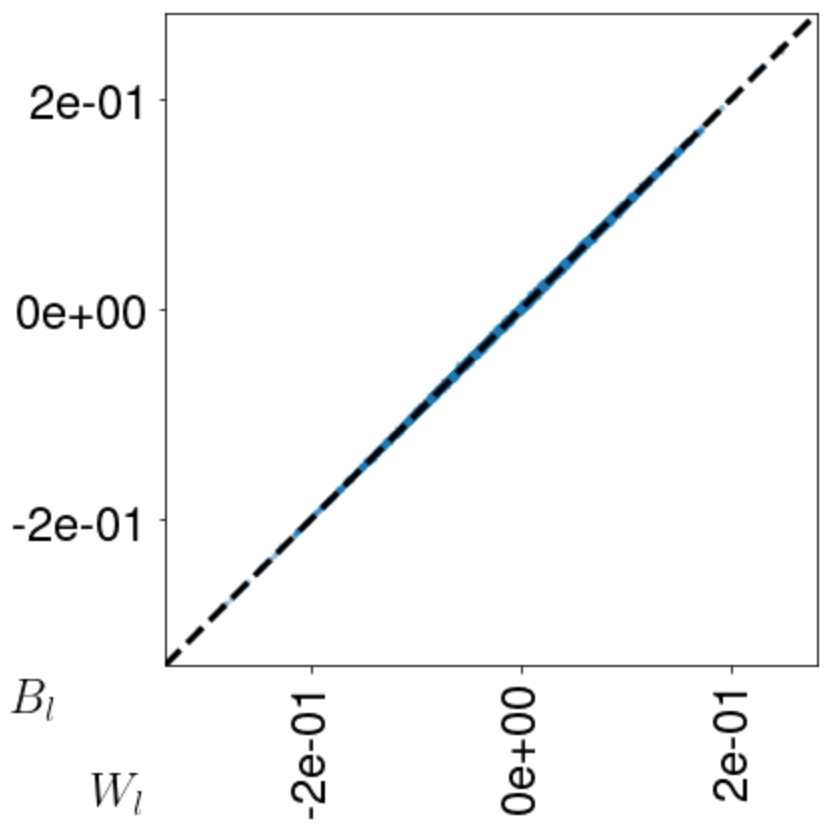}
    \caption{SA Epoch 2}
\end{subfigure}
\begin{subfigure}{0.3\columnwidth}
    \centering
    \includegraphics[width=\textwidth]{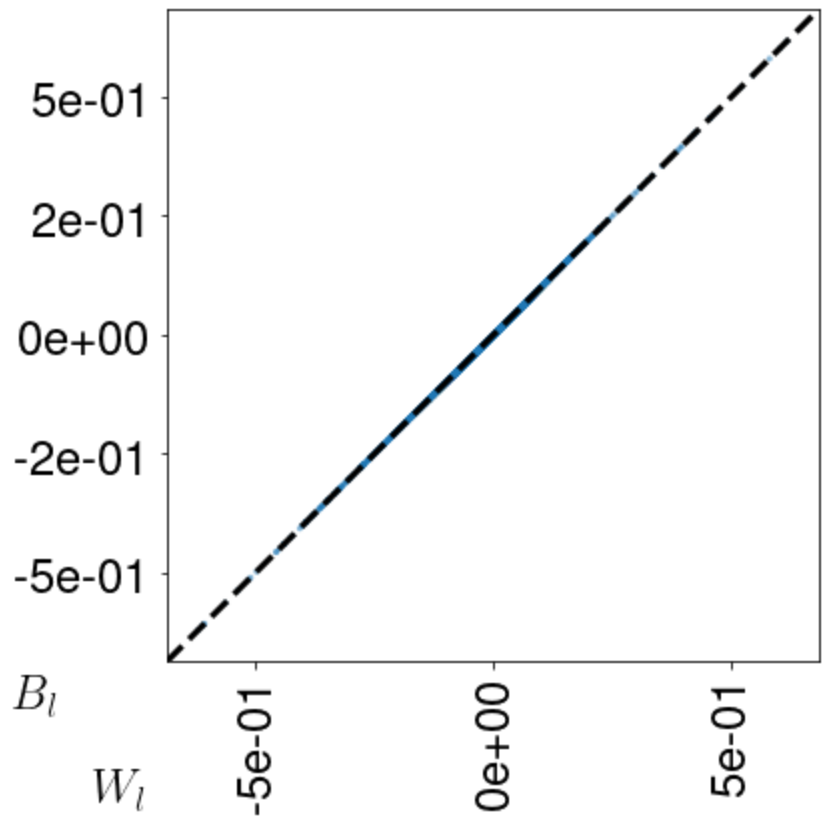}
    \caption{SA Epoch 90}
\end{subfigure}

\begin{subfigure}{0.3\columnwidth}
    \centering
    \includegraphics[width=\textwidth]{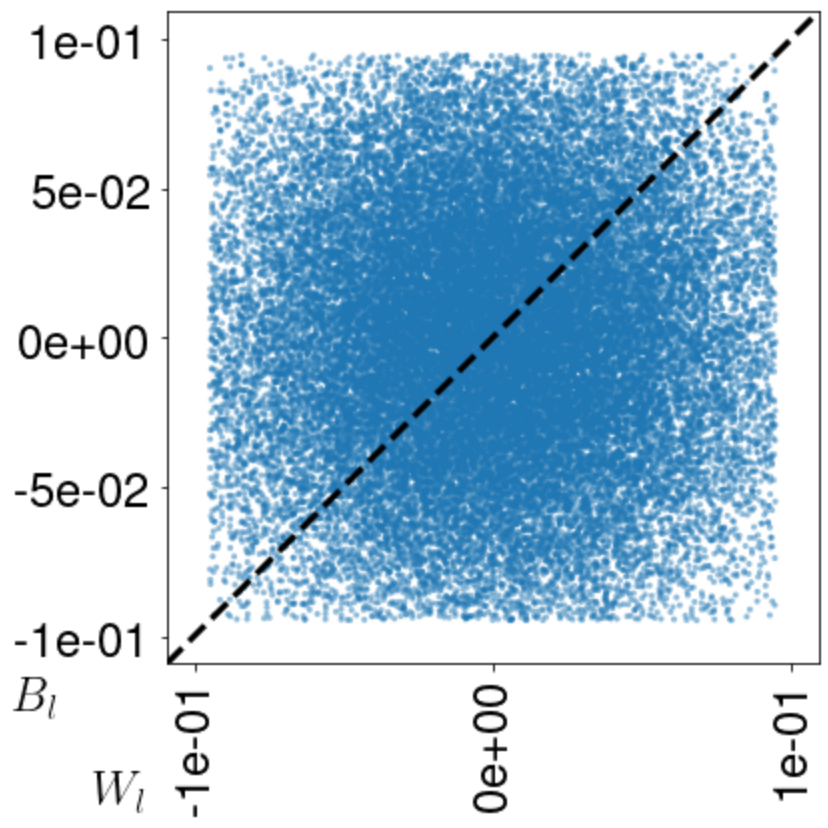}
    \caption{AA Epoch 0}
\end{subfigure}
\begin{subfigure}{0.3\columnwidth}
    \centering
    \includegraphics[width=\textwidth]{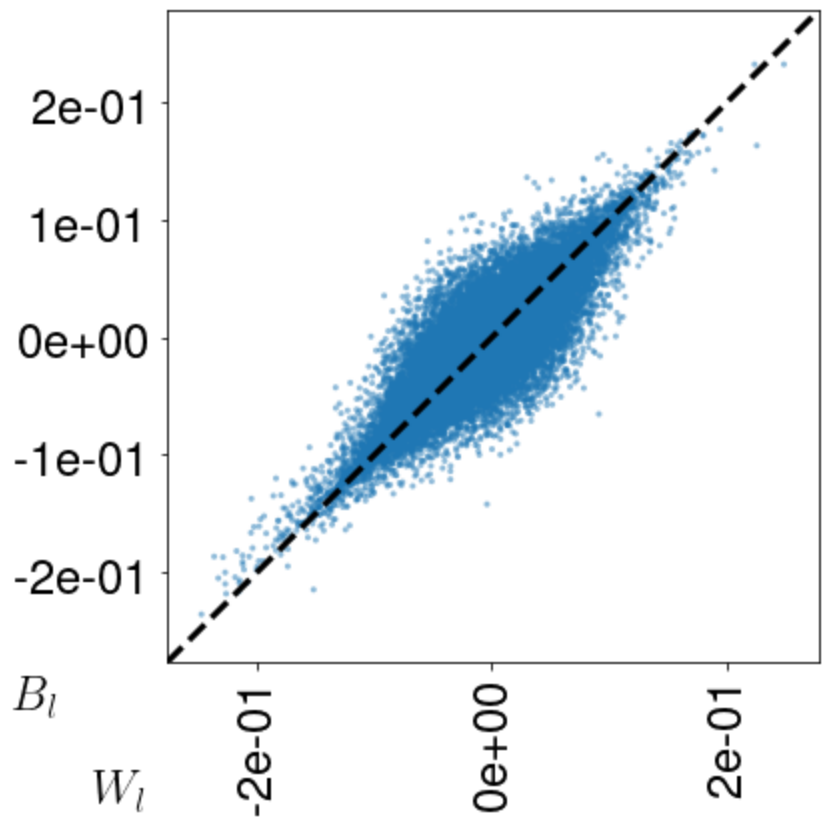}
    \caption{AA Epoch 2}
\end{subfigure}
\begin{subfigure}{0.3\columnwidth}
    \centering
    \includegraphics[width=\textwidth]{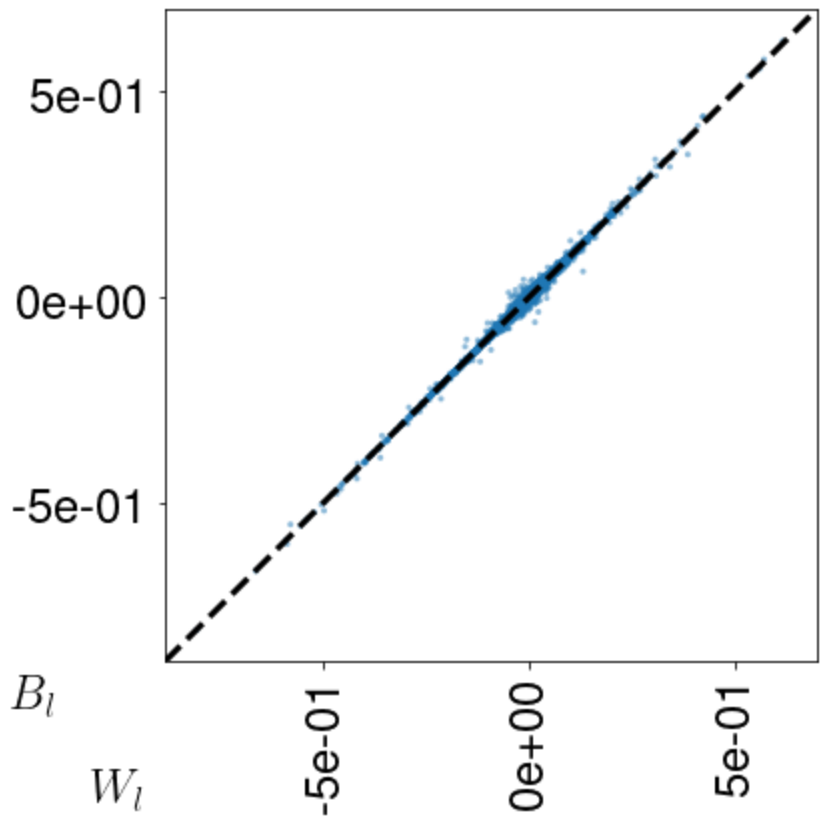}
    \caption{AA Epoch 90}
\end{subfigure}

\begin{subfigure}{0.3\columnwidth}
    \centering
    \includegraphics[width=\textwidth]{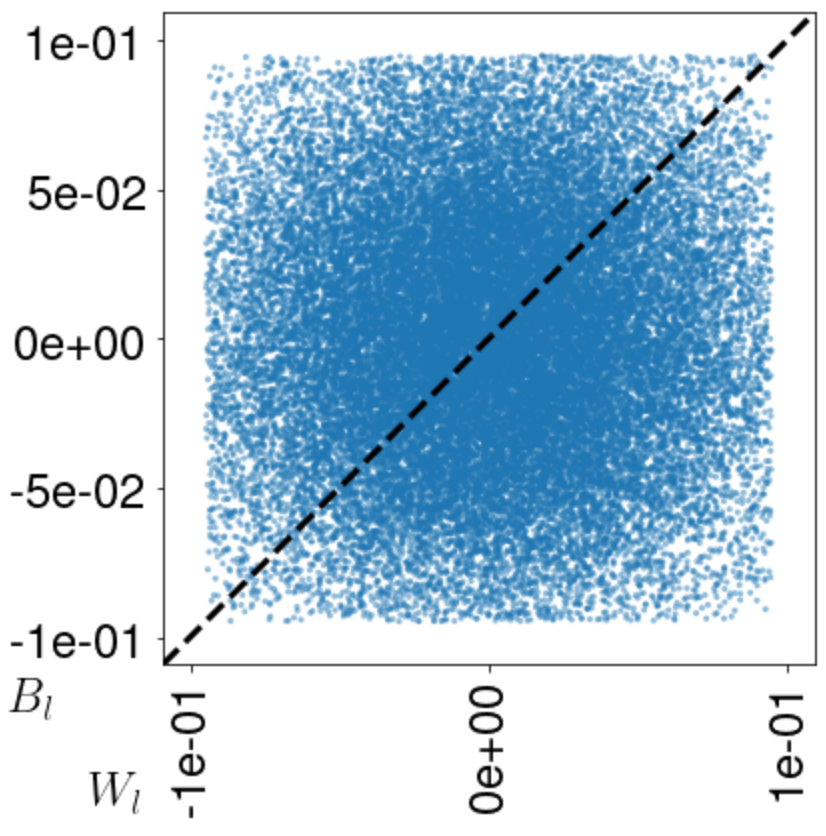}
    \caption{WM Epoch 0}
\end{subfigure}
\begin{subfigure}{0.3\columnwidth}
    \centering
    \includegraphics[width=\textwidth]{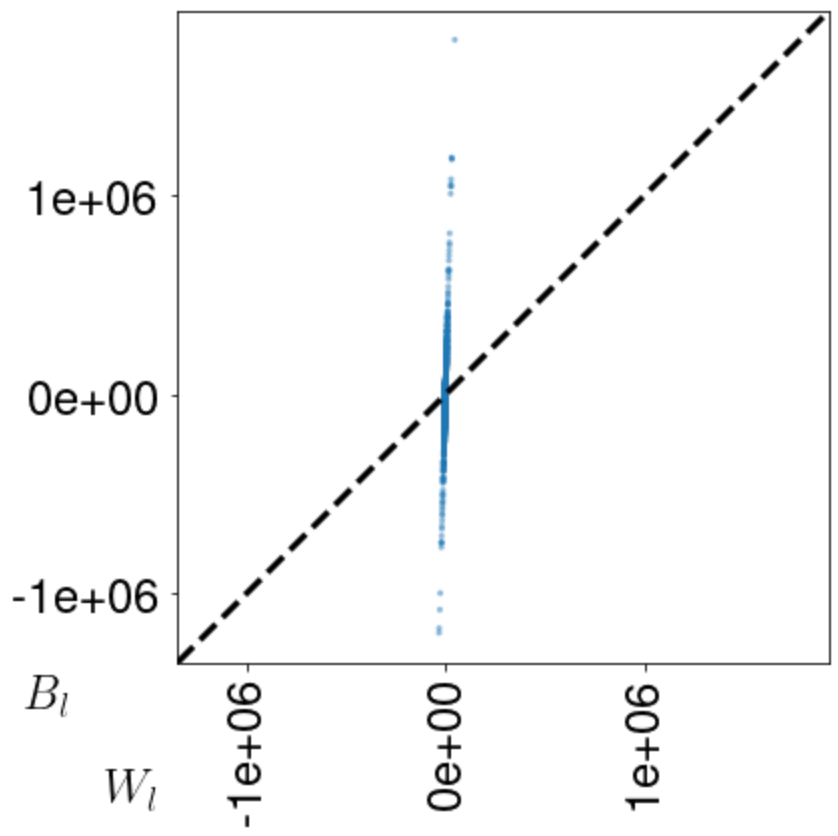}
    \caption{WM Epoch 2}
\end{subfigure}
\begin{subfigure}{0.3\columnwidth}
    \centering
    \includegraphics[width=\textwidth]{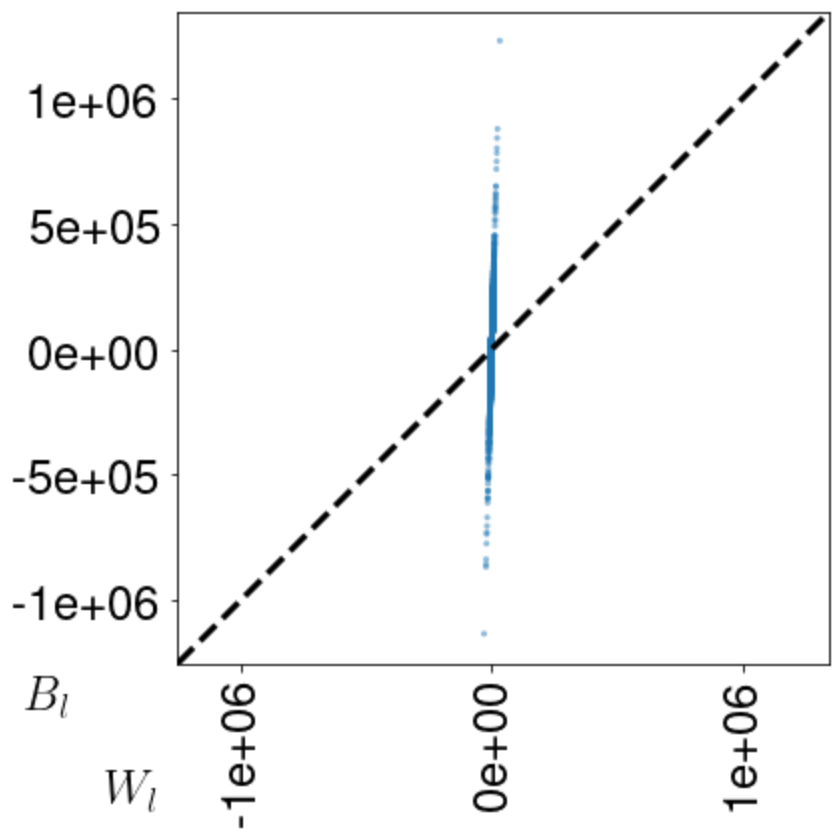}
    \caption{WM Epoch 90}
\end{subfigure}

\begin{subfigure}{0.3\columnwidth}
    \centering
    \includegraphics[width=\textwidth]{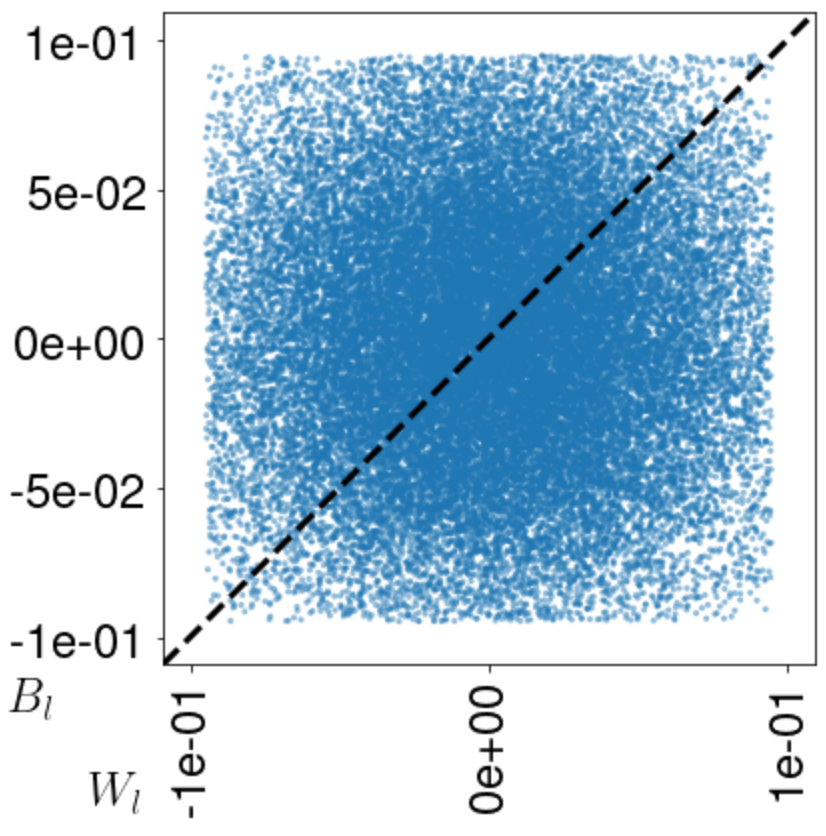}
    \caption{IA Epoch 0}
\end{subfigure}
\begin{subfigure}{0.3\columnwidth}
    \centering
    \includegraphics[width=\textwidth]{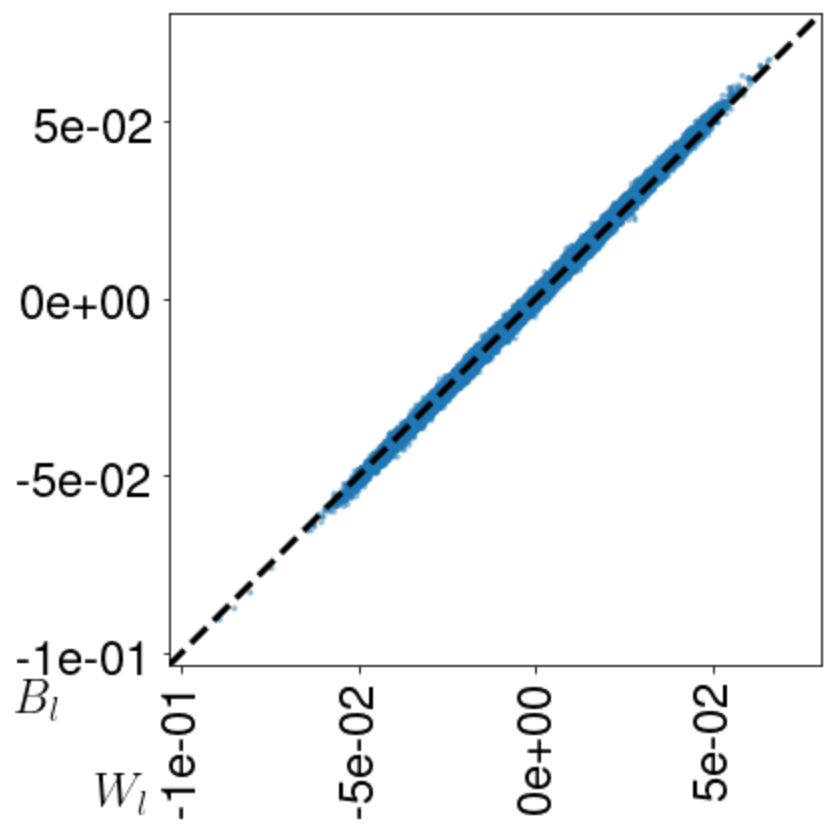}
    \caption{IA Epoch 2}
\end{subfigure}
\begin{subfigure}{0.3\columnwidth}
    \centering
    \includegraphics[width=\textwidth]{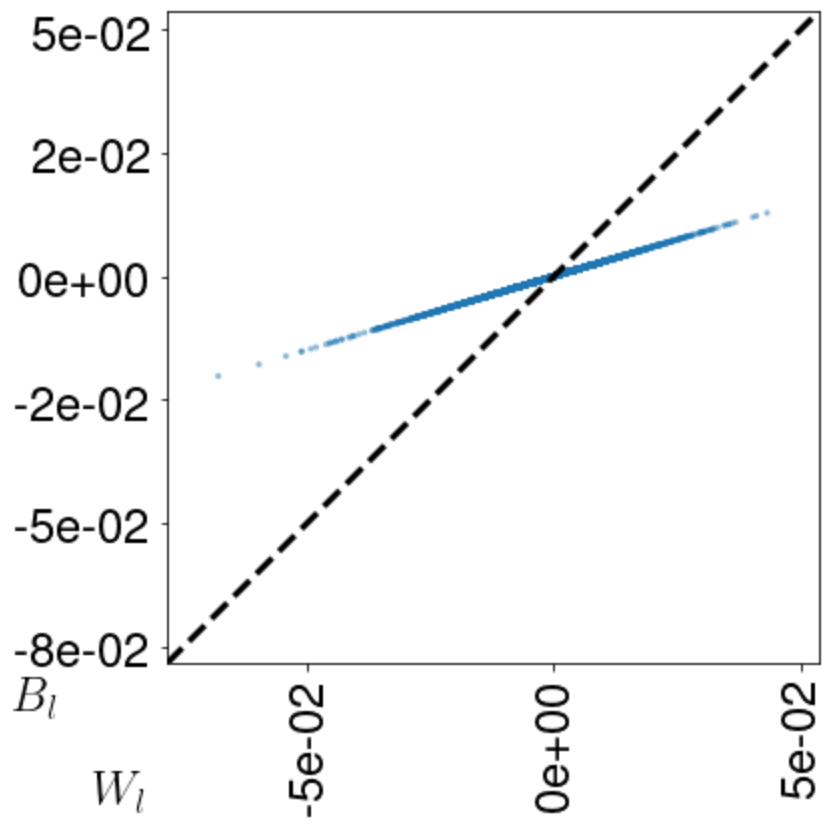}
    \caption{IA Epoch 90}
\end{subfigure}
\caption[Learning symmetry]{\textbf{Learning symmetry.} Weight values of the third convolutional layer in ResNet-18 throughout training with various learning rules.
Each dot represents an element in layer $l$'s weight matrix and its $(x,y)$ location corresponds to its forward and backward weight values, $(W_l^{(i,j)},B_l^{(j,i)} )$. The dotted diagonal line shows perfect weight symmetry, as is the case in backpropagation.
\label{lrperf:fig:full_weight_scatter}}
\end{figure}

\begin{figure*}
\begin{subfigure}{\columnwidth}
    \centering
    \includegraphics[width=0.46\textwidth]{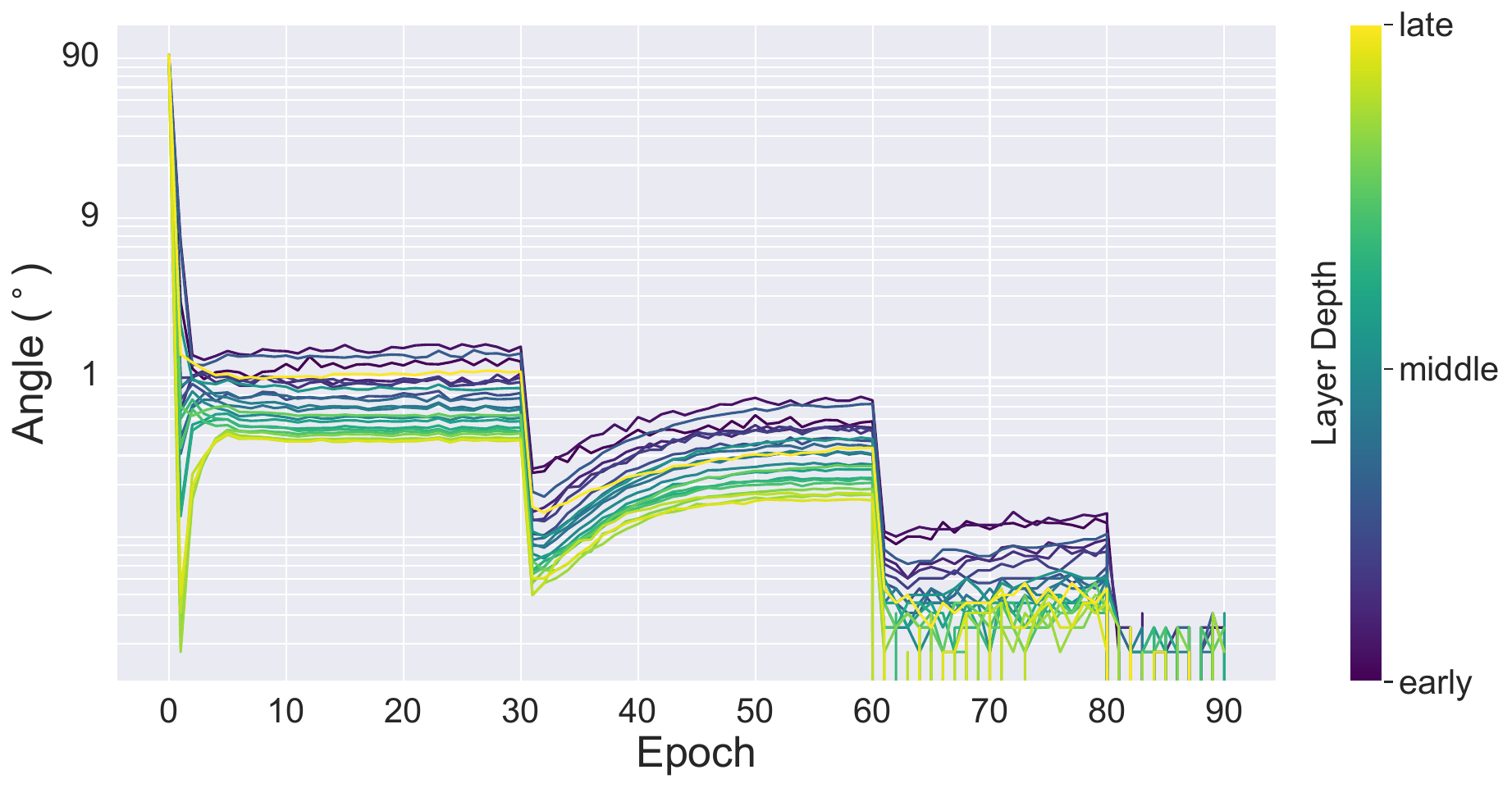}
    \includegraphics[width=0.49\textwidth]{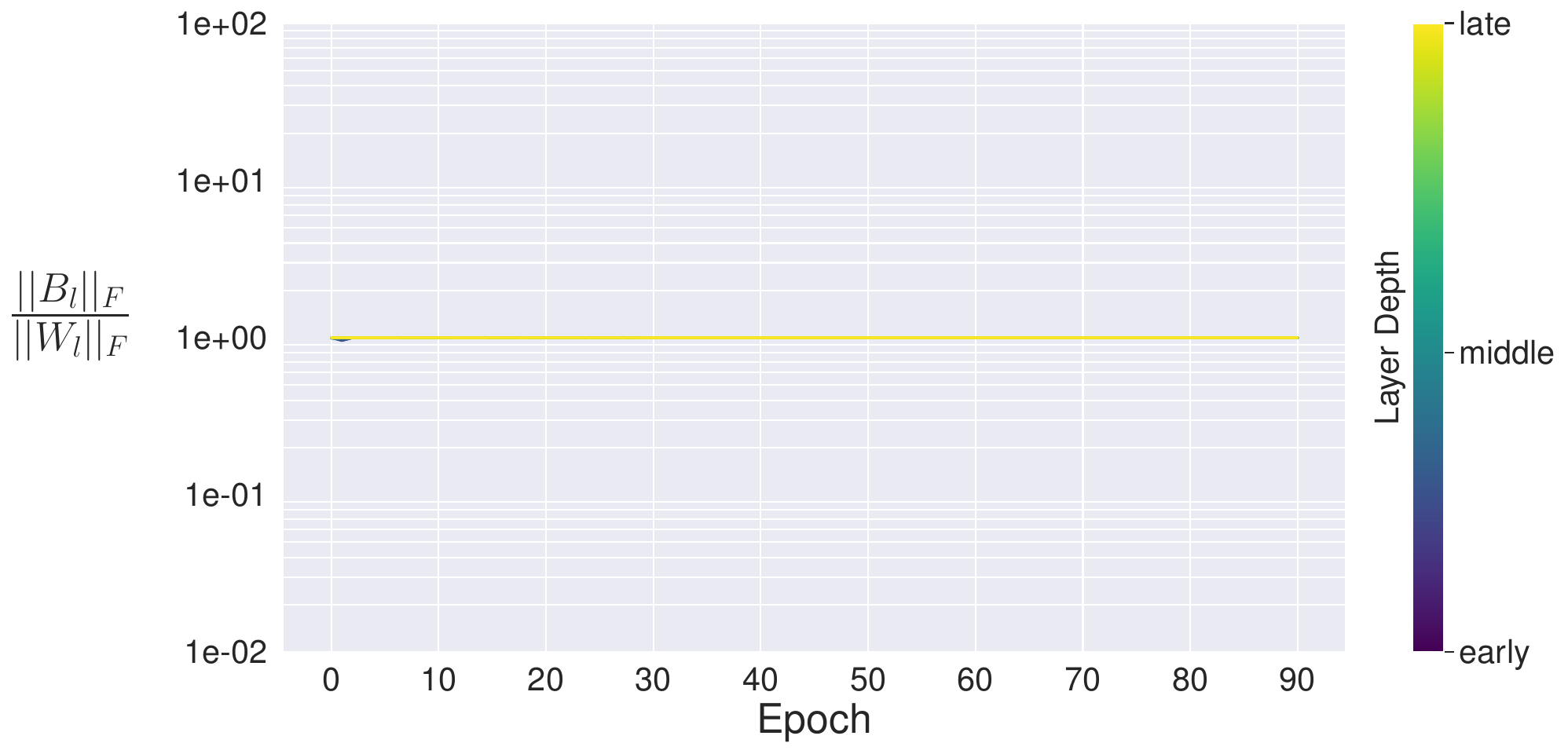}
    \caption{Symmetric Alignment}
\end{subfigure}
\begin{subfigure}{\columnwidth}
    \centering
    \includegraphics[width=0.46\textwidth]{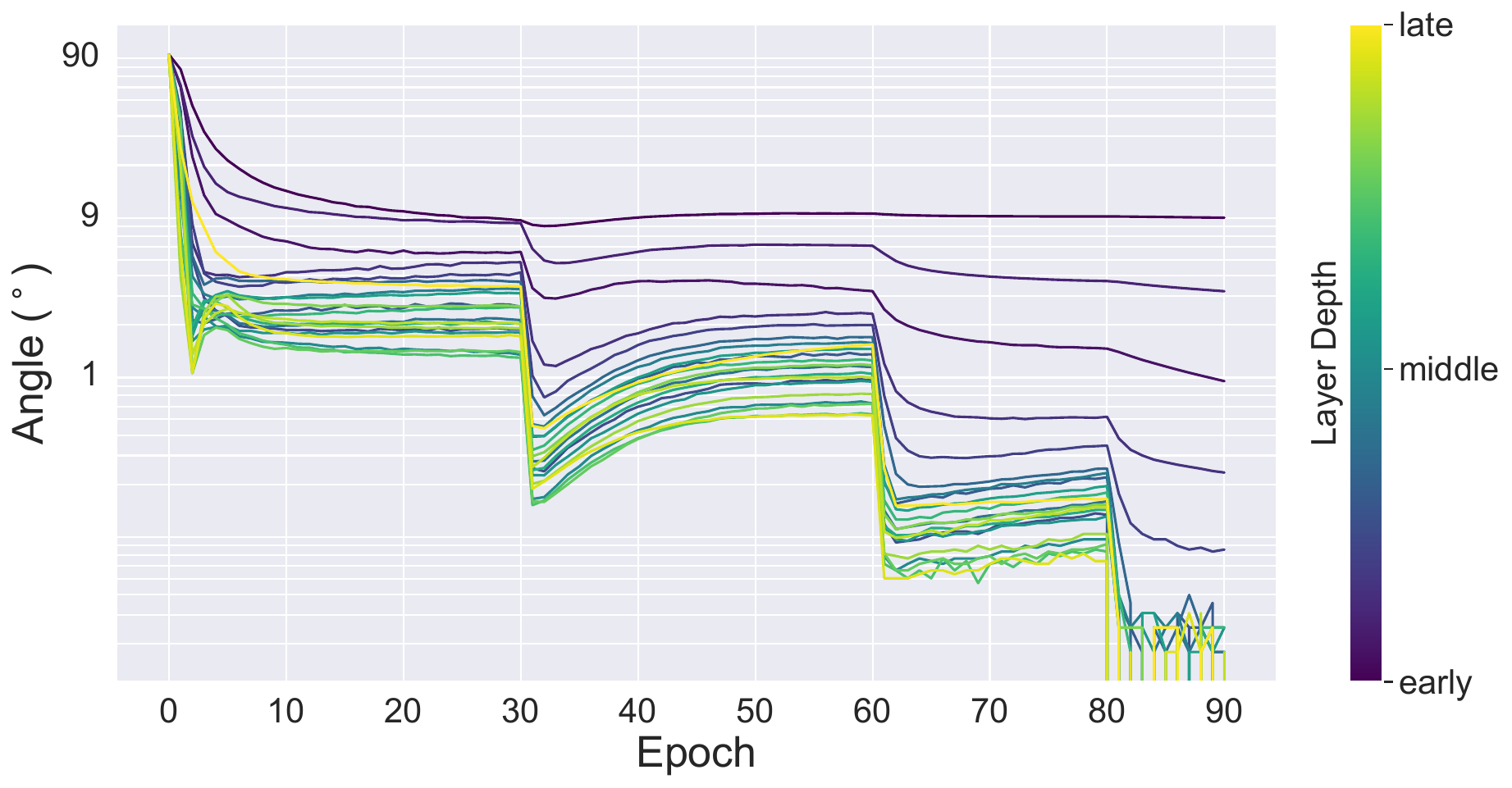}
    \includegraphics[width=0.49\textwidth]{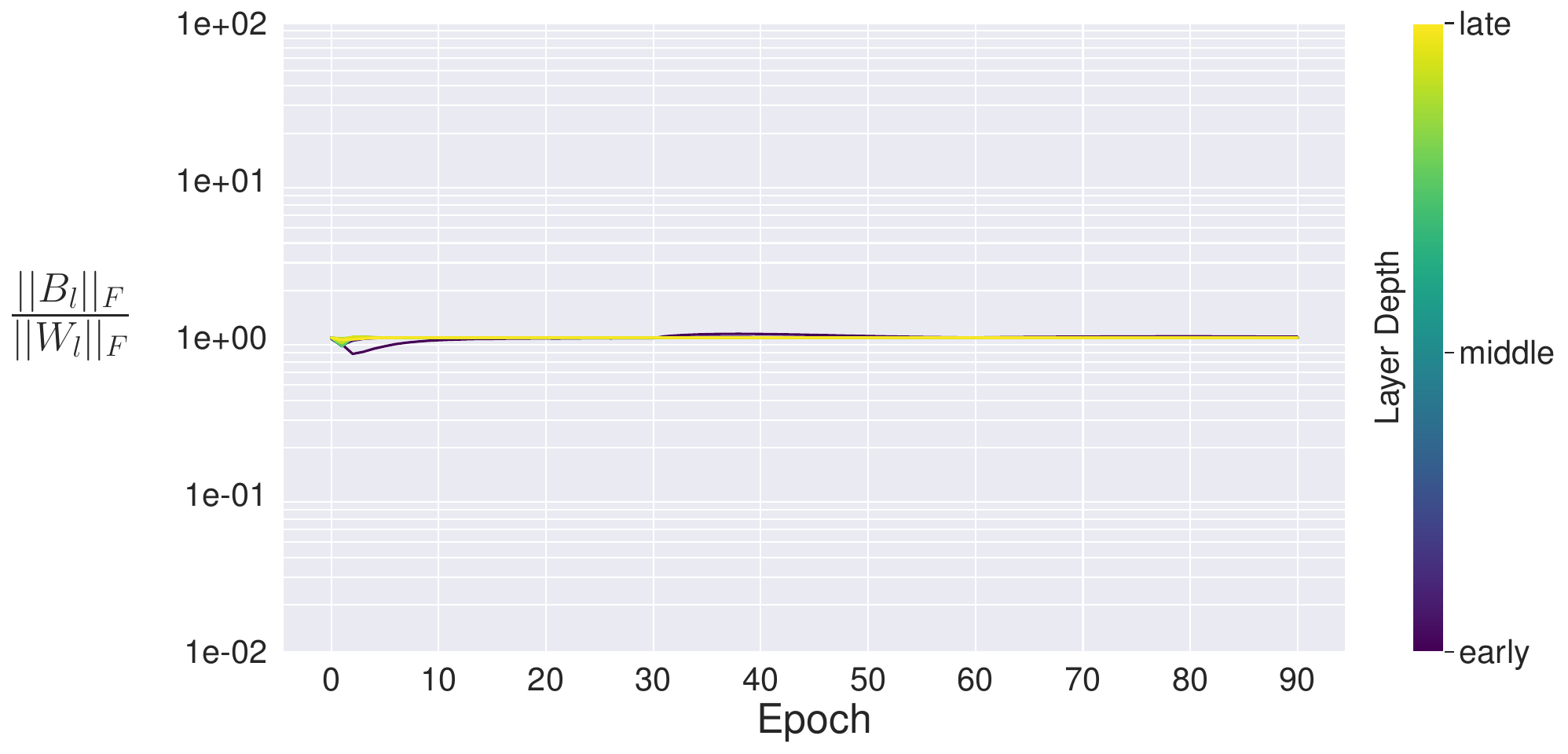}
    \caption{Activation Alignment}
\end{subfigure}
\begin{subfigure}{\columnwidth}
    \centering
    \includegraphics[width=0.46\textwidth]{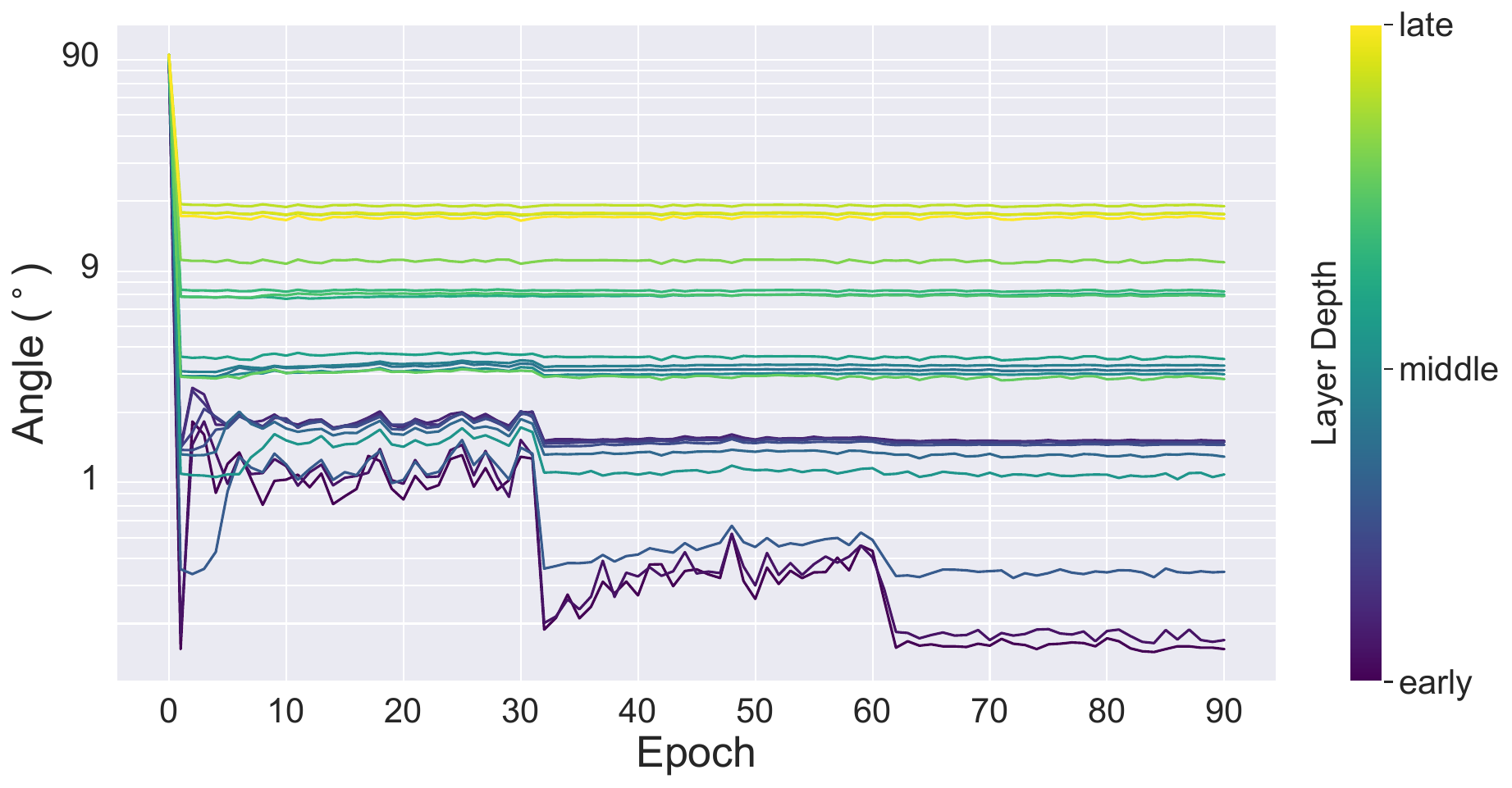}
    \includegraphics[width=0.49\textwidth]{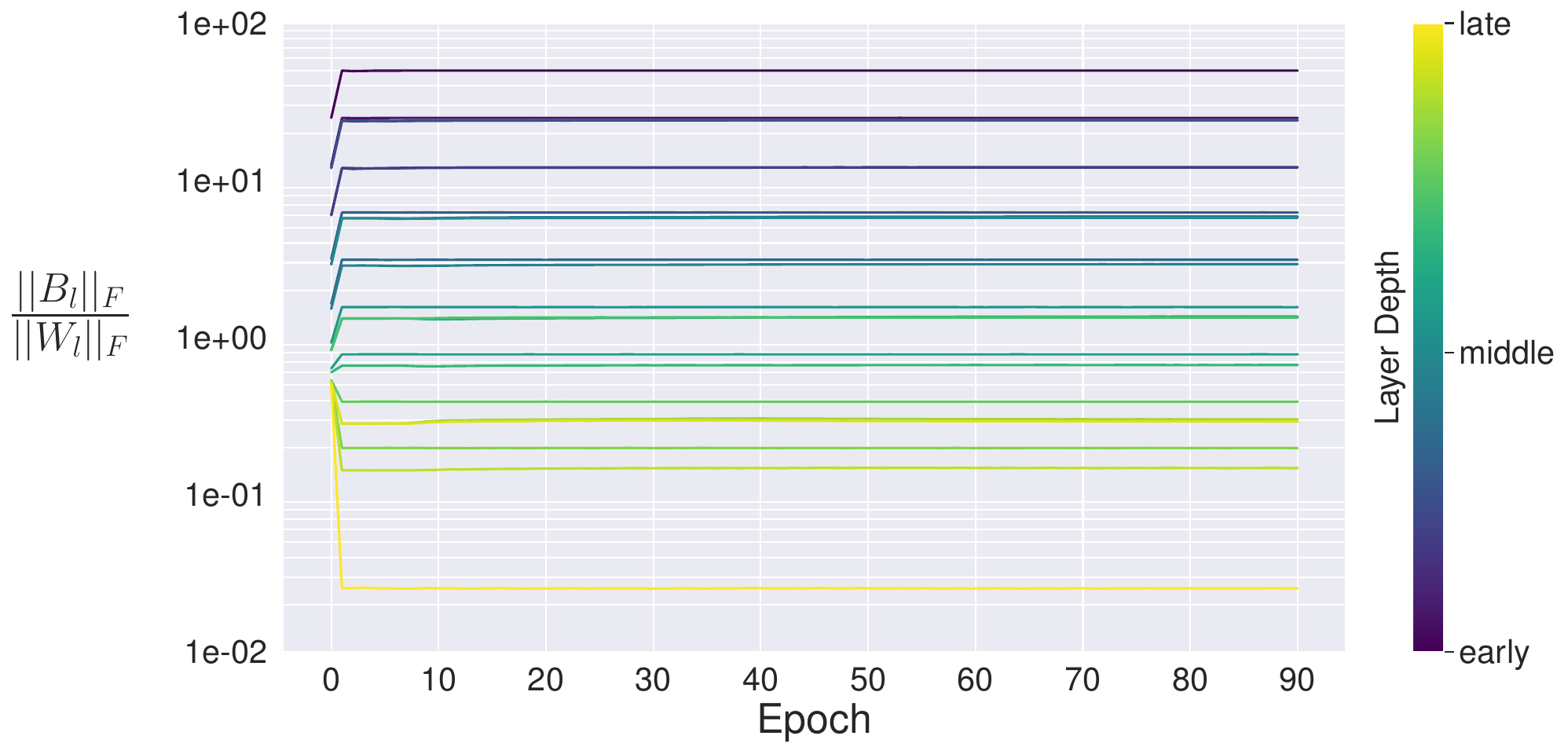}
    \caption{Weight Mirror}
\end{subfigure}
\begin{subfigure}{\columnwidth}
    \centering
    \includegraphics[width=0.46\textwidth]{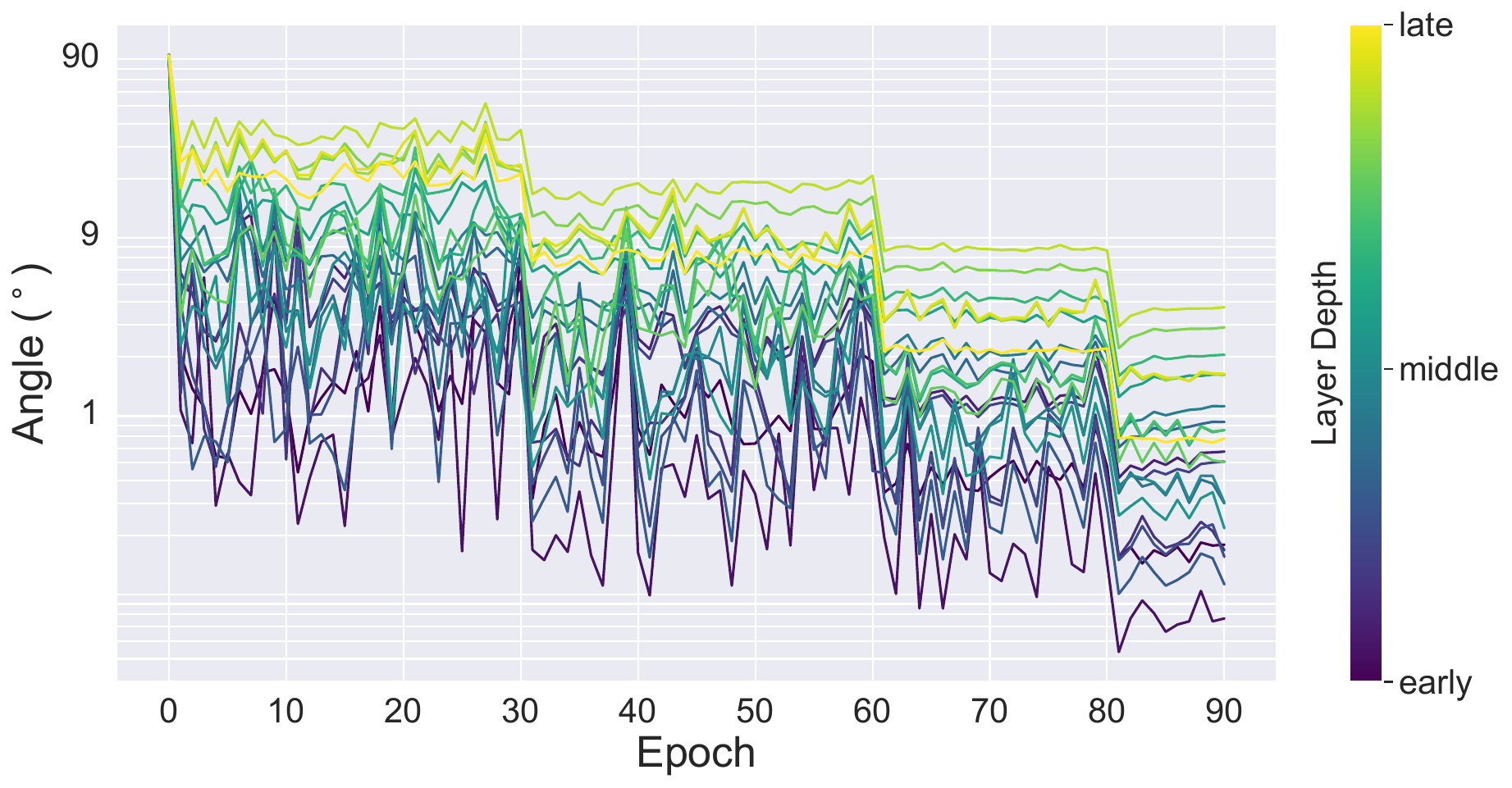}
    \includegraphics[width=0.49\textwidth]{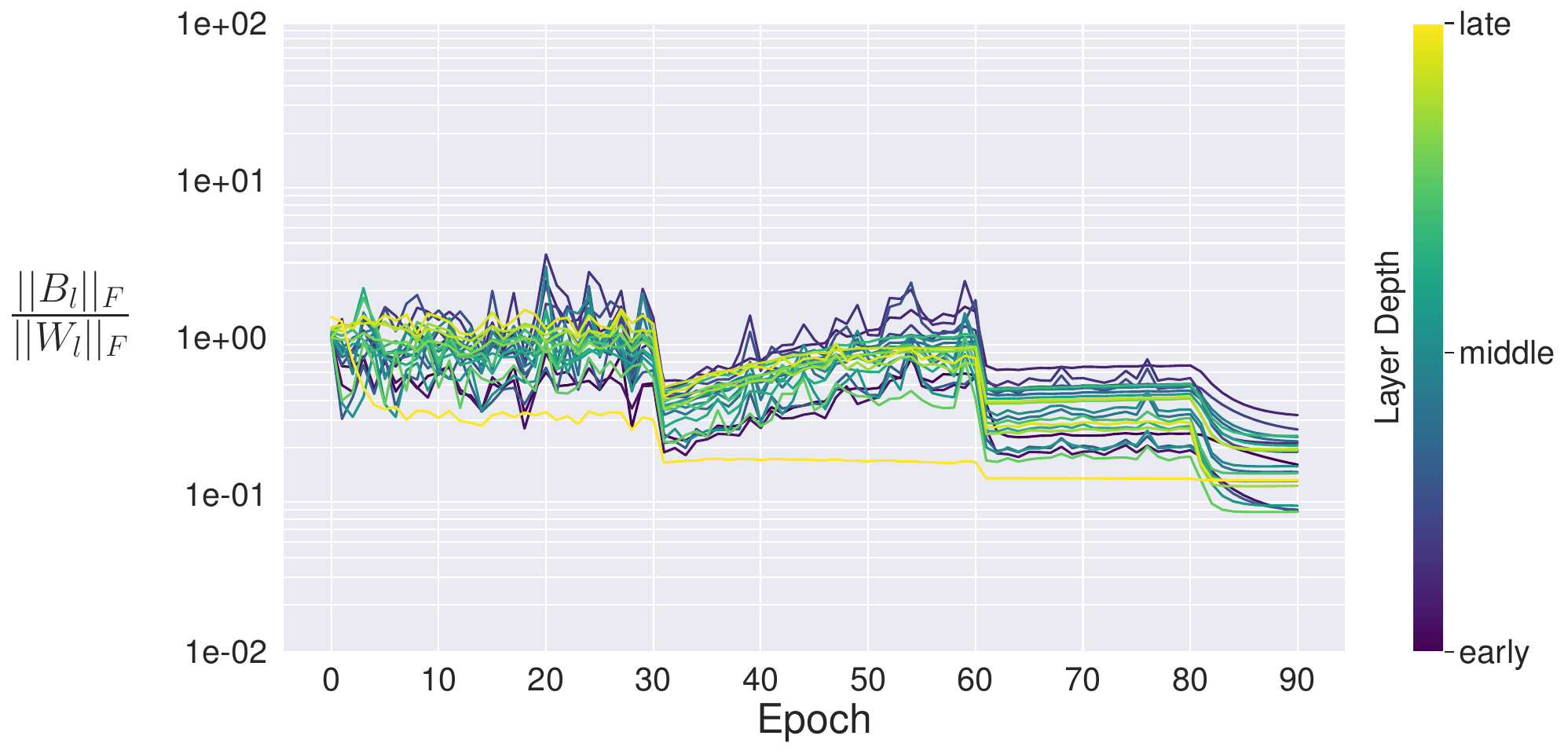}
    \caption{Information Alignment}
\end{subfigure}
\caption[Weight metrics during training]{\textbf{Weight metrics during training.} Figures on the left column show the angle between the forward and the backward weights at each layer, depicting their degree of alignment. 
Figures on the right column show the ratio of the Frobenius norm of the backward weights to the forward weights during training. 
For Symmetric Alignment \textbf{(a)} we can clearly see how the weights align very early during training, with the learning rate drops allowing them to further decrease. 
Additionally, the sizes of forward and backwards weight also remain at the same scale during training.
Activation Alignment \textbf{(b)} shows similar behavior to activation, though some of the earlier layers fail to align as closely as the Symmetric Alignment case.
Weight Mirror \textbf{(c)} shows alignment happening within the first few epochs, though some of the later layers don't align as closely.
Looking at the size of the weights during training, we can observe the unstable dynamics explained in Section~\ref{lrperf:sec:local-learning-rules} with exploding and collapsing weight values (Figure~\ref{lrperf:fig:stability}) within the first few epochs of training.
Information Alignment \textbf{(d)} shows a similar ordering in alignment as weight mirror, but overall alignment does improve throughout training, with all layers aligning within 5 degrees. 
Compared to weight mirror, the norms of the weights are more stable, with the backward weights becoming smaller than their forward counterparts towards the end of training. 
\label{lrperf:fig:weight_norms}}
\end{figure*}

In this section we present some visualizations which deepen the understanding of the weight dynamics and stability during training, as presented in Section~\ref{lrperf:sec:local-learning-rules} and Section~\ref{lrperf:sec:non-local-learning-rules}. 
By looking at the weights of the network at each validation point, we are able to compare corresponding forward and backward weights (see Figure~\ref{lrperf:fig:full_weight_scatter}) as well as to measure the angle between the vectorized forward and backward weight matrices to quantify their degree of alignment (see Figure~\ref{lrperf:fig:weight_norms}). 
Their similarity in terms of scale can also be evaluated by looking at the ratio of the Frobenius norm of the backward weight matrix to the forward weight matrix, $\|B_l\|_F / \|W_l\|_F$.
Separately plotting these metrics in terms of model depth sheds some insight into how different layers behave. 

\section{Further Analysis}
\label{lrperf:sup:analysis}

\subsection{Instability of Weight Mirror}
\label{lrperf:sup:stability-analysis}

As explained in Section~\ref{lrperf:sec:local-learning-rules}, the instability of weight mirror can be understood by considering the dynamical system given by the symmetrized gradient flow on $\mathcal{R}_{\text{SA}}$, $\mathcal{R}_{\text{AA}}$, and $\mathcal{R}_{\text{WM}}$ at a given layer $l$.  
By symmetrized gradient flow we imply the gradient dynamics on the loss $\mathcal{R}$ modified such that it is symmetric in both the forward and backward weights.  
We ignore biases and non-linearities and set $\alpha = \beta$ for all three losses.

When the weights, $w_l$ and $b_l$, and input, $x_l$, are all scalar values, the gradient flow for all three losses gives rise to the dynamical system,
$$\frac{\partial}{\partial t}\begin{bmatrix}
w_l\\
b_l
\end{bmatrix} = - A \begin{bmatrix}
w_l\\
b_l
\end{bmatrix},$$
For Symmetric Alignment and Activation Alignment, $A$ is respectively the positive semidefinite matrix
$$A_{\text{SA}} = 
\begin{bmatrix}
1 & -1\\
-1 & 1
\end{bmatrix}
\quad\text{ and }\quad
A_{\text{AA}} = 
\begin{bmatrix}
x_l^2 & -x_l^2\\
-x_l^2 & x_l^2
\end{bmatrix}.$$
For weight mirror, $A$ is the symmetric indefinite matrix
$$
A_{\text{WM}} = 
\begin{bmatrix}
\lambda_{\text{WM}} & -x_l^2\\
-x_l^2 & \lambda_{\text{WM}}
\end{bmatrix}.$$
In all three cases $A$ can be diagonally decomposed by the eigenbasis
$$
\left\{u,v\right\} = \left\{\begin{bmatrix}
1\\
1
\end{bmatrix},
\begin{bmatrix}
1\\
-1
\end{bmatrix}\right\},$$
where $u$ spans the symmetric component and $v$ spans the skew-symmetric component of any realization of the weight vector $\begin{bmatrix}
w_l &
b_l
\end{bmatrix}^\intercal$.

As explained in Section~\ref{lrperf:sec:local-learning-rules}, under this basis, the dynamical system decouples into a system of ODEs governed by the eigenvalues $\lambda_u$ and $\lambda_v$ associated with $u$ and $v$. For all three learning rules, $\lambda_v > 0$ ($\lambda_v$ is respectively $1$, $x^2$, and $\lambda_{\text{WM}} + x_l^2$ for SA, AA, and weight mirror).
For SA and AA, $\lambda_u = 0$, while for weight mirror $\lambda_u = \lambda_{\text{WM}} - x_l^2$.

\subsection{Beyond Feedback Alignment}
\label{lrperf:sec:analysis-beyond_fa}

An underlying assumption of our work is that certain forms of layer-wise regularization, such as the regularization introduced by Symmetric Alignment, can actually improve the performance of feedback alignment by introducing dynamics on the backward weights. 
To understand these improvements, we build off of prior analyses of backpropagation \cite{saxe_exact_2013} and feedback alignment \cite{baldi_learning_2018}.

Consider the simplest nontrivial architecture: a two layer scalar linear network with forward weights $w_1, w_2$, and backward weight $b$. 
The network is trained with scalar data $\{x_i,y_i\}_{i=1}^n$ on the mean squared error cost function
$$\mathcal{J} = \sum_{i=1}^n\frac{1}{2n}(y_i - w_2w_1x_i)^2.$$
The gradient flow of this network gives the coupled system of differential equations on $(w_1, w_2, b)$
\begin{align}
    \label{lrperf:dw1}
    \dot{w_1} &= b(\alpha - w_2w_1\beta) \\
    \label{lrperf:dw2}
    \dot{w_2} &= w_1(\alpha - w_2w_1\beta)
\end{align}
where $\alpha = \sum_{i=1}^n \frac{y_ix_i}{n}$ and $\beta = \sum_{i=1}^n \frac{x_i^2}{n}$. 
For backpropagation the dynamics are constrained to the hyperplane $b = w_2$, while for feedback alignment the dynamics are contained on the hyperplane $b = b(0)$ given by the initialization. 
For Symmetric Alignment, an additional differential equation
\begin{equation}
\label{lrperf:db}
\dot{b} = w_2 - b,
\end{equation}
attracts all trajectories to the backpropagation hyperplane $b = w_2$. 

To understand the properties of these alignment strategies, we explore the fixed points of their flow. 
From equation (\ref{lrperf:dw1}) and (\ref{lrperf:dw2}) we see that both equations are zero on the hyperbola
$$w_2w_1 = \frac{\alpha}{\beta},$$
which is the set of minima of $\mathcal{J}$. 
From equation (\ref{lrperf:db}) we see that all fixed points of Symmetric Alignment satisfy $b = w_2$. 
Thus, all three alignment strategies have fixed points on the hyperbola of minima intersected with either the hyperplane $b = b(0)$ in the case of feedback alignment or $b = w_2$ in the case of backpropagation and Symmetric Alignment.

In addition to these non-zero fixed points, equation (\ref{lrperf:dw1}) and (\ref{lrperf:dw2}) are zero if $b$ and $w_1$ are zero respectively.  
For backpropagation and Symmetric Alignment this also implies $w_2 = 0$, however for feedback alignment $w_2$ is free to be any value. 
Thus, all three alignment strategies have rank-deficient fixed points at the origin $(0,0,0)$ and in the case of feedback alignment more generally on the hyperplane $b = w_1 = 0$.

To understand the stability of these fixed points we consider the local linearization of the vector field by computing the Jacobian matrix\footnote{In the case that the vector field is the negative gradient of a loss, as in backpropagation, then this is the negative Hessian of the loss.}
$$J = \begin{bmatrix}
\partial_{w_1}\dot{w_1} & \partial_{w_1}\dot{w_2} & \partial_{w_1}\dot{b}\\
\partial_{w_2}\dot{w_1} & \partial_{w_2}\dot{w_2} & \partial_{w_2}\dot{b}\\
\partial_{b}\dot{w_1} & \partial_{b}\dot{w_2} & \partial_{b}\dot{b}
\end{bmatrix}.$$
A source of the gradient flow is characterized by non-positive eigenvalues of $J$, a sink by non-negative eigenvalues of $J$, and a saddle by both positive and negative eigenvalues of $J$.

On the hyperbola $w_2w_1 = \frac{\alpha}{\beta}$ the Jacobian matrix for the three alignment strategies have the corresponding eigenvalues:
\begin{table}[h]
\hspace*{-0.25cm}
\centering
\begin{tabular}{@{}lccc@{}} \toprule
    & $\lambda_1$ & $\lambda_2$ & $\lambda_3$ \\ \midrule
    Backprop. & $-\left(w_1^2 + w_2^2\right)x^2$ & $0$ & \\
    Feedback & $-\left(w_1^2 + bw_2\right)x^2$ & $0$ & $0$\\
    Symmetric & $-\left(w_1^2 + w_2^2\right)x^2$ & $0$ & $-1$ \\ \bottomrule
\end{tabular}
\end{table}
Thus, for backpropagation and Symmetric Alignment, all minima of the the cost function $\mathcal{J}$ are sinks, while for feedback alignment the stability of the minima depends on the sign of $w_1^2 + bw_2$. 

From this simple example there are two major takeaways:
\begin{enumerate}
    \item All minima of the cost function $\mathcal{J}$ are sinks of the flow given by backpropagation and Symmetric Alignment, but only some minima are sinks of the flow given by feedback alignment.
    \item Backpropagation and Symmetric Alignment have the exact same critical points, but feedback alignment has a much larger space of rank-deficient critical points.
\end{enumerate}
Thus, even in this simple example it is clear that certain dynamics on the backward weights can have a stabilizing effect on feedback alignment.

\subsection{Kolen-Pollack Learning Rule}
\label{lrperf:sec:kolen-pollack}

If we consider primitives that are functions of the pseudogradients $\widetilde{\nabla}_{l}$ and $\widetilde{\nabla}_{l+1}$ in addition to the forward weight $W_l$, backward weight $B_l$, layer input $x_l$, and layer output $x_{l+1}$, then the Kolen-Pollack algorithm, originally proposed by \citet{Kolen1994backpropagation} and modified by \citet{Akrout2019}, can be understood in our framework. 

\begin{figure}
\centering
\includegraphics[width=1.0\columnwidth]{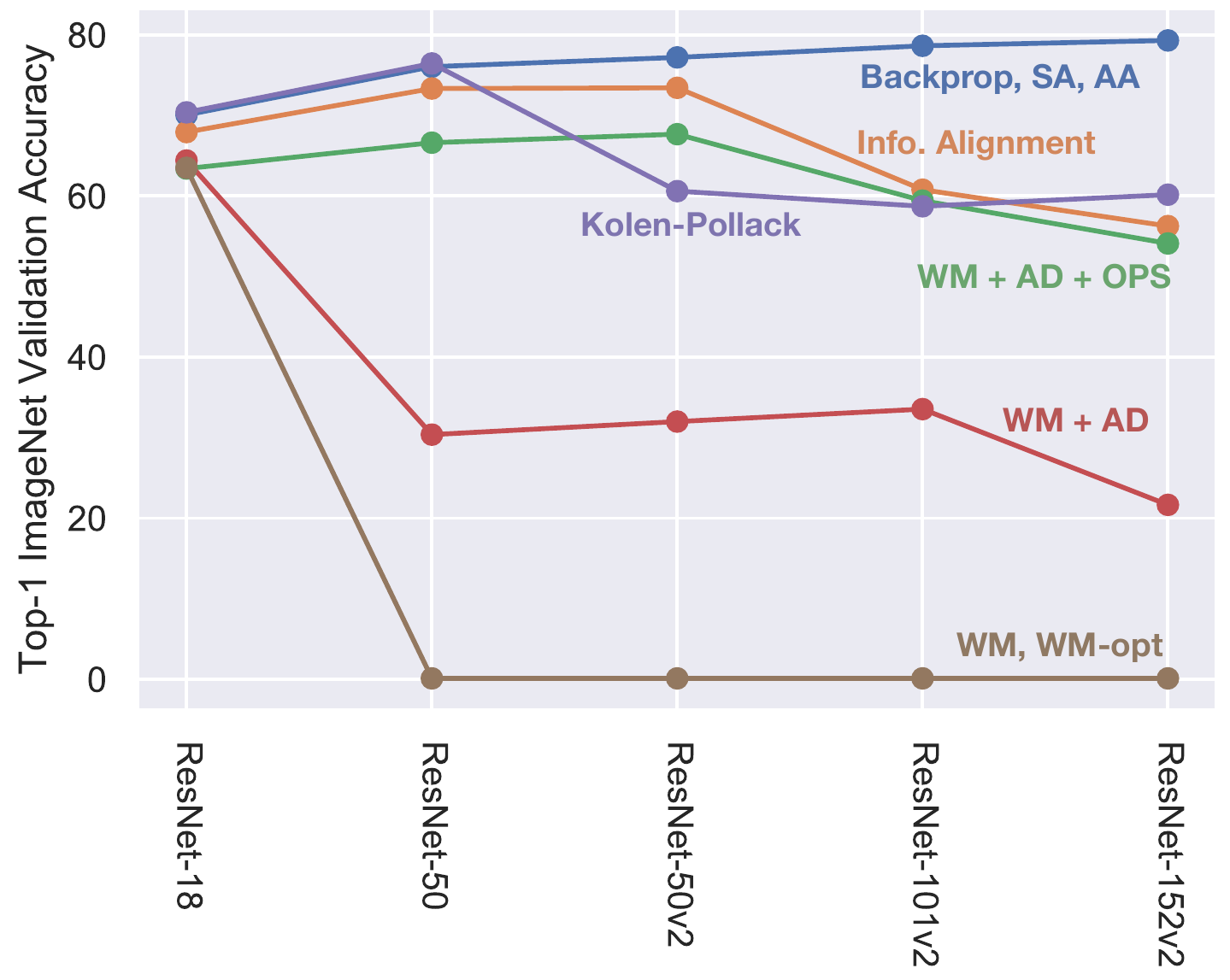}
\caption[Performance of Kolen-Pollack across architectures]{\textbf{Performance of Kolen-Pollack across architectures.} We fixed the categorical and continuous metaparameters for ResNet-18 and applied them directly to deeper and different ResNet variants (e.g. v2) as in Figure~\ref{lrperf:fig:hp-deeper}. The Kolen-Pollack learning rule, matched backpropagation performance for ResNet-18 and ResNet-50, but a performance gap emerged for different (ResNet-50v2) and deeper (ResNet-101v2, ResNet-152v2) architectures.}
\label{lrperf:fig:hp-deeper-kp}
\end{figure}

The Kolen-Pollack algorithm circumvents the weight transport problem, by instead transporting the weight updates and adding weight decay.
Specifically, the forward and backward weights are updated respectively by
\begin{align*}
    \Delta W_l &= -\eta \widetilde{\nabla}_{l+1}x_l^\intercal - \lambda_{\text{KP}} W_l,\\
    \Delta B_l &= -\eta x_l \widetilde{\nabla}_{l+1}^\intercal - \lambda_{\text{KP}} B_l,
\end{align*}
where $\eta$ is the learning rate and $\lambda_{\text{KP}}$ a weight decay constant.
The forward weight update is the standard pseudogradient update with weight decay, while the backward weight update is equivalent to gradient descent on 
$$\mathrm{tr}(x_l^\intercal B_l \widetilde{\nabla}_{l+1}) + \frac{\lambda_{\text{KP}}}{2\eta}||B_l||^2.$$
Thus, if we define the \textit{angle} primitive
$$\mathcal{P}_l^{\text{angle}} = \mathrm{tr}(x_l^\intercal B_l\widetilde{\nabla}_{l+1}) =  \mathrm{tr}(x_l^\intercal \widetilde{\nabla}_{l}),$$
then the \textbf{Kolen-Pollack (KP)} update is given by gradient descent on the layer-wise regularization function
$$\mathcal{R}_{\text{KP}} = \sum_{l \in \text{layers}} \alpha\mathcal{P}^{\text{angle}}_l + \beta\mathcal{P}^{\text{decay}}_l,$$
for $\alpha = 1$ and $\beta = \frac{\lambda_{\text{KP}}}{\eta}$.
The angle primitive encourages alignment of the forward activations with the backward pseudogradients and is local according to the criterion for locality defined in Section~\ref{lrperf:sec:framework-primitives}.
Thus, the Kolen-Pollack learning rule only involves the use of local primitives, but it does necessitate that the backward weight update given by the angle primitive is the exact transpose to the forward weight update at each step of training. 
This constraint is essential to showing theoretically how Kolen-Pollack leads to alignment of the forward and backward weights \cite{Kolen1994backpropagation}, but it is clearly as biologically suspect as exact weight symmetry.
To determine empirically how robust Kolen-Pollack is when loosening this hard constraint, we add random Gaussian noise to each update.
As shown in Figure~\ref{lrperf:fig:all-noise}, even with certain levels of noise, the Kolen-Pollack learning rule can still lead to well performing models.
This suggests that a noisy implementation of Kolen-Pollack that removes the constraint of exactness might be biologically feasible.

While Kolen-Pollack uses significantly fewer metaparameters than weight mirror or information alignment, the correct choice of these metaparameters is highly dependent on the architecture. 
As shown in Figure~\ref{lrperf:fig:hp-deeper-kp}, the Kolen-Pollack learning rule, with metaparameters specified by \citet{Akrout2019}, matched backpropagation performance for ResNet-18 and ResNet-50.
However, a considerable performance gap with backpropagation as well as our proposed learning rules (information alignment, SA, and AA) emerged for different (ResNet-50v2) and deeper (ResNet-101v2, ResNet-152v2) architectures, providing additional evidence for the necessity of the circuits we propose in maintaining robustness across architecture.

\chapter{Identifying Learning Rules}
\label{ch:lrobs}
\section{Chapter Abstract}
In the first part of this chapter, we study the problem of extracting dynamical rules from observations related to synaptic plasticity in biological data.
Cortical function relies on the balanced activation of excitatory and inhibitory neurons. 
However, little is known about the organization and dynamics of shaft excitatory synapses onto cortical inhibitory interneurons. 
Here, we use the excitatory postsynaptic marker PSD-95, fluorescently labeled at endogenous levels, as a proxy for excitatory synapses onto layer 2/3 pyramidal neurons and parvalbumin-positive (PV+) interneurons in the barrel cortex of adult mice.
Longitudinal \emph{in vivo} imaging under baseline conditions reveals that, while synaptic weights in both neuronal types are log-normally distributed, synapses onto PV+ neurons are less heterogeneous and more stable. 
Markov-model analyses suggest that the synaptic weight distribution is set intrinsically by ongoing cell type-specific dynamics, and substantial changes are due to accumulated gradual changes. 
Synaptic weight dynamics are multiplicative, i.e., changes scale with weights, although PV+ synapses also exhibit an additive component. 
These results reveal that cell type-specific processes govern cortical synaptic strengths and dynamics.

In the second part of this chapter, we study the problem of identifying learning rules in artificial neural networks to examine whether it is possible to use a different experimental observable than the synaptic strengths, which is generally difficult to measure.
It is an open question as to what specific experimental measurements would need to be made to determine whether any given learning rule is operative in a real biological system.
In this work, we take a ``virtual experimental'' approach to this problem.
Simulating idealized neuroscience experiments with artificial neural networks, we generate a large-scale dataset of learning trajectories of aggregate statistics measured in a variety of neural network architectures, loss functions, learning rule hyperparameters, and parameter initializations.
We then take a discriminative approach, training linear and simple non-linear classifiers to identify learning rules from features based on these observables.
We show that different classes of learning rules can be separated solely on the basis of aggregate statistics of the weights, activations, or instantaneous layer-wise activity changes, and that these results generalize to limited access to the trajectory and held-out architectures and learning curricula.
We identify the statistics of each observable that are most relevant for rule identification, finding that statistics from network activities across training are more robust to unit undersampling and measurement noise than those obtained from the synaptic strengths.
Our results suggest that activation patterns, available from electrophysiological recordings of post-synaptic activities on the order of several hundred units, frequently measured at wider intervals over the course of learning, may provide a good basis on which to identify learning rules.

\section{Part I: \emph{In vivo} Experimental Data}
\subsection{Introduction}
The formation, plasticity, and rewiring of synaptic connections are fundamental to brain function. Toward understanding these processes, the organization and dynamics of excitatory synapses onto cortical pyramidal (Pyr) neurons have been studied extensively using dendritic spines as a proxy~\citep{bhatt2009dendritic, grutzendler2002long, holtmaat2009experience, trachtenberg2002long}.
\emph{In vivo} two-photon imaging of morphologically identified dendritic spines has revealed several principles: their sizes are log-normally distributed and change in a multiplicative manner (i.e., the magnitude of spine size change is proportional to the spine size)~\citep{buzsaki2014log, loewenstein2011multiplicative, ziv2018synaptic}. 
Furthermore, the addition and elimination of such spiny synapses correlate with the acquisition of certain learned behaviors~\citep{hayashi2015labelling, hofer2009experience, holtmaat2009experience, johnson2016rule, xu2009rapid, zuo2005development}. 
These observations are foundational to our current understanding of brain function~\citep{buzsaki2014log}.
\emph{In vivo} imaging of spines, however, is limited in several aspects. 
First, the majority of excitatory synapses onto cortical inhibitory interneurons, such as parvalbumin-positive (PV+) interneurons, are shaft synapses~\citep{harris2015neocortical,huang2014toward,kim2017brain,lee2012activation} which cannot be identified using morphology under light microscopy~\citep{goldberg2003calcium, keck2011loss, sancho2018functional}. 
Even within pyramidal neurons, not all excitatory synapses reside on dendritic spines~\citep{knott2006spine, santuy2018volume}. 
Little is known about the distribution and dynamics of these synapses, particularly \emph{\emph{in vivo}}. 
Second, spines that protrude axially are difficult to distinguish under light-microscopy. 
Third, while spine size is correlated with synaptic size and strength, such correlations are not perfect~\citep{fortin2014live, harris1989dendritic, matsuzaki2001dendritic}.
To overcome these limitations, several studies have overexpressed an essential postsynaptic scaffolding protein at excitatory synapses, PSD-95, tagged with a fluorescent protein (FP), to visualize synapses \emph{in vivo}~\citep{cane2014relationship, gray2006rapid, sheng2011postsynaptic, villa2016inhibitory}.
However, the overexpression of PSD-95 results in aberrant synaptic function, and defective synaptic plasticity~\citep{beique2003psd, ehrlich2004postsynaptic, elias2008differential, schnell2002direct}.
Herein, we used a conditional knock-in strategy called endogenous labeling via exon duplication (ENABLED) to label PSD-95 at its endogenous levels with the yellow FP mVenus~\citep{fortin2014live} in layer 2/3 (L23) cortical spiny pyramidal neurons and aspiny PV+ inhibitory interneurons. 
We found that PSD-95 abundance provides an accurate assessment of synaptic weight in both cell types. 
Chronic \emph{in vivo} two-photon imaging in the mouse barrel cortex revealed that the distribution and dynamics of excitatory synapses were cell type-specific. 
Although excitatory synapses onto both neuronal types exhibited log-normal weight distributions, those onto PV+ dendrites were less heterogeneous and, on average, contained lower-levels of PSD-95 than those onto pyramidal neurons. 
PV+ synapses were also more stable than their pyramidal counterparts. 
Furthermore, although synaptic weight changes were largely multiplicative, changes in PV+ neurons also exhibited a significant additive, weight-independent component. 
Notably, our results indicated that large, rapid changes in synaptic weights were rare, and substantial changes were primarily the accumulation of incremental changes over time. 
These results document the organization and dynamic properties of shaft excitatory synapses onto inhibitory interneurons \emph{in vivo}, and reveal that cell type-specific processes govern cortical synaptic strengths and dynamics.

\subsection{Results}
\subsubsection{Synaptic weight dynamics are cell type-specific}
We examined the evolution of synaptic weights over time, which has been less characterized \emph{in vivo}. 
To do so, we pooled data of integrated PSD-95mVenus fluorescence from different dendrites and days together by normalizing individual synapses to the average fluorescence intensity of synapses within 40 – 60th percentile of the corresponding dendrite and time point.
This normalization was necessary to correct for inevitable variations in imaging conditions across dendrites, animals, and days, and appeared more robust than an alternative approach that normalized mVenus intensity to the cytosolic reporter tdTomato due to the differential bleaching of PSD-95mVenus and tdTomato.

\begin{figure}
  \centering
  \includegraphics[width=1.0\columnwidth]{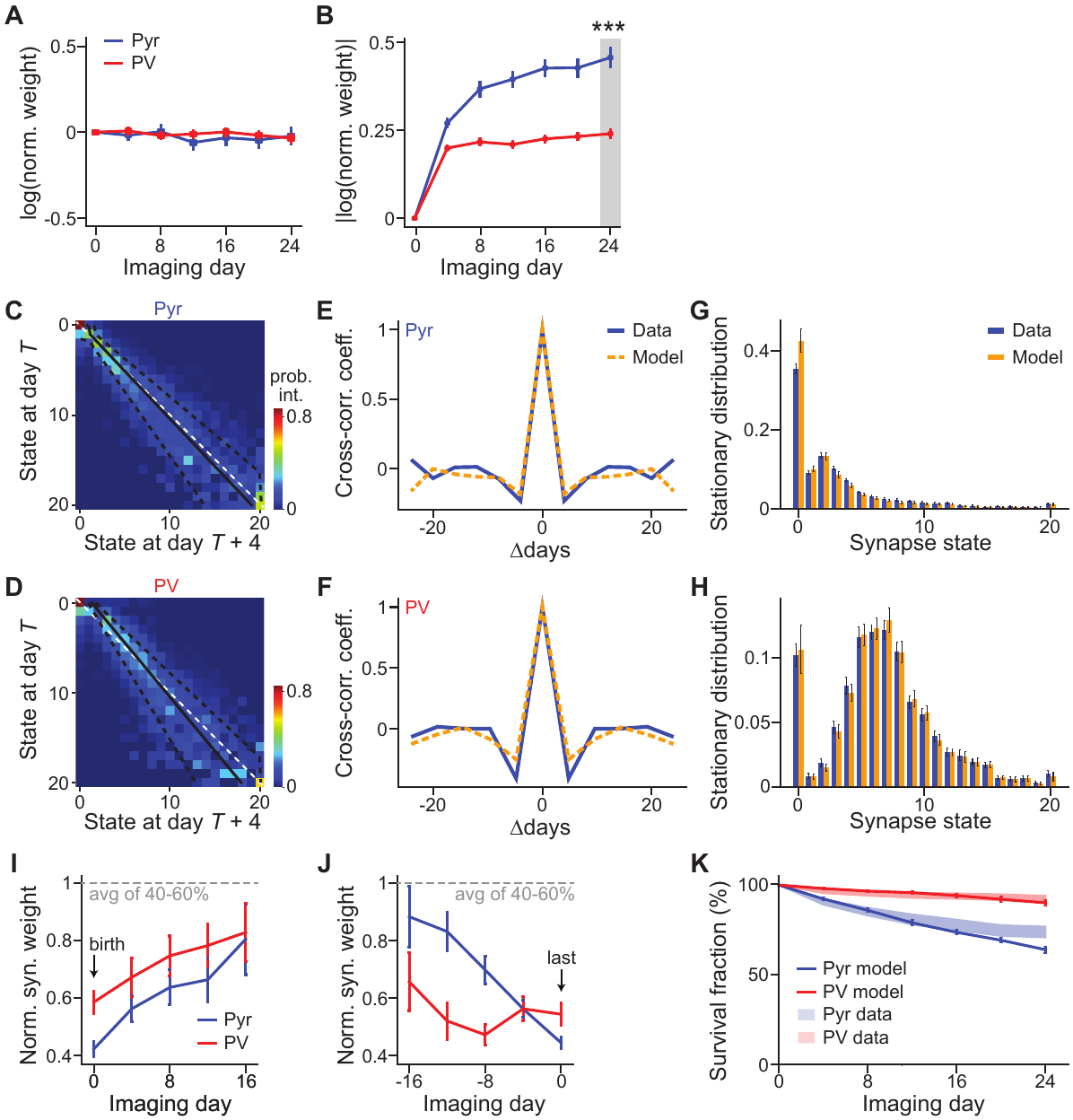}
  \caption[Markov-chain model reveals cell type-specific principles of synaptic weight dynamics]{\textbf{Markov-chain model reveals cell type-specific principles of synaptic weight dynamics.}
  \textbf{(a, b)} Average of signed \textbf{(a)} and absolute \textbf{(b)} synaptic weight changes relative to day 1 in the natural log scale. $n$ = 174 for Pyr and 190 for PV+.
  \textbf{(c, d)} The Markov state transition matrices of Pyr \textbf{(c)} and PV+ \textbf{(d)} synapses. 
  The unity line (white dashed) and fitted Kesten process with its $\pm$1 standard deviation (solid and dashed black lines, respectively) are superimposed. The Markov process was fit using 678 and 451 high-quality synapses for Pyr and PV+ dendrites, respectively.
  \textbf{(e, f)} Cross-correlation coefficient between synaptic weight changes from day $T$ to day $T+4$ and those from day $T+\Delta$ to day $T+\Delta+4$, averaged over all synapses and days.
  \textbf{(g, h)} Steady state distributions of synaptic weight for Pyr \textbf{(g)} and PV+ \textbf{(h)}, as computed from experimental data (orange) or as predicted by the Markov model (blue). 
  Error bars denote standard deviation across 30 bootstrap runs.
  \textbf{(i, j)} Addition-triggered \textbf{(i)} and elimination-triggered \textbf{(j)} averages of weight-trajectories, aligned to birth and death, respectively. $n$ (synapses) = 225 (Pyr birth), 271 (PV+ birth), 45 (Pyr death), and 50 (PV+ death).
  \textbf{(k)} The survival fraction as a function of time predicted by the Markovian transition model (solid lines) compared to experimental data (shades, s.e.m.) for both Pyr and PV+ synapses.
  $n$ (synapses) = 721 (Pyr model), 468 (PV+ model), 496 (Pyr data), and 416 (PV+ data).
}
 \label{synvivo:fig1}
\end{figure}

We asked whether synaptic weight changes were cell type-specific. Only synapses from the 25th to 75th percentiles were analyzed to minimize the effect of low signal-to-noise ratios associated with the smallest synapses. 
For both neuronal types, the average weight of synapses that persisted throughout all imaging time points did not change over time (Figure~\ref{synvivo:fig1}a), presumably due to balanced weight increases and decreases across synapses. 
Indeed, when the absolute weight changes were analyzed (Figure~\ref{synvivo:fig1}b), the weights of individual synapses deviated from their original weight substantially. 
The degree of weight change, however, was cell type-specific, with Pyr synapses exhibiting greater changes than PV+ synapses (Pyr: 0.46 $\pm$ 0.03 units on the natural log scale at day 24, corresponding to ~58\% changes, and PV+: 0.24 $\pm$ 0.02 units, corresponding to $\sim$27\% changes; Figure~\ref{synvivo:fig1}b).

\subsubsection{Markov model of weight dynamics predicts cell type-specific stationary synaptic weight distributions}
We then modeled the synaptic dynamics as a Markovian process.
Synaptic weights were binned into 21 states, including one zero-strength state, to yield a Markov state transition matrix (Figures~\ref{synvivo:fig1}c and~\ref{synvivo:fig1}d) between adjacent time points. 
Each row denoted the probability distribution of synaptic weights on the next observation day ($T + 4$) for synapses within a weight bin on the current day ($T$). 
To test the validity of our Markov assumption, changes in synaptic weights from day $T$ to day $T+4$ were correlated with those from day $T+\Delta$ to day $T+\Delta+4$. 
For both neuronal types, the cross-correlation of synaptic weight changes between two pairs of observations days dropped sharply as the separation $\Delta$ increased (Figures~\ref{synvivo:fig1}e and~\ref{synvivo:fig1}f), both in the data and in our model. 
In other words, knowledge of synaptic weights 8 days ago did not provide more information about the current weights, as measured via linear correlation, than did knowledge 4 days ago. 
These results indicate that the underlying dynamics of synaptic strengths can be modeled as a Markovian process on our experimental time scales.

\begin{figure}
  \centering
  \includegraphics[width=1.0\columnwidth]{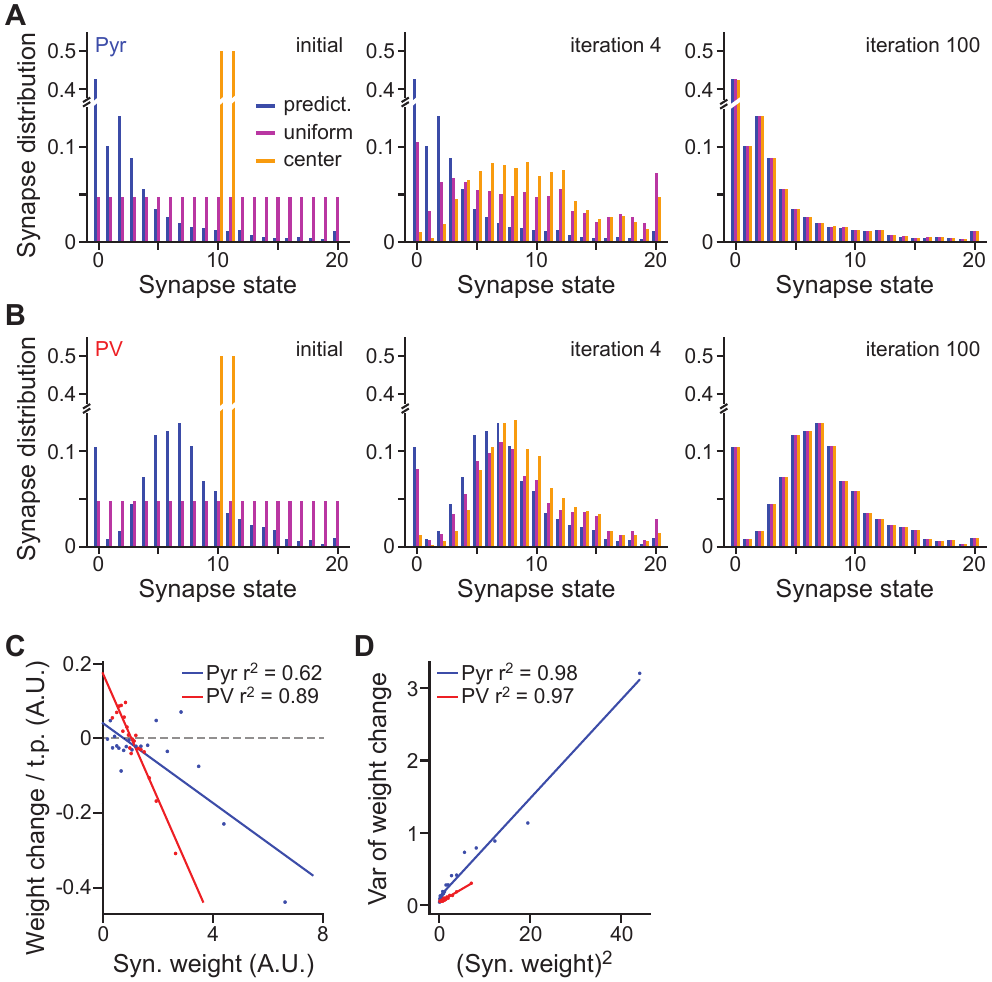}
  \caption[Synaptic weight dynamics are cell type-specific and determine the steady state synaptic weight distribution]{\textbf{Synaptic weight dynamics are cell type-specific and determine the steady state synaptic weight distribution.}
  \textbf{(a, b)} The evolution of three different starting distributions for Pyr \textbf{(a)} and PV+ \textbf{(b)} synapses by iterating through the corresponding Markovian transition matrix, as illustrated in Figures~\ref{synvivo:fig1}c and~\ref{synvivo:fig1}d. 
  Three distributions were tested for each neuronal type: 1) the predicted steady state distribution of the fit Markov chain (blue), 2) a hypothetical ``center'' distribution in which all synaptic weights are assigned to the center two states (orange), and 3) a uniform distribution across states (magenta) are included. 
  The predicted steady distributions of both neuronal type did not change significantly over iterations, but the two hypothetical distributions gradually evolved and converged to mimic the corresponding predicted steady state distribution.
  \textbf{(c)} For each cell type, synaptic strengths were binned into 20 bins with equal counts. 
  A line to the average (within bins) consecutive weight changes $w_{t+1} – w_{t}$ was fit using the average synaptic weight $\langle w_{t}\rangle$ as input (See Section~\ref{synvivo:methods:kesten} for details).
  Following the framework of \cite{ziv2018synaptic} that describes weight-dynamics as a combination of addition and weight-dependent multiplicative changes, these data suggest that excitatory synapses on PV+ dendrites have a stronger additive component in their dynamics than those on L23 pyramidal dendrites.
  \textbf{(d)} Same as \textbf{(c)}, but the variance of the consecutive weight changes $w_{t+1} – w_{t}$ was fit using the square of the average synaptic weight ${\langle w_t\rangle}^2$ as input.
}
 \label{synvivo:fig2}
\end{figure}

We initially found that synaptic weights adopted a cell type-specific log-normal distribution.
However, the underlying mechanism was poorly understood. 
Our Markov models predicted steady-state distributions of synaptic weights that were specific to the corresponding cell type. 
Indeed, when drastically different starting distributions were evolved by iterating through the experimentally-derived Markovian transition matrices, the distributions converged to the predicted stationary distribution of the corresponding neuronal type (Figures~\ref{synvivo:fig2}a and~\ref{synvivo:fig2}b). 
This predicted stationary distribution was very similar to the corresponding empirically measured distribution (Figures~\ref{synvivo:fig1}g and~\ref{synvivo:fig1}h). 
These results suggest that the synaptic weight distribution is intrinsically determined by the dynamics of synapses onto each neuronal type.

\subsubsection{Gradual accumulative changes predict synapse addition and elimination}
A noticeable feature of the Markov transition matrices was that synaptic changes were gradual, with very few large jumps, as evidenced by the concentration of higher transition probabilities near the diagonal (Figures~\ref{synvivo:fig1}c and~\ref{synvivo:fig1}d). 
One prediction of such Markovian dynamics is that synapses are both born into and die in states of weak strength. 
To test this prediction, we aligned synaptic weight dynamics by their addition or elimination day. 
For both cell types, new synapses were added with low synaptic weights, with those that survived gradually gaining strength over weeks (Figure~\ref{synvivo:fig1}i). 
At the same time, synapses were eliminated from a low synaptic weight state (Figure~\ref{synvivo:fig1}j). 
These results suggest that cortical synaptic changes, if they happen, are gradual, and large changes (e.g., the formation or loss of a strong synapse) arise primarily through the accumulation of multiple small events over the course of weeks.

To further test the ability of the Markov model to make predictions about synapse elimination over a prolonged period of time, we examined whether it could predict the survival function of synapses.
Iterations through the model over multiple observation periods resulted in a survival function well matched with the experimental data for each neuronal type (Figure~\ref{synvivo:fig1}k). 
Overall, these and above results indicate that our Markovian synaptic dynamics model accurately predicts fundamental cell type-specific properties that emerge over a month-long time scale.

\subsubsection{Synaptic weight changes are not always solely multiplicative}
Our data provides the opportunity to test two opposing models of synaptic weight dynamics~\citep{loewenstein2011multiplicative}. 
In the multiplicative model, unobserved synaptic activity leads to a synaptic change that is proportional to the synaptic weight. 
In contrast, the additive model hypothesizes that the magnitude of synaptic change is independent of the current weight. 
The Markov state transition matrix of both Pyr and PV+ dendrites exhibited an increasingly-widened central band as the previous day’s synaptic weights increased (Figures~\ref{synvivo:fig1}c and~\ref{synvivo:fig1}d). 
This indicates that the average magnitude of synaptic change grows with existing synaptic strength, suggesting a multiplicative component in both neuronal types. 
This conclusion is further supported by fitting the synaptic weights with a more constrained Kesten process consisting of both additive and multiplicative components~\citep{hazan2020activity,statman2014synaptic,ziv2018synaptic}.
Under such a framework, the intercept of the linear-fit to weight changes ($w_{t+1} – w_t$) would indicate an additive component, whereas a linear correlation between the variance of synaptic weight changes with the square of the current synaptic weight indicates a multiplicative component. 
The fit Kesten process produced a widening central band structure with increasing state (overlaid black lines in Figures~\ref{synvivo:fig2}c and~\ref{synvivo:fig2}d), indicating consistency between the Kesten and the Markov transition models. 
Synapses onto both cell types have a clear multiplicative component (Figure~\ref{synvivo:fig2}d). 
Interestingly, the Kesten fit also revealed a significant additive component for PV+ synapses, as evidenced by deviation from the diagonal line in Figure~\ref{synvivo:fig1}d and the positive $y$-intercept in Figure~\ref{synvivo:fig2}c. Furthermore, removing the additive component from the Kesten model resulted in a poor fit for PV+ synapses ($r^2$ = 0.03 and 0.89 for without and with the additive component, respectively), but only moderately affected the fit for Pyr synapses ($r^2$ = 0.53 and 0.62 for without and with the additive component, respectively). 
Together, these analyses suggest that, although excitatory synaptic weight changes are multiplicative, synapses onto PV+ dendrites also have a significant additive component.

\subsection{Discussion}
Here, we examined the \emph{in vivo} dynamics of an understudied cortical synapse type, excitatory shaft synapses onto inhibitory interneurons, and compared them to L23 Pyr synapses by using endogenously labeled PSD-95mVenus as a marker for excitatory synapses. 
We found that synaptic organization and plasticity are cell type-specific. 
Furthermore, we demonstrated several principles of synaptic dynamics that may have implications for both the experimental and computational understanding of synaptic plasticity. 
First, the stationary synaptic weight distribution is cell type-specific and is intrinsically associated with the day-to-day synaptic dynamics. 
Second, the majority of synaptic weight changes in the cortex is incremental. 
Third, although synaptic weight changes follow multiplicative dynamics in both cell types, there is a cell type-specific additive component present only in PV+ dendrites.

Our results suggest that weight dynamics are well described by small, analog changes, and larger weight changes, including synaptic additions and eliminations, are the cumulative result of these smaller events. 
Overall, our results call for attention to such analog changes in future studies.
We found that shaft excitatory synapses onto PV+ interneurons exhibit markedly different characteristics compared to the spiny synapses onto pyramidal neurons. 
Shaft excitatory synapses onto PV+ neurons are packed at a lower density, contain a lower PSD-95 content, and exhibit a narrower range of synaptic weights. 
They are also less dynamic than Pyr synapses. These cell type-specific properties may be intrinsic to the geometric constraints of the synapse type. 
Spines effectively increase the cylindrical volume along the dendrite, allowing more synapses to be packed per unit dendritic length. Spiny synapses also enable the dendrite to sample a larger space and interact with more presynaptic partners. Regardless of the underlying mechanism, these observations add to the notion that the plasticity of cortical synapses is specific to the postsynaptic cell. The uniformity and stability of synapses onto PV+ neurons are consistent with their function as maintainers of stable excitation and inhibition (E/I) balance in the brain~\citep{antoine2019increased, xue2014equalizing,zhou2014scaling}. 
In contrast, the larger range in synaptic weights and higher turnover of Pyr synapses may allow for the rewiring of cortical circuits when needed. Interestingly, the synaptic weight distribution is determined by the day-to-day dynamics of each synapse type. This may be the result of underlying homeostatic mechanisms~\citep{turrigiano2012homeostatic} and contribute to the cortical E/I balance.

Finally, classic models of synaptic plasticity often treat potentiation and depression events additively~\citep{gerstner1996neuronal,hopfield1982neural,song2000competitive}. 
However, recent studies~\citep{loewenstein2011multiplicative} suggest that in vivo spine size changes in L23 pyramidal neurons might be multiplicative. 
Here, by directly measuring the molecular content of the postsynaptic density and using it to assess synaptic weight, we find that that both Pyr and PV+ neurons exhibit multiplicative dynamics, suggesting that multiplicative scaling may be a general rule of synaptic plasticity. Interestingly, PV+ synapses, but not Pyr synapses, also exhibited an additive component, providing in vivo evidence that additive and multiplicative mechanisms are not mutually exclusive~\citep{ziv2018synaptic} and the degree of their co-existence is cell type-specific.


\subsection{Methods}
\label{synvivo:methods}

\subsubsection{Markov Chain Model}
\label{synvivo:methods:markov}
\underline{Fitting Procedure.} 
We binned the non-zero synaptic strengths into 40 equal-width bins, but group the last 21 bins into a single bin given that the counts were small in these last 20 bins. 
These bins constituted the ``states'' of the Markovian transition matrix, resulting in 21 total states (20 states of non-zero synaptic strength and one state reserved for a strength of 0). 
The number of bins were empirically determined. Other number of bins gave qualitatively similar results. The Markovian transition matrix was fit using a maximum likelihood estimator (MLE), in which for each pair of consecutive observation days ($t$ and $t+1$), we counted the number of times the synaptic strength changed from state $s_1$ on day $t$ to state $s_2$ on day $t+1$, for all pairs of states $s_1$ and $s_2$. 
We then normalized these counts to form a conditional probability distribution such that $\sum_{s_2}P(w_{t+1}=s_2\mid w_t=s_1) = 1$.
These transition probabilities were used as the elements of the Markov transition matrix.
Each row of this matrix indicates the initial state $s_1$ and each column indicates the final state $s_2$. 
Thus, each row is a conditional probability distribution of future state given current state and sums to 1. 
The model was fit to 678 synapses from 14 dendrites across 4 animals for Pyr neurons, and 451 synapses from 21 dendrites across 7 animals for PV+ neurons.

\underline{Stationary Distribution.} 
The stationary distribution was obtained from the right eigenvector of the Markov transition matrix with an eigenvalue of 1. This same stationary distribution can also be obtained from iterating the Markov dynamics many times from an arbitrary initial distribution. We compared this to the empirical stationary distribution fit using a maximum likelihood estimator (MLE), where the synaptic strengths were binned into the same states as used for the Markovian transition matrix, and the number of occurrences of synaptic strengths (aggregated across observation days) was counted in each bin, normalized by the total number of synapses.

To generate error bars, the original data were sampled with replacement from 30 bootstrap runs. In each bootstrap run, the data were fit with a Markov transition matrix to obtain the stationary distribution from the eigenvector (model, as described above), as well as fit by the empirical stationary distribution (experimental).

To generate Figures~\ref{synvivo:fig2}a and~\ref{synvivo:fig2}b, we started from a distribution over the 21 Markov states (including the 0 state), and then iterated the Markov chain to produce the next distribution over states.

\underline{Cross-correlation Coefficient of Weight Changes.} 
Once the Markovian transition matrix had been established, we sampled from the chain starting at the stationary distribution. 
Although the Markovian transition matrix returned discrete states at each time point, we converted those states back to continuous synaptic strengths by uniformly sampling between the minimum and maximum synaptic strength values within that state bin. 
This procedure resulted in a continuous valued trajectory of synaptic strengths $S$ for each synapse in the original dataset.

We then computed the cross-correlation coefficient on both $S$ and the original dataset $W$ (both of which were matrices of size, number of synapses $N$ $\times$ total observation days $T$). 
First, the consecutive weight changes were computed in $W$, given by $v_t = w_t - w_{t-1}$, resulting in the matrix $V$, and in $S$, ${\hat{v}}_t = s_t - s_{t-1}$, resulting in the matrix $\hat{V}$. 
The cross correlation was then computed from these consecutive weight change matrices. 
For a given increment of observation time $\Delta$,
\begin{equation*}
C_f(\Delta) = \frac{1}{N}\frac{1}{T}\sum_{s=1}^N\sum_{t=1}^T f[s,t]*f[s,t+\Delta],
\end{equation*}
where $f$ can either be the matrix $V$ or the matrix $\hat{V}$. 
The ``cross-correlation coefficient'' was given by $CC_f(\Delta) = C_f(\Delta)/C_f(0)$, in order to attain a maximum value of 1.

\underline{Survival Fraction.} 
For the simulated survival fraction in Figure~\ref{synvivo:fig1}k, we generated a continuous valued trajectory of synaptic strengths $S$ from the Markovian transition matrix (as described above), but starting from the distribution of synaptic strengths from the first day to match the original data. 
The survival fraction was then computed both in the model and in the data as the fraction of synapses on each observation day that persisted from the original set of synapses on the first day. 
In the survival fraction simulations, synapses could only enter and exit the ``0'' state once to simulate synapse addition and elimination, respectively.

\subsubsection{Kesten Model}
\label{synvivo:methods:kesten}
A Kesten process is given by $w_{t+1} = \varepsilon_t\cdot w_t + \eta_t$, which can be rewritten as $w_{t+1} = w_t + (\varepsilon_t – 1)\cdot w_t + \eta_t$, where $w_t$ is the synaptic size at time $t$ and $\varepsilon_t$ and $\eta_t$ are random variables that can be drawn from any distribution (we set them to be Gaussians, as explained below). 
Set $\alpha_t = \varepsilon_t – 1$, so that
\begin{equation*}
\begin{split}
\Delta w_t &= (\varepsilon_t – 1)\cdot w_t + \eta_t \\
&= \alpha_t\cdot w_t + \eta_t.
\end{split}
\end{equation*}
We assumed that $\alpha_t$ and $\eta_t$ were random processes as they were the results of unobserved pre- and post-synaptic activities. 
We simply modeled them as Gaussians, whereby $\alpha_t \sim \mathcal{N}(a_t, b_t)$ and $\eta_t\sim \mathcal{N}(c_t, d_t)$. 
Here, $\mathcal{N}(a,b)$ denotes a Gaussian distribution with mean $a$ and variance $b$. 
As a result, 
\begin{equation*}
\Delta w_t \sim \mathcal{N}(a_t\cdot w_t + c_t, b_t\cdot w_t^2 + d_t). 
\end{equation*}
We binned the weights $w_t$ in order to get an approximate distribution for the synaptic strengths at day $t$, and then plotted the average value of the consecutive weight change $\langle \Delta w_t\rangle = a_t\cdot w_t + c_t$, and its variance $Var(\Delta w_t) = b_t\cdot w_t^2 + d_t$, versus the average (across synapses in that bin), $w_t$, and the square of this average, $w_t^2$, respectively. 
This allowed us to first determine respectively if $a_t$, $b_t$, $c_t$, and $d_t$ are indeed independent of momentary synaptic strength $w_t$ (and can therefore be treated as constants in a linear regression), as well as their corresponding values.

To perform this linear regression, we binned the (nonzero) synaptic strengths, across synapses and observation days, into 20 bins with roughly equal numbers of elements, which for synapses onto the pyramidal cell type resulted in $\sim$165 synaptic strength values per bin, and for those onto the PV cell type resulted in $\sim$125 synaptic strength values per bin. 
We then computed $\Delta w_t = w_{t+1} - w_t$, for all synaptic strength values $w_t$ in that bin, from which we further computed $\langle w_t\rangle$, ${\langle w_t\rangle}^2$, $\langle\Delta w_t\rangle$, and $Var(\Delta w_t)$ per bin. 
These four values per bin were then linearly regressed, namely $\langle \Delta w_t\rangle$ vs. $\langle w_t\rangle$ and $Var(\Delta w_t)$ vs. ${\langle w_t\rangle}^2$ -- yielding the coefficients $a$, $b$, $c$, and $d$, as described above.

\section{Part II: \emph{In silico} Experimental Data From Artificial Neural Networks}
\subsection{Introduction}
\label{lrobs:sec:intro}
One of the tenets of modern neuroscience is that the brain modifies its synaptic connections during learning to improve behavior \citep{hebb1949organization}.
However, the underlying plasticity rules that govern the process by which signals from the environment are transduced into synaptic updates are unknown.
Many proposals have been suggested, ranging from Hebbian-style mechanisms that seem biologically plausible but have not been shown to solve challenging real-world learning tasks \citep{bartunov_assessing_2018}; to backpropagation \citep{rumelhart_learning_1986}, which is effective from a learning perspective but has numerous biologically implausible elements \citep{grossberg_competitive_1987, crick_recent_1989}; to recent regularized circuit mechanisms that succeed at large-scale learning while remedying some of the implausibilities of backpropagation \citep{Akrout2019, tworoutes2020}. 

A major long-term goal of computational neuroscience will be to identify which of these routes is most supported by neuroscience data, or to convincingly identify experimental signatures that reject all of them and suggest new alternatives. 
However, along the route to this goal, it will be necessary to develop practically accessible experimental observables that can efficiently separate between hypothesized learning rules. 
In other words, what specific measurements -- of activation patterns over time, or synaptic strengths, or paired-neuron input-output relations -- would allow one to draw quantitatively tight estimates of whether the observations are more consistent with one or another specific learning rule? 
This in itself turns out to be a substantial problem, because it is difficult on purely theoretical grounds to identify which patterns of neural changes arise from given learning rules, without also knowing the overall network architecture and loss function target (if any) of the learning system.

In this work, we take a ``virtual experimental'' approach to this question, with the goal of answering whether it is even possible to generically identify which learning rule is operative in a system, across a wide range of possible learning rule types, system architectures, and loss targets; and if it is possible, which types of neural observables are most important in making such identifications.
We simulate idealized neuroscience experiments with artificial neural networks trained using different learning rules, across a variety of architectures, tasks, and associated hyperparameters.
We first demonstrate that the learning rules we consider can be reliably separated \textit{without} knowledge of the architecture or loss function, solely on the basis of the trajectories of aggregate statistics of the weights, activations, or instantaneous changes of post-synaptic activity relative to pre-synaptic input, generalizing as well to unseen architectures and training curricula.
We then inject realism into how these measurements are collected in several ways, both by allowing access to limited portions of the learning trajectory, as well as subsampling units with added measurement noise.
Overall, we find that measurements temporally spaced further apart are more robust to trajectory undersampling.
We find that aggregated statistics from recorded activities across training are most robust to unit undersampling and measurement noise, whereas measured weights (synaptic strengths) provide reliable separability as long as there is very little measurement noise but can otherwise be relatively susceptible to comparably small amounts of noise.

\subsection{Related Work}
\label{lrobs:sec:related}
\cite{lim2015inferring} infer a Hebbian-style plasticity rule from IT firing rates recorded in macaques to novel and familiar stimuli during a passive and an active viewing task.
By assuming that the plasticity rule is a separable function of pre-synaptic and post-synaptic rates acting on a non-linear recurrent rate model, their result demonstrates that one can infer the \emph{hyperparameters} of a learning rule from post-synaptic activities alone, given a specific architecture (e.g. recurrent excitatory to excitatory connections) and \emph{single class} of learning rule (e.g. Hebbian).

Since \cite{grossberg_competitive_1987} introduced the \textit{weight transport problem}, namely that backpropagation requires exact transposes to propagate errors through the network, many credit assignment strategies have proposed circumventing the problem by introducing a distinct set of feedback weights to propagate the error backwards.
Broadly speaking, these proposals fall into two groups: those that encourage symmetry between the forward and backward weights \citep{lillicrap_random_2016,nokland_direct_2016,liao_how_2016,xiao_biologically-plausible_2019,moskovitz_feedback_2018,Akrout2019,tworoutes2020}, and those that encourage preservation of information between neighboring network layers \citep{bengio_how_2014, lee_difference_2015,tworoutes2020}.

However, \cite{tworoutes2020} show that temporally-averaged post-synaptic activities from IT cortex (in adult macaques during passive viewing of images) are not sufficient to separate many different \emph{classes} of learning rules, even for a fixed architecture.
This suggests that neural data most useful for distinguishing between classes of learning rules in an \emph{in vivo} neural circuit will likely require more temporally-precise measurements during learning, but does not prescribe what quantities should be measured.

In artificial networks, every unit can be measured precisely and the ground truth learning rule is known, which may provide insight as to what experimentally measurable observables may be most useful for inferring the underlying learning rule.
In the case of single-layer linear perceptrons with Gaussian inputs, the error curves for a mean-squared error loss can be calculated exactly from the unit-averaged weight updates \citep{baldi1989neural, heskes1991learning,werfel2004learning}.
Of course, it is not certain whether this signal would be useful for multi-layered networks across multiple loss functions, architectures, and highly non-Gaussian inputs at scale.

In order to bear on future neuroscience experiments, it will be important to identify useful aggregate statistics (beyond just the mean) and how robust these observable statistics are when access to the full learning trajectory is no longer allowed, or if they are only computed from a subset of the units with some amount of measurement noise.

\subsection{Approach}
\label{lrobs:sec:approach}

\begin{figure}
    \centering
    \includegraphics[width=1.0\columnwidth]{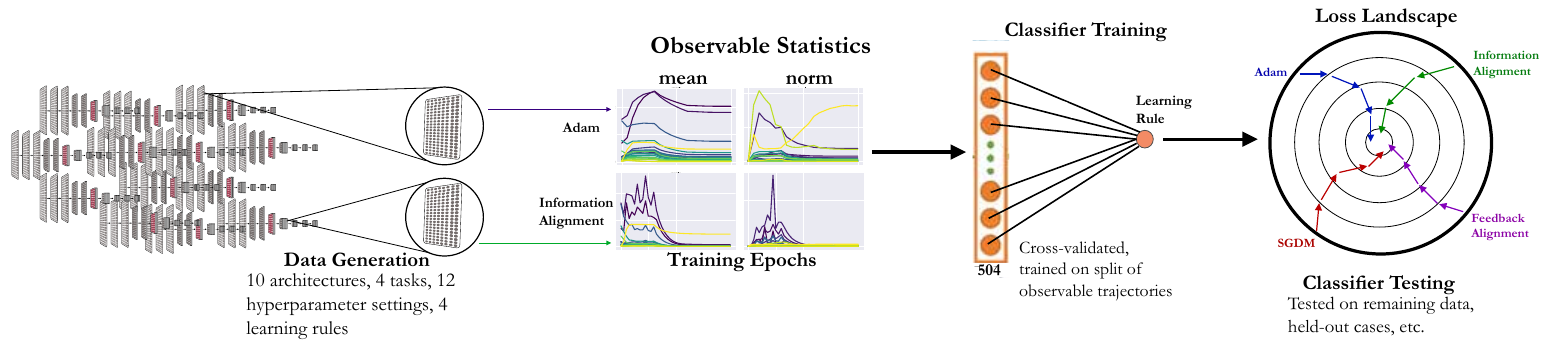}
    \caption[Observable statistic dataset generation]{\textbf{Overall approach.} Observable statistics are generated from each layer of 1,056 neural networks, through the model training process for each learning rule. Qualitatively, trajectories show patterns that are common across the observable statistics and distinctive among learning rules.
    We take a quantitative approach whereby a classifier is cross-validated and trained on a subset of these trajectories and evaluated on the remaining data.}
    \label{lrobs:fig:approachschema}
\end{figure}

\textbf{Defining features.}
The primary aim of this study is to determine the general separability of classes of learning rules. 
Figure~\ref{lrobs:fig:approachschema} indicates the general approach of doing so, and the first and second panels illustrate feature collection.
In order to determine what needs to be measured to reliably separate classes of learning rules, we begin by defining representative features that can be drawn from the course of model training.
For each layer in a model, we consider three measurements: \textbf{weights} of the layer, \textbf{activations} from the layer, and \textbf{layer-wise activity change} of a given layer's outputs relative to its inputs.
We choose artificial neural network weights to analogize to synaptic strengths in the brain, activations to analogize to post-synaptic firing rates, and layer-wise activity changes to analogize to \emph{paired} measurements that involve observing the change in post-synaptic activity with respect to changes induced by pre-synaptic input.
The latter observable is motivated by experiments that result in LTP induction via tetanic stimulation \citep{wojtowicz1985correlation}, in the limit of an infinitesimally small bin-width.

For each measure, we consider three functions applied to it: identity, absolute value, and square.
Finally, for each function of the weights and activations, we consider seven statistics: Frobenius norm, mean, variance, skew, kurtosis, median, and third quartile.
For the layer-wise activity change observable, we only use the mean statistic because instantaneous layer-wise activity change is operationalized as the gradient of outputs with respect to inputs (definable for \emph{any} learning rule, regardless of whether or not it uses a gradient of an error signal to update the weights).
The total derivative across neurons can be computed efficiently, but computing the derivative for every output neuron in a given layer is prohibitively computationally complex.
This results in a total of 45 continuous valued observable statistics for each layer, though 24 observable statistics (listed in Figure~\ref{lrobs:fig:featimp}) are ultimately used for training the classifiers, since we remove any statistic that has a divergent value during the course of model training.
We also use a ternary indicator of \textbf{layer position} in the model hierarchy: ``early'', ``middle'', or ``deep'' (represented as a one-hot categorical variable).

\textbf{Constructing a varied dataset.}
We use the elements of a trained model other than learning rule as factors of variation. 
Specifically, we vary architectures, tasks, and learning hyperparameters.
The tasks and architectures we consider are those that have been shown to produce good representations to (mostly primate) sensory neural and behavioral data \citep{yamins2014performance, kell2018task, nayebi2018task, Schrimpf2018, cadena2019deep, feather2019metamers}.
Specifically, we consider the tasks of \textbf{supervised 1000-way ImageNet categorization} \citep{Deng2009}, \textbf{\emph{self-supervised} ImageNet} in the form of the recent competitively performing ``SimCLR'' contrastive loss and associated data augmentations \citep{chen2020simple}, \textbf{supervised 794-way Word-Speaker-Noise (WSN) categorization} \citep{feather2019metamers}, and \textbf{supervised ten-way CIFAR-10 categorization} \citep{krizhevsky2010cifar}.
We consider six architectures on ImageNet, SimCLR, and WSN, namely, ResNet-18, ResNet-18v2, ResNet-34, and ResNet-34v2 \citep{He2016}, as well as Alexnet \citep{Krizhevsky2012} both without and with local response normalization (denoted as Alexnet-LRN).
On CIFAR-10, we consider four shallower networks consisting of either 4 layers with local response normalization, due to Krizhevsky \citep{krizhevsky2012cuda}, or a 5 layer variant (denoted as KNet4-LRN and KNet5-LRN, respectively), as well as without it (denoted as KNet4 and KNet5).
The latter consideration of these networks on CIFAR-10 is perhaps biologically interpretable as expanding scope to shallower \emph{non-primate} (e.g. mouse) visual systems \citep{harris2019hierarchical}.

The dependent variable is the learning rule (across hyperparameters), and we consider four: \textbf{stochastic gradient descent with momentum (SGDM)} \citep{Sutskever2013}, \textbf{Adam} \citep{kingma2014adam}, \textbf{feedback alignment (FA)} \citep{lillicrap_random_2016}, and \textbf{information alignment (IA)} \citep{tworoutes2020}.
Learning hyperparameters for each model under a given learning rule category are the Cartesian product of three settings of batch size (128, 256, and 512), two settings of the random seed for architecture initialization (referred to as ``model seed''), and two settings of the random seed for dataset order (referred to as ``dataset randomization seed'').
The base learning rate is allowed to vary depending on what is optimal for a given architecture on a particular dataset and is rescaled accordingly for each batch size setting.
All model training details can be found in Appendix~\ref{lrobs:supp:model-training}.

We select these learning rules as they span the space of commonly used variants of backpropagation (SGDM and Adam), as well as potentially more biologically-plausible ``local'' learning rules that efficiently train networks at scale to varying degrees of performance (FA and IA) and avoid weight transport.
We do not explicitly train models with more classic learning rules such as pure Hebbian learning or weight/node perturbation \citep{widrow199030,jabri1992weight}, since for the multi-layer non-linear networks that we consider here, those learning rules are either highly inefficient in terms of convergence (e.g. weight/node perturbation \citep{werfel2004learning}) or they will readily fail to train due to a lack of stabilizing mechanism (e.g. pure Hebbian learning).
However, the local learning rules rules we do consider (FA and IA) incorporate salient aspects of these algorithms in their implementation -- IA incorporates Hebbian and other stabilizing mechanisms in its updates to the feedback weights and FA employs a pseudogradient composed of random weights reminiscent of weight/node perturbation.

To gain insight into the types of observables that separate learning rules in general, we use statistics (averaged across the validation set) generated from networks trained across 1,056 experiments as a feature set on which to train simple classifiers, such as a linear classifier (SVM) and non-linear classifiers (Random Forest and a Conv1D MLP) on various train/test splits of the data.
This is illustrated in the third and fourth panels of Figure~\ref{lrobs:fig:approachschema}, with the feature and parameter selection procedure detailed in Appendices~\ref{lrobs:supp:cls-feats} and ~\ref{lrobs:supp:cls-params}.

\subsection{Results on Full Dataset}
\label{lrobs:sec:results}

In this section, we examine learning rule separability when we have access to the entire training trajectories of the models, as well as noiseless access to \emph{all} units.
This will allow us to assess whether the learning rules can be separated independent of loss function and architecture, what types of observables and their statistics are most salient for separability, and how robust these conclusions are to certain classes of input types to explore strong generalization.

\begin{figure}
    \centering
    \begin{subfigure}{0.49\columnwidth}
        \centering
        \includegraphics[width=\textwidth]{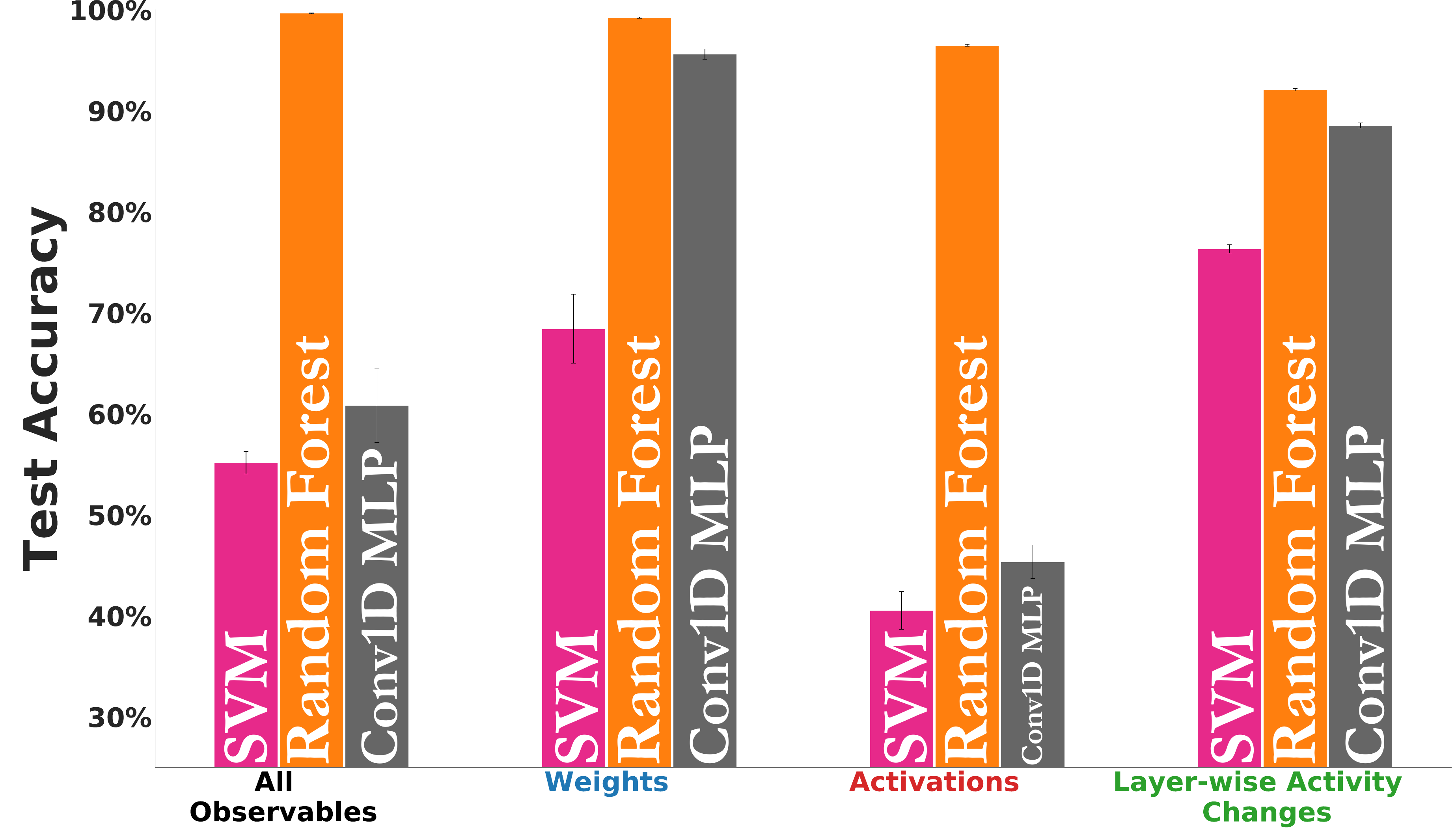}
        \caption{General separability is feasible}
    \end{subfigure}
    \begin{subfigure}{0.49\columnwidth}
        \centering
        \includegraphics[width=\textwidth]{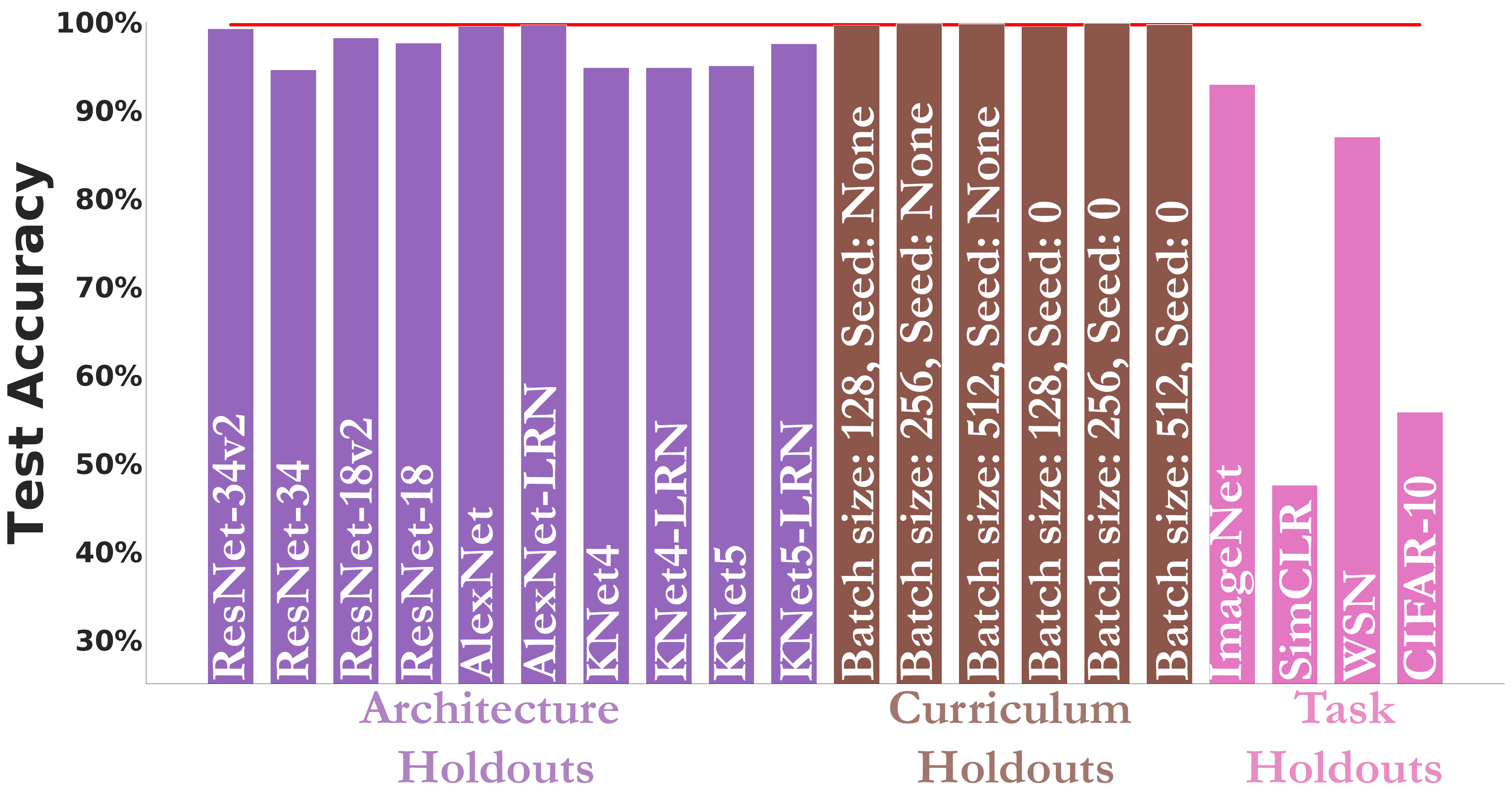}
        \caption{Heldout architecture, training curriculum, and task}
    \end{subfigure}
    \begin{subfigure}{\columnwidth}
        \centering
        \includegraphics[width=\textwidth]{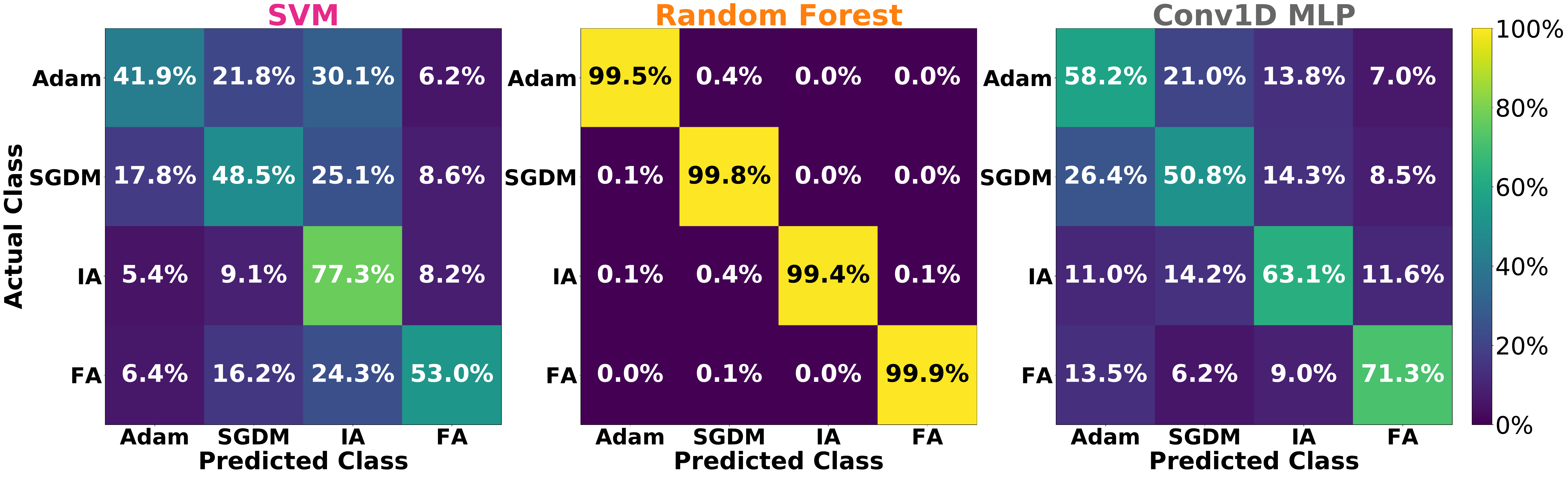}
        \caption{Differences in learning rate policy (Adam vs. SGDM) are more difficult to distinguish}
    \end{subfigure}
    \caption[Quantifying successful learning rule separation]{\textbf{Quantifying successful separation.}
    \textbf{(a)} shows the test accuracy of each classifier, with mean and s.e.m. across ten category-balanced 75\%/25\% train/test splits, using the observable measures in Section~\ref{lrobs:sec:approach}.
    \textbf{(b)} Let the red line indicate the mean and s.e.m. test accuracy of the Random Forest trained on all observable measures in (a).
    The left-most violet bars indicate Random Forest performance when holding out \emph{all} examples from each of the ten architectures we consider, for each observable measure.
    The middle brown bars indicate Random Forest performance when holding out \emph{all} examples from each of the six combinations of batch size and dataset randomization seed pair.
    The right-most pink bars indicate Random Forest performance when holding out \emph{all} examples from each of the four tasks.
    \textbf{(c)} Confusion matrices on the test set (1,296 examples per class), averaged across the ten category-balanced 75\%/25\% train/test splits, for each of the three classifiers when trained on all observable measures in (a).
    Chance performance in these settings is 25\% test accuracy.
    }
    \label{lrobs:fig:allcls_accuracy}
\end{figure}

\textbf{General separability problem is tractable. }
For each observable measure, we train the classifier on the concatenation of the trajectory of statistics identified in Section~\ref{lrobs:sec:approach}, generated from each model layer.
Already by eye (Figure~\ref{lrobs:fig:approachschema}), one can pick up distinctive differences across the learning rules for each of the training trajectories of these metrics.
Of course, this is not systematic enough to clearly judge one set of observables versus another, but provides some initial assurance that these metrics seem to capture some inherent differences in learning dynamics across rules.

As can be seen from Figure~\ref{lrobs:fig:allcls_accuracy}a, across all classes of observables, the Random Forest attains the highest test accuracy, and all observable measures perform similarly under this classifier, even when distinguishing between any given pair of learning rules (Figure~\ref{lrobs:fig:clspairs}).
The Conv1D MLP never outperforms the Random Forest, and only slightly outperforms the SVM in most cases, despite having the capability of learning additional non-linear features beyond the input feature set.

In fact, from the confusion matrices in Figure~\ref{lrobs:fig:allcls_accuracy}c, the Random Forest hardly mistakes one learning rule from any of the others.
However, when the classifiers make mistakes, they tend to confuse Adam vs. SGDM more so than IA vs. FA, suggesting that they are able to pick up more on differences, reflected in the observable statistics, due to high-dimensional direction of the gradient tensor than the magnitude of the gradient tensor (the latter directly tied to learning rate policy).

\begin{figure}
    \centering
    \includegraphics[width=\columnwidth]{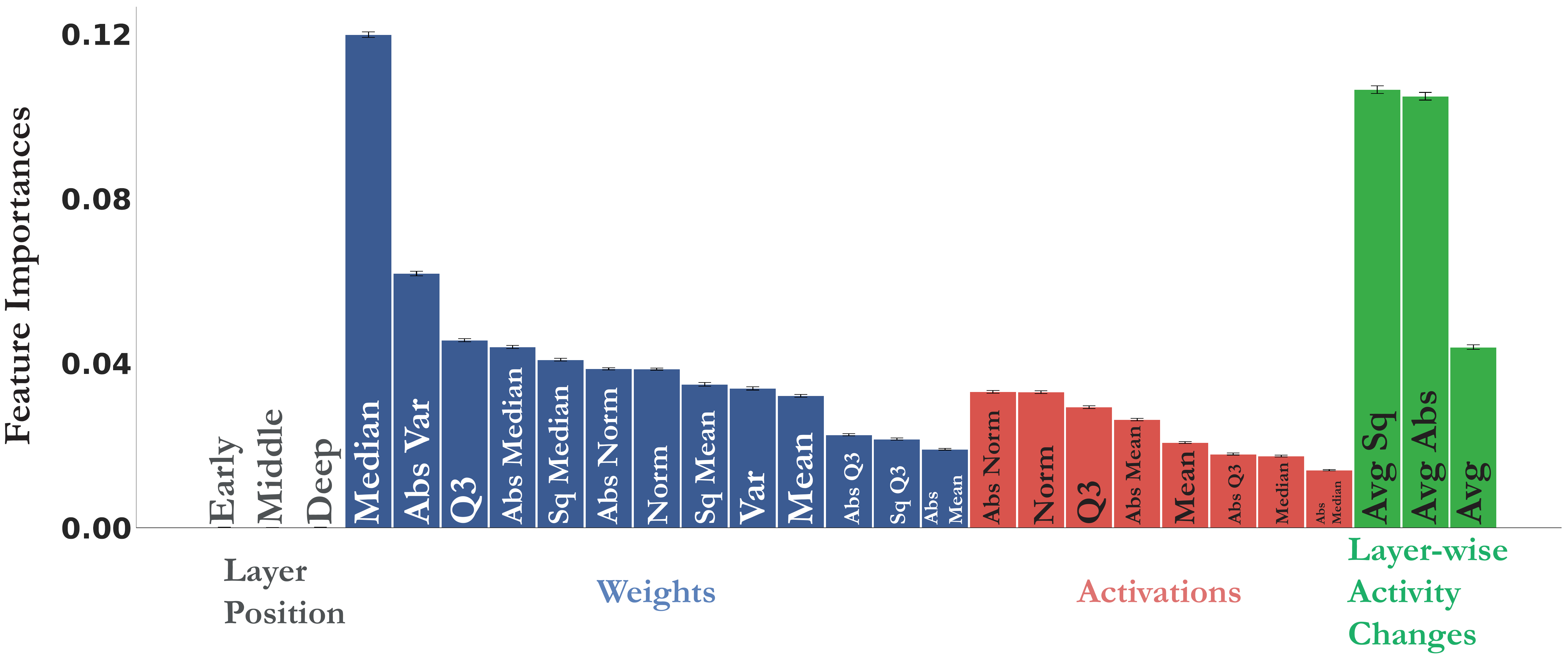}
    \caption[Relative importances of observable statistics]{\textbf{Relative importances of observable statistics.} We show the Gini impurity feature importance (summed across the learning trajectory) of each observable statistic for the Random Forest classifier trained on all observable measures in Figure~\ref{lrobs:fig:allcls_accuracy}a. 
    The colors indicate observable measures, demonstrating the prevalence and importance of observable statistics from a given measure.
    Mean and s.e.m. across trees in the Random Forest and ten category-balanced 75\%/25\% train/test splits.
    ``Sq'' and ``Abs'' indicate squaring or taking the absolute value across all units prior to computing the statistic, respectively.
    ``Q3'' is the abbreviation for the third quartile statistic.
    }
    \label{lrobs:fig:featimp}
\end{figure}

In terms of linear separability, the SVM can attain its highest test accuracy using either the weight or layer-wise activity change observables.
Interestingly, the layer-wise activity change observable has the least number of statistics (three) associated with it compared to the weight and activation observables.
The class of learning rule, however, is least able to be \emph{linearly} determined from the activation observables.
The latter observation holds as well when distinguishing between any given pair of learning rules (Figure~\ref{lrobs:fig:clspairs}).

To further gain insight into drivers of learning rule discriminability, we consider two additional sets of controls, namely that of task performance and the scale of observable statistics.
In Figure~\ref{lrobs:fig:gensep_ctrl}a, we find that where defined, task performance is a confounded indicator of separability. 
SGDM and IA are easily separated by classifiers trained on any observable measure, despite yielding similar top-1 validation accuracy on ImageNet across ResNet architectures and hyperparameters.

In Figure~\ref{lrobs:fig:gensep_ctrl}b, we train classifiers on observable statistics averaged over the whole learning trajectory.
We find that averaging the trajectory hurts performance, relative to having the full trajectory, for the (linear) SVM classifier across all classes of observables. 
This is also the case for the Random Forest, but far less for the weight observable statistics, suggesting this type of non-linear classifier has enough information from the trajectory-averaged weight statistics during the course of learning to decode the identity of the learning rule.
Overall, this result suggests that the scale of the observable statistics is not sufficient in all cases to distinguish learning rules, though it certainly contributes in part to it.

\textbf{Generalization to entire held-out classes of input types.}
The results so far were obtained across ten category-balanced splits of examples with varying tasks, model architectures, and training curricula.
However, it is not clear from the above whether our approach will generalize to \emph{unseen} architectures and training curricula.
In particular, if a new ``animal'' were added to our study, we ideally do not want to be in a position to have to completely retrain the classifier or have the conclusions change drastically, nor do we want the conclusions to be highly sensitive to training curricula (the latter analogized in this case to batch size and dataset randomization seed).

As seen in Figure~\ref{lrobs:fig:allcls_accuracy}b, the performance of the Random Forest for the held out conditions of architecture and training curriculum is not much worse than its corresponding test set performance in Figure~\ref{lrobs:fig:allcls_accuracy}a, indicated by the red line.
This relative robustness across architectures and training curricula holds even if you train on either the weight or activation observable measures individually (Figure~\ref{lrobs:fig:rfcls_weightactgrad_genvar}a,b), but not as consistently with the layer-wise activity changes (Figure~\ref{lrobs:fig:rfcls_weightactgrad_genvar}a) or with a linear classifier (Figure~\ref{lrobs:fig:svmcls_weightactgrad_genvar}a,b).

We also tested held-out task to quantify differences in learning dynamics between different tasks, although it is not necessarily reflective of a real experimental condition where task is typically fixed.
In particular, generalizing to deep models trained on ImageNet or WSN works the best, despite them being different sensory modalities.
However, generalizing from the supervised cross-entropy loss to the self-supervised SimCLR loss performs the lowest, as well as generalizing from deep architectures on ImageNet, SimCLR, and WSN to shallow architectures on CIFAR-10 (in the latter case, both task and architecture have changed from the classifier's train to test set). 
This quantifies potentially fundamental differences in learning dynamics between supervised vs. self-supervised tasks and between deep networks on large-scale tasks vs. shallow networks on small-scale tasks.

Taken together, Figure~\ref{lrobs:fig:allcls_accuracy} suggests that a simple non-linear classifier (Random Forest) can attain consistently high generalization performance when trained across a variety of tasks with either the weight or activation observable measures \emph{alone}.

\textbf{Aggregate statistics are not equally important.}
We find from the Random Forest Gini impurity feature importances for each measure's individual statistic that not all the aggregate statistics within a given observable measure appear to be equally important for separating learning rules, as displayed in Figure~\ref{lrobs:fig:featimp}.
For the weights, the median is given the most importance, even in an absolute sense across all other observable measures.
For the activations, the norm statistics are given the most importance for that measure.
For the averaged (across output units) layer-wise activity changes, the magnitude of this statistic either in terms of absolute value or square are assigned similar importances. 
On the other hand, the first three features comprising the ternary categorical ``layer position'' are assigned the lowest importance values.

Of course, we do not want to over-interpret these importance values given that Gini impurity has a tendency to bias continuous or high cardinality features.
The more computationally expensive permutation feature importance could be less susceptible to such bias.
However, these results are suggestive that for a given observable measure, there may only be a small subset of statistics most useful for learning rule separability, indicating that in a real experiment, it might not be necessary to compute very many aggregate statistics from recorded units so long as a couple of the appropriate ones for that measure (e.g. those identified from classifiers) are used.
While no theory of neural networks yet allows us to derive optimal statistics mathematically (motivating our empirical approach), ideally in the future we can combine better theory with our method to sharpen feature design.

\subsection{Access to Only Portions of the Learning Trajectory}
\label{lrobs:sec:subtraj}

The results in Section~\ref{lrobs:sec:results} were obtained with access to the entire learning trajectory of each model.
Often however, an experimentalist collects data throughout learning at regularly spaced intervals \citep{holtmaat2005transient}.
We capture this variability by randomly sampling a fixed number of samples at a fixed spacing (``subsample period'') from each trajectory, as is done in Figure~\ref{lrobs:fig:rfcls_subsample}a.
See Appendix \ref{lrobs:supp:cls-trajsubsample} for additional details on the procedure.

We find across observable measures that robustness to undersampling of the trajectory is largely dependent on the subsample period length (number of epochs between consecutive samples).
As the subsample period length increases (middle and right-most columns of Figure~\ref{lrobs:fig:rfcls_subsample}a), the Random Forest classification performance\footnote{The advantage of larger subsample period lengths holds for a linear classifier such as the SVM (Figure~\ref{lrobs:fig:svmcls_subsample}), but as expected, absolute performance is lower.} increases compared to the \emph{same} number of samples for a smaller period (left-most column).

To further demonstrate the importance of subsampling widely \emph{across} the learning trajectory, we train solely on a consecutive third of the trajectory (``early'', ``middle'', or ``late'') and separately test on each of the remaining consecutive thirds in Figure~\ref{lrobs:fig:rfcls_subsample}b.
This causes a reduction in generalization performance relative to training and testing on the \emph{same} consecutive portion (though there is some correspondence between the middle and late portions).
For a fixed observable measure, even training and testing on the same consecutive portion of the learning trajectory does not give consistent performance across portions (Figure~\ref{lrobs:fig:allcls_trajslice}).
The use of a non-linear classifier is especially crucial in this setting, as the (linear) SVM often attains chance test accuracy or close to it (Figure~\ref{lrobs:fig:svmcls_trajslicegen}).

Taken together, these results suggest that data that consists of measurements collected temporally further apart across the learning trajectory is more robust to undersampling than data collected closer together in training time.
Furthermore, across \emph{individual} observable measures, the \emph{weights} are overall the most robust to undersampling of the trajectory, but with enough frequency of samples we can achieve comparable performance with the activations.

\begin{figure}
  \centering
    \begin{subfigure}{\columnwidth}
        \centering
        \includegraphics[width=\textwidth]{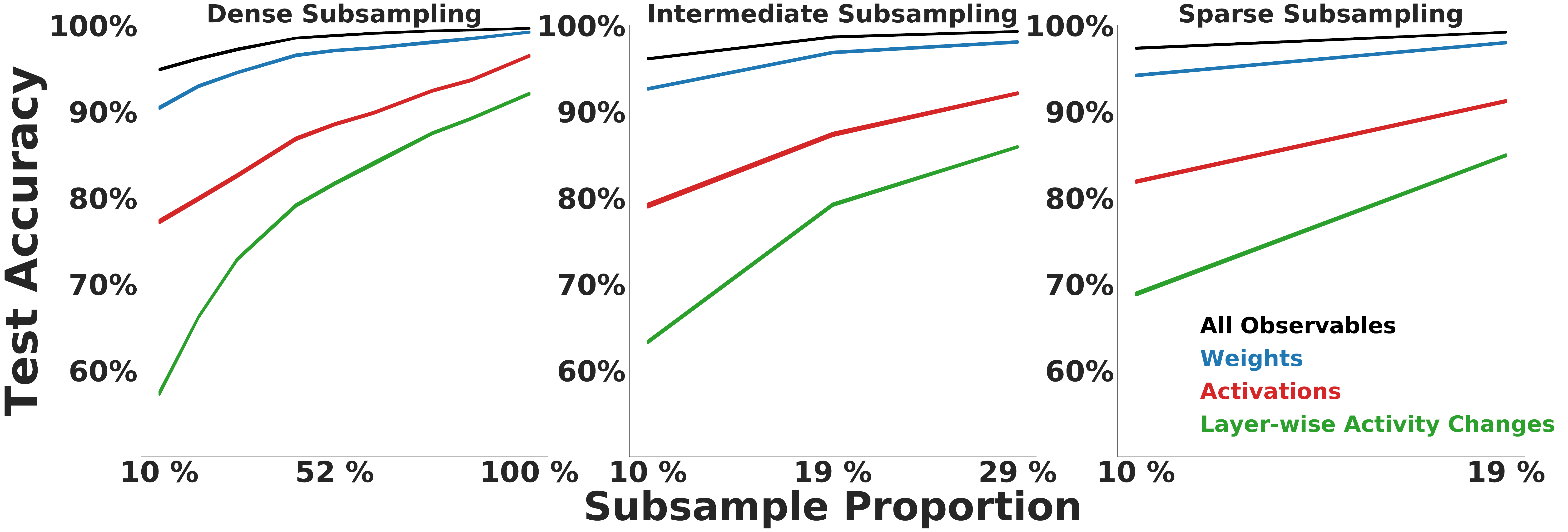}
        \caption{Sparse subsampling \emph{across} the learning trajectory is robust to trajectory undersampling.}
    \end{subfigure}
    \begin{subfigure}{\columnwidth}
        \centering
        \includegraphics[width=\textwidth]{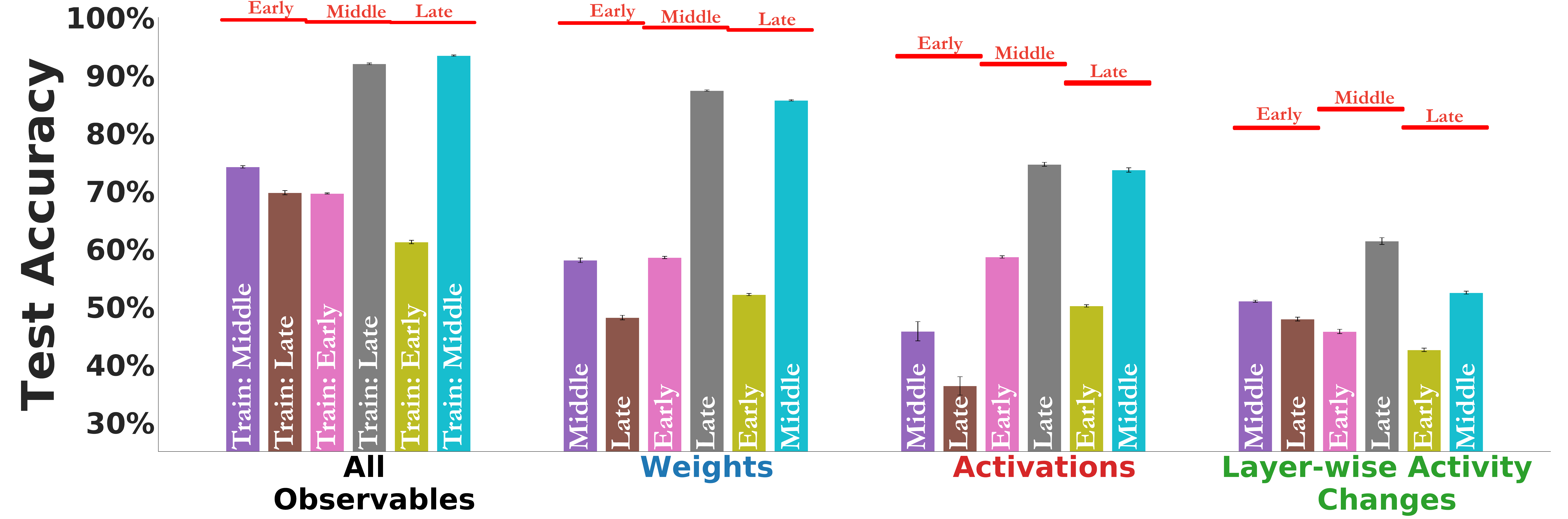}
        \caption{Training solely on consecutive portions of the learning trajectory is \emph{not} robust to trajectory undersampling.}
    \end{subfigure}
    \caption[Trajectory subsampling]{\textbf{Trajectory subsampling.} \textbf{(a)} Sparse subsampling, where the epochs between nearby trajectory samples are the furthest apart (right panel; 25 epochs apart), requires far fewer samples to achieve the same performance for the Random Forest as the full trajectory (left panel; 5 epochs apart). 
    ``Subsample Proportion $Y$\%'' refers to the number of samples chosen for the trajectory subsample relative to that of the full trajectory (21 total samples).
    \textbf{(b)} We train the Random Forest on 75\% of examples with access to only one consecutive third of the full trajectory (7 consecutive samples each out of 21 total samples), testing on the remaining 25\% of examples from one of the other two consecutive thirds. 
    Red lines denote Random Forest performance when tested on 25\% of examples from the \emph{same} portion of the trajectory as in classifier training for each observable measure, reported in the bottom right row of Figure~\ref{lrobs:fig:allcls_trajslice}.
    Mean and s.e.m. in all cases are across ten category-balanced train/test splits.
    Chance performance in these settings is 25\% test accuracy.
    \label{lrobs:fig:rfcls_subsample}}
\end{figure}

\subsection{Unit Undersampling and Measurement Noise Robustness of Observables}
\label{lrobs:sec:subnoise}

\begin{figure}
    \centering
    \includegraphics[width=1.0\columnwidth]{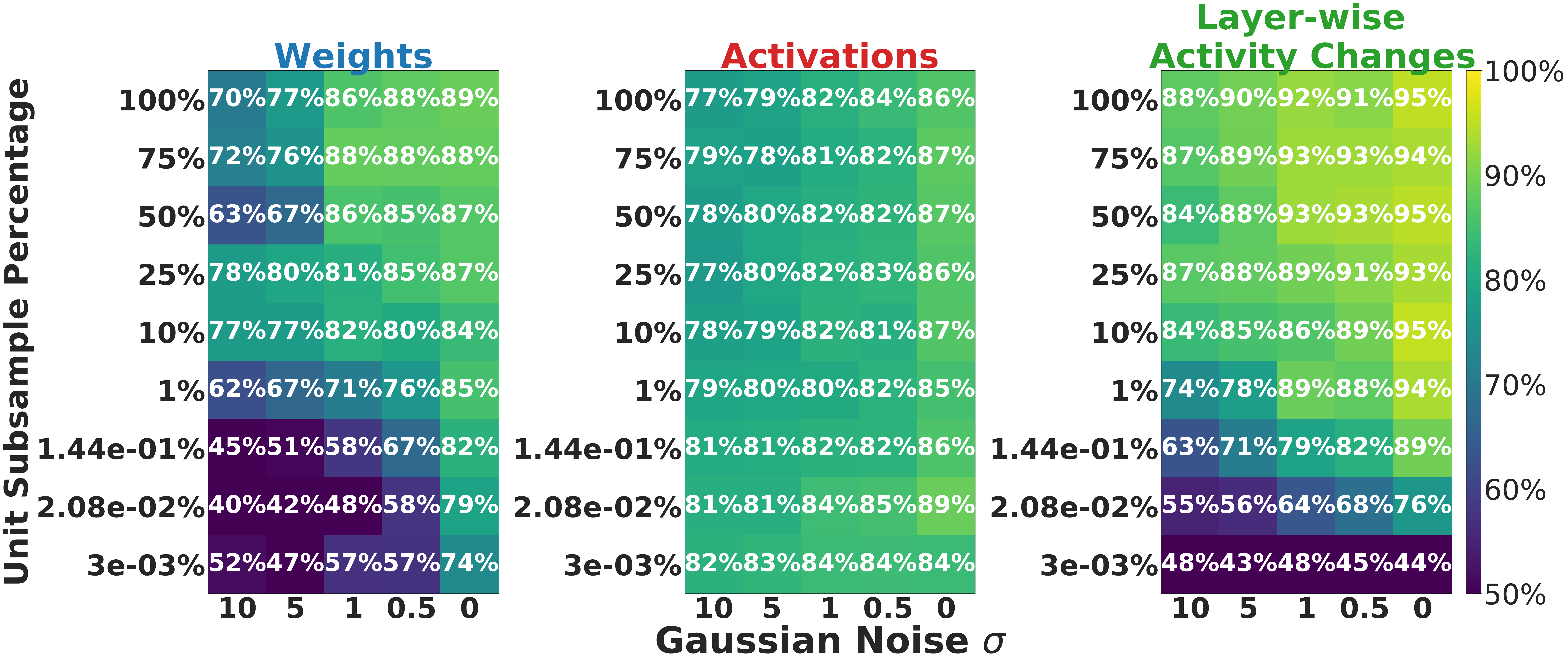}
    \caption[Activations are the most robust to measurement noise and unit undersampling]{\textbf{Activations are the most robust to measurement noise and unit undersampling.} 
    For each observable measure, the Random Forest test accuracy separating FA vs. IA on ResNet-18 across the ImageNet and SimCLR tasks, averaged across random seed and ten category-balanced 75\%/25\% train/test splits.
    The $x$-axis from the left to right of each heatmap corresponds to decreasing levels of noise, and the 
    $y$-axis from bottom to top corresponds to increasing levels of sampled units.
    Top right corner of each heatmap corresponds to the perfect information setting of Section~\ref{lrobs:sec:results}.
    Chance performance in this setting is 50\% test accuracy.}
    \label{lrobs:fig:rfcls_unitsubsamplenoise}
\end{figure}

The aggregate statistics computed from the observable measures have thus far operated under the idealistic assumption of noiseless access to every unit in the model.
However, in most datasets, there is a significant amount of unit undersampling as well as non-zero measurement noise.
How do these two factors affect learning rule identification, and in particular, how noise and subsample-robust are particular observable measures?
Addressing this question would provide insight into the types of experimental paradigms that may be most useful for identifying learning rules, and predict how certain experimental tools may fall short for given observables.

To assess this question, we consider an anatomically reasonable feedforward model of primate visual cortex, ResNet-18, trained on both supervised ImageNet and the self-supervised SimCLR tasks (known to give neurally-plausible representations \citep{nayebi2018task, Schrimpf2018, Zhuang2021}).
We train the network with FA and IA, for which there have been suggested biologically-motivated mechanisms \citep{guerguiev_towards_2017, lansdell2019spiking, tworoutes2020}.

We model measurement noise as a standard additive white Gaussian noise (AWGN) process applied independently per unit with five values of standard deviation (including the case of no measurement noise).
We subsample units from a range that is currently used by electrophysiological techniques\footnote{Based on estimates of the total neuronal density in visual cortex being on the order of hundreds of millions \citep{DiCarlo_2012}, compared to the number of units typically recorded on the order of several hundred \citep{Majaj2015, kar2019evidence}.} up to the case of measuring all of the units.
See Appendix \ref{lrobs:supp:cls-subnoise} for additional details on the procedure.

From Figure~\ref{lrobs:fig:rfcls_unitsubsamplenoise}, we find that both the weight and layer-wise activity change observables are the most susceptible to unit undersampling and noise in measurements.
The activations observable is the most robust across all noise and unit subsampling levels (regardless of classifier, see Figure~\ref{lrobs:fig:svmcls_unitsubsamplenoise} for the SVM).

This data suggests that if one collects experimental data by imaging synaptic strengths, it is still crucial that the optical imaging readout not be very noisy, since even with the amount of units typically recorded currently (e.g. several hundred to several thousand synapses, $\sim 2.08\times 10^{-2}$\% or less), a noisy imaging strategy of synaptic strengths may be rendered ineffective.
Instead, current electrophysiological techniques that measure the \emph{activities} from hundreds of units ($\sim 3\times 10^{-3}$\%) could form a good set of neural data to separate learning rules.
Recording more units with these techniques can improve learning rule separability from the activities, but it does not seem necessary, at least in this setting, to record a \emph{majority} of units to perform this separation effectively.

\subsection{Discussion}
\label{lrobs:sec:discussion}
In this work, we undertake a ``virtual experimental'' approach to the problem of identifying measurable observables that may be most salient for separating hypotheses about synaptic updates in the brain.
We have made this dataset publicly available\footnote{\url{https://github.com/neuroailab/lr-identify}}, enabling others to analyze properties of learning rules without needing to train neural networks.

Through training 1,056 neural networks at scale on both supervised and self-supervised tasks, we find that across architectures and loss functions, it is possible to reliably identify what learning rule is operating in the system only on the basis of aggregate statistics of the weights, activations, or layer-wise activity changes.
We find that with a simple non-linear classifier (Random Forest), each observable measure forms a relatively comparable feature set for predicting the learning rule category.

We next introduced experimental realism into how these observable statistics are measured, either by limiting access to the learning trajectory or by subsampling units with added noise.
We find that measurements temporally spaced further apart are more robust to trajectory undersampling across all observable measures, especially compared to those taken at consecutive portions of the trajectory.
Moreover, aggregate statistics across units of the network's activation patterns are most robust to unit undersampling and measurement noise, unlike those obtained from the synaptic strengths alone.

Taken together, our findings suggest that \emph{in vivo} electrophysiological recordings of post-synaptic activities from a neural circuit on the order of several hundred units, frequently measured at wider intervals during the course of learning, may provide a good basis on which to identify learning rules.
This approach provides a computational footing upon which future experimental data can be used to identify the underlying plasticity rules that govern learning in the brain.

\subsection{Model Training Details}
\label{lrobs:supp:model-training}
We use TensorFlow version 1.13.1 to conduct all model training experiments.
All model function code can be found here: \url{https://github.com/neuroailab/lr-identify/tree/main/tensorflow/Models}.
The names of the model layers from which we generate observable statistics can be found here: \url{https://github.com/neuroailab/lr-identify/blob/main/tensorflow/Models/model_layer_names.py}.

The v2 variants of ResNets use the pre-activation of the weight layers rather than the post-activation used in the original versions \citep{He2016}.
Furthermore, the v2 variants of ResNets apply batch normalization \citep{ioffe2015batch} and ReLU to the input \emph{prior} to the convolution, whereas the original variants apply these operations after the convolution.
We use the TF Slim architectures for these two variants provided here: \url{https://github.com/tensorflow/models/tree/master/research/slim}.

Following \cite{tworoutes2020}, we replace tied weights in backpropagation with a regularization loss on untied forward and backward weights in order to train the alternative learning rules of FA \citep{lillicrap_random_2016} and IA \citep{tworoutes2020} at scale. 
Each model layer consists of forward weights that parametrize the task objective function and the backward weights specify a descent direction, as implemented here: \url{https://github.com/neuroailab/neural-alignment}.
The total network loss is defined as the sum of the task objective function $\mathcal{J}$ and the local alignment regularization $\mathcal{R}$.
Setting $\mathcal{R}\equiv 0$ results in FA, and we use the same $\mathcal{R}$ and corresponding metaparameters for IA that were specified in \cite{tworoutes2020}.
SGDM refers to stochastic gradient descent with Nesterov momentum of 0.9 in all cases.

\subsubsection{ImageNet}
\label{lrobs:supp:imagenet}
This dataset \citep{Deng2009} consists of 1,218,167 training images and 50,000 validation images from 1,000 total categories.
The architectures trained on this task are AlexNet, AlexNet-LRN, ResNet-18, ResNet-18v2, ResNet-34, and ResNet-34v2.
All models, except for the v2 ResNets, are trained on the standard ResNet preprocessing with $224\times 224\times 3$ sized ImageNet images.
As used in the TF Slim repository linked above, the v2 ResNets are trained with Inception preprocessing, which uses larger $299\times 299\times 3$ sized images.
The ResNets used an L2 regularization of $1 \times 10^{-4}$ and the AlexNets used an L2 regularization of $5\times 10^{-4}$.

The base learning rate for SGDM and IA for the ResNet models is set to 0.125 for a batch size of 256, and linearly rescaled for the remaining batch sizes of 128 and 512.
For AlexNet and AlexNet-LRN, this base learning rate is set an order of magnitude lower to 0.0125 since it does not employ batch normalization layers.
For FA and Adam across all models, this base learning rate is 0.001 (in the case of FA, the Adam optimizer operates on the pseudogradient as is also done by \cite{tworoutes2020}).
Each base learning rate is linearly warmed up to its corresponding value for 6 epochs followed by $90\%$ decay at 30, 60, and 80 epochs, training for 100 epochs total, as prescribed by \cite{buchlovsky2019tf}.

The AlexNet model function code can be found here: \url{https://github.com/neuroailab/lr-identify/blob/main/tensorflow/Models/alexnet.py}.
The ResNet model function code for both variants can be found here: \url{https://github.com/neuroailab/lr-identify/blob/main/tensorflow/Models/resnet_model_google.py}.

\subsubsection{SimCLR}
\label{lrobs:supp:simclr}
The architectures trained on this task are the same as in Section~\ref{lrobs:supp:imagenet}, namely, AlexNet, AlexNet-LRN, ResNet-18, ResNet-18v2, ResNet-34, and ResNet-34v2.
Following \cite{chen2020simple}, for all models we use the same ImageNet preprocessing with $224\times 224\times 3$ sized images, L2 regularization of $1 \times 10^{-6}$, and two contrastive layers after dropping the final ImageNet categorization layer of the model, adapting their implementation provided here: \url{https://github.com/google-research/simclr/}.
The only difference is that given that our batch sizes are small (between 128-512), we do not use the LARS optimizer \citep{you2017large} nor do we aggregate batch examples and batch norm statistics across TPU shards, in order to keep the learning rule and model architectures comparable across the other datasets.
The learning rule hyperparameters are otherwise the same as in Section~\ref{lrobs:supp:imagenet}.

The SimCLR model function code can be found here: \url{https://github.com/neuroailab/lr-identify/blob/main/tensorflow/Models/simclr_model.py}.

\subsubsection{Word-Speaker-Noise (WSN)}
\label{lrobs:supp:audionet}
This is the word recognition task described in \cite{kell2018task}, but with an updated dataset from \cite{feather2019metamers} that consists of 793 word class labels (and a null class when there is no speech), with the inputs being cochleagrams generated from waveforms of speech segments (from the Wall Street Journal \citep{paul1992design} and Spoken Wikipedia Corpora \citep{kohn2016mining}) that are superimposed on AudioSet \citep{gemmeke2017audio} background noises.
There are 5,810,600 training cochleagrams and 369,864 test set cochleagrams, where each cochleagram is of dimensions $203\times 400 \times 1$.
The architectures trained on this task are the same as in Section~\ref{lrobs:supp:imagenet} (with the exception that their final readout layer consists of 794 output units as opposed to 1000 output units), namely, AlexNet, AlexNet-LRN, ResNet-18, ResNet-18v2, ResNet-34, and ResNet-34v2.
The learning rule hyperparameters and L2 regularizations are otherwise the same as in Section~\ref{lrobs:supp:imagenet}.

The WSN model function code can be found here: \url{https://github.com/neuroailab/lr-identify/blob/main/tensorflow/Models/audionet_model.py}.

\subsubsection{CIFAR-10}
\label{lrobs:supp:cifar10}
This dataset \citep{krizhevsky2010cifar} consists of 50,000 training images and 10,000 test set images, all of which are of dimensions $32\times 32\times 3$.
We applied standard CIFAR-10 data augmentations used by the Keras library of random height and width shift, followed by a random horizontal flip.
The architectures trained on this task are KNet4, KNet4-LRN, KNet5, and KNet5-LRN.
KNet4-LRN is due to \cite{krizhevsky2012cuda}, and the 5 layer variant KNet5 we use adds an extra convolution layer ($5\times 5$ kernel size with 128 output channels) with a ReLU followed by average pooling with a kernel size of $3\times 3$ and a stride of $2\times 2$.
The models use an L2 regularization of $1\times 10^{-5}$ for all learning rules, except for IA which used an L2 regularization of $1 \times 10^{-12}$ (L2 regularization in IA is an important metaparameter for its learning stability).

The base learning rate for SGDM and IA is set to 0.01 for a batch size of 256, and linearly rescaled for the remaining batch sizes.
For FA and Adam across all models, this base learning rate is 0.001.
Unlike the other datasets, we found the best top-1 performance by setting the base learning rate to be constant for the entire 100 epochs of training with no learning rate warmup.

The KNet model function code can be found here: \url{https://github.com/neuroailab/lr-identify/blob/main/tensorflow/Models/knet.py}.

\subsection{Classifier Training Details}
\label{lrobs:supp:cls-training}

\subsubsection{Feature Selection Procedure}
\label{lrobs:supp:cls-feats}
As there are 1,056 training experiments across all learning rules, each model is trained for 100 epochs on its own Tensor Processing Unit (TPUv2-8 and TPUv3-8), saving checkpoints every 5 epochs, resulting in 21 samples in a single trajectory.
Once all the models are trained, we generate on GPU their observable statistics (per layer) from their saved checkpoints, on the respective dataset's validation set (averaged across all validation stimuli).
The resultant dataset of generated observables therefore consists of 20,736 total examples.
For each example, an observable statistic is \emph{not} used if and only if it had a divergent value during the course of training for any task, architecture, or learning rule hyperparameter, since all network parameters and units (as well as loss function and, when applicable, categorization performance) were well defined across these.
Since each example in this resultant dataset is the concatenation of each observable statistic's trajectory, this corresponds to a total of at most 504 feature dimensions.

We additionally include the ternary indicator layer position (``early'', ``middle'', or ``deep''), determined by which third of the total number of layers for a given model each of its layers belongs to.
For SimCLR, the two contrastive head layers are always assigned to ``deep''.
This mapping for each model can be found in the \texttt{group\_layers()} function defined here: \url{https://github.com/neuroailab/lr-identify/blob/main/tensorflow/Models/model_layer_names.py}.

\subsubsection{Parameter Selection Procedure}
\label{lrobs:supp:cls-params}
We run all classifiers on ten category-balanced train/test splits, which are always (with the exception of architecture, learning curriculum, and task holdouts in Figs.~\ref{lrobs:fig:allcls_accuracy}b,~\ref{lrobs:fig:rfcls_weightactgrad_genvar},~\ref{lrobs:fig:svmcls_weightactgrad_genvar}) 75\%/25\% in proportion.
For each train/test split, we perform five-fold stratified cross-validation for each classifier's parameters.
The SVM and Random Forest use \texttt{scikit-learn} 0.20.4.

For the SVM, we use the \texttt{svm.LinearSVC} function in \texttt{scikit-learn}.
PCA can be applied to the trajectories of the continuous valued observables as a preprocessing step (namely, excluding the ternary feature of layer position), and the binary decision to use it is chosen at the cross-validation stage.
Specifically, when PCA is applied, the number of components $\in \{10, 20, 30, 40, 50, 100, 200, 300, 400, 500, 600\}$.
The strength of the regularization parameter $C \in \{1.0, 50, 500, 5\times 10^{3}, 5\times 10^{4}, 5\times 10^{5}, 5\times 10^{6}\}$.

For the Random Forest, we use the \texttt{ensemble.RandomForestClassifier} function in \texttt{scikit-learn}.
We allow for the number of trees in each forest to be $\in \{20, 50, 100, 500, 1000\}$.
We allow the number of features to consider at each split of an internal node to be $\in \{n_{feats}, \log_2(n_{feats}), \sqrt{n_{feats}}\}$, where $n_{feats}$ is the total number of input features.

For the Conv1D MLP, we use TensorFlow 1.13.1 to train a two-layer neural network that consists of a learned 1-layer 1D convolution, followed by a ReLU and (optional) pooling prior to the fully connected categorization layer.
We use the Adam optimizer \citep{kingma2014adam} to train the classifier on a separate Titan X GPU per split for 400 epochs.
The Adam learning rate is set to be $\in \{1\times 10^{-3}, 1\times 10^{-4}\}$.
The train batch size is set to be $\in \{512, 1024\}$.
The kernel size of the 1D convolution is set to be a fraction of the trajectory length $\in \{3\times 10^{-3}, 7\times 10^{-3}, 5\times 10^{-2}, 0.25, 0.5, 1.0\}$.
The strides of the 1D convolution is set to be $\in \{1, 2, 4\}$.
The number of output filters of the 1D convolution is set to be $\in \{20, 40\}$.
The binary choice of a pooling layer between the 1D convolution and fully connected categorization layer is chosen during cross-validation and if used, can be either 1D max pooling or 1D average pooling.
L2 regularization is set to be $\in \{1\times 10^{-4}, 0\}$.

The classifier code can be found in: \url{https://github.com/neuroailab/lr-identify/blob/main/fit_pipeline.py}.
The classifier cross-validation parameter ranges can be found in:
\url{https://github.com/neuroailab/lr-identify/blob/main/cls_cv_params.py}.

\subsubsection{Variability of Generalization}
\label{lrobs:supp:cls-gen}
The second aim of this study is to understand whether the approach described above generalizes to selected variations in the available data. 
We introduce two variations: (1) \textbf{trajectory subsampling}, in which data is taken for part of the model learning trajectory, and (2) \textbf{holdouts}, in which all but one ``animals'', ``curricula'', and ``tasks'' are used only during training and the remaining one is used only during testing. 
(1) investigates whether our discriminative approach is successful when only part of the model training process is available, and (2) investigates where it transfers successfully to unseen classes of input types.
These two variations address the fact that neural and behavioral data is usually \emph{not} collected throughout the developmental trajectory of the animal's entire lifespan as well as generalization (particularly) to new animals and curricula.

\subsubsection{Trajectory Subsampling}
\label{lrobs:supp:cls-trajsubsample}
Trajectory subsampling is performed to represent data taken at a low frequency, data taken for short periods of time, and data taken at different points in the training process. 
All of these are defined relative to our ``full'' model learning trajectory, which is generated from epochs $E = \{0, 5, 10,\ldots, 95, 100\}$ spanning the entire 100 epochs, where $|E| = 21$ samples since we sample every 5 epochs during model training.
We note that this is inherently subsampled, but we find that it is a good proxy for the entire model learning trajectory while avoiding the prohibitive computational complexity of generating observable statistics for each step across the 100 epochs of every model training process.

We therefore consider a subsampled dataset to be generated from epochs $E' \subset E$. 
We define three quantities to represent this limitation.
\textbf{Subsample start position} is the epoch in the model learning trajectory from which the subsampled dataset starts.
\textbf{Subsample period} is the space between two consecutive samples of the learning trajectory (in units of model training epochs).
We consider three subsample periods of 5 epochs (the original value), 15 epochs, and 25 epochs, which correspond to ``Dense Subsampling'', ``Intermediate Subsampling'', and ``Sparse Subsampling'' in Figure~\ref{lrobs:fig:rfcls_subsample}a, respectively.
Finally, \textbf{subsample proportion} is the proportion of the number of samples chosen for the trajectory subsample relative to that of the overall model learning trajectory (21 total samples).
In Figure~\ref{lrobs:fig:rfcls_subsample}a, we represent this as a percentage of the total trajectory, but the number of samples this corresponds to in the ``Dense Subsampling'' regime is 2, 4, 6, 9, 11, 13, 16, 18, and 21 samples (the latter is the full trajectory); in the ``Intermediate Subsampling'' regime it is 2, 4, and 6 samples; and in the ``Sparse Subsampling'' regime it is 2 and 4 samples.  
We use a different \emph{single} random seed (corresponding to a potentially different subsample start position) for each example to keep the size of the dataset the same (in terms of number of examples) as when the full model learning trajectory is present.

\subsubsection{Holdouts}
\label{lrobs:supp:cls-holdouts}
We hold out certain parts of the dataset from training and use them for testing to understand the transfer capability of a classifier that can separate certain learning rules when encountering new learning rules. 
We start by considering two types of holdouts: \textbf{animals} and \textbf{curricula}. 
We consider an ``animal'' to be a certain architecture (one of ten) and a ``curriculum'' to be a certain combination of batch size and dataset randomization seed (one of six), and hold them out by using the remaining animals or curricula in the classifier training set and the held-out one in the classifier testing set. 
Specifically, for each of the ten architectures, we train the classifier on the data for the other nine architectures and test on the remaining one.
We do the same for the six batch size/dataset randomization seed combinations.
We also consider \textbf{task} holdouts to get a sense of the differences in learning dynamics across different tasks, holding out all examples from each one of the four tasks.
We then measure the test accuracy for each holdout.

\subsubsection{Realistic Data Measurements: Unit Subsampling and Noise Robustness}
\label{lrobs:supp:cls-subnoise}
Thus far we have assumed that whenever data is collected, it can be collected perfectly: no noise, no loss of neurons. 
In the final aim of this study, we address how our approach handles more realistic situations under which neuroscience data is collected, we analyze the noise robustness and sample efficiency of the observable statistics. 
We consider the ResNet-18 architecture trained on supervised ImageNet and self-supervised SimCLR with either the FA or IA learning rules, using a batch size of 256, model seed of None, and dataset randomization seed of None.
For noise robustness, we add Gaussian noise ($\sigma = 0, 0.5, 1, 5, 10$) to the (forward) weight/activation/layer-wise activity change matrix of each layer before generating observable statistics. 
For sample efficiency, we randomly subsample the units from the matrix at rates of $3\times 10^{-3}$\%, $2.08\times 10^{-2}$\%, $1.44\times 10^{-1}$\%, 1\%, 10\%, 25\%, 50\%, 75\%, and 100\%. 
We consider test accuracy for the Cartesian product of these two axes.
For each Cartesian product, we use the \emph{same} random seed for unit subsampling and Gaussian noise.
The number of random seeds depends on the unit subsample fraction based on how many are needed until test performance stabilizes when averaged across random seeds and ten category-balanced 75\%/25\% train/test splits.
Thus, smaller unit subsample fractions will require more random seeds (and are computationally more efficient to generate data from) than larger fractions, which will require fewer.
Specifically, we use six random seeds for $3\times 10^{-3}$\%, four random seeds each for $2.08\times 10^{-2}$\% and $1.44\times 10^{-1}$\%, three random seeds for 1\%, and two random seeds for each of the remaining unit subsample fractions.

\subsection{Supplementary Figures}

\begin{figure}[h]
  \centering
    \begin{subfigure}{0.49\columnwidth}
        \centering
        \includegraphics[width=\textwidth]{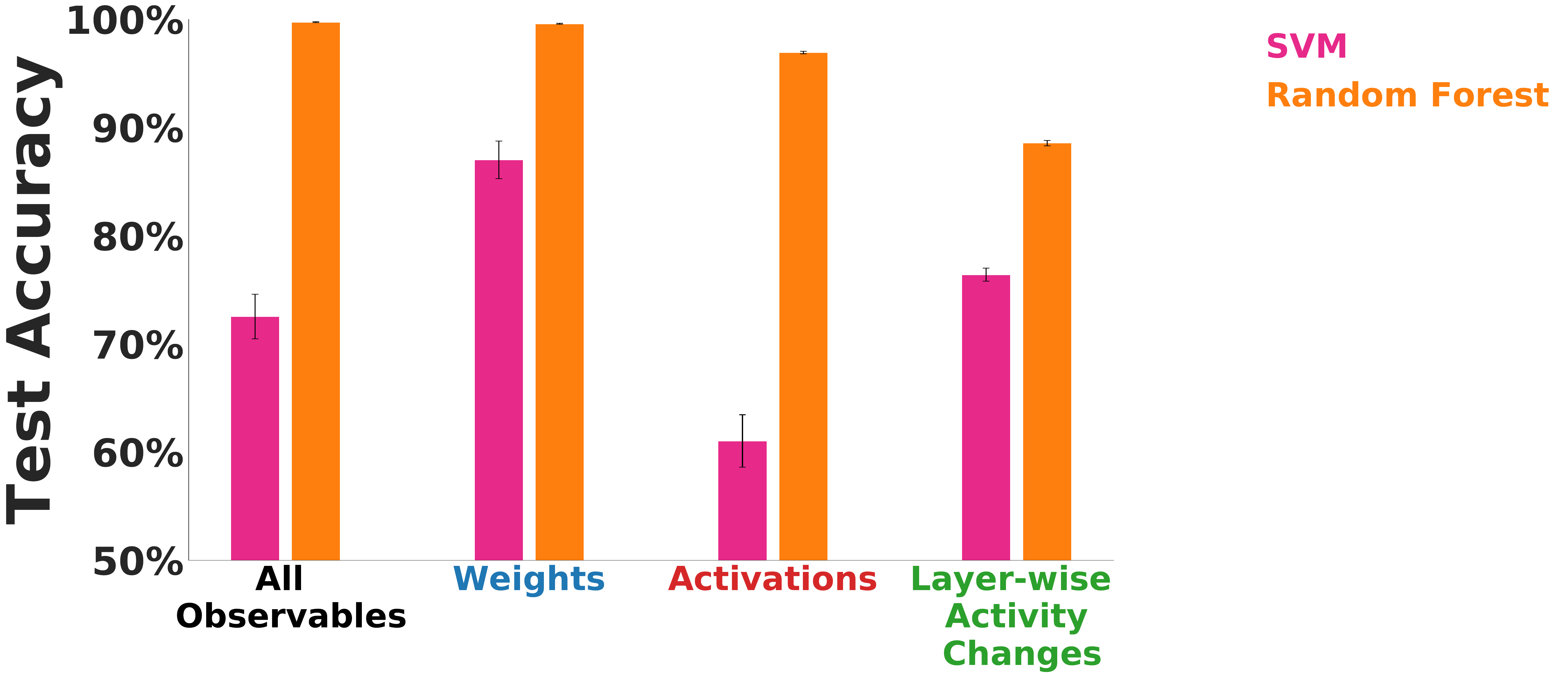}
        \caption{Adam vs. SGDM}
    \end{subfigure}
    \begin{subfigure}{0.49\columnwidth}
        \centering
        \includegraphics[width=\textwidth]{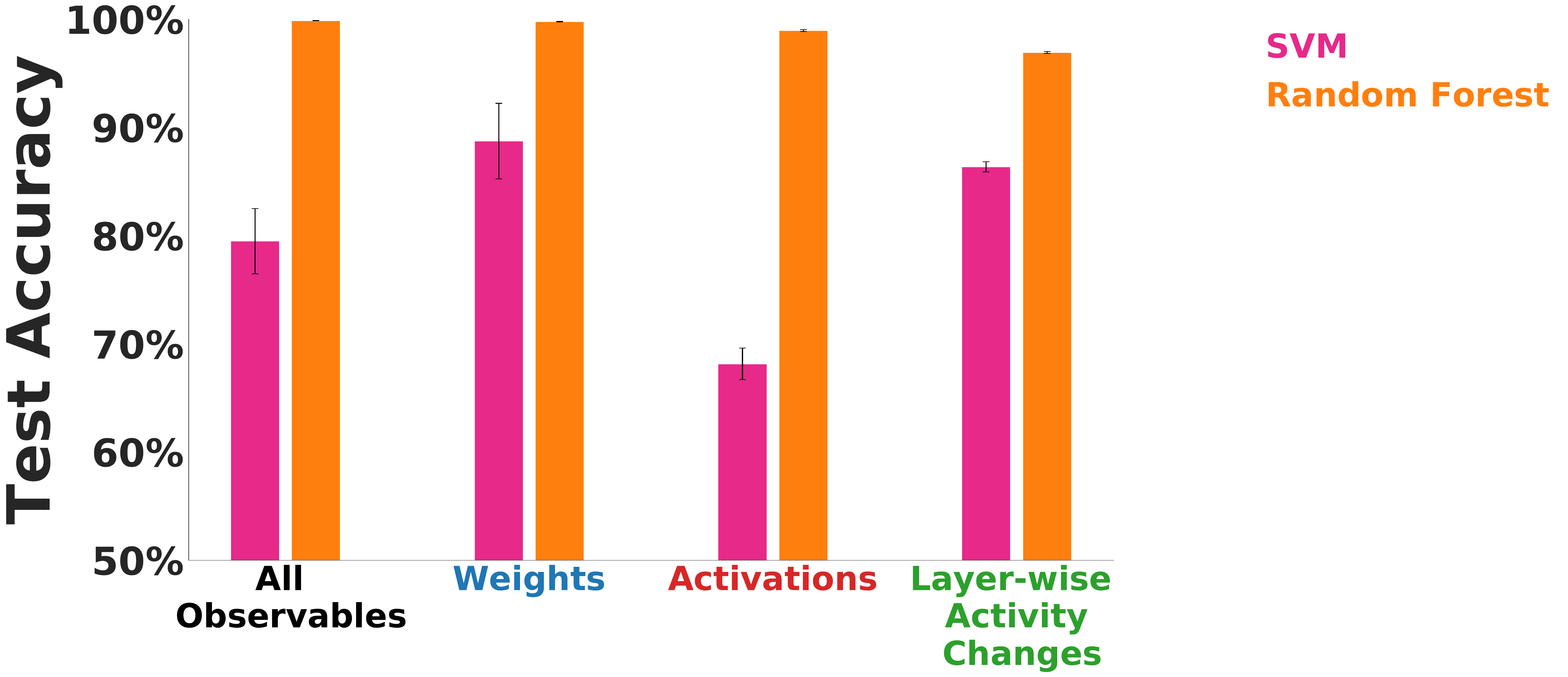}
        \caption{Adam vs. IA}
    \end{subfigure}
    \begin{subfigure}{0.49\columnwidth}
        \centering
        \includegraphics[width=\textwidth]{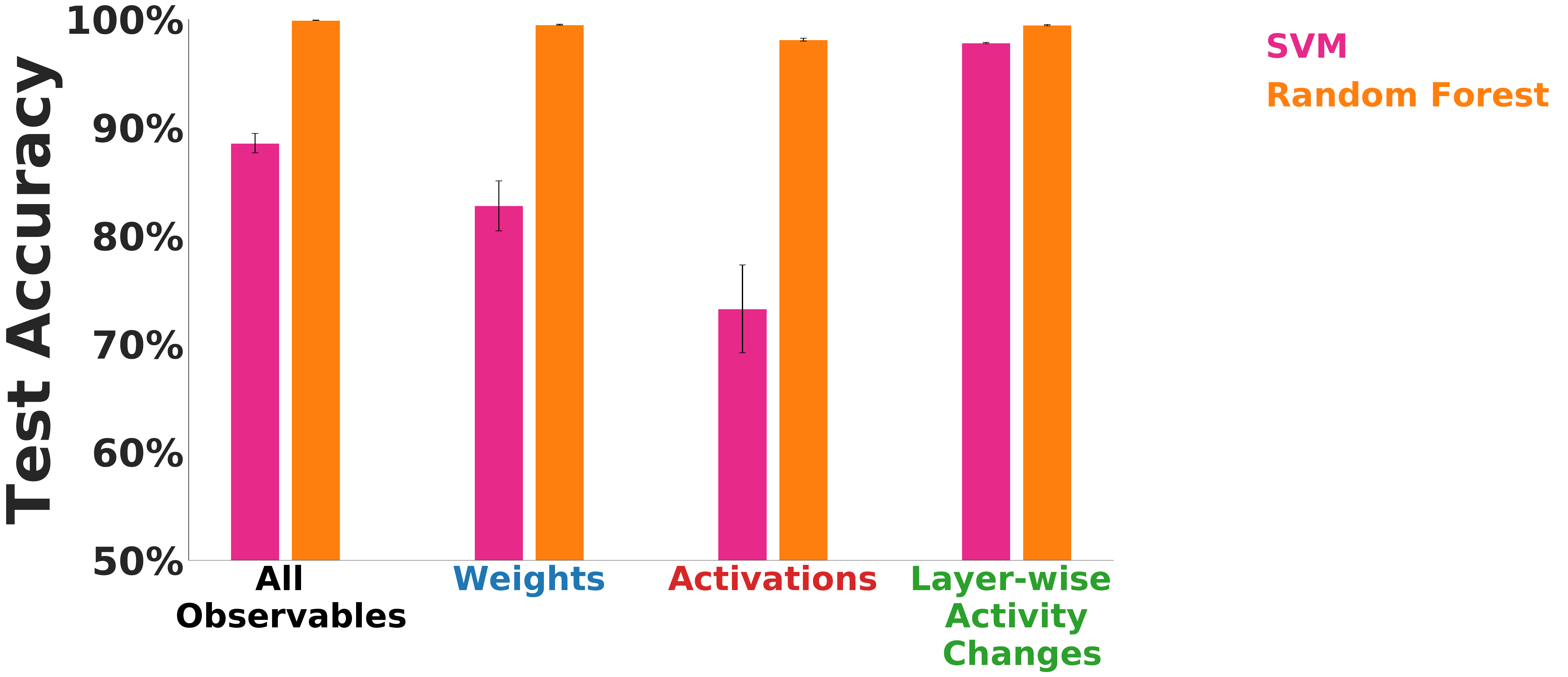}
        \caption{Adam vs. FA}
    \end{subfigure}
    \begin{subfigure}{0.49\columnwidth}
        \centering
        \includegraphics[width=\textwidth]{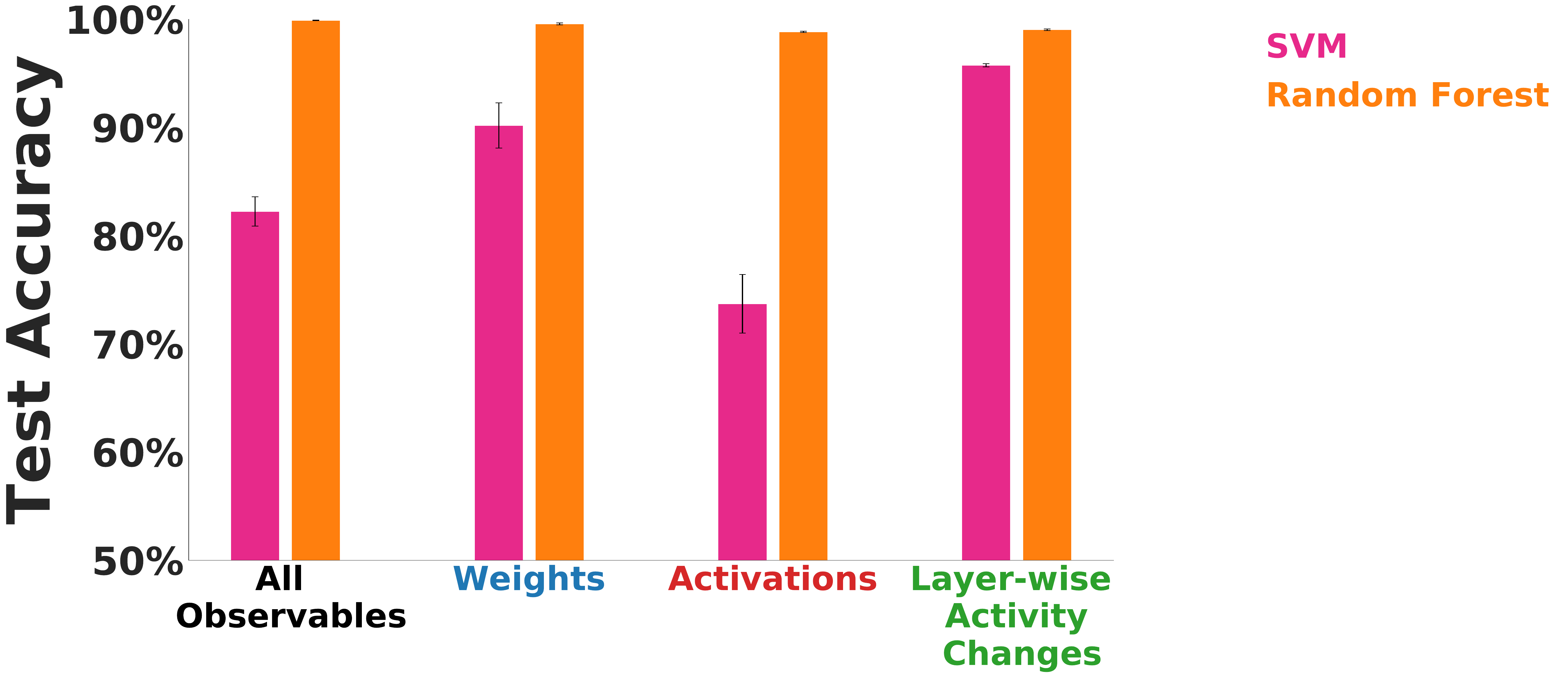}
        \caption{SGDM vs. FA}
    \end{subfigure}
    \begin{subfigure}{0.49\columnwidth}
        \centering
        \includegraphics[width=\textwidth]{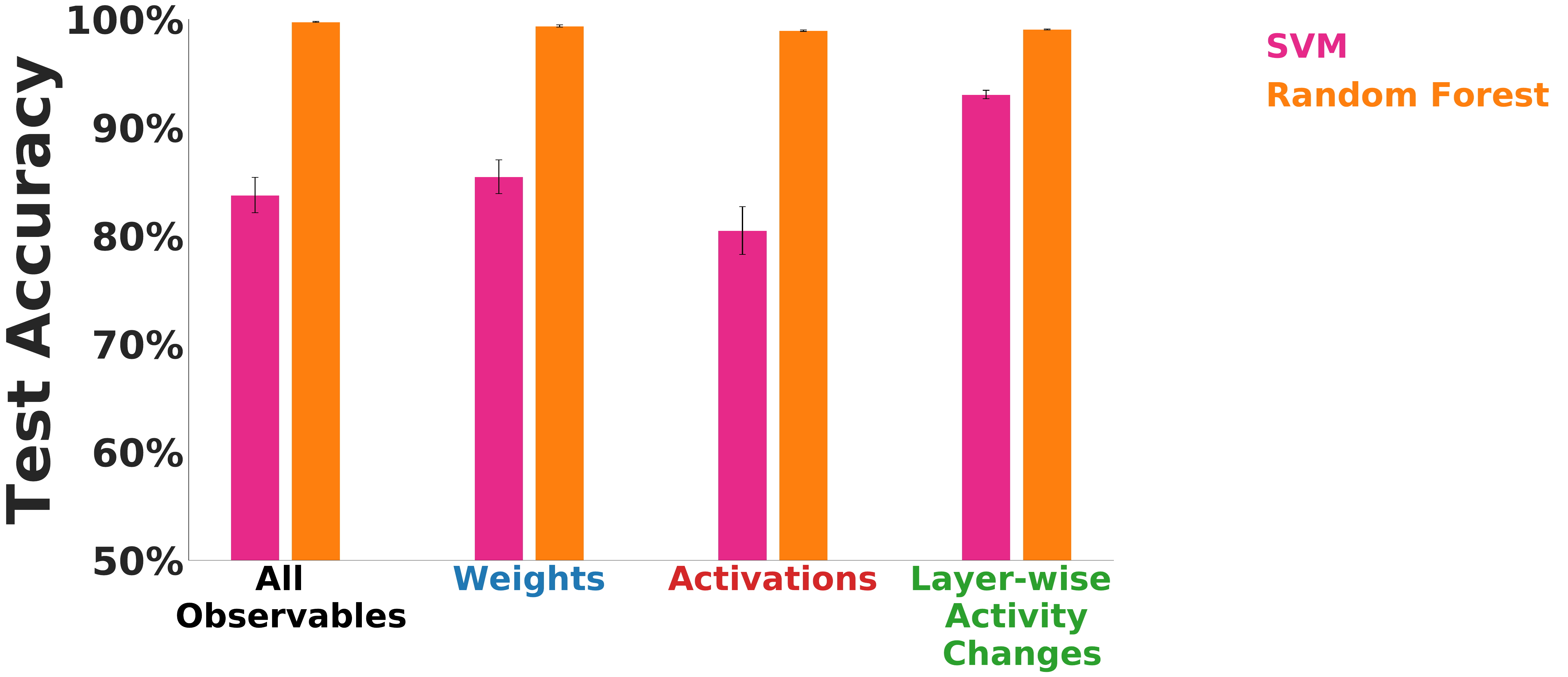}
        \caption{FA vs. IA}
    \end{subfigure}
    \begin{subfigure}{0.49\columnwidth}
        \centering
        \includegraphics[width=\textwidth]{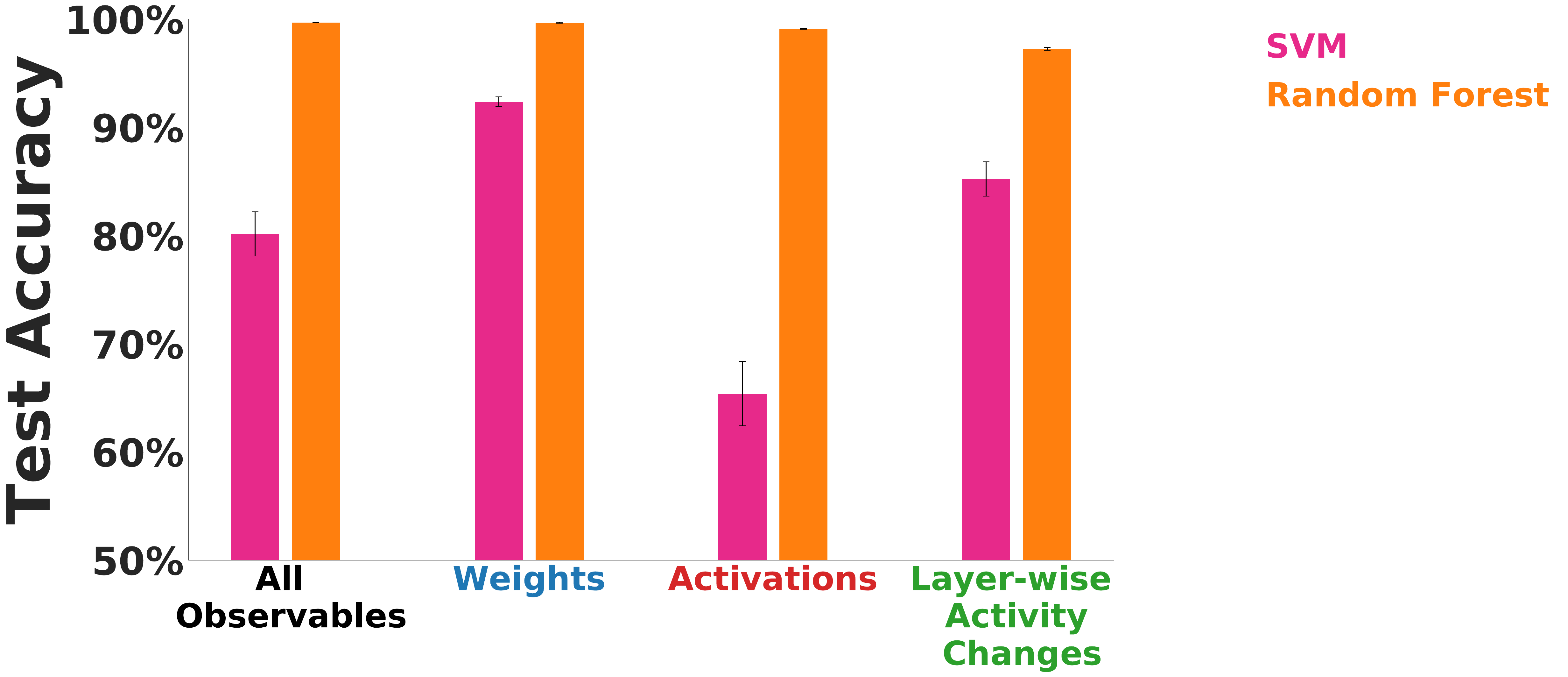}
        \caption{SGDM vs. IA}
    \end{subfigure}
  \caption[Learning rule is least able to be \emph{linearly} determined from the activations for any pair of learning rules]{\textbf{Learning rule is least able to be \emph{linearly} determined from the activations for any pair of learning rules.}
  For all six pairs of learning rules, we train a separate classifier to distinguish each learning rule pair per observable measure.
  Mean and s.e.m. are across ten category-balanced 75\%/25\% train/test splits.
  Chance performance in these settings is 50\% test accuracy.
  \label{lrobs:fig:clspairs}}
\end{figure}

\begin{figure}[tb]
    \centering
    \begin{subfigure}{0.49\columnwidth}
    \includegraphics[width=\textwidth]{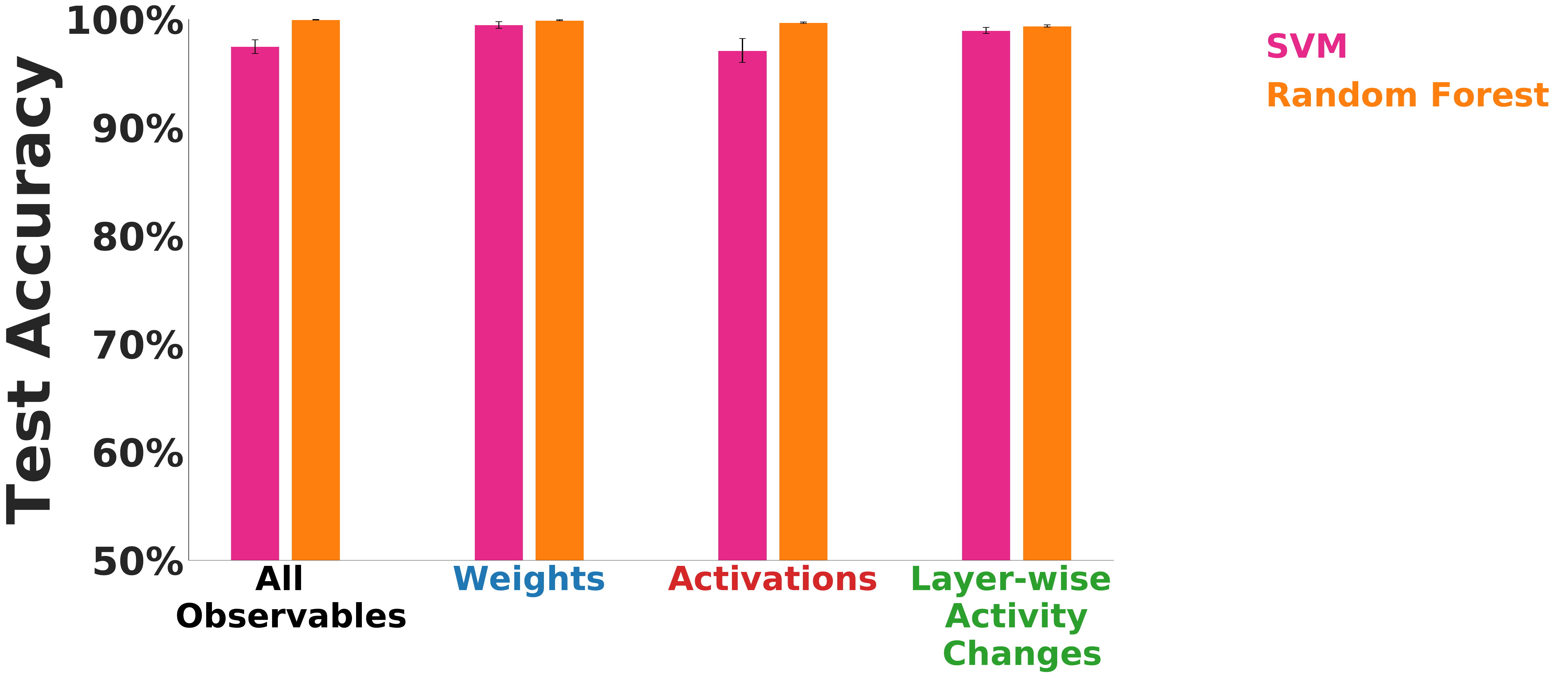}
    \caption{Task performance as a confounded indicator of learning rule separability.}
    \end{subfigure}
    \begin{subfigure}{0.49\columnwidth}
    \includegraphics[width=\textwidth]{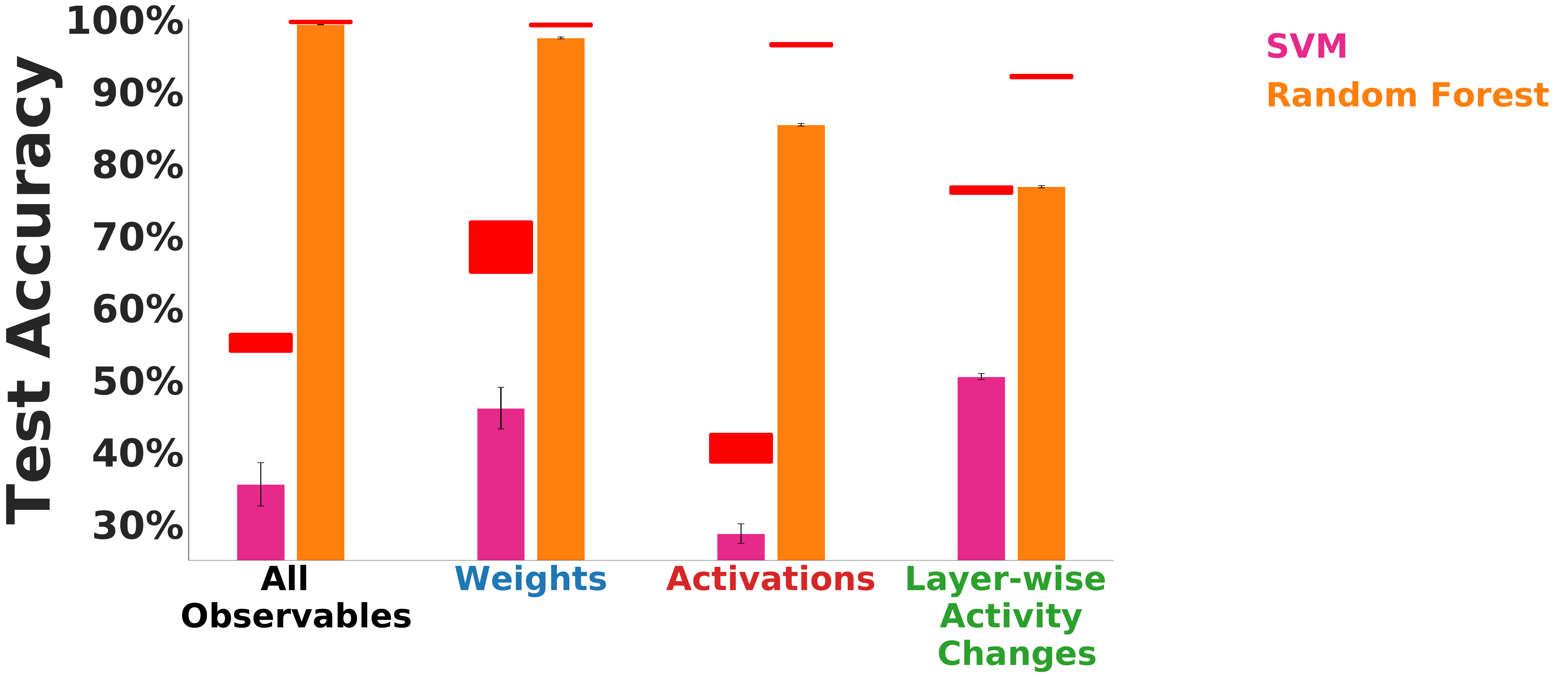}
    \caption{Averaging across the trajectory can hurt generalization.}
    \end{subfigure}
    \caption[Learning rule separability controls (task performance and averaging)]{\textbf{Learning rule separability controls.} \textbf{(a)} We train classifiers to separate IA vs. SGDM when their top-1 validation accuracy is similar on ImageNet, namely across all learning hyperparameters for ResNet-18, ResNet-18v2, ResNet-34, and ResNet-34v2.
    Mean and s.e.m. are across ten category-balanced 75\%/25\% train/test splits.
    Chance performance in this setting is 50\% test accuracy.
    \textbf{(b)} We train classifiers to separate all four learning rules after averaging the statistics across the trajectory.
    Red lines indicate test set performance of each classifier when trained on the entire trajectory for each respective observable measure, as reported in Figure~\ref{lrobs:fig:allcls_accuracy}a.
    Mean and s.e.m. are across ten category-balanced 75\%/25\% train/test splits.
    Chance performance in this setting is 25\% test accuracy.
    \label{lrobs:fig:gensep_ctrl}}
\end{figure}

\begin{figure}[tb]
    \centering
    \begin{subfigure}{0.8\columnwidth}
        \centering
        \includegraphics[width=\textwidth]{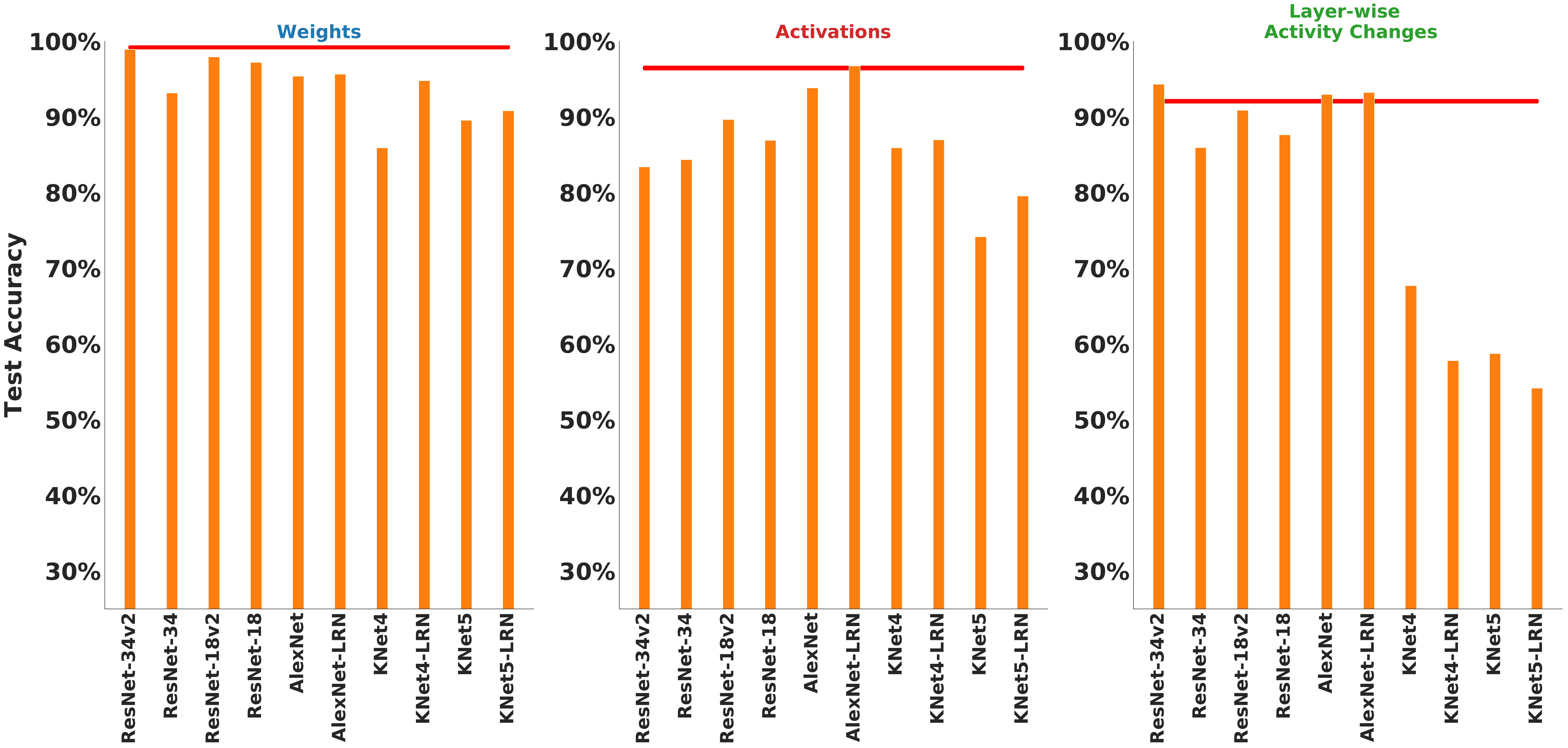}
        \caption{Heldout architecture}
    \end{subfigure}
    \begin{subfigure}{0.8\columnwidth}
        \centering
        \includegraphics[width=\textwidth]{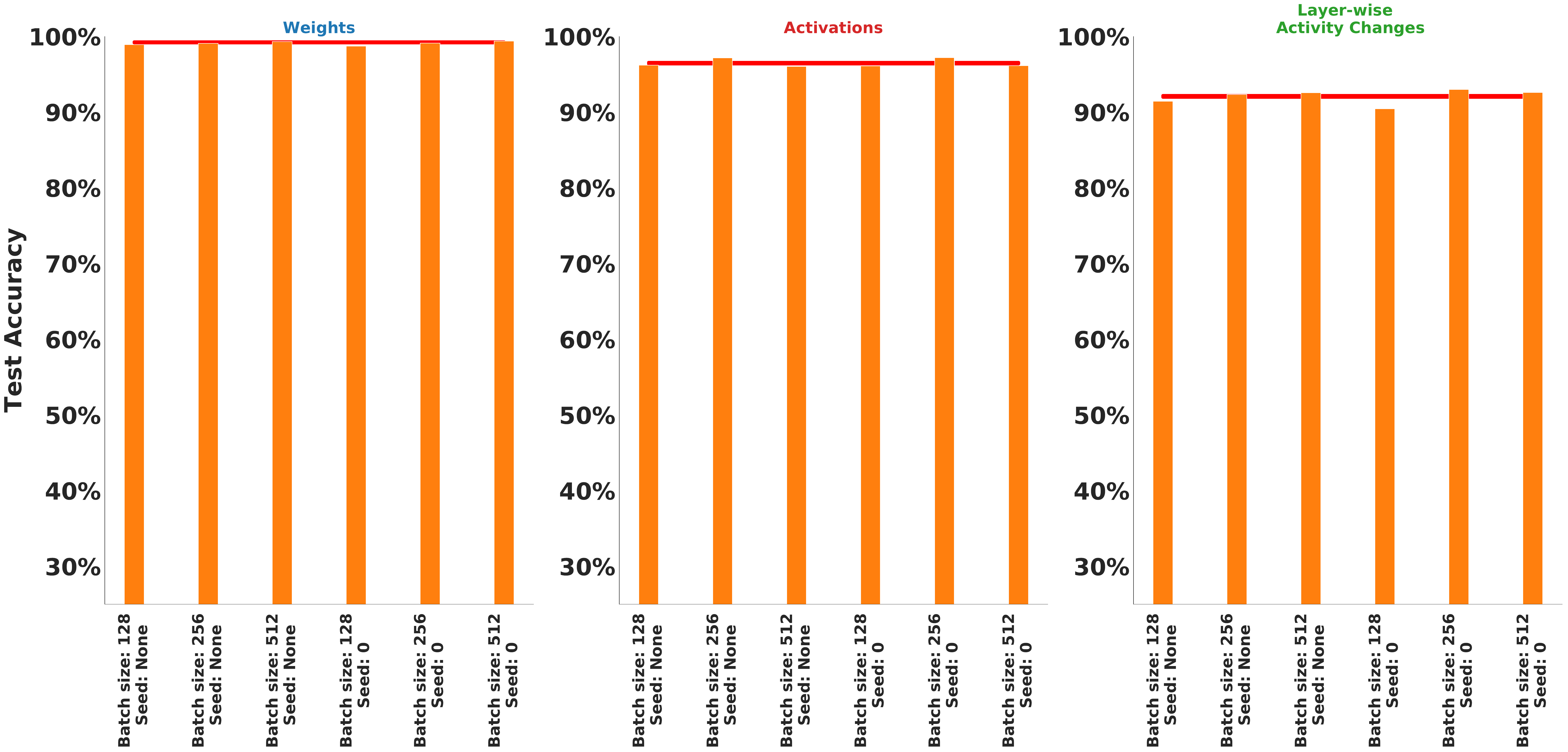}
        \caption{Heldout training curriculum}
    \end{subfigure}
    \begin{subfigure}{0.8\columnwidth}
        \centering
        \includegraphics[width=\textwidth]{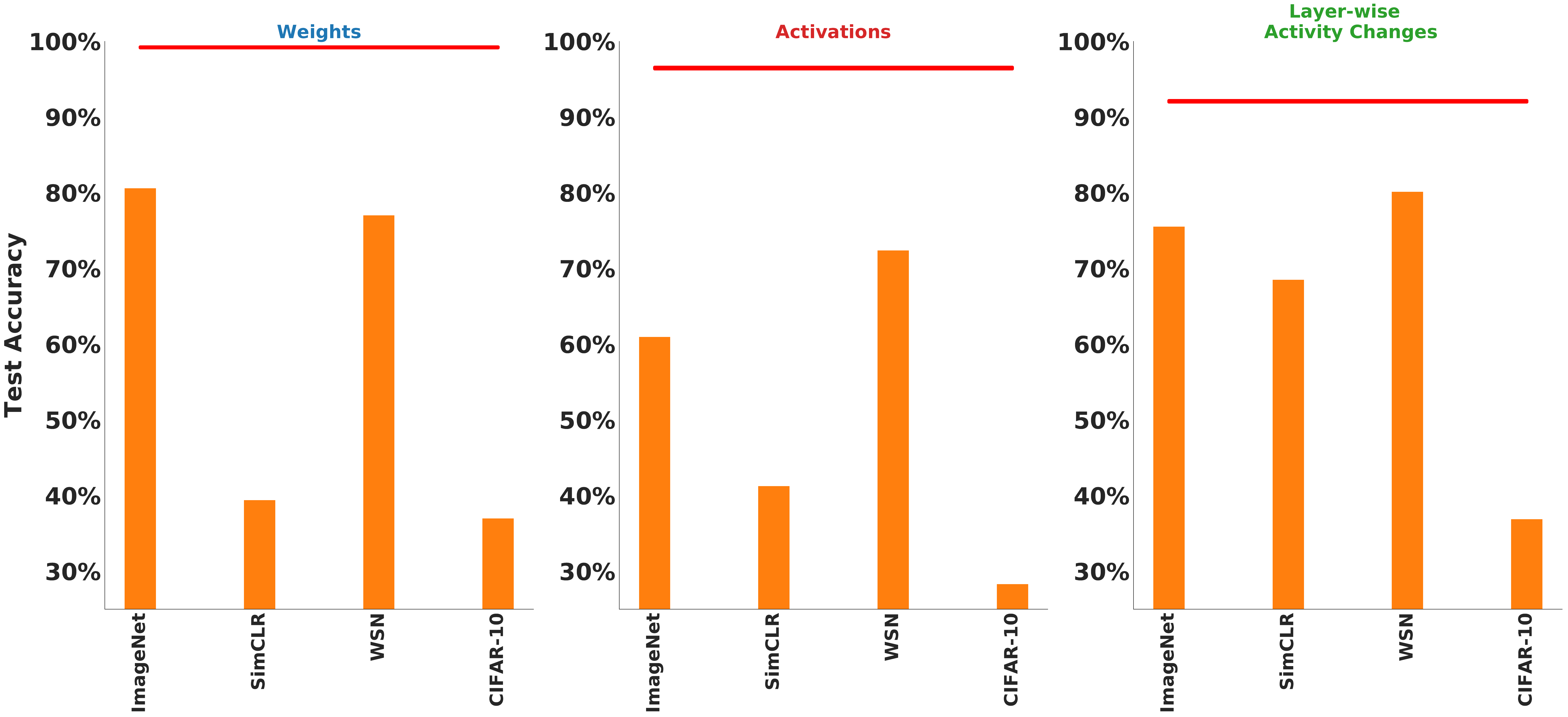}
        \caption{Heldout task}
    \end{subfigure}
    \caption[Assessing robustness of generalization for individual observable measures (Random Forest)]{\textbf{Assessing robustness of generalization for individual observable measures (Random Forest).} Let the red line indicate the mean and s.e.m. test accuracy from Figure~\ref{lrobs:fig:allcls_accuracy}a for the Random Forest trained on each observable measure. 
    \textbf{(a)} We hold out \emph{all} examples from each of the ten architectures.
    \textbf{(b)} We hold out \emph{all} examples from each of the six combinations of batch size and dataset randomization seed pair.
    \textbf{(c)} We hold out \emph{all} examples from each of the four tasks.
    Chance performance is 25\% test accuracy.}
    \label{lrobs:fig:rfcls_weightactgrad_genvar}
\end{figure}

\begin{figure}[tb]
    \centering
    \begin{subfigure}{0.8\columnwidth}
        \centering
        \includegraphics[width=\textwidth]{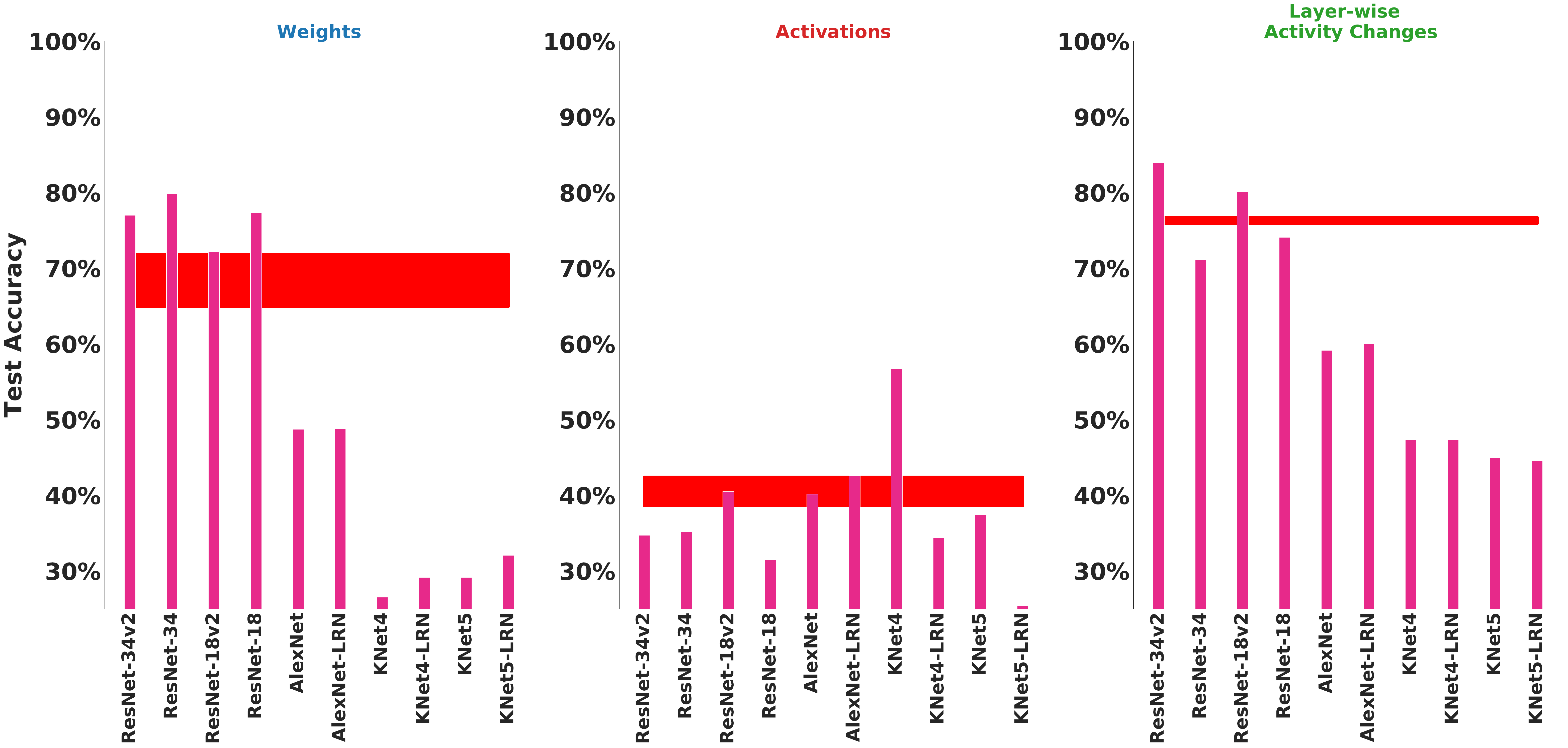}
        \caption{Heldout architecture}
    \end{subfigure}
    \begin{subfigure}{0.8\columnwidth}
        \centering
        \includegraphics[width=\textwidth]{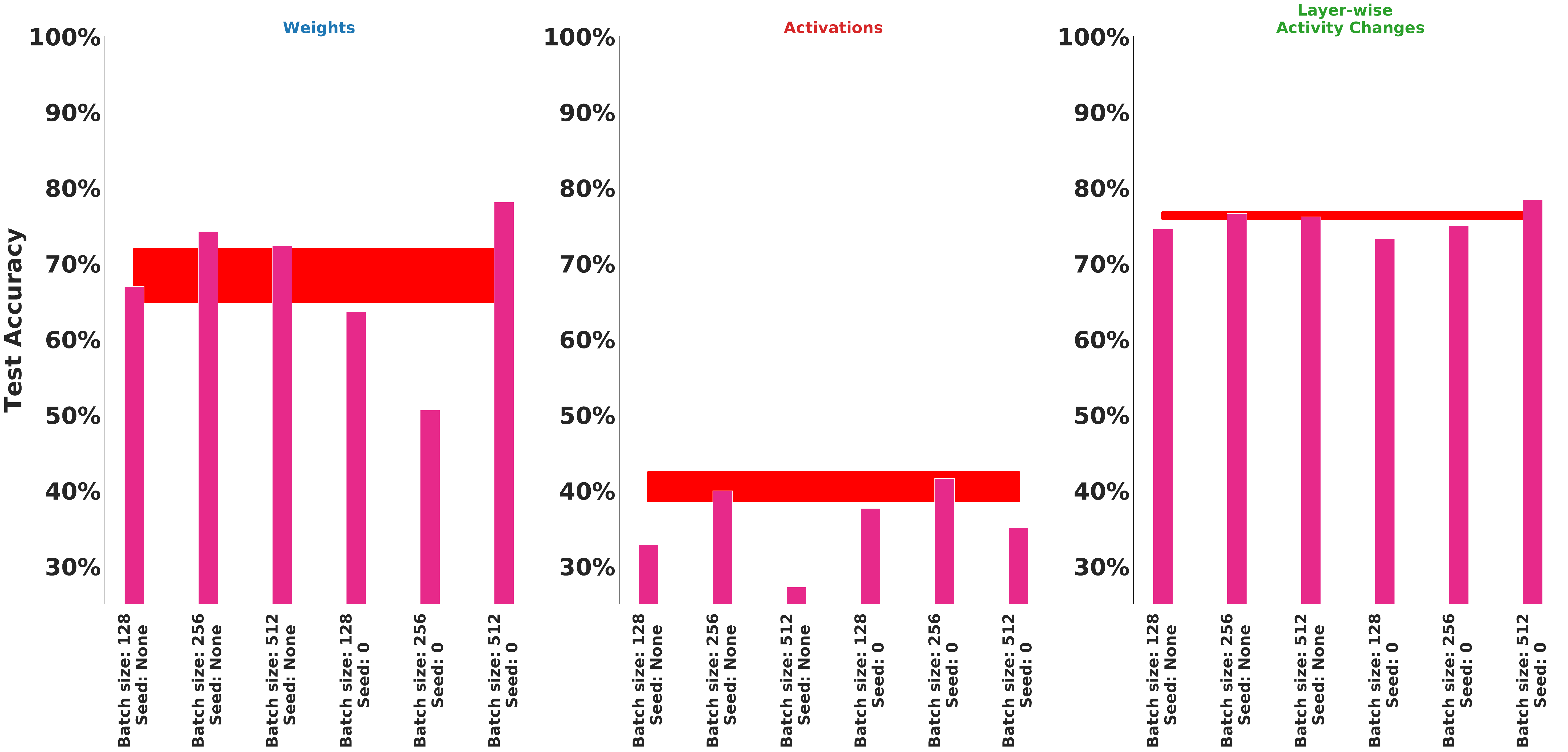}
        \caption{Heldout training curriculum}
    \end{subfigure}
    \begin{subfigure}{0.8\columnwidth}
        \centering
        \includegraphics[width=\textwidth]{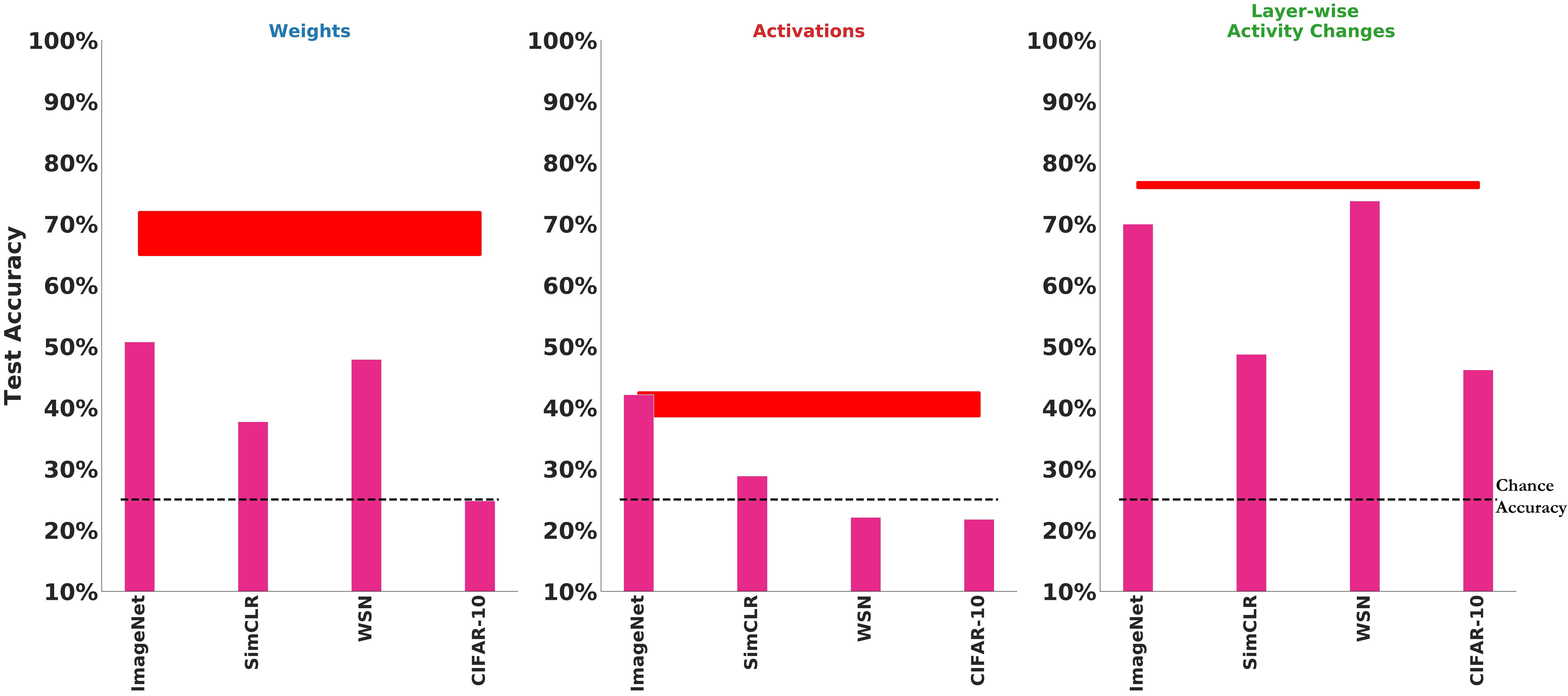}
        \caption{Heldout task}
    \end{subfigure}
    \caption[Assessing robustness of generalization for individual observable measures (SVM)]{\textbf{Assessing robustness of generalization for individual observable measures (SVM).} Let the red line indicate the mean and s.e.m. test accuracy from Figure~\ref{lrobs:fig:allcls_accuracy}a for the SVM trained on each observable measure. 
    \textbf{(a)} We hold out \emph{all} examples from each of the ten architectures.
    \textbf{(b)} We hold out \emph{all} examples from each of the six combinations of batch size and dataset randomization seed pair.
    \textbf{(c)} We hold out \emph{all} examples from each of the four tasks.
    Chance performance is 25\% test accuracy.}
    \label{lrobs:fig:svmcls_weightactgrad_genvar}
\end{figure}

\begin{figure}[tb]
    \centering
    \includegraphics[width=\columnwidth]{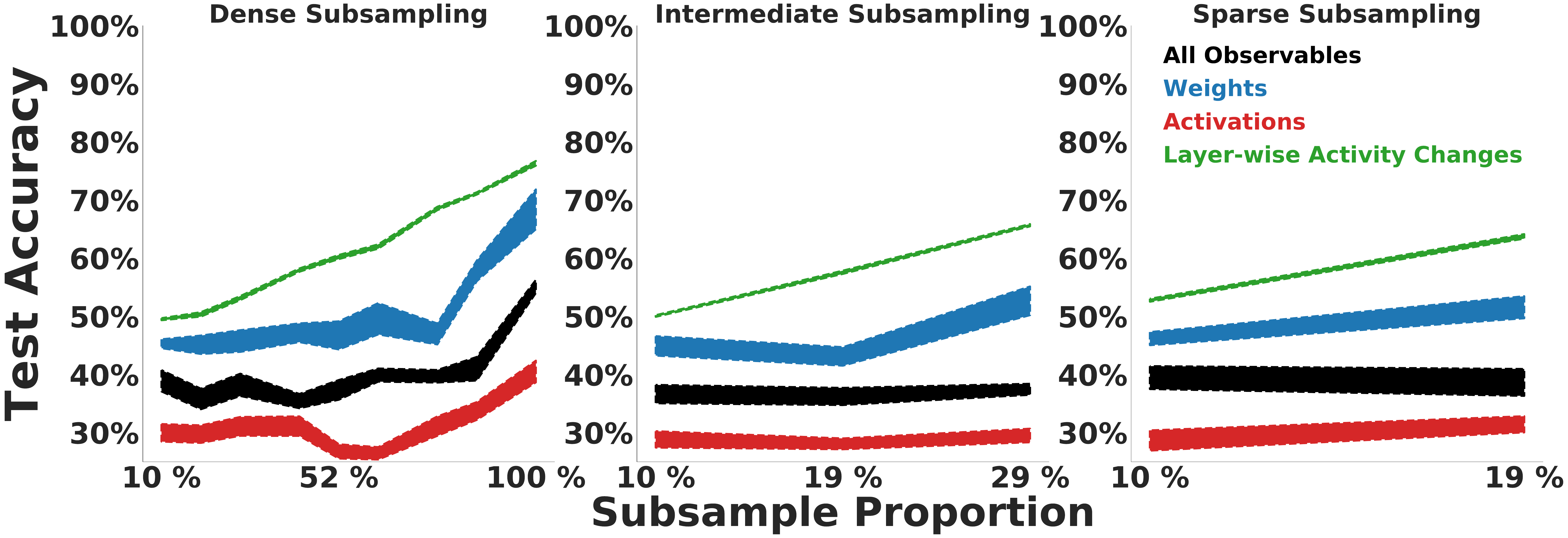}
    \caption[Sparse subsampling \emph{across} the learning trajectory is robust to trajectory undersampling (SVM)]{\textbf{Sparse subsampling \emph{across} the learning trajectory is robust to trajectory undersampling (SVM).} For each observable measure, the test accuracy of the SVM on a random 25\% split of the data.
    Mean and s.e.m. are across the ten category-balanced 75\%/25\% train/test splits.
    Each``full'' training trajectory contains 21 samples (corresponding to epochs $0, 5, 10, \cdots, 100$).
    ``Subsample Proportion $Y$\%'' refers to the number of samples chosen for the trajectory subsample relative to that of the full trajectory (21 total samples).
    Chance performance in these settings is 25\% test accuracy.}
    \label{lrobs:fig:svmcls_subsample}
\end{figure}

\begin{figure}[tb]
    \centering
    \includegraphics[width=1.0\columnwidth]{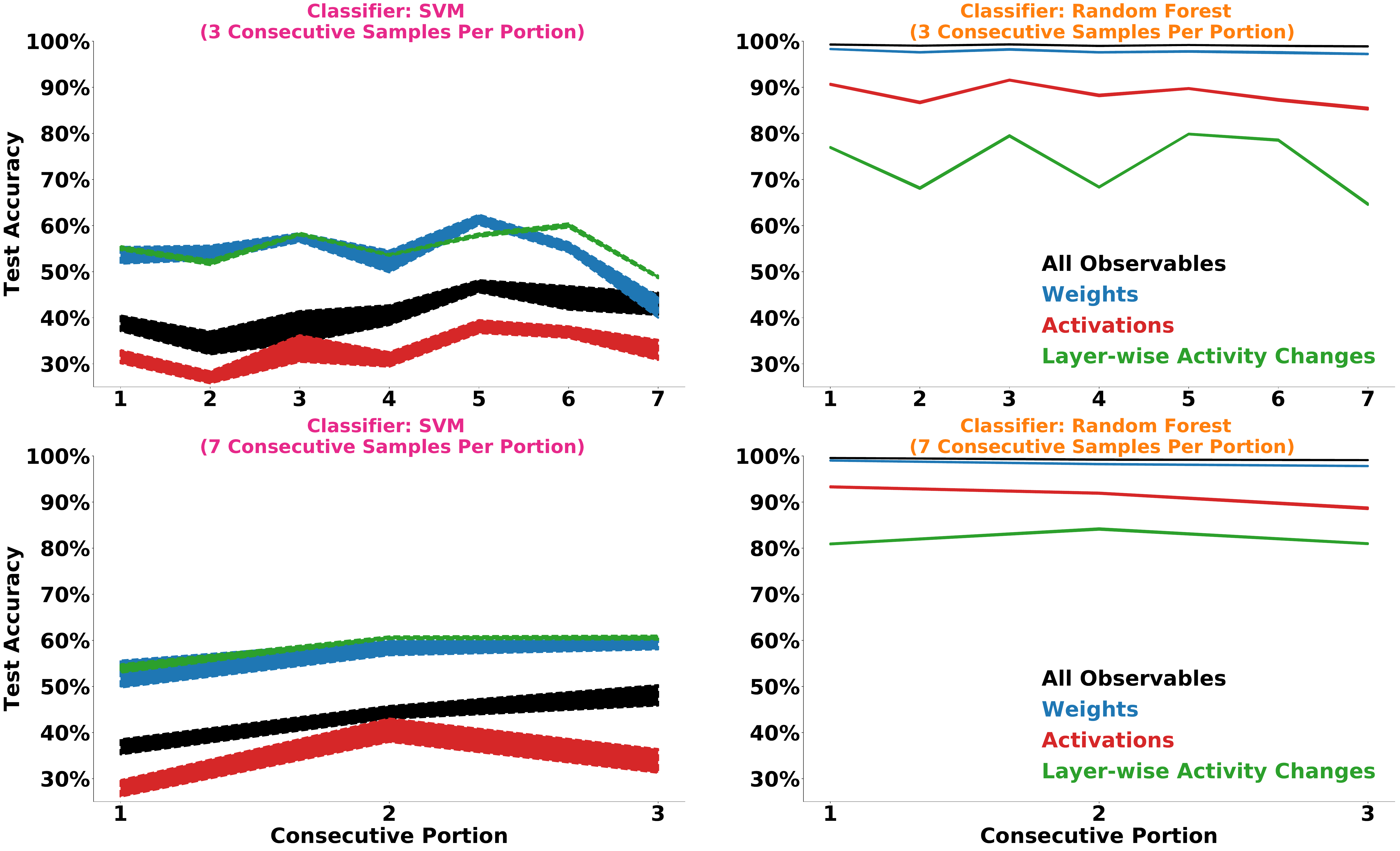}
    \caption[eneralization performance can vary when trained solely on a consecutive portion of the learning trajectory, regardless of classifier]{\textbf{Generalization performance can vary when trained solely on a consecutive portion of the learning trajectory, regardless of classifier.} For each observable measure, the test accuracy on a random 25\% split of the data.
    Mean and s.e.m. are across the ten category-balanced 75\%/25\% train/test splits.
    Each training sample contains each consecutive seventh of the trajectory, totalling 7 consecutive portions each consisting of 3 samples (top row), or consecutive third of the trajectory, totalling 3 consecutive portions each consisting of 7 samples (bottom row).
    Chance performance in these settings is 25\% test accuracy.}
    \label{lrobs:fig:allcls_trajslice}
\end{figure}

\begin{figure}[tb]
    \centering
    \includegraphics[width=1.0\columnwidth]{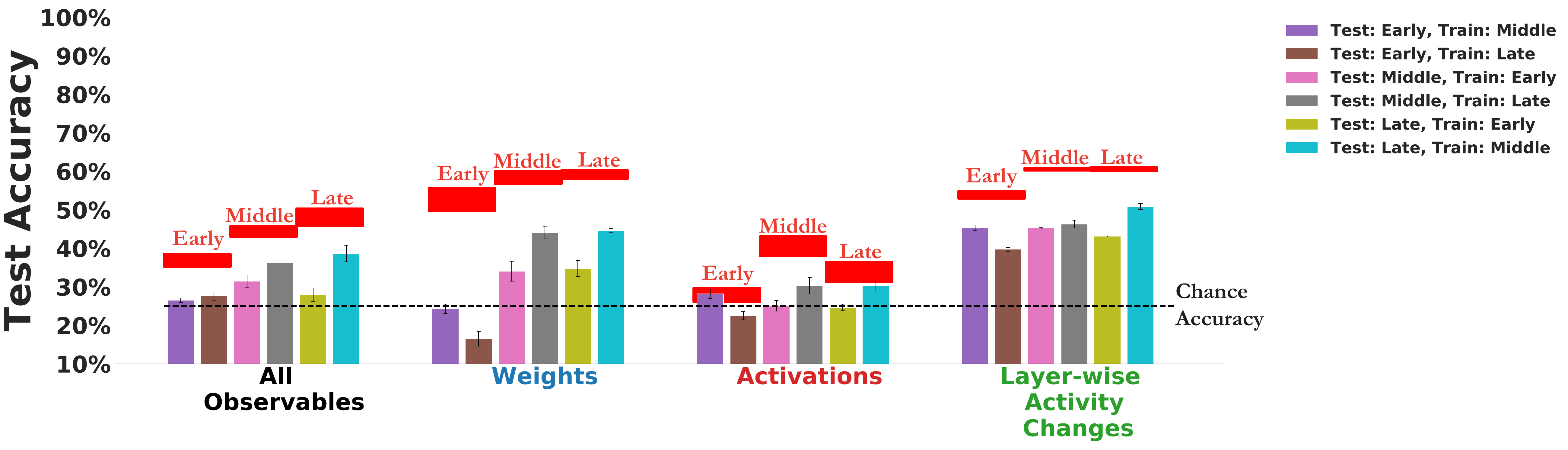}
    \caption[Training solely on consecutive portions of the learning trajectory is \emph{not} robust to trajectory undersampling (SVM)]{\textbf{Training solely on consecutive portions of the learning trajectory is \emph{not} robust to trajectory undersampling (SVM).} We train the SVM on 75\% of examples with access to only one consecutive third of the full trajectory (7 consecutive samples each out of 21 total samples), testing on the remaining 25\% of examples from one of the other two consecutive thirds. 
    Red lines denote SVM performance when tested on 25\% of examples from the \emph{same} portion of the trajectory as in classifier training for each observable measure, reported in the bottom left row of Figure~\ref{lrobs:fig:allcls_trajslice}.
    Mean and s.e.m. in all cases are across ten category-balanced train/test splits.
    Chance performance in this setting is 25\% test accuracy (denoted by the dotted black line).}
    \label{lrobs:fig:svmcls_trajslicegen}
\end{figure}

\begin{figure}[tb]
    \centering
    \includegraphics[width=1.0\columnwidth]{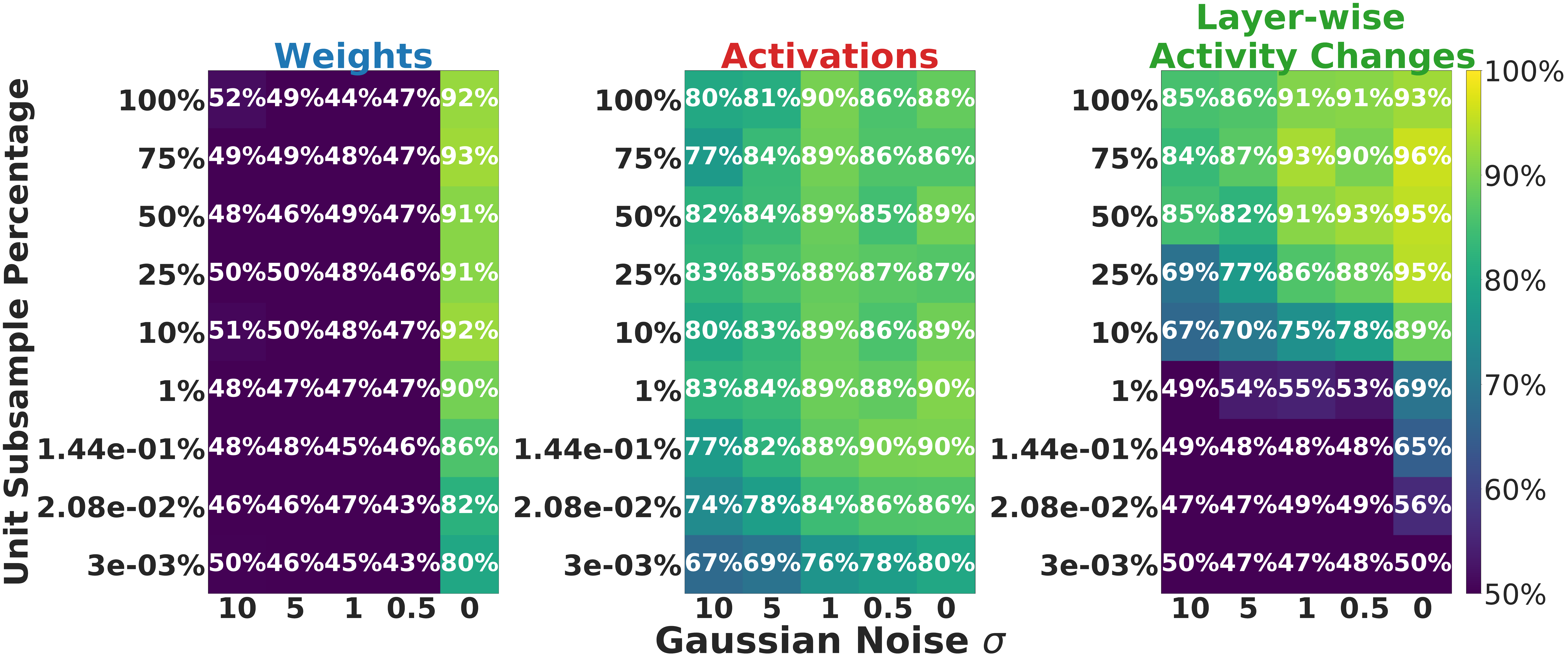}
    \caption[Activations are the most robust to measurement noise and unit undersampling (SVM)]{\textbf{Activations are the most robust to measurement noise and unit undersampling (SVM).}
    For each observable measure, the SVM test accuracy separating FA vs. IA on ResNet-18 across the ImageNet and SimCLR tasks, averaged across random seed and ten category-balanced 75\%/25\% train/test splits.
    The $x$-axis from the left to right of each heatmap corresponds to decreasing levels of noise, and the 
    $y$-axis from bottom to top corresponds to increasing levels of sampled units.
    Top right corner of each heatmap corresponds to the perfect information setting of Section~\ref{lrobs:sec:results}.
    Chance performance in this setting is 50\% test accuracy.}
    \label{lrobs:fig:svmcls_unitsubsamplenoise}
\end{figure}

\chapter{Conclusions}
\label{conclusions}
In this dissertation, we apply goal-driven models to improve our understanding of the role of multiple brain areas, across several species and modalities.
In Chapter~\ref{ch:convrnn}, we find quantitative evidence for the role of recurrent connections in the primate ventral visual stream (VVS) during core object recognition as mediating a tradeoff between task performance and network size.
In Chapter~\ref{ch:mouse}, we study the rodent visual system, motivated chiefly by the ubiquity of this model organism in systems neuroscience.
We build models that combine ``architecture'', ``objective function'', and ``data stream'', to show that the most quantitatively accurate description of mouse visual cortex is a low-resolution, shallow network that makes best use of these limited resources to be a general-purpose visual system.
Finally, in Chapter~\ref{ch:mec}, we apply this approach to understand the heterogeneity of cell types in rodent medial entorhinal cortex (MEC).
We show that these cell types can be encompassed in a unified framework of biological performance optimization, demonstrating that they form a continuum of cells within a single framework that encompasses grid, border, and heterogeneous cells.
This study indicates goal-driven modeling is a promising new path forward for studying MEC beyond cells with stereotyped, interpretable response profiles, that otherwise form a small fraction of the overall circuit.
In Chapter~\ref{ch:lrperf}, we turn to addressing the biological implausibilities of backpropagation.
We demonstrate that suitably chosen recurrent circuits facilitate effective, error-driven learning \emph{without} strict weight symmetry, making them strong candidates for investigating learning rules in biological circuits.
In Chapter~\ref{ch:lrobs}, we examine the inverse problem of inferring the learning rule from ``virtual'' experimental measurements, showing that temporally dependent measurements from network activations can robustly separate learning rules, where the number of units recorded is within reach of current electrophysiological techniques.

There are several immediate next steps stemming from this work, towards a more complete goal-driven theory of (primate and rodent) visual cortex and medial entorhinal cortex.
For the work with visual cortex in both species, a short-term target would be a recurrent, unsupervised model (in the primate work, the models were recurrent \emph{but} supervised; and in the rodent work, the models were unsupervised \emph{but} feedforward).
The question of biologically plausible learning targets is similarly linked to biologically plausible mechanisms for learning such objective functions.
One could additionally train these networks with more biologically plausible learning rules, though the question of how these representations are acquired through development (versus what is endowed by the genome) would be the subject of a much longer-term aim.
For the work with medial entorhinal cortex, we are really at the beginning when it comes to goal-driven modeling.
Perhaps the most exciting direction in addressing gaps in our understanding is with regard to the loss function of the network.
The networks that explain almost all of the response variability were supervised to path integrate an intermediate (simulated) place cell representation.
While this target does not seem unreasonable given the connections between MEC and HPC, we do not yet have a more concise rendering of what the system is optimized for.
Namely, what tasks yield representations that when trained end-to-end, give a strong correspondence both the MEC and HPC, across a range of environments?
Does the answer to this question involve architectures that can represent both space but also other semantic variables?
Finally, does the resultant architecture and loss target yield an improvement in the ability of artificial agents to plan over long timescales, in the midst of interference and other environmental non-stationarities?

Overall, the methods covered in this dissertation operate under a unified, ``reverse engineering'' framework by which to better understand the principles of biological intelligence, but there is much more that remains to be done.
In the near term, we aim to build integrated models of both sensory and non-sensory areas, especially as we engage artificial agents in richer, more ethologically relevant environments.
One implication from our work is that not only should experimental neuroscience datasets naturally involve more neurons and animals, but in order to be effective at separating models, they should involve the organism under study to encounter varied environments and behavioral conditions.
As these datasets become available, we not only will have more effective estimates of the inter-animal consistency ceiling, but also will be better at discerning competing modelling hypotheses.
And as our integrated agents become more autonomous and explore their environments similarly to the animals under investigation (dealing with non-stationarities in an online and unsupervised way), these will not only provide interpretable hypotheses about such large-scale datasets, but eventually more realistically stand-in as ``virtual animals'', providing a closed loop between the collection of new data and building improved models.
While evolution gave rise to our brains -- we have used ours to design \emph{in silico} versions, and through this iterative process, would have come to better understand our own.

% ...
\bibliographystyle{plainnat}
\bibliography{ref}
\end{document}